\documentclass[11pt]{book}
\usepackage{amsthm} 

\usepackage{amsmath} 

\usepackage{amssymb} 

\usepackage{color}        

\usepackage{hyperref}
\hypersetup{colorlinks=true,linkcolor=black,urlcolor=blue, citecolor=blue}

\usepackage{graphicx}  

\usepackage{eurosym}

\usepackage{pgfplots} % For plotting functions
\pgfplotsset{compat=1.11}

\usepackage{enumitem}
\setlist[1]{itemsep=-5pt}

\usepackage{soul}

\usepackage{subcaption} %  for subfigures environments 

\usepackage{tcolorbox}  % for exercise environments
	
\usepackage{authblk}

\usepackage{multirow}

\usepackage{algorithm}
\usepackage{algpseudocode}

\definecolor{CommentColor}{RGB}{9,181,106}

\renewcommand{\Comment}[2][.75\linewidth]{%
  \leavevmode\hfill\makebox[#1][l]{$\triangleright$\color{CommentColor} ~#2}}
  
\newcommand{\Commentt}[2][.75\linewidth]{%
  \leavevmode\hfill\makebox[#1][l]{\phantom{$\triangleright$}\color{CommentColor}~#2}}
  
\renewcommand{\comment}[1]{\Statex {$\triangleright$ \color{CommentColor} #1}}
\newcommand{\commentt}[1]{\Statex {\phantom{$\triangleright$} \color{CommentColor} #1}}
\usepackage{longtable}

\usepackage{dsfont}

\usepackage{listings}

\usepackage[scr=boondox]{mathalpha}
\newcommand{\mi}{\mathit}   

\newcommand{\mc}{\mathcal}

\newcommand{\al}{\alpha}

\newcommand{\ag}{ag} %agent
\newcommand{\numAgents}{n}
\newcommand{\off}{\omega} %offer
\newcommand{\Off}{\Omega} %offer set

\newcommand{\pareto}[1]{Par(#1)}
\newcommand{\OffPareto}{\pareto{\Off}} % the set of Pareto-optimal offers.

\newcommand{\offComp}[1]{x_#1} %component of a multi-issue offer.
\newcommand{\dom}{D} % negotiation domain.

\newcommand{\opp}{\mathit{opp}}

\newcommand{\ac}{a} %action
\newcommand{\Ac}{A} % set of actions.

\newcommand{\actype}{\eta} % 'action type' e.g. propose or accept
\newcommand{\prop}{\mathfrak{p}} % the 'propose' illoc. particle
\newcommand{\acc}{\mathfrak{a}} % the 'accept' illoc. particle

\newcommand{\pr}{\pi} %proposal

\newcommand{\hist}{h} % the history
\newcommand{\Hist}{\mathcal{H}} % the set of all histories.
\newcommand{\obsHist}[1]{h^o_{#1}}
\newcommand{\issue}[1]{I_{#1}}
\newcommand{\numIssues}{m} % number of issues in a domain.
\newcommand{\numOptions}{s} % the number of options in a single issue.
\newcommand{\util}{u} %utility function
\newcommand{\Util}{\mc{U}} %set of possible utility functions (in the context of Bayesian learning)
\newcommand{\eval}{v} %evaluation function (utility of a single issue)
\newcommand{\rv}{rv} %reservation value.
\newcommand{\df}{\delta} %discount factor

\newcommand{\offMax}{\off^{max}} % most preferred offer.
\newcommand{\offMin}{\off^{min}} % least preferred offer.
\newcommand{\utilMax}{\util^{max}}% utility of the most preferred offer.
\newcommand{\utilMin}{\util^{min}} % utility of the least preferred offer.

\newcommand{\offNext}{\off_{next}} % the next offer to propose
\newcommand{\offRec}{\off_{rec}} % the last offer received from the opponent

\newcommand{\dead}{T} %temporal deadline
\newcommand{\maxRounds}{N} %maximum number of rounds.

\newcommand{\dis}{\xi} % disagreement
\newcommand{\extOff}{\Off^{+}} %extended offer space (i.e. offer space + disagreement)

\newcommand{\hard}{hardheaded}
\newcommand{\soft}{conceding}

\newcommand{\asp}{\lambda} %aspiration level
\newcommand{\targ}{\beta} %target value.
\newcommand{\om}{\mc{M}} % opponent model
\newcommand{\safe}{\epsilon} %'safety term' to add to the target value to represent our uncertainty about the optimal target value.
\newcommand{\tftThresh}{\theta} %threshold value used in tit-for-tat strategies.

\newcommand{\con}{c}
\newcommand{\conGain}{\Delta \con}
\newcommand{\Rec}{\Off^{rec}} % set of received offers
\newcommand{\Pro}{\Off^{prop}} % set of proposed offers (by us)
\newcommand{\candidates}{C} % set off offers that are candidates to be proposed next.

\newcommand{\est}{\hat}%to indicate estimated quantities

\newcommand{\hypo}{y} %hypothesis, in Bayesian learning.
\newcommand{\Hypo}{Y} %set of hypotheses, in Bayesian learning.
\newcommand{\obs}{o} %observation, in Bayesian learning.
\newcommand{\Obs}{O} %the set of all possible observations, in Bayesian learning.
\newcommand{\numObs}{k} %number of observations, in Bayesian learning
\newcommand{\tri}{\Lambda} %triangular function.

\newcommand{\expt}{\overline} %to indicate expected quantities

\newcommand{\freq}{f} % frequency (in frequency analysis)
\newcommand{\numRec}{k} % number of received proposals.

\newcommand{\covmat}{\mathbf{K}} % covariance-matrix for Gaussian processes.
\newcommand{\kernel}{\kappa} %kernel function for Gaussian processes.

\newcommand{\br}{\mathit{BR}} %best response
\newcommand{\ms}{\mathscr{m}} %mixed strategy
\newcommand{\MS}{\mathcal{M}} %set of all mixed strategies
\newcommand{\emptyTuple}{\varepsilon}

\newcommand{\strat}{\sigma} %strategy for an extensive-form game.

\newcommand{\Strat}{\mathcal{S}} %Set of all strategies for an extensive-form game.
\newcommand{\act}{pl} % the 'active player' in an extensive-form game.

\newcommand{\term}{T} % indicates a subset of terminal tuples e.g. Y^T
\newcommand{\node}{\nu}

\newcommand{\nfgame}{G} % normal form game
\newcommand{\ff}{\mc{F}_A} % frequency function.
\newcommand{\ttgame}{\Gamma} % turn-taking game

\newcommand{\kleene}{\mathbb{N}}

\newcommand{\nat}{0} % `nature' i.e. the random player.
\newcommand{\obsfunc}{f^{obs}}
\newcommand{\delay}{\epsilon}
\newcommand{\utilSpace}{\Upsilon}

\newcommand{\pf}{\utilSpace^p} % pareto frontier
\newcommand{\kf}{\mathit{ks}} % kalai function

\newcommand{\cand}{C} % Candidate set of equilibria
\newcommand{\sfs}{S} % safe set.
\newcommand{\fs}{F} % full set

\newcommand{\numDomains}{m}

\newcommand{\numVars}{k}
\newcommand{\numReps}{R}

\newcommand{\tx}{\underline} % tounament index i.e. the index of an agent in a given tournament.
\newcommand{\ts}{U} % tournament score
\newcommand{\prot}{\Pi} % negotiation protocol
\newcommand{\scen}{sc}
\newcommand{\Scen}{Sc}
\newcommand{\Ag}{Ag} % set of agents.
\newcommand{\Dom}{\mathcal{D}} % set of negotiation domains.
\newcommand{\ind}{\mathds{1}} % the indicator function
\newcommand{\numSucc}{\mathit{NA}} % the number of agreements made
\newcommand{\uua}{\mathit{UA}} %utility under agreement
\newcommand{\ar}{\mathit{AR}} % agreement rate.
\newcommand{\se}{\mathit{se}} %standard error

\newcommand{\rvX}{\mathcal{X}}
\newcommand{\rvY}{\mathcal{Y}}
\newcommand{\rvZ}{\mathcal{Z}}
\newcommand{\varDomain}{S}
\newcommand{\diff}{\est{\mu}}

\newcommand{\tstat}{\mathfrak{t}} % t-statistic
\newcommand{\tdist}{t} % t-distribution

\newcommand{\rej}{\mathfrak{r}} % the 'reject' illoc. particle
\newcommand{\pa}{pa} % participating agents.
\newcommand{\feas}{\mathit{Fea}}

\newcommand{\partoff}[1]{\off^{(#1)}}
\newcommand{\Partoff}[1]{\Off^{(#1)}}
\newcommand{\bids}{B} % The BIDS algorithm
\newcommand{\precision}{q}
\newcommand{\range}[2]{R_{#1,#2}}
\newcommand{\domin}{dom}
\newcommand{\close}{cl}
\newcommand{\cluster}[1]{\mathit{Cl}_{#1}}
\newcommand{\approxPareto}[1]{\widetilde{Par}(#1)}

\newcommand{\infeas}{C}	% a hard constraint (i.e. a set of infeasible offers) on a cartesian domain.

\newcommand{\hypCube}{\mathit{hc}}
\newcommand{\HypCube}{\mathit{HC}}
\newcommand{\subIssue}[1]{J_{#1}}
\newcommand{\vol}{V} % the volume of a hypercube.
\newcommand{\bid}{\mc{B}}		% A collection of hypercubes.
\newcommand{\dist}{d} % Manhattan distance
\newcommand{\offSet}{\mathcal{O}} % a set of offers

\newcommand{\later}[1]{\textcolor{orange}{#1}}
\renewcommand{\later}[1]{}

\DeclareMathOperator*{\argmax}{arg\,max}
\DeclareMathOperator*{\argmin}{arg\,min}

\newtheorem{definition}{Definition}[section]
\newtheorem*{observation}{Observation}
\newtheorem{lemma}{Lemma}[section]
\newtheorem{theorem}[lemma]{Theorem}

\tcbuselibrary{breakable} % to ensure that the 'exercise' and 'opinion' environments can be broken across multiple pages.

\theoremstyle{definition}
\newtheorem{exer}{Exercise}
\usepackage{environ}
\NewEnviron{exercise}
  {\begin{tcolorbox}
  	\begin{exer}
        \BODY
     \end{exer}
    \end{tcolorbox}
  }

\theoremstyle{definition}
\newtheorem*{opin}{\textit{Opinion}}
\usepackage{environ}
\NewEnviron{opinion}
  {\begin{tcolorbox}[breakable]
  	\begin{opin}
        \BODY
     \end{opin}
    \end{tcolorbox}
  }

\begin{document}

\title{Introduction to Automated Negotiation}
\author{Dave de Jonge\\
IIIA-CSIC, Barcelona, Spain\\
Universitat Aut\`onoma de Barcelona, Spain
}

%\affil[1]{IIIA-CSIC, Barcelona, Spain}
%\affil[2]{Universitat Autonoma de Barcelona, Spain}

%\date{\today\\v0.7}
\date{17 November 2026\\v0.7}

\maketitle

%\clearpage
%
\vspace*{\fill}

\noindent \textcopyright\ 2025 Dave de Jonge. \\
This work is licensed under CC BY-NC-ND 4.0. To view a copy of this license, visit \url{https://creativecommons.org/licenses/by-nc-nd/4.0/}

%\clearpage

\tableofcontents

\chapter*{Preface}
This book is targeted towards computer science students who are completely new to the topic of automated negotiation. It does not require any prerequisite knowledge, except for elementary mathematics and basic programming skills. I have made this book available for free, so feel free to share it with anyone you like.

Please note that this book is meant as an organic document that keeps expanding over time. Therefore, I recommend to regularly check  the website of this book to see if there is any updated version available. Also note that since this is still only a preliminary version of the final book, some notations or definitions may change in future versions of this book, or may have changed with respect to earlier versions.
 
This book comes with an simple toy-world negotiation framework implemented in Python that can be used by the readers to implement their own negotiation algorithms and perform experiments with them. This framework is small and simple enough that any reader who does not like to work in Python should be able to re-implement it very quickly in any other programming language of their choice. It can be downloaded from the website of this book:

\url{https://www.iiia.csic.es/~davedejonge/intro_to_nego}

If you have any questions or comments on this book, please send me an e-mail: \url{davedejonge@iiia.csic.es}. I am more than happy to hear your suggestions so that I can improve this work. Especially, if you feel that something is not clearly explained, or that something important is missing, please let me know! 

\newpage

\noindent To cite this book, please use the following BibTeX:

\begin{verbatim}
@book{deJonge2025IntroToNego,
        title = "Introduction to Automated Negotiation",
        author = "de Jonge, Dave",
        year = "2025",
        publisher = "IIIA-CSIC",
        address = "Barcelona, Spain",
        doi = "10.48550/arXiv.2511.08659",
        url = "https://www.iiia.csic.es/~davedejonge/intro_to_nego"
    }
\end{verbatim}

\chapter*{Summary of Notation}

\begin{tabular}{lp{5in}}
\multicolumn{2}{l}{\textbf{Basic Negotiations}} \\
$\mathbb{R}$ & The set of real numbers.\\
$\mathbb{R}^+$ & The set of positive real numbers.\\
$\mathbb{N}$ & The set of natural numbers (including 0). \\
$\ag_i$ & Agent $i$ \\
$\Off$ & The set of all offers in a given negotiation domain. \\
$\off$ & Offer \\
$\issue{j}$ & Issue \\
$\numOptions_j$ & The size of issue $\issue{j}$, i.e. $\numOptions_j := |\issue{j}|$.\\
$\offComp{j}$ & Option\\
$\offComp{{j,l}}$ & The $l$-th option of issue $\issue{j}$.\\
$t$ & Time \\
$(i,\prop, \off, t)$ & Agent $\ag_i$ proposes offer $\off$ at time $t$.\\
$(i,\acc, \off, t)$ & Agent $\ag_i$ accepts offer $\off$ at time $t$.\\
$\actype$ & Action type, i.e. a variable that can either adopt the value $\prop$ or the value $\acc$. \\
$\dead$ & Deadline \\
$\maxRounds$ & The maximum number of rounds in a negotiation.\\
$\delay$ & Delay, i.e. the difference between the time a proposal or acceptance was sent by one agent and the time it was received by the other agent.\\
$\hist$ & Negotiation history or action history\\
$\obsHist{i}$ & Observed negotiation- or action- history (observed by agent $\ag_i$)\\
$\util_i$ & Utility function of agent $\ag_i$.\\
$\offMax_i$ & The most preferred offer by agent $\ag_i$ \\
$\offMin_i$ & The least preferred offer by agent $\ag_i$ \\
$\utilMax_i$ & The utility value, for $\ag_i$ of $\ag_i$'s most  preferred offer. \\
$\utilMin_i$ & The utility value, for $\ag_i$ of $\ag_i$'s least  preferred offer. \\
$\eval_i^j$ & Evaluation function for agent $\ag_i$ and issue $\issue{j}$ \\
$w_i^j$  & Weight for agent $\ag_i$ and issue $\issue{j}$\\
$\rv_i$ & Reservation value of agent $\ag_i$ (Def.~\ref{def:res_val}).\\
$\dis$ & The conflict outcome, a.k.a. disagreement.\\
$\extOff$ & The extended offer space (the union of  offer space and the conflict outcome)\\
$\df$ & Discount factor.\\
$\dom$ & Negotiation domain (Def.~\ref{def:nego_domain})\\
$\vec{\util}(\off)$ & Utility vector of offer $\off$.\\
$\OffPareto$ & Pareto set, i.e. the set of all Pareto-optimal offers (Def.~\ref{def:pareto_set}).\\
$\opp(\dom)$ & The amount of `opposition' of a domain $\dom$.\\
\end{tabular}

\begin{tabular}{lp{4in}}
\multicolumn{2}{l}{\textbf{Negotiation Strategies}} \\
$\offRec$ & The last offer that our agent has received from the opponent.\\
$\offNext$ & The offer that our agent is about to propose next.\\
$\om$ & Opponent model \\
$\asp(t)$ & Aspiration level at time $t$.\\
$\est{\util}_2$ & Estimation of $\ag_2$'s utility function, as estimated by $\ag_1$'s opponent modeling algorithm. \\
$\Pro_t$ & The set off all offers that have already been proposed by $\ag_1$ before time $t$. \\
$\alpha$ & Initial value of the aspiration function, i.e. $\asp(0)$.\\
$\targ$ & Target value, i.e. the final value of the aspiration function: $\asp(\dead)$.\\
$\gamma$ & The concession parameter.\\
$\dead'$ & Target time.\\
$\targ^*$ & Optimal target value for our agent, based on predictions of our opponent's future proposals.\\ 
$\Rec_t$ & The set of offers that have been \textit{received} by agent $\ag_1$ up until time $t$.\\
$\con_i$ & Function that measures the amount of concession made by agent $\ag_i$.\\
$2^\Off$ & Power set of the set of offers (i.e. the set of all subsets of $\Off$).\\
$\conGain_t$ & `concession gain' of agent 1 at time $t$.\\
$\tftThresh_{min}$ & minimum required concession gain for tit-for-tat strategy.\\
$\tftThresh_{max}$ & maximum required concession gain for tit-for-tat strategy.\\
\end{tabular}

\begin{tabular}{lp{4in}}
\multicolumn{2}{l}{\textbf{Opponent Modeling}} \\
$\Util$ & Some set of possible utility functions.\\
$\pr_j$ & Proposal\\
$\vec{\pr}$ & Sequence of proposals. \\ 
$P(\util | \pr_1, \pr_2, \dots, \pr_\numObs)$ & The probability that our opponent has utility function $\util$, given that our agent has received proposals $\pr_1, \pr_2, \dots, \pr_\numObs$ from that opponent.\\
$\hypo$ & Hypothesis\\
$\Hypo$ & Set of possible hypotheses.\\
$\obs$ & Observation\\
$\Obs$ & Set of possible observations.\\
$\vec{\obs}$ & Sequence of observations.\\
$P(\hypo | \vec{\obs})$ & Probability that hypothesis $\hypo$ holds, given the sequence of observations $\vec{\obs}$.\\
$P(o|\hypo)$ & Probability of making observation $o$ when hypothesis $\hypo$ holds.\\
$\tilde{P}$ & Unnormalized probability. \\
$\mc{N}(r | \mu, \sigma)$ & Probability of drawing the number $r$ from a Gaussian probability distribution with mean $\mu$ and standard deviation $\sigma$.\\
$\tri_j^n$ & Triangular function over issue $\issue{j}$, with peak at option $\offComp{{j,n}}$ (see Eq.~(\ref{eq:triangular})).\\
$\expt{w}^j$ & Expectation value for the weight that the opponent assigns to issue $\issue{j}$.\\
$\expt{\eval}^{j}$ & Expected evaluation function that the opponent applies to issue $\issue{j}$.\\
$\expt{\util}$ & Expected utility function for the opponent.\\
$\freq_\hist(\offComp{{j,l}})$ & The number of times the opponent has proposed an offer containing option $\offComp{{j,l}}$.\\
$z_j$ & Shorthand for the utility offered to us by the opponent in her $j$-th proposal to us, i.e.: $z_j := \util_1(\off_j)$.\\
$\vec{z}$ & Sequence of offered utilities, i.e. $\vec{z} = (z_1, z_2, \dots)$ \\
$\mathbf{I}$ & Identity matrix.\\
$\covmat$ & Covariance matrix.\\
$K_{i,j}$ & Element of the covariance matrix at row $i$ and column $j$.\\
$\kernel$ & Kernel function.\\
$P_\acc(z)$ & Probability that $\ag_2$ would accept an offer $\off$ with utility $u_1(\off) = z$.\\
\end{tabular}

\begin{longtable}{lp{4in}}
\multicolumn{2}{l}{\textbf{Game Theory}} \\
$\ac$ & Action\\
$\Ac_i$ & The set of actions available to player $i$.\\
$\nfgame$ &  Normal-form game.\\
$\br_j(\ac_i)$ & The set of actions that are a best response for agent $j$, against some action $\ac_i$ of its opponent.\\
$\ms$ & Mixed strategy \\
$\MS_i$ & The set of all mixed strategies for player $i$\\
$\vec{\ms}$ & Strategy profile (of mixed strategies). \\
$\br_j(\ms_i)$ & The set of mixed strategies that are a best response for agent~$j$, against some mixed strategy $\ms_i$ of its opponent.\\
% $(a,b)^\dagger$ & The `reflection' of a tuple $(a,b)$, i.e.: $(a,b)^\dagger := (b,a)$\\
$\mi{NE}$ & The set of all Nash equilibria of a normal-form game.\\
%$\mi{DNE}$ & The set of degenerate Nash equilibria of some normal-form game.\\
%$\mi{NDNE}$ & The set of non-degenerate Nash equilibria of some normal-form game, i.e. $\mi{NDNE} := \mi{NE} \setminus \mi{DNE}$. \\
$\mc{G}$ & A set of 2-player normal-form games. \\
$\ff$ & A `role frequency function', i.e. $\ff(\nfgame,i)$ is a number that represents the relative frequency in which an agent is going to be (or is expected to be) playing game $\nfgame$ in the role of player $i$.\\
$\mc{U}_A$, $\mc{U}_B$ & Meta-utility functions of agents $\ag_A$ and $\ag_B$.\\
$X^*$ & The set of tuples over some set $X$.\\
$X^n$ & The $n$-fold Cartesian product of some set $X$, i.e. $X^1 = X$, \quad $X^2 = X \times X$, \quad $X^3 = X\times X \times X$, \quad etc...\\
$\emptyTuple$ & The `empty tuple'.\\
$\circ$ & The concatenation operator for tuples, e.g. $(a,b) \circ (c,d,e) = (a,b,c,d,e)$.\\
$Y^\term$ & The set of terminal tuples among some set of tuples $Y$.\\
$\node$ & Tree node.\\
$d$ & Depth of a tree node.\\
$\ttgame$ & Turn-taking game.\\
$\Hist$ & The set of all legal action histories of a turn-taking game.\\
$\act$ & The active player function of a turn-taking game.\\
$\Hist_i$ & The set of all non-terminal histories after which player $i$ is the active player.\\
$\Ac_\hist$ & The set of actions that the active player is allowed to choose after history $\hist$.\\
$\strat$ & Strategy for a turn-taking game.\\
$\vec{\strat}$ & Strategy profile for a turn-taking game.\\
$\hist_{\vec{\strat}}$ & The unique terminal history generated by  strategy profile $\vec{\strat}$.\\
$\Strat_i$ & The set of all possible strategies for player $i$ in some turn-taking game.\\
$\ttgame_\hist$ & The subgame of $\ttgame$ at history $\hist$. \\
$\Hist_\hist$ & The set of legal action histories of the subgame $\ttgame_\hist$.\\
$\obsfunc_i$ & The observation function of player $i$.\\
$\Obs_i$ & The set of all possible observed histories after which it is player $i$'s turn. \\
$\Ac_i^\dom$ & The set of negotiation actions for player~$i$ in negotiation domain $\dom$.\\
$\utilSpace_{\dom}$ & The utility space of a negotiation domain $\dom$.\\
$\pf_\dom$ & The Pareto frontier of a negotiation domain $\dom$.\\
$\kf$ & The `Kalai-Smorodinsky function' (Eq.~\ref{eq:kalai_function}).\\
$L(\dom)$ & The line from the point $(\rv_1 , \rv_2)$ to the `utopian' point $(\utilMax_1, \utilMax_2)$.
\end{longtable}

\begin{longtable}{lp{4in}}
\multicolumn{2}{l}{\textbf{Evaluation of Negotiation Algorithms}} \\
$\prot$ & Negotiation protocol.\\
$\Dom$ & Set of negotiation domains. \\
$\Ag$ & Set of agents.\\
$\ag_{\tx{i}}$ & The $i$-th agent from the set of agents $\Ag$ in a tournament.\\
$\mc{\numReps}$ & Function that maps each possible negotiation scenario to its number of repetitions in the tournament.\\
$\numReps^{d,\tx{i},\tx{j}}$ & Shorthand for $\mc{\numReps}(\dom_d, \ag_{\tx{i}}, \ag_{\tx{j}})$.\\
$\util_{1}^{d,\tx{i},\tx{j},r}$ & Utility obtained by agent $\ag_k$ in the $r$-th negotiation between agents $\ag_{\tx{i}}$ and $\ag_{\tx{j}}$ over domain $\dom_d$.\\
$\ts_{\tx{i}}^{d,\tx{j}}$ & Average utility  obtained by agent $\ag_{\tx{i}}$ against opponent $\ag_{\tx{j}}$ when negotiating over domain $\dom_d$. \\
$\ts_{\tx{i}}$ & Tournament score of agent $\ag_{\tx{i}}$.\\
$\ind_{agr(d,\tx{i},\tx{j},r)}$ & Indicator function that has value 1 if the $r$-th repetition of scenario $(\prot, \dom_d, \ag_{\tx{i}}, \ag_{\tx{j}})$ ended with agreement, and 0 otherwise.\\
$\numSucc^{d,\tx{i},\tx{j}}$ & Number of times a negotiation in the scenario $(\prot, \dom_d, \ag_{\tx{i}}, \ag_{\tx{j}})$ ended with agreement.\\
$\ar_{\tx{i}}$ & Agreement rate of agent $\ag_{\tx{i}}$.\\
$\uua_{\tx{i}}$ & Utility-under-agreement of agent $\ag_{\tx{i}}$. \\
$\rvX$, $\rvY$, $\rvZ$ & Random variables. \\
$\varDomain_{\rvX}$ & The set of possible values of the random variable $\rvX$. \\
$P_{\rvX}$ & Probability distribution of $\rvX$.\\
$\mu_{\rvX}$ & Mean of $\rvX$.\\
$\mathit{Var}_{\rvX}$ & Variance of $\rvX$.\\
$\sigma_{\rvX}$ & Standard deviation of $\rvX$.\\
$\est{\mu}_{\rvX}$ & Estimated mean of $\rvX$.\\
$\est{\mathit{Var}}_{\rvX}$ & Estimated variance of $\rvX$ (a.k.a. sample variance).\\
$\est{\sigma}_{\rvX}$ & Estimated standard deviation of $\rvX$ (a.k.a. sample standard deviation).\\
$\se_{\mu_{\rvX}}$ & Standard error on the mean of $\rvX$.\\
$\est{\se}_{\mu_{\rvX}}$ & Estimated standard error on the  mean of $\rvX$.\\
$\scen$ & Negotiation scenario\\
$\Scen$ & Set of negotiation scenarios.\\
$\util_{\tx{i},s,r}$ & The $r$-th observation from random variable $X_{\tx{i},s}$.\\
$\est{\mu}_l^{d,\tx{i}, \tx{j}}$  & Average utility  obtained by agent $\ag_{l}$ in the negotiations between $\ag_{\tx{i}}$ and $\ag_{\tx{j}}$ over domain $\dom_d$. \\
$\tstat$ & t-statistic \\
$\tdist_{k-1}$ & The t-distribution with $k-1$ degrees of freedom.\\
$p$ & $p$-value
\end{longtable}

\begin{longtable}{lp{4in}}
\multicolumn{2}{l}{\textbf{Multilateral Negotiations}} \\
$\est{\util}_{opp}(\off)$ & The minimum estimated utility of $\off$ among all opponents.\\
$m_i$ & The number of unique offers that were either proposed or accepted by agent $\ag_i$.\\
$\ag_0$ & The `buyer' agent in a one-to-many negotiation.\\
$\Off_{i,j}$ & The offer space for negotiations between buyer $\ag_i$ and seller $\ag_j$.\\
$\pa$ & The participating agents function.\\
$\feas$ & The feasibility function.\\
$[\numAgents]$ & The set of integers from 1 to $\numAgents$.\\
$\rej$ & Symbol representing rejection. \\
$\obsHist{not}$ & The history as observed by the notary. \\ 
$\hslash$ & The history as observed by the notary (alternative notation to $\obsHist{not}$). \\ 
$rej_{\hist}(i,\off,t)$ & Predicate indicating that agent $\ag_i$ has rejected offer $\off$ at some time after time $t$.\\
$acc_{\hist}(i, \off)$ & Predicate indicating that agent $\ag_i$ has accepted offer $\off$ but has not rejected it afterwards.\\
$acc_{\hist}(\off)$ & Predicate indicating $acc_{\hist}(i, \off)$ holds for all negotiating agents.\\
$Agr_\hslash$ & The set of offers that have become binding agreements after the notary has observed history $\hslash$.
\end{longtable}

\begin{longtable}{lp{4in}}
\multicolumn{2}{l}{\textbf{Large and Complex Negotiation Domains}} \\
$\partoff{k}$ & Partial offer of size $k$. \\
$\Partoff{k}$ & The set of all partial offers of size $k$ (in some given negotiation domain). \\
$\bids_k$ & Function representing the BIDS algorithm.\\
$\tilde{\asp}$ & Value close to some given aspiration value $\asp$. \\
$\util_i^{min,k}$ & The minimum possible utility value for agent $\ag_i$, for any partial offer of size $k$ in some given negotiation domain.\\
$\util_i^{max,k}$ & The maximum possible utility value for agent $\ag_i$, for any partial offer of size $k$ in some given negotiation domain.\\
$\precision$ & The precision parameter of the BIDS algorithm or the IPS algorithm.\\
$\range{\precision}{k}$ & Discretization of the interval $[\util_1^{min,k} , \util_1^{max,k}]$ into $\precision$ equally spaced points. \\
$\domin(x, x')$ & A predicate representing the fact that some object $x$ is dominated by another object $x'$.\\
$\pareto{S}$ & The Pareto-set of $S$ (i.e. the subset of $S$ consisting of those elements of $S$ that are Pareto-optimal \textit{within} $S$).\\
$\close$ & Function that assigns to any partial offer $\partoff{k}$, the element of $\range{\precision}{k}$ that is closest to $\util_1(\partoff{k})$.\\
$\cluster{\tilde{\util},k}$ & The set of all partial offers $\partoff{k}$ for which $\close(\partoff{k}) = \tilde{\util}$.\\
$\approxPareto{\Partoff{k}}$ & Approximation, made by the IPS algorithm, of the set of Pareto-optimal partial offers of size $k$.\\
$\infeas$ & A hard constraint (i.e. a set of infeasible offers).\\
$\hypCube$ & A hypercube.\\
$w_{\hypCube}$ & The weight of hypercube $\hypCube$.\\
$\delta_{\hypCube}$ & Hypercube membership function. \\
$\HypCube$ & A set of hypercubes.\\
$\subIssue{j}$ & A subset of issue $\issue{j}$.\\
$\vol(\hypCube)$ & The volume of a hypercube $\hypCube$.\\
$\bid$ & A `bid', i.e. a set of hypercubes chosen from an agent's own set of hyper cubes. \\	
$\overline{\bid}$ & The intersection of all hypercubes in bid $\bid$.\\
$\bid_{\off}^i$ & The set of all hypercubes in $\HypCube_i$, that contain $\off$.\\
$Q(\hypCube)$ & The `quality' of a hypercube. \\ 
$dist_t(\off)$ & The `distance' of an offer $\off$.\\
$div_t(\off)$ & The `diversity' of an offer $\off$.\\
$\dist(\off, \off')$ & Manhattan distance between two offers $\off$ and $\off'$ in a hypercube domain.\\
$\offSet$ & A set of offers returned by a search algorithm.\\
\end{longtable}

%% If you want to cite this book, please do so as follows:

\chapter{Introduction}

\section{Characteristics of Negotiation}
Whenever we talk about `negotiation' we are referring to any form of communication between multiple `agents' (which can be either humans or software) with the goal of coordinating their actions, so that they can achieve a better outcome for themselves than what they could possibly achieve without such coordination.

A simple example is the scenario of a group of friends that want to go to the cinema together. In order to achieve that goal, they have to make a number of decisions together: which cinema to go to, which movie to watch, and at what time. If they do not manage to come to an agreement on all these decisions, then they will not be able to go to the cinema together. Clearly, coordination is essential to achieve the desired outcome. Another typical example is the case that a salesperson and a customer are negotiating the price of some product, such as a car.

In particular, we say that agents are negotiating whenever the following conditions are satisfied:
\begin{enumerate}
\item There is more than one agent.
\item These agents are able to communicate with each other.
\item The agents need to make one or more choices out of a number of options.
\item Each agent has its own individual preferences over the options.
\item Each agent is autonomous.
\end{enumerate}
%\todo{I'm not entirely happy about this. Maybe I should mention that there is a 'default' option, and that no agent can enforce any of the other options onto the other agents.}

The need for the first three of these conditions should be obvious. The fourth condition is also essential, because if an agent does not have its own preferences, then it would not have any reason to participate in the negotiations. It could simply let all the other agents make the decisions.  The fifth condition means that each agent has at least some partial freedom to do whatever it wants. If any of the agents does not have such freedom, then that agent would essentially be a slave to the others and it would not have any negotiation power. For example, a car seller cannot force the buyer to buy the car. The buyer has the autonomy to refuse any offer he or she doesn't like. Similarly, the buyer cannot force the seller to sell the car either. The seller too has the autonomy to reject any offer from the buyer. As a counter example, we can imagine a swarm of robots that are searching through the ruins of a collapsed building in order to find survivors. If these robots are fully controlled by a central computer, then there is no need for negotiation. The central computer simply dictates what all the robots should do.

Regarding the condition that all agents have their own preferences, we should note, however, that this does not mean their preferences need to be \textit{different}. For example, suppose two friends called Alice and Bob want to choose a movie to watch together. Even if they each want to see the same movie, they may still need to communicate this preference to one another in order to ensure that they are each \textit{aware} of this fact. For example, Alice could propose to Bob to see \textit{The Godfather}, and then Bob could accept that proposal. In other words, they still need a short negotiation, to establish their decision. The key point here, is that the two agents a priori do not know that their preferences are the same.

%% I HAVE COMMENTED OUT THIS PARAGRAPH, BECAUSE I'M NOT SURE THAT WE CAN REALLY CALL THIS A 'NEGOTIATION'. 
%Furthermore, even if they already know each other's preferences, they may still need to negotiate. Imagine, for example, that they need to choose between two different cinemas where they could watch their movie and that they are each completely indifferent between the two options. Even though they both have exactly the same preferences, and they each \textit{know} that they have the same preferences, they still need to come to an agreement about which one to pick. For example, Alice could flip a coin to decide on the cinema, and then still needs to propose it to Bob. 
%(let say they are called \textit{Rialto} and \textit{Paradiso}) 

Nevertheless, in the rest of this book we will almost always assume that there is some amount of conflict among the agents' preferences. After all, a scenario in which all agents exactly agree on their preferences is not a very interesting test case for scientific research. The scenario of the salesperson and the customer that are negotiating the price of a car is probably the most obvious example of a negotiation in which the two negotiators have conflicting interests. In this case the agents' preferences are diametrically opposed: the seller wants to sell the car for the highest possible price, while the buyer wants to buy it for the lowest possible price. Despite their conflicting interests, the two agents still aim to find a compromise that is acceptable to each of them individually.

It should be noted that there are many situations in daily life in which the above conditions hold, and therefore can be seen as a type of negotiation, even though we normally wouldn't think of them as a negotiation. In fact, any time two or more people make a joint decision, it is essentially a negotiation. For example, whenever you ask someone a question like ``shall we eat at 19:00?" or ``Do you want to go the cinema?" you are essentially starting a negotiation.

Another nice example of a negotiation scenario that we typically do not think of as a negotiation, is when you do your groceries at the supermarket. In this scenario there are indeed multiple agents, namely the customer and the supermarket. These two agents jointly aim to come to an agreement about which products the supermarket will sell to the customer. Each of these agents has a certain amount of autonomy: the supermarket can choose which products it offers and for what price. The customer, on the other had, can choose which of those products he or she will buy. Furthermore, each agent has their own preferences: the supermarket aims to make the highest possible financial profit, while the customer has preferences over which products he or she wants to buy, and prefers to buy them for the lowest possible price. Among the five conditions we listed above, the presence of communication is probably the least obvious in this example. However, the supermarket is, in fact, communicating to the customer, by means of labels and price tags on their products. Every time the costumer sees a label saying something like ``\textit{1~kg of beef, \$12}" this can be seen as a \textit{proposal} made by the supermarket to the customer. The customer can then either \textit{accept} that proposal by taking the product from the shelf and adding it to their shopping cart, or \textit{reject} it by walking along without taking the product. This is, essentially, a form of negotiation. Of course, it is a somewhat limited form of negotiation since the supermarket is the only agent here that can make proposals, while the customer can only accept or reject those proposals, but cannot make any counter-proposals to the supermarket.

In the literature one sometimes distinguishes between \textit{negotiation} and \textit{bargaining}. The exact definitions differ per author, where `bargaining' is often used exclusively to refer to the exchange of proposals that can be accepted or rejected, while `negotiation' often refers to a more general process in which the agents may use a broader form of communication that allows them to express their respective interests, or allows them to convince the other agents to change their points of view. In the rest of this book, however, we will not distinguish between the two concepts and simply always use the term `negotiation' even were some authors might argue that `bargaining' would be the more appropriate term.

\section{What this Book is (not) About}
Of course, this book is not just about negotiation. It is about  \textit{automated} negotiation. That is, we discuss how to implement a software agent that can conduct negotiations with other agents. 

%That is, the study of how to develop computer programs that can perform negotiations autonomously, either with other computer programs or with humans (although in this book we will focus mainly on negotiations between computers only).

It is important to understand that when we implement such a negotiating agent, this agent is supposed to defend the interests of only \textit{one} party in the negotiation. For example, if we implement an agent that is going to be used by a car seller to negotiate the price of a car with a potential customer, then our agent should try to sell the car for the highest possible price. In particular, we have to see the other agents that it is going to negotiate with, as its `opponents', and we have to keep in mind that, in general, our agent will not have any knowledge about the implementations or strategies used by those opponents, in the same sense that a chess computer should be able to play chess against any arbitrary opponent. In other words, when we implement an agent, we should consider each of its negotiation partners as a `black box' that we don't have access to.

In general, the research topic of automated negotiation may be about software agents that negotiate with other software agents (agent-to-agent negotiations), or about software agents that negotiate with \textit{humans} (human-to-agent negotiations). In this book, however, we will focus exclusively on agent-to-agent negotiations and we will not discuss negotiations with humans (at least, not in the \textit{current} version of this book). This is not because human-to-agent negotiations aren't important, but rather because human-to-agent negotiations are fundamentally different from agent-to-agent negotiations. After all, when conducting negotiations with humans, the psychology of humans becomes a major factor, which does not play any role in agent-to-agent negotiations. 

For example, two software agents could easily exchange thousands of proposals in a matter of seconds, while we cannot expect a human negotiator to be able to process that many different proposals. So, when negotiating with humans, we have to keep in mind that the opponent will get tired quickly. Furthermore, a human negotiator might get frustrated or even upset if their negotiation partner is only making very small concessions. They might accuse their opponent of being too greedy or stubborn and  leave the negotiations for emotional reasons, even if that only hurts themselves. These considerations imply that human-to-agent negotiations require an entirely different approach from agent-to-agent negotiations.

So, while human-to-agent negotiation is also a very interesting and important topic, we have chosen not to go into that area, and focus exclusively on agent-to-agent negotiations, which do not involve psychology.

Another important topic within the field of automated negotiation that, for now, we will not discuss in this book, is the topic of \textit{argumentation-based} negotiation (ABN). In argumentation-based negotiation one does not only focus on making the right proposals, but also on the use of argumentation to change the other agent's mind.

\section{The History of Automated Negotiation}

The topic of automated negotiation dates back to the 1950's, starting with the work of John Nash~\cite{Nash1950}. Back in those days, however, automated negotiation was mainly studied from a purely theoretical point of view, rather than from an algorithmic point of view. The typical approach followed by Nash and other researchers of his time, would be to 
argue that the outcome of a negotiation should satisfy  a certain set of mathematical axioms. They would then formally prove that there exists a unique outcome satisfying those axioms. Several different solution concepts were proposed in this way, based on different sets of axioms~\cite{KalaiSmorodinsky,herrero1989nash,conley1996nonConvex}.

This changed in 1998 with the seminal paper by Faratin et al.~\cite{Faratin1998}. Rather then trying to find theoretically optimal outcomes, they took a more practical approach and proposed a number of possible negotiation strategies, which we will discuss in Chapter~\ref{sec:negotiation_strategies}. This was a great step forwards towards realistic applications of automated negotiation, because it takes into account that real agents typically would not have complete domain knowledge and would not be willing to share strategic information with each other.

Another pivotal event in the history of automated negotiation was the inception of the Automated Negotiating Agents Competition (ANAC) in 2010~\cite{anac2010} and the development of the Genius framework \cite{lin2014genius} on which ANAC was run. Since then, ANAC has been held almost every year at major A.I. conferences such as IJCAI and AAMAS and has greatly boosted the number of papers published on the topic of automated negotiation. Furthermore, ANAC has led to the development of hundreds of negotiating agents and a plethora of different opponent modeling techniques, which are still used by many researchers today, as a baseline against which they can test new negotiation algorithms.

Initially, most research on automated negotiation focused on the most basic type of negotiations with two agents negotiating over a small set of possible agreements with linear utility functions~\cite{anac2010}.  However, over the years, more and more researchers have started investigating more complex negotiation scenarios. For example, several researchers have studied negotiation domains with non-linear utility functions and with an extremely large number of possible agreements~\cite{ito2008nonlinear, marsaMaestre2009nonlinear}. This was later taken even a step further by considering negotiations in which the evaluation of just a single proposal was already computationally complex problem~\cite{deJonge2015nb3,deJonge2017dbrane,deJonge2022multiObjectiveVRP}. 

Other researchers have focused on multilateral negotiations (negotiations between 3 or more agents)~\cite{Nguyen2004MultipleConcurrentNegotiations,Endriss2006,deJonge2015nb3,aydougan2017saop}, or the use of  machine learning algorithms such as deep learning and reinforcement learning to train negotiation algorithms~ \cite{Sengupta2021,Bakker2019RLBOA}.

Most of these developments have also been closely mirrored by the various editions of ANAC. For example, ANAC 2014 involved negotiations with non-linear utility functions and extremely large search spaces~\cite{anac2014}, while from 2015 to 2018 ANAC focused on multilateral negotiations~\cite{anac2015}. Then, in 2019 and 2020 the focus shifted back to small, bilateral negotiations, but in which each agent only had \textit{partial} knowledge about its own preferences~\cite{anac2019}. After that, several editions focused on the use of machine learning to allow the agents to learn the characteristics of their opponents, from earlier negotiations~\cite{renting}. Furthermore, from 2017 onward the ANAC competition was divided into a `main league' and one or more sub-leagues focusing on more specialized negotiation problems, such as high computational complexity in the game of Diplomacy~\cite{dejonge2019challengeOfNegotiation}, multilateral negotiations in a supply-chain environment~\cite{Mohammad2019SCML} negotiations between computers and humans~\cite{anac2017HumanAgentLeague}, and negotiations in the game of 
Werewolves~\cite{anac2019}. 

%However, throughout the years this competition has shifted focus to many different types of negotiation settings such as negotiation with learning from previous negotiations~\cite{anac2013, renting}, negotiations with non-linear utility functions and astronomically large search spaces~\cite{anac2014} negotiations between three or more agents~\cite{anac2015}, negotiations with \textit{partial} knowledge about your own utility function~\cite{anac2019}, negotiation in the game of Diplomacy~\cite{dejonge2019challengeOfNegotiation}, in the game of Werewolves~\cite{anac2019}, or in a supply chain environment \cite{anac2019}, or negotiations between computers and humans~\cite{anac2017HumanAgentLeague}.

For a long time, the Genius framework, which was written in Java, was the main platform that researchers used for their experiments in the field of automated negotiation. It was especially useful because it included a large set of hand-crafted test-domains that were used in the ANAC competitions and a large set of agents that participated in those competitions. This immediately gave researchers access to a vast library of benchmark test cases and baseline algorithms for their experiments.

However, it has recently been shown, both experimentally \cite{deJonge2022MiCRO} and theoretically \cite{deJonge2024theoreticalPropertiesMicro}, that a very simple negotiation strategy called MiCRO is able to achieve near-optimal results on the Genius test domains even without using any form of machine learning or opponent modeling. It was therefore argued that those hand-crafted test cases should no longer be used.

The Genius framework is no longer maintained, and has now been superseded by the NegMas framework \cite{Mohammad2020NegMas} as the main platform for research on automated negotiation. It is written in Python, but it still includes the possibility to run the Java agents from the Genius framework. Furthermore, it allows generating random test domains which are harder to tackle than the hand-crafted ones from Genius. Another framework, called GeniusWeb, was also developed by the makers of Genius, but this framework never gained much traction.

%\essential{
%These agents and test-cases have therefore become the standard benchmark for automated negotiations research.
%}

%\essential{
%It has been followed-up by GeniusWeb, which however never gained much traction. Instead, the NegMas framework \cite{Mohammad2020NegMas} nowadays seems to be the most commonly used framework for experiments in automated negotiation. It is written in Python, but it still includes the possibility to run the Java agents from the Genius framework.
%}

%
%\essential{
%For a long time the hand-crafted negotiation test scenarios that were included in the Genius framework have been used as the standard benchmark to test new negotiation scenarios. However, it was recently shown both experimentally \cite{deJonge2022MiCRO} and theoretically \cite{deJonge2024theoreticalPropertiesMicro} that a very simple algorithm called MiCRO is able to achieve near-optimal results on such domains even without any form of machine learning or opponent modeling. It has since then become more common to test algorithms on randomly generated domains that are harder to tackle.
%}

\later{Mention Colored Trails}

\later{Mention Argumentation-Based Negotiation?}

%Furthermore, the ANAC competitions have led to an increased attention to the field of automated negotiation and to development of several software platforms for the development of, and the  experimentation with automated negotiating agents, such as the Genius framework~\cite{lin2014genius}, GeniusWeb~\cite{geniusWeb}, and NegMas~\cite{Mohammad2020NegMas}.

\later{Discuss relation to cooperative game theory}

\later{Discuss the future of negotiation, e.g. with LLM's}

\chapter{Basic Negotiations}\label{sec:basic_negotiations}
In this chapter we discuss the basic ideas of automated negotiation. For now we will focus mainly on \textbf{bilateral} negotiation. That is,  negotiations between exactly two agents, as opposed to \textbf{multilateral} negotiation, which takes place between more than two agents. The only exception is that some of the mathematical definitions below will be given for arbitrary numbers of agents, because it would not simplify anything if we presented them for only two agents.

We here focus on bilateral negotiation because they are the simplest to explain, because they have been studied much more extensively in the literature and because they are sufficient to explain the most basic aspects of automated negotiation. We will discuss multilateral negotiations later on in Chapter~\ref{sec:multilateral}.

\section{Informal Description}

Imagine there are two agents, which we will  call the `buyer' and the `seller' respectively, that are negotiating the price of a second-hand car. The negotiations start with one agent proposing an offer to the other agent. For example, the seller might start by proposing a price of \$10,000. Next, the buyer can do two things: to accept the proposal, or to reject it. If the buyer accepts the proposal, then it becomes a formally binding agreement and the negotiations are over. Otherwise, if she rejects the proposal, then she can make a counter-proposal. For example, she might propose a price of \$5,000. Next, it is again the seller's turn. The seller now also has the choice between accepting the last proposal, or rejecting it and making a new proposal. For example, she could then propose a price of \$9,500. This will continue until they come to an agreement, or one of the agents decides to withdraw from the negotiations, or a given deadline has passed, or when a fixed maximum number of proposals has been made.

In this example we assumed the agents negotiated according to the so-called \textbf{alternating offers protocol} (AOP) \cite{rosen94}, meaning that the agents take turns making proposals. Specifically, it means that an agent is not allowed to make two proposals in a row. After making a proposal the agent first needs to wait for the other agent to respond and make a counter-proposal before she can make a new proposal herself. While this is certainly not the only protocol for automated negotiation, it does seem to be the one that is most commonly used in the literature.

In the field of automated negotiation we typically assume there is a fixed set of possible offers that the agents can propose to one another. This set is called the \textbf{offer space} (or sometimes \textbf{agreement space}). In the example of the car sale, the offer space consisted of every possible price that the seller could possibly ask, or that the buyer could possibly offer. So, this could be the set of all positive integers. One important thing to notice about this example, is that the agents were negotiating over just one issue: the price of the car. This is what we call a \textbf{single-issue} negotiation. In many cases in the literature, however, one studies \textbf{multi-issue negotiations}. That is, negotiations in which each proposal may involve multiple different components. For example, suppose there are two friends, Alice and Bob, that want to go to the cinema together. They need to agree on three different issues: 
\begin{enumerate}
\item Which movie they will see.
\item Where they will see this movie (in which cinema).
\item When they will see this movie (which day of the week and at which time).
\end{enumerate}
One way to conduct such multi-issue negotiations would be to negotiate each issue separately, one by one. However, a more common approach in the literature is to just negotiate all issues at the same time. This means that each proposal indicates a value for all three issues at the same time. For example, Alice might start by proposing to see \textit{The Godfather} in cinema \textit{Rialto} on Friday at 20:00. Bob might then reject this proposal, and instead propose to see \textit{Casablanca}, in cinema \textit{Paradiso}, on Saturday at 18:00, etcetera.

%For example, rather than just negotiating the price of the car, they agents could also negotiate about \textit{which} car exactly the customer wants to buy, and also other extras such as an annual service contract, and 

%An example could be that a buyer and a seller are negotiating the details of a health insurance contract. That is, they are not only negotiating the price of the contract (i.e. the monthly fee), but also what is included in the contract, such as dental care, pregnancy and physio therapy, as well as the height of any excess fees. 

We should remark that in this book we will use the term \textbf{offer} to refer to a potential outcome of a negotiation. That is, something that can be proposed or accepted or rejected.  So, in the scenario of the car sale, the price of \$10,000 would be an example of an offer, while in the scenario of the two friends who are going to the cinema, the tuple (\textit{The Godfather}, Rialto, Fri 20:00) would be an example of an offer. Furthermore we will use the term \textbf{proposal} to refer to the \textit{action} of proposing an offer. Finally, we use the term \textbf{agreement} to refer to an offer that has been accepted as the final outcome of the negotiation between the two agents.  We should note however, that the literature is not very consistent on this matter. Other authors may use these terms in  different ways, or they may use alternative terms such as \textbf{deal}, \textbf{contract}, or \textbf{bid} with their meanings being different for each author.

\section{Formal Model}

In order to be able to implement an agent that can negotiate, we first need to have a formalization of what `negotiation' means exactly. We will here discuss this formal model. We assume there are exactly two \textbf{agents}, which we denote by $\ag_1$ and $\ag_2$ respectively.

\subsection{The Offer Space}\label{sec:offer_space}
In order to implement a negotiating agent, the first thing we need to know is which offers the agents can possibly propose. This is known as the \textbf{offer space} or \textbf{agreement space} and is usually denoted by $\Off$. In the example of a single-issue car sale, the set of possible offers was the set of all positive integers $\mathbb{N}$, where each number $k\in \mathbb{N}$ represents a proposal to trade the car for a price of $k$ dollars. A single offer from the offer space is usually denoted by $\off$.

%Of course, there is only a small subset of integers that actually represents a realistic price. Therefore, we could just as well reduce it to the set of only 

In the case of a multi-issue negotiation, the offer space can be written as the cartesian product of smaller sets that we call \textbf{issues}:
\[\Off = \issue{1} \times \issue{2} \times \dots \times \issue{\numIssues}\]
so each offer $\off$ is a tuple:
\[\off = (\offComp{1}\ ,\ \offComp{2}\ ,\ \dots \ ,\ \offComp{\numIssues})\]
where each $\offComp{j} \in \issue{j}$. For each issue, we will refer to its elements as its \textbf{options}.

For example, the scenario in which two friends are planning to see a movie together, can be modeled as a negotiation over the following three issues, representing the movie, the cinema, and the time slot, respectively:
\begin{eqnarray*}
\issue{1} &=& \{\mathit{The\ Godfather}, \mathit{Casablanca}, \mathit{The\ Big\ Lebowski}\}\\
\issue{2} &=& \{\mathit{Rialto}, \mathit{Paradiso}\}\\
\issue{3} &=& \{\text{Fri\ 18:00}, \text{Fri\ 20:00}, \text{Fri\ 22:00}, \text{Sat\ 18:00}, \text{Sat\ 20:00}, \text{Sat\ 22:00}  \}
\end{eqnarray*}
We see that the issue `movie' has 3 options, the issue `cinema' has 2 options, and the issue `time slot' has 6 options. So, the offer space contains $3 \times 2 \times 6 = 36$ possible offers.

Note that issues may or may not have a natural ordering. For example, the issue $\issue{3}$ above, representing the time slot, is naturally ordered from early to late. On the other hand, the other two issues $\issue{1}$ and $\issue{2}$ do not have any ordering (of course, we could put them in any order we like, such as an alphabetical order, but that is not very meaningful for the negotiations). 

Furthermore, note that the division of an offer space into separate issues can sometimes be a bit arbitrary. For example, rather than having one issue representing the time slot, we could instead have defined two separate issues: one issue for the day of the week, and one issue for the time. So, we could have defined the offer space as a product of the following 4 issues:
\begin{eqnarray*}
\issue{1} &=& \{\mathit{The\ Godfather}, \mathit{Casablanca}, \mathit{The\ Big\ Lebowski}\}\\
\issue{2} &=& \{\mathit{Rialto}, \mathit{Paradiso}\}\\
\issue{3} &=& \{\text{Fri}, \text{Sat} \}\\
\issue{4} &=& \{\text{18:00}, \text{20:00}, \text{22:00} \} 
\end{eqnarray*}
This would not have made any difference. This also works in the other direction: if we wanted, we could have just ignored the separate issues altogether and model the entire domain as one single issue containing 36 different options, without any structure. However, as we will see in Section \ref{sec:linear_util_functions}, decomposing the offer space into separate issues has the advantage that it allows us to define simple utility functions that are linear combinations of smaller functions that are each defined over a single issue.

Also note that in a real-world scenario there may exist constraints among the issues. For example, Cinema Rialto might only screen \textit{The Godfather} on Saturdays, and Cinema Paradiso might not screen any movie at all on Friday at 18:00. So, in that case not \textit{every} combination of options would be possible, and the offer space $\Off$ would only be a \textit{subset} of the Cartesian product $\issue{1} \times \issue{2} \times \dots \times \issue{\numIssues}$. However, in most of the literature such constraints are not taken into account and one typically assumes that all possible combinations of options are allowed.

Another example of a multi-agent negotiation scenario, is the case that two collectors of, say, rare coins, negotiate a trade of coins. For example, say Alice and Bob each have 3 different coins they are willing to trade. We will denote these coins as $c_1, c_2, \dots, c_6$, where the first three coins ($c_1$, $c_2$, and $c_3$) belong to Alice, and the last three coins ($c_4$, $c_5$, and $c_6$) belong to Bob. However, Alice may not be interested in each of Bob's coins, and vice versa. So, they aim to find an agreement in which Alice gives some subset of her coins to Bob, and in return Bob gives some subset of his coins to Alice. This can be modeled as a negotiation domain with 6 issues (one for each coin) and where each issue $\issue{i}$ consists of 2 options: $\issue{i} = \{Alice, Bob\}$, corresponding to who will be the new owner of coin $c_i$. So, the offer:
\[\off = (Bob, Bob, Bob, Alice, Alice, Alice)\]
represents the deal in which Alice gives all her coins to Bob (so Bob ends up owning the first three coins) and in return, Bob gives all his coins to Alice (so Alice ends up owning the last three coins). Furthermore, the offer 
\[\off' = (Bob, Alice, Alice, Bob, Bob, Alice)\]
corresponds to the deal in which Alice just gives 
coin $c_1$ to Bob, and in return Bob just gives coin $c_6$ to Alice, so Alice ends up with coins $c_2$, $c_3$, and $c_6$, and Bob ends up with coins $c_1$, $c_4$, and $c_5$. Note that they may even agree to not exchange any coins at all, which is represented by the offer:
\[\off'' = (Alice, Alice, Alice, Bob, Bob, Bob)\]
where each collector just ends up with the same coins they originally already owned. Moreover, note that this domain also contains trades in which only one collector gives any coins to the other, and receives nothing in return. While one may argue that such trades would be unrealistic, that does not mean that we need to exclude them from the offer space. After all, if either of the two agents is unhappy with such an offer, they can simply refuse to ever propose or accept it.

\subsection{The Alternating Offers Protocol}\label{sec:aop}

The next thing we need to specify is the \textbf{negotiation protocol}. That is, the rules that determine when which agent is allowed to propose or accept which offer, and when a proposal will be considered a formally binding agreement. 

The most commonly used protocol for \textit{bilateral} negotiations, is the alternating offers protocol (AOP) which we have already seen above. In this protocol the agents take turns, so the protocol needs to specify which of the two agents will make the first proposal. In this section we will, without loss of generality, assume that this is always agent $\ag_1$.

At the start of the negotiations, agent $\ag_1$ can choose any $\off \in \Off$ from the offer space and propose it to $\ag_2$. Next it is agent $\ag_2$'s turn. Agent $\ag_2$ can now either accept the previous proposal from $\ag_1$, or propose an alternative offer $\off' \in \Off$. If $\ag_2$ accepts the previous offer $\off$ then the negotiations are over and $\off$ will be considered a formally binding agreement. Otherwise, if $\ag_2$ does not accept $\off$ and instead makes a new proposal, then we  say that $\ag_2$ \textbf{rejects} the offer $\off$. Next, it is again $\ag_1$'s turn. This time, $\ag_1$ can choose between accepting the previously received proposal $\off'$, or rejecting it and proposing a new offer $\off''$ from the offer space $\Off$. 

This continues until one of the following stopping criteria is satisfied:
\begin{enumerate}
\item A proposal is accepted.
\item A given temporal deadline $\dead$ has passed.
\item A maximum number of rounds $\maxRounds$ have passed.
%\item One of the two agents withdraws from the negotiations.
\end{enumerate}
In the first case we say the negotiations have \textbf{succeeded}, while in the other two cases we say the negotiations have \textbf{failed}, meaning that the agents did not manage to come to any agreement. When we say that a `\textit{round}' has passed, we mean that an agent has proposed or accepted an offer. So, if $\maxRounds = 10$ it means that each agent can make at most 5 proposals (or 4 proposals and an acceptance).

We should remark here, that many authors assume there is only a temporal deadline, but no maximum number of rounds, or vice versa. However, if there is no temporal deadline then we can equivalently just say that $\dead=\infty$. Similarly, if there is no maximum number of rounds, then this is equivalent to saying that $\maxRounds=\infty$. So, we can always say---without loss of generality---that there is a temporal deadline as well as a maximum number of rounds, as long as we allow these values to be infinite.

In the rest of this book we will use the notation $(i, \prop, \off, t)$ to indicate that agent $\ag_i$ proposes offer $\off$ at time $t$, and we will use the notation $(i, \acc, \off, t)$ to indicate that agent $\ag_i$ accepts offer $\off$ at time $t$. We follow the convention that $t=0$ represents the time at which the negotiations start.
\begin{definition}\label{def:action}
We define a \textbf{negotiation action} to be a tuple 
\[(i, \actype, \off, t) \ \ \in  \ \  \{1,2\}\times \{\prop, \acc\} \times \Off \times \mathbb{R}^+\]
where $i$ represents the index of the agent performing the action, and $\actype$ represents the \textbf{type} of the action, which can be either the symbol $\prop$ (`propose'), or the symbol $\acc$ (`accept'). Furthermore, $\off$ is the offer that is being proposed or accepted, and $t$ is the time at which the agent proposes or accepts the offer. We define a \textbf{proposal} to be a negotiation action for which $\actype = \prop$ and we define an \textbf{acceptance} as a negotiation action for which $\actype = \acc$.
\end{definition}

Some authors also include a third type of action, besides `propose' and `accept', which is called `\textit{withdraw}'. If an agent withdraws, it means that the agent chooses to end the negotiations immediately, without agreement. So, this also adds a fourth stopping criterion to the three that we mentioned above. However, since this type of action does not play an important role in the rest of this book, we prefer not to include it here, to keep the formalization simple.

Whenever two agents are negotiating with each other, they obviously need to be connected to each other through some communication channel such as the Internet or a local network. This means that whenever one agent proposes an offer, it will take some time, due to network latency, for the other agent to receive that proposal. Since this delay is typically unpredictable, we will model it as a random variable denoted $\delay$. This motivates the following definition. %Later on in this book it will turn out to be important to include this randomness explicitly in our model.

\begin{definition}\label{def:history}
A \textbf{negotiation history} $\hist$ is a finite list that alternates between negotiation actions $\ac_j$ and positive real numbers $\delay_j \in \mathbb{R}^+$:
\[\hist  = \Big((i_1, \actype_1, \off_1, t_1) \ , \ \delay_1 \ , \  (i_2, \actype_2, \off_2, t_2) \ ,\ \delay_2 \ , \  (i_3, \actype_3, \off_3, t_3) \ , \ \delay_3 \ ,\ \dots \Big) \]
such that the negotiation actions appear in chronological order (i.e. for all $j$ we have $t_j \leq t_{j+1}$).
%such that for any $j\in \mathbb{N}$ we have $t_j + \delay_j < t_{j+1}$ and all $\eps_j$ are independent random variables with $\eps_j > 0$.
\end{definition}
In this definition, each $\delay_j$ represents the time it takes for the 
action $(i_j, \actype_j, \off_j, t_j)$ to be received by the other agent. So, a proposal made at time $t_j$ will be received by the other agent at time $t_j + \delay_j$. Each $\delay_j$ is assumed to be drawn independently from some probability distribution.

So, a negotiation history is a list containing negotiation actions, which themselves are defined as 4-tuples. Furthermore, in the case of multi-issue negotiations, the offers $\off$ inside those tuples are also tuples. For example, a negotiation history with 10 negotiation actions could look as follows:
\begin{eqnarray*}
\hist &=& \Big(\ac_1 \ ,\ \delay_1 \ ,\ \ac_2 \ ,\ \delay_2 , \quad \dots \quad, \ \ac_{9}  \ ,\ \delay_9 \ ,\  \ac_{10} \ ,\ \delay_{10} \ \Big) \\
	  &=& \Big((1, \prop, \off_1, t_1),\ \delay_1,\ (2, \prop, \off_2, t_2), \ \delay_2, \ \ \dots \ \ ,\ (1, \prop, \off_9, t_9), \ \delay_9,\  (2, \acc, \off_{9}, t_{10}), \ \delay_{10} \Big) \\
	   &=& \Big((1, \prop, (\offComp{{1}^1}, \offComp{1}^2, \offComp{1}^3), t_1),\ \delay_1,\ \ (2, \prop, (\offComp{{2}^1}, \offComp{2}^2, \offComp{2}^3), t_2),\ \delay_2,\ \ \\
	   & & \quad \dots \ \ ,\ \ (1, \prop, (\offComp{{9}^1}, \offComp{9}^2, \offComp{9}^3), t_9),\ \delay_9,\ \ (2, \acc, (\offComp{{9}^1}, \offComp{9}^2, \offComp{9}^3), t_{10}),\ \delay_{10}  \Big) \\
\end{eqnarray*}
where each $\ac_k$ is a negotiation action and each $\offComp{k}^j \in \issue{j}$ is an option from the $j$-th issue in the $k$-th proposal. In this example we assumed that the domain has three issues. Note that in the 10-th action agent $\ag_2$ accepts the offer $\off_9$ that was proposed by $\ag_1$ directly before that. 
%Furthermore, note that since the AOP dictates that the negotiations are over as soon as one of the two agents accepts a proposal, the symbol $\acc$ can only appear in the last action (if the history obeys the AOP).

We can now formally define the AOP by means of the following two definitions.
\begin{definition}\label{def:aop}
We say a negotiation history $\hist$ satisfies the AOP (with deadline $\dead$ and maximum number of rounds $\maxRounds$) if and only if all of the following conditions hold:
\begin{enumerate}
\item For any two consecutive negotiation actions $\ac_j = (i_j, \actype_j, \off_j, t_j)$ and $\ac_{j+1} = (i_{j+1}, \actype_{j+1}, \off_{j+1}, t_{j+1})$ in $h$, we have:
\begin{enumerate}
\item $i_j\neq i_{j+1}$, and
\item $t_j + \delay_j < t_{j+1}$
\end{enumerate}
% that $t_j + \delay_j < t_{j+1}$.
%\item For any two consecutive negotiation actions $\ac_j = (i_j, \actype_j, \off_j, t_j)$ and $\ac_{j+1} = (i_{j+1}, \actype_{j+1}, \off_{j+1}, t_{j+1})$ in $h$, we have that $i_j\neq i_{j+1}$.
\item The first negotiation action must be a proposal (i.e. cannot be an acceptance): $\actype_1 = \prop$.
\item A negotiation action with $\actype = \acc$ can only appear as the last action in the negotiation history.
\item If $(i, \actype, \off, t)$ and $(i', \actype', \off', t')$ are the respective second-last and last action of the negotiation history and $\actype' = \acc$, then we must have $\off = \off'$.
\item For all negotiation actions $(i, \actype, \off, t)$ in $\hist$ we have $t\leq \dead$.
\item The history $\hist$ can contain at most $\maxRounds$ negotiation actions.
\end{enumerate}
\end{definition}
The first condition states that the two agents have to alternate turns and that an agent can only propose or accept an offer after it has received the previous proposal from the other agent. The third condition states that the negotiations are over as soon as any of the two agents accepts an offer. The fourth condition states that an agent can only accept the offer from the \textit{previous} proposal and not from any earlier proposals. The fifth condition states that the negotiations are over when the deadline $\dead$ has passed, and the last condition states that the negotiations are over as soon as $\maxRounds$ negotiation actions have been made.

\begin{definition}\label{def:agreement}
Let $\hist$ be a negotiation history that satisfies the AOP and let $\ac_k = (i_k, \actype_k, \off_k, t_k)$ be the last negotiation action of this history. Then, the AOP defines that the negotiation has ended in \textbf{agreement} if $\actype_k = \acc$ and $t_k + \epsilon_k < \dead$. In that case we say that $\off_k$ is the \textbf{accepted offer}. Otherwise, we say the negotiations have \textbf{failed}.
\end{definition}
Note that this definition states that even if an agent accepts an offer before the deadline, the negotiations may still fail if the other agent does not \textit{receive} this acceptance message before the deadline. This is a detail that is often not specified in the literature, but we add it here to be as precise as possible.
%
%\later{Alternatively, we could assume there is a `notary' agent and $\delay$ represents the time it takes for the message to arrive at that agent.}

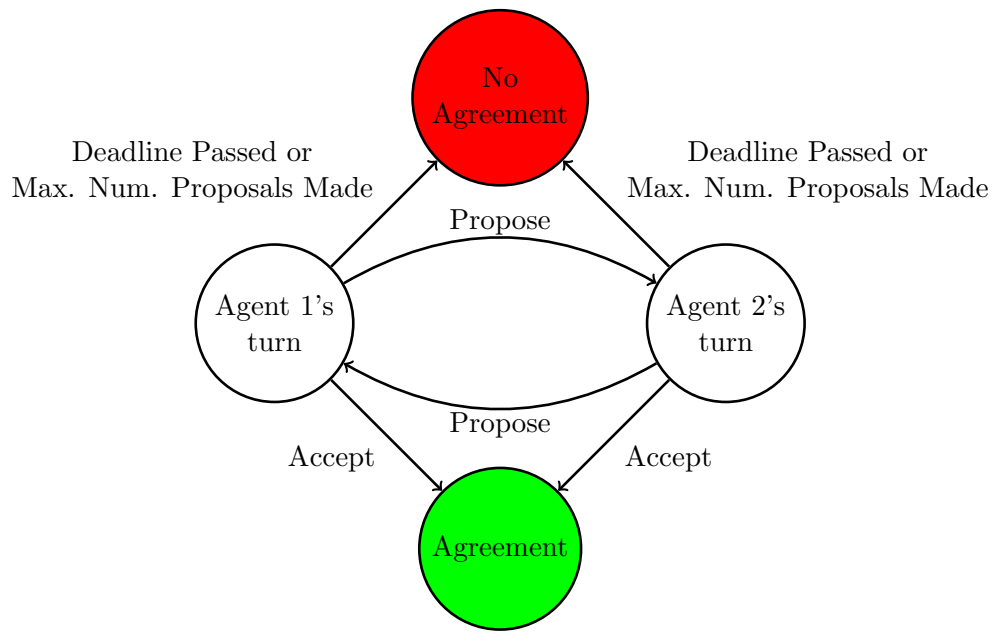
\begin{figure}
\tikzset{every picture/.style={line width=1pt}} %set default line width to 1pt        
\begin{center}
\begin{tikzpicture}[scale=1.5]

\node (NoAgr) [draw, fill=red, align=center, circle]  at (0,2) {No \\ Agreement};

\node (Ag1) [draw, align=center, circle]  at (-2,0) {Agent 1's \\ turn};

\node (Ag2) [draw, align=center, circle]  at (2,0) {Agent 2's \\ turn};

\node (Agr) [draw, fill=green, align=center, circle]  at (0,-2) {Agreement};

\draw[->] (Ag1.north east)  -- (NoAgr.south west) node[pos=0.5,anchor=south east, align=center] {Deadline Passed or  \\ Max. Num. Proposals Made};

\draw[->] (Ag2.north west)  -- (NoAgr.south east) node[pos=0.5,anchor=south west, align=center] {Deadline Passed or  \\ Max. Num. Proposals Made};

\draw[->] (Ag1.south east)  -- (Agr.north west) node[pos=0.5,anchor=north east] {Accept};

\draw[->] (Ag2.south west)  -- (Agr.north east) node[pos=0.5,anchor=north west] {Accept};

\node at (0,0.9) {Propose};
\draw[->, bend left] (Ag1) to (Ag2);

\draw[->, bend left] (Ag2) to (Ag1);
\node at (0,-0.9) {Propose};

%node[pos=0.5,anchor=north] {\tiny Propose};

\end{tikzpicture}
\end{center}
\caption{The alternating offers protocol as a finite-state machine.}\label{fig:aop}
\end{figure}

The alternating offers protocol is displayed as a finite-state machine in Figure~\ref{fig:aop}.

It is important to note that the agents individually cannot observe the delays. That is, if agent 1 proposes an offer, then he will only know the time $t$ at which he proposed the offer, but he will not know the time $t + \delay$ at which the offer was received by agent 2. On the other hand, agent 2 will only observe the time $t + \delay$ at which she received that proposal, but she will not know the exact time $t$ at which it was sent. In other words, each of the agents only has a partial view of the negotiation history, and neither of them knows the full history $\hist$. This motivates the following definition.

\begin{definition}\label{def:observed_history}
An \textbf{observed negotiation history} is a list of negotiation actions, sorted in chronological order (i.e. in order of increasing values of $t$). Specifically, if $\hist$ is a negotiation history:
\[\hist  = \Big((1, \actype_1, \off_1, t_1) \ , \ \delay_1 \ , \  (2, \actype_2, \off_2, t_2) \ ,\ \delay_2 \ , \  (3, \actype_3, \off_3, t_3) \ , \ \delay_3 \ ,\ \dots \Big) \]
then the corresponding observed negotiation history $\obsHist{1}$ for agent 1, is:
\[\obsHist{1} = \Big((1, \actype_1, \off_1, t_1) \ , \  (2, \actype_2, \off_2, t_2 + \delay_2) \ , \  (3, \actype_3, \off_3, t_3) \ , \ \dots \Big) \]
while the corresponding observed negotiation history $\obsHist{2}$ for agent 2 is:
\[\obsHist{2}  = \Big((i_1, \actype_1, \off_1, t_1 + \delay_1) \ , \  (i_2, \actype_2, \off_2, t_2)  \ , \  (i_3, \actype_3, \off_3, t_3 + \delay_3) \ ,\ \dots \Big) \]
\end{definition}
So, if $\hist$ is the true negotiation history, then agents 1 and 2 will only be aware of their respective \textit{observed} histories $\obsHist{1}$ and $\obsHist{2}$. 

In the rest of this book we will often just use the term `history' or `negotiation history' when we actually mean an \textit{observed} negotiation history, because it should be clear from the context what we mean. 
%Also, we will often just use the symbol $\hist$ to denote an observed negotiation history.

%In the rest of this book we may sometimes abuse notation and terminology and consider $\hist$ to be a \textit{set} of negotiation actions rather than a list, so we may write, for example, $\ac_k \in \hist$. 

%This does not really matter, since we have defined a negotiation action to include the time $t$ at which it was proposed or accepted, which means that for each set of negotiation actions there is a unique, chronologically ordered list corresponding to it (note that in the AOP two agents cannot make a proposal at the same time, so there cannot be two actions with the same value of $t$). 

%\essential{Include `withdraw' in the formal definition? Idea: just mention that a withdrawal does not need to be modeled explicitly by an action, because it just corresponds to an agent not making any action at all until the deadline. On the other hand, that doesn't work with discounted reservation values.}

%\begin{definition}
%Let $\hist$ be a negotiation history that obeys the AOP and let $(i_k, \actype_k, \off_k, t_k)$ be its last element. Then we say that $\hist$ ended with \textbf{agreement} if and only if $\actype_k = \acc$. In that case we say that $\off_k$ is the \textbf{accepted offer}.
%\end{definition}

\subsubsection{Notary Agents}\label{sec:notary_agents}
In Definition \ref{def:agreement} we stated that the `accept' message must be \textit{received} before the deadline. This, however, is by no means a standard convention. Most papers in the literature are ambiguous about this, because they only mention that an offer must be \textit{accepted} before the deadline, but they don't specify whether that means the accept message must be \textit{sent} before the deadline, or be \textit{received} before the deadline. In order to make our formalization precise, we therefore had to make a choice and chose the convention that it should be \textit{received} before the deadline. 

In practice, however, most researchers implicitly use an alternative convention, involving the presence of a third agent, which we will here call the `\textit{notary}'~\cite{deJonge2015nb3}. This notary agent does not take part in the negotiations themselves, but only \textit{observes} the actions by the negotiators (i.e. all propose- and accept- messages from the negotiators are also received by the notary). This notary is only there to judge whether or not the acceptance was made before the deadline. In this model, the negotiations end with agreement if and only if an accept message was \textit{received by the notary} before the deadline. While most researchers never actually mention the presence of such a notary agent, it is in fact the solution that is applied by software frameworks such as Genius and NegMas, because those frameworks \textit{themselves} essentially act as the notary.

The downside of this solution, however, is that it makes the formalization of the protocol more complicated, so for this reason why decided not to use it here.

\subsubsection{Proposals and Acceptances}\label{sec:proposals_and_acceptances}
Technically speaking, it was not really necessary to introduce two different action types (`propose' $\prop$ and `accept' $\acc$). Instead, we could have just defined a `proposal' to be a triple $(i, \off, t)$ and we could have defined an `acceptance' as a special case of a proposal, that occurs whenever an agent replies to the previous proposal by proposing exactly the same offer again. 

After all, when you think about it, `accepting' an offer is essentially the same as `proposing' an offer. Both actions are a way for an agent to signal that he agrees with that offer. The only difference between a `proposal' and an `acceptance' is that a proposal comes first and that an acceptance is a direct reply to a proposal.

Nevertheless, we here make an explicit distinction between `proposals' and `acceptances' because this is the standard approach in the literature, and because we also believe that it is more intuitive to think of them as different actions. At least, in the case of \textit{bilateral} negotiations. However, when we discuss \textit{multilateral} negotiations later on, we will see that this distinction is not so clear anymore.

\subsubsection{Some Further Remarks about the AOP}

Some authors model the action of rejecting a proposal and the action of making a counter-proposal as two separate actions. However, since in the AOP a counter-proposal is always preceded by a rejection, this distinction is not really necessary. So, in this book we follow the convention that the act of rejecting the previous proposal and the act of making a new proposal are modeled as one single action.

 %(technically, this offer $\off'$ could be the same as the previous offer $\off$, but that would not make much sense. After all, if $\ag_2$ really wants $\off$ then she might as well accept it)

At first sight, it may seem a bit unrealistic to assume that there is a single deadline $\dead$ which is imposed upon the two agents. After all, in a real-world negotiation, who would impose such a deadline onto the two agents? However, we can imagine that in a real-world scenario each agent $\ag_i$ \textit{itself} has its own individual deadline $\dead_i$, which may be determined by various external factors. In that case, we can simply define the global deadline $\dead$ as the individual deadline that comes earliest. That is:
$\dead := \min \{\dead_1, \dead_2\}$. We can imagine that before the negotiations begin each agent announces his personal deadline $T_i$ to the other agent, so that both agents will be aware of the global deadline $\dead$.

Arguably less realistic, is the assumption that the agents have a maximum number of proposals $\maxRounds$ that they can make. The main advantage of this assumption is that it makes it easier to analyze the negotiations using  mathematical or game-theoretical techniques that require a fixed and commonly-known number of rounds.  However, one major disadvantage of including a maximum number of proposals, is that it implies an asymmetry between the two negotiators, since the agent that has the last turn will not be able to make any new proposals, and thus will be forced to either accept the last proposal or to end the negotiations without agreement.

\begin{opinion}
I personally have never been a fan of negotiations with a fixed maximum number of proposals $\maxRounds$. This is because I can't really imagine any real-world situation in which the two negotiators would face such a constraint and in which the number $\maxRounds$ would be known to both negotiators in advance. The only similar scenario I can imagine, is that either of the negotiators is human and therefore would get tired after rejecting a certain number of proposals and give up. However, even in that case I don't think there would be a clearly fixed number $\maxRounds$ that is known by both negotiators in advance. Instead, I think it would be more realistic to model this with a random variable that assigns a probability $P(N)$ to every possible value of $N$, to represent the probability that the human would be too tired to continue after $N$ rounds.  
\end{opinion}

Finally, we should remark that according to the definition of the AOP that we used here, an agent is only allowed to accept the \textit{last} proposal it received from its opponent. That is, an agent is not allowed to accept any proposals that it received from the opponent in any of the \textit{earlier} rounds. So, if an agent does not immediately accept a certain offer $\off$ proposed by the opponent, then the possibility of accepting that offer may be lost forever. While this may seem overly strict, in practice this rule is not much of a restriction because if an agent $\ag$ does want to accept an offer $\off$ that was proposed by the opponent $\ag'$ in an earlier round, then instead agent $\ag$ can simply propose that offer again itself. Since the opponent $\ag'$  already proposed it earlier, there are good reasons to believe that $\ag'$  will now be willing to accept it (more about this later in Section \ref{sec:reproposing}).

\subsection{Utility Functions}

The negotiation protocol defines what the agents are \textit{allowed} to do, but does not specify anything about how an agent would choose between its various legal actions. That is, it does not specify the agents' \textit{preferences}. Such preferences are typically modeled by means of \textit{utility functions}. If we see negotiations as a game, and we see the negotiation protocol as the rules of the game, then the utility functions specify, for each agent, its \textit{goal} in the game.

Clearly, each agent has its own preferences over the set of possible agreements. For example, in the case of a negotiation between a buyer and seller over the price of a car, the seller prefers to sell the car for the highest possible price, while the buyer prefers to buy the car for the lowest possible price. To model these preferences we assume that each agent has its own personal \textbf{utility function} $\util_i$, which is a map from the set of offers to the set of real numbers:
\[\util_i : \Off \rightarrow \mathbb{R}\]
A higher utility value represents a more desired outcome. So, each agent aims to make an agreement for which his utility value is as high as possible. In the example of the car sale, the seller would have a utility function that strictly increases as a function of the price, while the buyer has a utility function that strictly decreases as a function of the price.

In the rest of this book it will turn out useful to use the notation $\offMax_i$ for the offer most preferred by agent $\ag_i$, and the notation $\offMin_i$ for the offer least preferred by agent $\ag_i$:
\begin{equation}\label{eq:offMax}
\offMax_i \quad := \quad \argmax_{\off \in \Off} \{u_i(\off) \}
\end{equation}
\begin{equation}\label{eq:offMin}
\offMin_i \quad := \quad \argmin_{\off \in \Off} \{u_i(\off) \}
\end{equation}
Furthermore, we will use the notation $\utilMax_i$ and $\utilMin_i$ to denote the corresponding utility values of the most preferred and least preferred offers:
\begin{equation}\label{eq:utilMax}
\utilMax_i \quad := \quad \util_i(\offMax_i) \quad = \quad \max_{\off \in \Off} \{u_i(\off) \}
\end{equation}
\begin{equation}\label{eq:utilMin}
\utilMin_i \quad := \quad \util_i(\offMin_i) \quad = \quad \min_{\off \in \Off} \{u_i(\off) \}
\end{equation}

\subsubsection{Von Neumann-Morgenstern Utilities}\label{sec:neumann_morgenstern}
When we only look at a single negotiation, the interpretation of the utility functions is clear: they represent the agents' respective preferences over the possible outcomes of that negotiation. However, you typically do not implement a negotiation algorithm to use it only one time and then throw it away. Ideally, it should be possible to use the same negotiation algorithm more than once. But then, how do we interpret the utility functions? After all, if we use the algorithm, say, five times, then it may make five different agreements. But how do we determine which combination of five agreements is the best?

While there are many possibilities, the most obvious and most commonly used interpretation is that the agent would prefer those outcomes that maximize the \textit{sum} of their utility values (or equivalently: the \textit{average}). That is, if the algorithm is used $n$ times, then the agent $\ag_i$ aims to maximize $\sum_{k=1}^n \util_i(\off_k)$, where $\util_i$ is the utility function of the agent and $\off_k$ the agreement reached in the $k^{th}$ negotiation. Utility functions that are interpreted in this way are called \textbf{von Neumann - Morgenstern utilities}. In the rest of this book we will always assume that utility functions are such von Neumann-Morgenstern utilities, unless specified otherwise.

One important aspect of von Neumann-Morgenstern utilities is that we can add any arbitrary constant to them or multiply them with any arbitrary positive constant, without changing the actual preferences. In other words, if $a$ and $b$ are two arbitrary real numbers (but with $a>0$) and $\util_i$ is the utility function of our agent, then it does not make any difference if we use the utility function $\util_i' = a\cdot \util_i + b$ instead of $\util_i$. Any set of agreements that is optimal under $\util_i$ will also be optimal under $\util_i'$.

\begin{definition}\label{def:invariance_lin_trans}
The principle of \textbf{Invariance under Linear Transformations} says that if an agent $\ag_i$ has a von Neumann Morgenstern utility function $\util_i$, then that agent should not behave any differently if instead it had a utility function $\util_i' = a \util_i + b$, where $a,b \in \mathbb{R}$ can be any arbitrary real numbers, as long as $a > 0$.
\end{definition}

\subsubsection{Self-interested Agents}\label{sec:self_interested}
In the rest of this book, we will assume that agents are always \textit{purely self-interested} with respect to their utility functions. This means that each agent only aims to maximize its own utility function, and does not care at all if its opponents also receive high utility values.

Of course, the point of automated negotiation is that agents need to compromise. An agent that only makes proposals that yield high utility for itself and low utility for its opponent will never be able to come to an agreement and therefore only end up with low utility. So, in negotiations, even a purely-self interested agent still needs to take the other agents' preferences into account as well. However, the point is that when an agent makes a concession to its opponent, it does that not because it \textit{wants} the opponent to receive more utility, but rather only because it \textit{needs} to concede, to secure an agreement.

Now, this may \textit{sound} like we are only trying to model very selfish and anti-social agents that do not care about each others' welfare. However, it is extremely important to understand that this is not the case. That is, \textit{`self-interested' does not mean the same as `selfish'}. 

For example, suppose that we have two agents $\ag_1$ and $\ag_2$ with respective utility functions $\util_1$ and $\util_2$. Furthermore, suppose that agent $\ag_1$ is a social agent that cares just as much about the opponent's utility as it cares about its own. So, it aims to maximize the sum $\util_1 + \util_2$ of the two utility functions (this is also known as the \textit{social welfare}). Now, note that we can simply define a new utility function $\util_1'$ for agent $\ag_1$ as follows:
\[\util_1' \quad := \quad \util_1 + \util_2\]
We now see that, even though $\ag_1$ is a very social agent, we can at the same time say that, \textit{with respect to utility function} $\util_1'$, it is purely self-interested. In other words, the question whether or not an agent is self-interested depends entirely on how we define its utility function and has nothing to do with the question whether or not it is \textit{selfish}.

\later{Discuss some common misconceptions about utility functions.}

\subsubsection{Linear Utility Functions}\label{sec:linear_util_functions}

In the case of multi-issue negotiations, one often assumes \textbf{linear utility functions}. We say a utility function is linear, if it is composed as a linear combination of several smaller functions, each one defined over one of the issues of the domain. That is:
\begin{equation}\label{eq:linear_util_func_no_weights}
\util_i(\off) = \sum_{j=1}^{\numIssues} \eval_i^j(\offComp{j}) \end{equation}
where:
\[\off \ \ = \ \ (\offComp{1}, \offComp{2}, \dots, \offComp{\numIssues})\ \ \in \ \ \issue{1} \times \issue{2} \times \dots \times \issue{\numIssues}\]
and each $\eval_i^j$ is a function that maps issue $\issue{j}$ to the real numbers: $\eval_i^j : \issue{j} \rightarrow \mathbb{R}$. We will call these functions $\eval_i^j$ the \textbf{evaluation functions}. The superscript $j$ refers to the issue $\issue{j}$ for which it is defined, while the subscript $i$ refers to the agent $\ag_i$ to which it belongs.

Alternatively, linear utility functions are often written as:
\begin{equation}\label{eq:linear_util_func}
\util_i(\off) = \sum_{j=1}^{\numIssues} w_i^j \cdot \eval_i^j(\offComp{j})
\end{equation}
\sloppy where the $w_i^j$ are the so-called \textbf{weights}, which typically sum to one: ${\sum_{j=1}^{\numIssues} w_i^j = 1}$. However, this expression is not fundamentally different from Equation~(\ref{eq:linear_util_func_no_weights}), as the weights can simply be `absorbed' inside the evaluation functions $\eval_i^j$. That is, to re-write Eq.~(\ref{eq:linear_util_func}) into the form of Eq.~(\ref{eq:linear_util_func_no_weights}), we can simply define ${\eval_i^j}' := w_i^j \cdot \eval_i^j$.

Nevertheless, the expression of Eq.~(\ref{eq:linear_util_func})  is often preferred, because it allows to emphasize that an agent might consider some issues more important than other issues, by giving them a higher weight. Furthermore, in this form it is easier to define utility functions that are normalized, because all you need to do is choose the weights and evaluation functions such that the following conditions are met:
\begin{itemize}
\item All evaluation functions $\eval_i^j$ are mapped into the interval $[0,1]$.
\item Each issue $\issue{j}$ has at least one option $\offComp{j} \in \issue{j}$ for which $\eval_i^j(\offComp{j}) = 0$.
\item Each issue $\issue{j}$ has at least one option $\offComp{j} \in \issue{j}$ for which $\eval_i^j(\offComp{j}) = 1$.
\item The weights sum to one: $\sum_{j=1}^{\numIssues} w_i^j = 1$
\end{itemize}
Just be careful not to confuse the notation $w$ for weights, with the notation $\off$ for offers.

One should realize, that when we say a utility function is linear, it only refers to the fact that it is a linear combination of evaluation functions $\eval_i^j$, while those evaluation functions themselves may still be non-linear. In fact, it often does not even make sense to ask if a certain evaluation function is linear or not, unless its options are numerical. For example, say that Alice's preferences over which movie to watch are given as follows:
\begin{eqnarray*}
\eval_{Alice}^1(\mi{The\ Godfather}) &=& 0\\
\eval_{Alice}^1(\mi{Casablanca}) &=& 1\\
\eval_{Alice}^1(\mi{The\ Big\ Lebowski}) &=& 0.7\\
\end{eqnarray*}
There is no way to tell if this function is linear or not. This is because the options of this issue (\textit{The Godfather}, \textit{Casablanca} and \textit{The Big Lebowski}) are non-numerical. For the same reason it normally does not make sense to ask if a utility function is linear if that function is defined over an offer space that only consists of a single issue.

In the rest of this book, we will sometimes abuse notation and write $\eval_i^j(\off)$ when we actually mean $\eval_i^j(\offComp{j})$, where $\offComp{j}$ is the $j$-th component of $\off$. That is:
\[\eval_i^j(\offComp{1}, \offComp{2}, \dots, \offComp{j}, \dots, \offComp{\numIssues}) \quad := \quad  \eval_i^j(\offComp{j})\]

\subsection{Reservation Values}
In many real negotiation scenarios it may happen that some proposals are so bad that you would rather not to make any agreement at all than to accept such a proposal.

For example, in the example of a car sale, if the seller asks a ridiculously high price, then the buyer would prefer not to buy the car at all than to pay that price. This could be because the buyer knows she can get a better deal elsewhere, ro because she simply doesn't have that amount of money, or because she would just prefer not to own a car at all, rather than to pay that much.

This means that a negotiating agent should not only be able to compare the various possible offers with each other, but should also be able to compare them with the situation that the negotiations end without agreement. For this, we define the \textit{reservation value}.

\begin{definition}\label{def:res_val}
An agent's \textbf{reservation value} is the amount of utility it receives when the negotiations end without  agreement.
\end{definition}
This definition implies that a rational agent would never accept any proposal that yields a utility value smaller than that agent's reservation value. After all, the agent \textit{by definition} prefers to not make any agreement at all than to accept that proposal. 
Another way to look at it, is to say that the reservation value $\rv_i$ is the minimum amount of utility that the agent $\ag_i$ is guaranteed to get. After all, $\ag_i$ can always choose to withdraw from the negotiations, or reject any proposals it receives. Therefore, a rational agent would only propose or accept any offers that yield more utility than its reservation value.

Here is another example. Suppose two friends, Alice and Bob, want to go out for dinner together and they are discussing where to go.  They have three options: a Chinese restaurant, an Italian restaurant, or a Mexican restaurant. Let us denote this as follows: 
\[\Off = \{\mi{CHI}, \mi{ITA}, \mi{MEX}\}.\] 
Unfortunately, they have different preferences, so they will have to find a compromise. If they can't agree about where they will eat, then they will each just have to stay home and eat alone. Let's suppose that Alice assigns the following utility values to the options:
\[\util_{Alice}(\mi{CHI}) = 1, \quad \util_{Alice}(\mi{ITA}) = 4, \quad  \util_{Alice}( \mi{MEX}) = 5\]
and that her reservation value is 3, which we denote as:
\[\rv_{Alice} = 3\]
The fact that she assigns the lowest utility to Chinese food means that this is her least preferred option. In fact, the utility she assigns to Chinese food is even lower than her reservation value. This means that she dislikes Chinese food so much, that she would prefer to just eat alone at home than to eat Chinese food with Bob. Furthermore, we see that she prefers Mexican food over Italian food. However, the utility she assigns to Italian food is still higher than her reservation value, which means that she still prefers to eat Italian food with Bob, than to stay at home.

One thing you may be wondering now, is what an agent should do when it receives an offer $\off$ for which the utility is exactly \textit{equal} to the reservation value, i.e. $\util_i(\off) = \rv_i$. We argue that in that case the agent should also reject the offer. After all, if he accepts the offer he will certainly receive $\rv_i$, while if he rejects it, he is also guaranteed to obtain at least $\rv_i$, but on top of that he also still has the possibility to get a better deal later and thus obtain more utility. 
\begin{observation}
A rational agent $\ag_i$ should never accept any offer $\off$ for which his utility $\util_i(\off)$ is smaller than or equal to his reservation value $\rv_i$.
\end{observation}

The situation that the negotiations end without agreement is often called the \textbf{conflict outcome}, or \textbf{disagreement}, and we will use the symbol $\dis$ to denote this outcome. Some authors define the \textbf{extended offer space} $\extOff$ as the union of the offer space together with the conflict outcome: $\extOff := \Off \cup \{\dis\}$ and then define the utility functions $\util_i$ to be functions over this extended offer space $\util_i : \extOff \rightarrow \mathbb{R}$. In that case, the utility functions implicitly also define the reservation values, because we would have, by definition, $\rv_i := \util_i(\dis)$. Furthermore, note that in some domains the disagreement outcome is already represented as an offer in the offer space, so we have $\dis \in \Off$ and therefore $\extOff = \Off$. See, for example, the coin exchange domain we discussed in Section \ref{sec:offer_space}.

We should warn the reader that the term `reservation value' is not always used in the same way by all authors. Especially in the field of economics, the term `reservation value' is often used to refer to the minimum utility value that a negotiator is willing to accept. This is what we will later call the \textit{target value} (in Section \ref{sec:bidding_strategies}), and is, in general, not the same as the utility he receives without agreement. Authors who use the term `reservation value' in that way, often use the term `BATNA' for what we call the reservation value. This stands for `Best Alternative To Negotiated Agreement'. Such inconsistencies in terminology are unfortunate, but hard to prevent. Since in the field of automated negotiation the term `reservation value' is now commonly used as a synonym for BATNA, we will stick with that convention.

\subsection{Normalized Utility Functions}\label{sec:normalized_util_funcs}
In the literature it is common to assume the agents have utility functions for which the offer with highest utility has utility value $\utilMax_i = 1$ and the offer with lowest utility has utility value $\utilMin_i = 0$. We call such functions \textbf{normalized utility functions}. 

Note that this assumption can be made without loss of generality, because the principle of Invariance under Linear Transformation (Def.~\ref{def:invariance_lin_trans}) implies that if the utility functions are not normalized, then we can simply apply a linear transformation to normalize them, without changing the essence of the negotiation domain.

Indeed, if $\util_i$ is some arbitrary utility function, then it is easy to check that the utility function $\util_i'$ defined as follows is a \textit{normalized} utility function, which is obtained from $\util_i$ by a linear transformation.
\[\forall \off \in \Off: \quad \util_i'(\off) \ \ := \ \  \frac{\util_i(\off) - \utilMin_i}{\utilMax_i - \utilMin_i}\]
%So, any von Neumann-Morgenstern utility function over a finite offer space can be normalized, without making any essential change to the negotiation domain. 

However, we should not forget that the reservation value is also a utility value, so whenever we normalize a utility function, we should apply the same linear transformation also to the corresponding reservation value:
\[\rv_i' \ \ := \ \ \frac{\rv_i - \utilMin_i}{\utilMax_i - \utilMin_i}\]
Finally, we should remark that, instead, one could 
apply an alternative kind of normalization, using the following transformations:
\[\forall \off \in \Off: \quad \util_i'(\off) \ \ :=  \ \  \frac{\util_i(\off) - \rv_i}{\utilMax_i - \rv_i}\]
\[\rv_i' \ \ := \ \ 0\]
This ensures that $\utilMax_i = 1$ and $\rv_i = 0$, and that any offer that is worse for $\ag_i$ than no agreement at all, will have a negative utility value.

\subsection{Discount Factors}\label{sec:discount_factors}
In the literature, many authors have studied models of negotiation in which the utility obtained by the agents does not only depend on the agreement they make, but also on the time at which they make that agreement. That is, the faster they make the agreement, the higher their respective utilities. This is typically modeled by introducing so-called \textbf{discount factors}. In a negotiation with discount factors, when the agents come to an agreement $\off$ each agent receives a \textbf{discounted utility} $\util_i(\off, t)$ defined as:
\[\util_i(\off, t) := \util_i(\off) \cdot \df^t\]
where $\df \in (0, 1]$ is called the discount factor, $t$ is the time at which the agents come to an agreement and the function $\util_i$ on the right-hand side is the ordinary utility function as defined previously, which in this context is also referred to as the \textbf{undiscounted utility}. Note that since $\df$ is between 0 and 1, the discounted utility decreases over time. Furthermore, note that if $\df=1$ then the discounted utility is just the same as the undiscounted utility, so this is equivalent to saying that there is no discount factor at all.

Furthermore, when studying negotiations with discount factors, it is sometimes also assumed that the reservation values are discounted as well. This means that if one of the two agents decides to withdraw from the negotiations at time $t$, then each agent $\ag_i$ receives its respective \textbf{discounted reservation value} $\rv_i \cdot \df_i^t$. In that case it may indeed be beneficial for an agent to withdraw from the negotiations early, if it seems unlikely that they will come to a good deal. This is why some authors include a `withdraw' action in the AOP, as we briefly discussed in Section~\ref{sec:aop}.

\begin{opinion}
I personally feel that the presence of discount factors is a somewhat unrealistic assumption. It seems to me that most researchers only make this assumption in order to obtain more interesting results, rather than because it yields a realistic model of negotiation. For example, Rubinstein~\cite{rubinstein1982perfect} used discount factors because it enabled him to find a mathematically optimal solution for certain negotiation scenarios. More generally, the advantage of discount factors is that they force the agents to concede quicker. After all, without discount factors an agent could simply refuse to make any concessions until very close to the deadline.

Some people might argue that discount factors could be used to model a human's impatience. However, that argument of course only holds in the case that you are modeling negotiations with humans. Furthermore, I don't think it is very obvious that a human's impatience is indeed accurately modeled by an exponentially decreasing discount factor. 

Another argument that some people might use in favor of discount factors, is that they can model the fact that certain goods such as fish or flowers are perishable, so their value quickly decreases over time. However, I don't think that that is a strong argument, since the typical time scale for the decay of such products is several days, which is still much longer than the time span of a typical negotiation involving such products, which might take place in a matter of seconds, or at most minutes.
\end{opinion}

\subsection{Knowledge}
The final ingredient that is still missing before we can fully specify a negotiation scenario, is the question how much knowledge each agent has about the other agents' utility functions, reservation values and discount factors (if present).

Authors that mainly focus on the theoretical aspects of negotiation, often assume full knowledge about the utility functions and reservation values because it is typically much harder to derive formal mathematical results under partial knowledge.

On the other hand, authors that focus more on algorithms and experiments often assume that each agent only knows its own utility function and reservation value, while it does not know anything about its opponent's utility function or reservation value, except maybe that the opponent's utility function is linear. Furthermore, they may sometimes assume that some of the issues are ordered, and that each agent knows, for each such issue, whether the opponent's preference over the options of that issue are increasing or decreasing w.r.t the ordering (e.g. Alice knows that Bob prefers to go to the cinema as late as possible).

Of course, for many commercial applications it would be unrealistic to assume the agents know each other's utility functions. After all, each agent would aim to exploit the other one as much as possible and would therefore try to hide its utility function. Nevertheless, theoretical research that does assume full knowledge is still very valuable, since it allows us to determine a theoretical `upper bound' to what an agent could hypothetically achieve in the ideal case of full knowledge (for example, \textit{the Nash bargaining solution} \cite{Nash1950} which we will discuss later on in this book). This, in turn, allows us to quantify how well practical algorithms are able to approach that upper bound~\cite{deJonge2024theoreticalPropertiesMicro}.

Furthermore, one can argue that the assumption of having no knowledge about the opponent's utility at all, is also unrealistic. For example, a car dealer knows that some cars are  more valuable than other cars and understands that the customer's preference is largely determined by his budget. 

\begin{opinion}
I would argue that for many real-world negotiation scenarios the most realistic model lies somewhere in between `full-knowledge' and `zero-knowledge'. A real negotiator would not know the \textit{exact} utility function of its opponent, but it would have at least some background knowledge about the negotiation domain, from which it could make some basic assumptions about the opponent's preferences. 
\end{opinion}
A good example of a real-world scenario where the agents' knowledge about the opponent lies somewhere in the middle of `zero-knowledge' and `full-knowledge', is given in \cite{deJonge2021multiAgentVehicleRouting} and \cite{deJonge2022multiObjectiveVRP}. In their model, two logistics companies negotiate the exchange of truck loads. Their utility functions depend on expenses like fuel price and truck driver salaries. While neither company knows exactly how much the other company pays for fuel and salaries, they do know that these prices cannot be radically different between the two companies. So, they can each make an educated guess about the opponent's utility function.

\subsection{Negotiation Domains}

\begin{definition}\label{def:nego_domain}
A \textbf{negotiation domain} $\dom$ for $\numAgents$ agents consists of the following components:
\begin{itemize}
\item An offer space $\Off$.
%\item For each agent $\ag_i \in \{\ag_1, \ag_2, \dots, \ag_\numAgents\}$:
\item For each $i \in \{1, 2, \dots, \numAgents\}$:
\begin{itemize}
\item a utility function $\util_i : \Off \rightarrow \mathbb{R}$
\item a reservation value $\rv_i \in \mathbb{R}$
\item a discount factor $\df_i \in (0, 1]$
\end{itemize}
\end{itemize}
\end{definition}
A negotiation domain with two agents (i.e. $n=2$) is called a \textbf{bilateral negotiation domain} and a negotiation domain with more than two agents (i.e. $n>2$) is called a \textbf{multilateral negotiation domain}.

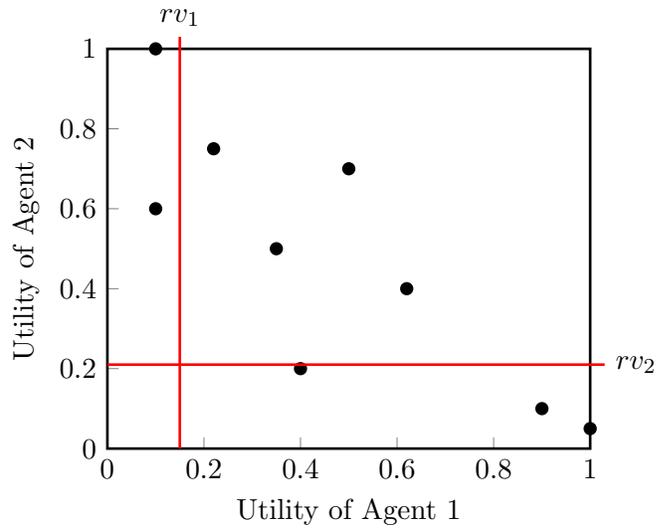
\begin{figure}
\tikzset{every picture/.style={line width=1pt}} %set default line width to 1pt        
\pgfplotsset{width=8cm}
\begin{center}
\begin{tikzpicture}[scale=1]
\begin{axis}[xmin=0, xmax=1,ymin=0,ymax=1,xlabel={Utility of Agent 1}, ylabel={Utility of Agent 2}, 
		clip=false, %ensures that you can draw outside the boundaries.
		xtick pos=left,
		ytick pos=left		
		]
	%%OFFERS:
    \filldraw [black] (0.1,1) circle [radius=2pt]  
    					(0.1,0.6) circle [radius=2pt]  
                     (0.5,0.7) circle [radius=2pt] %node[anchor=south west]{$\off_1$}
                     (0.4,0.2) circle [radius=2pt]
                     (0.35,0.5) circle [radius=2pt]
                     (0.22,0.75) circle [radius=2pt]
                     (0.62,0.4) circle [radius=2pt]
                     (0.9,0.1) circle [radius=2pt]
                     (1.0,0.05) circle [radius=2pt]
                     ;
    %% RESERVATION VALUES:
    \draw[red] (0,0.21) -- (1.03,0.21) node[anchor=west,color=black] {$\rv_2$}; %rv 1
    \draw[red] (0.15,0) -- (0.15,1.03) node[anchor=south,color=black] {$\rv_1$}; %rv 2
\end{axis}
\end{tikzpicture}
\end{center}
\caption{Utility space diagram. Every dot is the utility vector of one offer $\off$ in the offer space $\Off$. The red lines represent the reservation values of the two respective agents.}\label{fig:util_space_diagram}
\end{figure}

\begin{definition}
In a negotiation domain for $\numAgents$ agents, each offer $\off$ corresponds to an $\numAgents$-tuple which we call the \textbf{utility vector} and which consists of the utility values of all agents:
\[(\util_1(\off), \util_2(\off), \dots, \util_\numAgents(\off) )\]
We may also denote this vector as $\vec{\util}(\off)$.
\end{definition}
It is often instructive (in the case of bilateral negotiations) to plot the utility vectors of a given negotiation domain in a diagram such as in Figure~\ref{fig:util_space_diagram}. We will call this a \textbf{utility space diagram} or simply a \textbf{utility diagram}. In such diagrams, each black dot represents one offer. For example, if an offer $\off$ yields utility values $\util_1(\off) = 0.3$ and $\util_2(\off) = 0.6$ for the two agents respectively, then that offer is represented by a black dot with coordinates $(0.3\ ,\ 0.6)$. Furthermore, in such diagrams we may draw the reservation values of the agents with a horizontal line and a vertical line respectively. For example, if agent $\ag_1$ has a reservation value of $\rv_1 = 0.1$, then we draw a vertical line at $x=0.1$ and if agent $\ag_2$ has a reservation value of $\rv_1 = 0.2$, then we draw a horizontal line at $y=0.2$.

Whenever we refer to such diagrams we may use somewhat sloppy language and use the term `offer' or the symbol $\off$ when we technically mean the \textit{utility vector} of that offer.

Of course, it is important to remember that we often assume that neither of the two agents knows the utility function of the other and therefore neither of the two agents would be able to draw such a diagram. In other words, such diagrams are typically only meaningful to you, as the researcher, but not to the agents themselves.

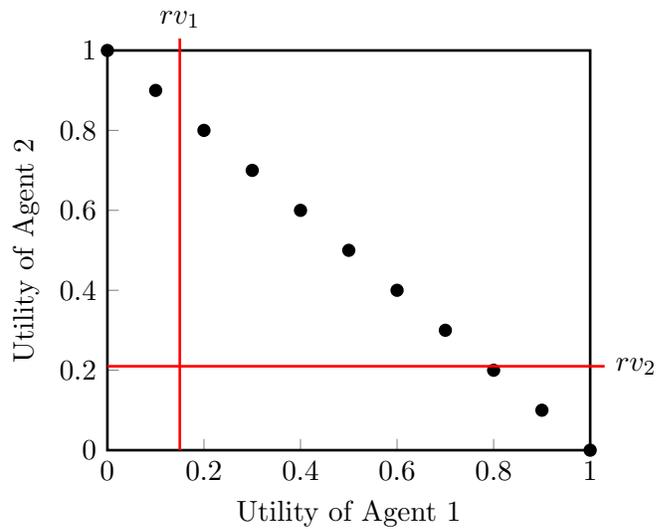
\begin{figure}
\tikzset{every picture/.style={line width=1pt}} %set default line width to 1pt        
\pgfplotsset{width=8cm}
\begin{center}
\begin{tikzpicture}[scale=1]
\begin{axis}[xmin=0, xmax=1,ymin=0,ymax=1,xlabel={Utility of Agent 1}, ylabel={Utility of Agent 2}, 
		clip=false, %ensures that you can draw outside the boundaries.
		xtick pos=left,
		ytick pos=left		
		]
	%%OFFERS:
    \filldraw [black] (0.0,1.0) circle [radius=2pt]
    					(0.1,0.9) circle [radius=2pt]  
    					(0.2,0.8) circle [radius=2pt]  
                     (0.3,0.7) circle [radius=2pt] 
                     (0.4,0.6) circle [radius=2pt]
                     (0.5,0.5) circle [radius=2pt]
                     (0.6,0.4) circle [radius=2pt]
                     (0.7,0.3) circle [radius=2pt]
                     (0.8,0.2) circle [radius=2pt]
                     (0.9,0.1) circle [radius=2pt]
                     (1.0,0.0) circle [radius=2pt]
                     ;
    %% RESERVATION VALUES:
    \draw[red] (0,0.21) -- (1.03,0.21) node[anchor=west,color=black] {$\rv_2$}; %rv 1
    \draw[red] (0.15,0) -- (0.15,1.03) node[anchor=south,color=black] {$\rv_1$}; %rv 2
\end{axis}
\end{tikzpicture}
\end{center}
\caption{Utility space diagram of a split-the-pie domain. Note that all utility vectors lie on the line $y=1-x$.}\label{fig:split_the_pie}
\end{figure}

A bilateral negotiation domain is called a \textbf{split-the-pie} domain if it satisfies $\forall \off \in \Off: \util_1(\off) + \util_2(\off) = 1$. It is called this way, because it is as if the two agents are negotiating about how to divide a pie among them. The size of the pie is 1, and each agent's utility is proportional to the size of the pie she gets. So, if $\ag_1$ gets, say, 40\% of the pie then her utility is 0.4 and therefore $\ag_2$ gets 60\% of the pie, corresponding to a utility of 0.6. Another example of split-the-pie domain is the scenario of the seller and the buyer that are negotiating the price of a car. A utility diagram of a split-the-pie domain is displayed in Figure~\ref{fig:split_the_pie}.

\subsubsection{Single-Issue Domains vs. Multi-Issue Domains}
%This section should come after explaining linear utility functions.

It is sometimes argued that multi-issue negotiations are more complex than single-issue negotiations, because they involve making trade-offs between the various different issues. However, this is somewhat misleading.

Of course, if you compare a single-issue domain $\dom_1$ that contains 10 different offers, with a multi-issue domains $\dom_2$ that contains 3 issues with 10 options each, then indeed a negotiation over the multi-issue domain will be more complex because it involves $10^3 = 1,000$ offers in total. However, this is not because there are multiple issues, but rather because the domain simply contains more offers.

In fact, if we compare domain $\dom_2$ with a single-issue domain $\dom_3$ of the same size (i.e. with 1,000 offers), then one could even say that the single-issue domain $\dom_3$ is more complex, especially if the utility functions of $\dom_2$ are linear. After all, in that case, to describe the utility functions of $\dom_2$ we only need 33 parameters (three weights, plus 10 numbers for each issue $I_j$ to represent the values $\eval_i^j(x_j)$). On the other hand, to describe the utility functions in the single-issue domain $\dom_3$ we need 1,000 parameters: one for each offer to describe its utility value $\util_i(\off)$. As we will see later on in Chapter \ref{sec:opponent_modeling}, this means that for many opponent modeling algorithms it is much easier to learn the opponent's utility function in the multi-issue domain. In fact, many existing opponent modeling algorithms would not even work on single-issue domains, because they are based on the assumption that the utility functions are linear over the issues.

\begin{observation}
If a single-issue domain and a multi-issue domain each have the same size, then one could argue that, typically, the single-issue domain would be more complex than the multi-issue domain.
\end{observation}

One exception to this rule, however, would be if we assume that all issues are ordered and that we know, for each issue, the opponent's preference ordering over that issue. In that case a single-issue domain would be easier to handle, because we would have a full preference ordering over all offers in such a domain.

\section{Pareto Optimality and Individual Rationality}\label{sec:pareto_optimal_offers}
In this section we discuss two important properties that any agreement between two agents should ideally satisfy: \textit{individual rationality}, and \textit{Pareto optimality}.

As mentioned before, a rational agent would never accept an offer that yields a utility value lower than or equal to its reservation value. This motivates the definition of individual rationality.
\begin{definition}\label{def:individually_rational}
In any negotiation domain an offer $\off$ is said to be \textbf{rational for agent} $\ag_i$ if that agent's utility for that offer is strictly greater than that agent's reservation value:
\[\util_i(\off) > \rv_i\]
Furthermore, we say an offer $\off$ is \textbf{individually rational} if it is rational for \textit{all} agents:
\[\forall i\in \{1, 2, \dots, \numAgents\} \ : \ \util_i(\off) > \rv_i\]
\end{definition}
You may find this terminology a bit confusing, since \textit{individual} rationality actually refers to \textit{all} agents, but this is an established term in the literature.

The importance of individual rationality, is that in a bilateral negotiation only the individually rational offers could ever become an agreement. After all, if an offer is not individually rational, then at least one of the two agents would never accept or propose it (unless, of course, the agent is very badly programmed). 

In a multilateral negotiation, on the other hand, this depends on the details of the protocol. If the protocol prescribes that \textit{all} agents need to agree with an offer for it to become an agreement, then again we have that only individually rational offers can become agreements. However, there are scenarios and protocols in which it is possible for \textit{subsets} of agents to make agreements. In such cases, of course, an agreement only needs to be rational for that subset of agents.

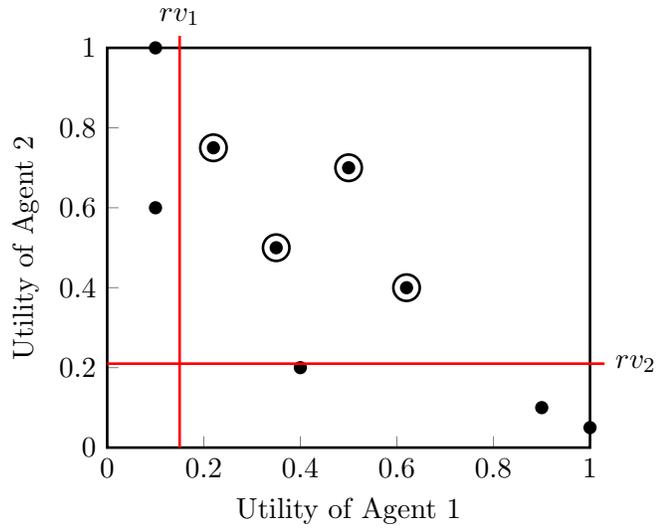
\begin{figure}
\tikzset{every picture/.style={line width=1pt}} %set default line width to 1pt        
\pgfplotsset{width=8cm}
\begin{center}
\begin{tikzpicture}[scale=1]
\begin{axis}[xmin=0, xmax=1,ymin=0,ymax=1,xlabel={Utility of Agent 1}, ylabel={Utility of Agent 2}, 
		clip=false, %ensures that you can draw outside the boundaries.
		xtick pos=left,
		ytick pos=left		
		]
	%%OFFERS:
	\draw [black] 	(0.22,0.75) circle [radius=5pt]
					(0.5,0.7) circle [radius=5pt]
					(0.35,0.5) circle [radius=5pt]
					(0.62,0.4) circle [radius=5pt] ;
    \filldraw [black] 	(0.1,1) circle [radius=2pt]  
    						(0.1,0.6) circle [radius=2pt]  
    					  	(0.22,0.75) circle [radius=2pt]
    						(0.35,0.5) circle [radius=2pt]
    						(0.4,0.2) circle [radius=2pt]
                     (0.5,0.7) circle [radius=2pt] 
                     (0.62,0.4) circle [radius=2pt]
                     (0.9,0.1) circle [radius=2pt]
                     (1.0,0.05) circle [radius=2pt];
    %% RESERVATION VALUES:
    \draw[red] (0,0.21) -- (1.03,0.21) node[anchor=west,color=black] {$\rv_2$}; %rv 1
    \draw[red] (0.15,0) -- (0.15,1.03) node[anchor=south,color=black] {$\rv_1$}; %rv 2
\end{axis}
\end{tikzpicture}
\end{center}
\caption{The individually rational offers are those for which their utility vector lies above the horizontal line representing $\rv_2$ and to the right of the vertical line representing $\rv_1$. Here these utility vectors are all drawn with a circle around them. }\label{fig:indiv_rat}
\end{figure}

The set of individually rational offers can be  visualized easily in a utility diagram, since it is the set of all offers that lie above the horizontal line representing $\rv_2$, as well as to the right of the vertical line representing $\rv_1$. See Figure~\ref{fig:indiv_rat}.

\begin{figure}
\tikzset{every picture/.style={line width=1pt}} %set default line width to 1pt        
\pgfplotsset{width=8cm}
\begin{center}
\begin{tikzpicture}[scale=1]
\begin{axis}[xmin=0, xmax=1,ymin=0,ymax=1,xlabel={Utility of Agent 1}, ylabel={Utility of Agent 2}, 
		clip=false, %ensures that you can draw outside the boundaries.
		xtick pos=left,
		ytick pos=left		
		]
	%%OFFERS:
    \filldraw [black] (0.1,1) circle [radius=2pt]  
    					(0.1,0.6) circle [radius=2pt]  
                     (0.5,0.7) circle [radius=2pt] node[anchor=south west]{$\off$}
                     (0.4,0.2) circle [radius=2pt]
                     (0.35,0.5) circle [radius=2pt] node[anchor=south east,color=black] {$\off'$}
                     (0.22,0.75) circle [radius=2pt]
                     (0.62,0.4) circle [radius=2pt]
                     (0.9,0.1) circle [radius=2pt]
                     (1.0,0.05) circle [radius=2pt]
                     ;
    %% RESERVATION VALUES:
    \draw[red] (0,0.21) -- (1.03,0.21) node[anchor=west,color=black] {$\rv_2$}; %rv 1
    \draw[red] (0.15,0) -- (0.15,1.03) node[anchor=south,color=black] {$\rv_1$}; %rv 2
    %% HELP LINES
    \draw[black,dashed] (0,0.5) -- (1,0.5);
    \draw[black,dashed] (0.35,0) -- (0.35,1);
\end{axis}
\end{tikzpicture}
\end{center}
\caption{Example of domination. The offer $\off$ lies to the top-right of $\off'$ and we therefore say that $\off$ dominates $\off'$. }\label{fig:domination}
\end{figure}
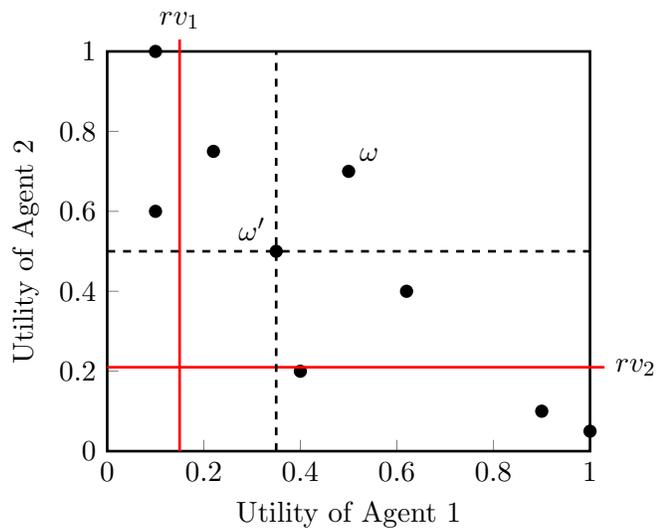

Before we can define the concept of Pareto optimality, we first have to define the concept of \textit{domination}. Suppose that we have two offers, $\off$ and $\off'$, such that each agent prefers $\off$ over $\off'$. We then say that $\off$ \textit{dominates} $\off'$, or that $\off'$ is \textit{dominated} by $\off$. We can give a precise definition as follows.
\newpage
\begin{definition}
We say that an offer $\off$ \textbf{dominates} another offer $\off'$ if:
\[\forall i\in \{1, 2, \dots, \numAgents\} \ : \ \util_i(\off) \geq \util_i(\off')\]
and there is at least one agent for which this inequality is strict:
\[\exists i\in \{1, 2, \dots, \numAgents\} \ : \ \util_i(\off) > \util_i(\off')\]
We say an offer $\off'$ \textbf{is dominated by} $\off$, if $\off$ dominates $\off'$.
\end{definition}
In a utility diagram, this can be visualized as follows: first, draw a vertical line through the point representing $\off'$, next, draw a horizontal line through $\off'$. Now, if $\off$ lies on or above the horizontal line, and also lies on or to the right of the vertical line, then $\off$ dominates $\off'$. See Figure \ref{fig:domination}.

Clearly, if the agents agree upon an offer $\off'$ that is dominated by some other offer $\off$, then this outcome would not be optimal, since at least one agent would actually prefer $\off$ as the final agreement and none of the other agents would have any objection against $\off$ instead of $\off'$. So, ideally, agents would only agree upon offers that that are not dominated by any other offer. Such offers are called \textit{Pareto-optimal}.
\begin{definition}\label{def:pareto_optimal}
An offer $\off$ is \textbf{Pareto optimal} if it is not dominated by any other offer.
\end{definition}
However, unlike individual rationality, Pareto optimality is hard to guarantee in practice, if the agents don't know each other's utility functions. So, many negotiation algorithms  still often make deals that are not Pareto optimal.

\begin{figure}
\tikzset{every picture/.style={line width=1pt}} %set default line width to 1pt        
\pgfplotsset{width=8cm}
\begin{center}
\begin{tikzpicture}[scale=1]
\begin{axis}[xmin=0, xmax=1,ymin=0,ymax=1,xlabel={Utility of Agent 1}, ylabel={Utility of Agent 2}, 
		clip=false, %ensures that you can draw outside the boundaries.
		xtick pos=left,
		ytick pos=left		
		]
	%%OFFERS:
    \filldraw [black] 	(0.1,1) circle [radius=2pt]  
    						(0.1,0.6) circle [radius=2pt]  
    					  	(0.22,0.75) circle [radius=2pt]
    						(0.35,0.5) circle [radius=2pt]
    						(0.4,0.2) circle [radius=2pt]
                     (0.5,0.7) circle [radius=2pt] node[anchor=south west]{$\off$}
                     (0.62,0.4) circle [radius=2pt]
                     (0.9,0.1) circle [radius=2pt]
                     (1.0,0.05) circle [radius=2pt];
    %% RESERVATION VALUES:
    \draw[red] (0,0.21) -- (1.03,0.21) node[anchor=west,color=black] {$\rv_2$}; %rv 1
    \draw[red] (0.15,0) -- (0.15,1.03) node[anchor=south,color=black] {$\rv_1$}; %rv 2
        %% HELP LINES
    \draw[black,dashed] (0.5,0) -- (0.5,1); %vertical
    \draw[black,dashed] (0,0.7) -- (1,0.7); %horizontal
\end{axis}
\end{tikzpicture}
\end{center}
\caption{The offer $\off$ is Pareto optimal because it is not dominated by any other offer. We can see this because the area that lies above the horizontal dashed line and to the right of the vertical dashed line is empty. }\label{fig:pareto_optimal}
\end{figure}
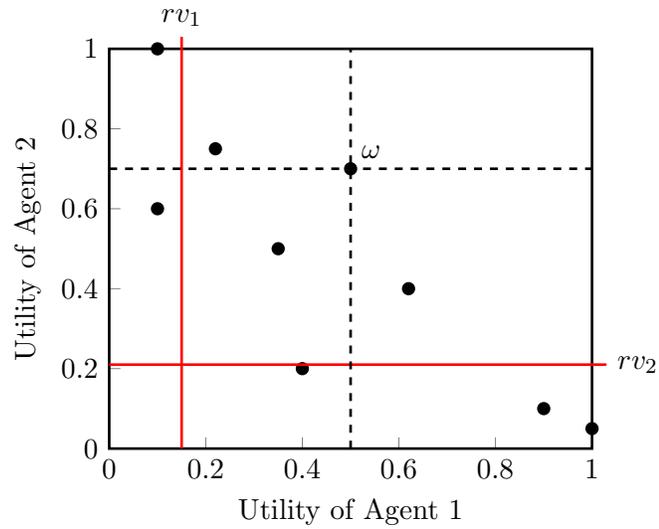

To visualize Pareto optimality, again draw a horizontal line and a vertical line through a given offer $\off$. The lines divide the space into for quadrants. If the top-right quadrant (including the lines themselves) is empty, then $\off$ is Pareto optimal. See Figure~\ref{fig:pareto_optimal}.

%This is especially the case if there are many possible offers, while the time is limited. For example, suppose agent $\ag_1$ is a time-based agent and, just like $\ag_2$ it prefers $\off$ over $\off'$. Since the agent does not have the time to propose all the possible offers during the negotiations, its aspiration function may tell it to propose offer $\off'$, while it has never proposed $\off$.

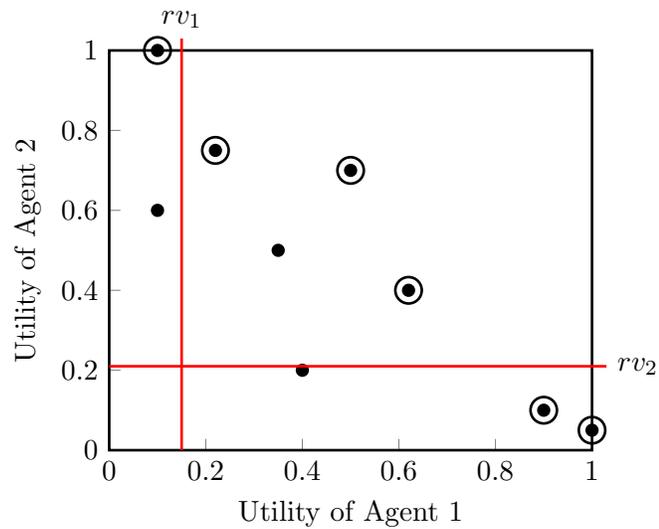
\begin{figure}
\tikzset{every picture/.style={line width=1pt}} %set default line width to 1pt        
\pgfplotsset{width=8cm}
\begin{center}
\begin{tikzpicture}[scale=1]
\begin{axis}[xmin=0, xmax=1,ymin=0,ymax=1,xlabel={Utility of Agent 1}, ylabel={Utility of Agent 2}, 
		clip=false, %ensures that you can draw outside the boundaries.
		xtick pos=left,
		ytick pos=left		
		]
	%%OFFERS:
	\draw [black] 	(0.1,1) circle [radius=5pt]
					(0.22,0.75) circle [radius=5pt]
					(0.5,0.7) circle [radius=5pt]
					(0.62,0.4) circle [radius=5pt]
					(0.9,0.1) circle [radius=5pt]
					(1.0,0.05) circle [radius=5pt] ;
    \filldraw [black] 	(0.1,1) circle [radius=2pt]  
    						(0.1,0.6) circle [radius=2pt]  
    					  	(0.22,0.75) circle [radius=2pt]
    						(0.35,0.5) circle [radius=2pt]
    						(0.4,0.2) circle [radius=2pt]
                     (0.5,0.7) circle [radius=2pt] 
                     (0.62,0.4) circle [radius=2pt]
                     (0.9,0.1) circle [radius=2pt]
                     (1.0,0.05) circle [radius=2pt];
    %% RESERVATION VALUES:
    \draw[red] (0,0.21) -- (1.03,0.21) node[anchor=west,color=black] {$\rv_2$}; %rv 1
    \draw[red] (0.15,0) -- (0.15,1.03) node[anchor=south,color=black] {$\rv_1$}; %rv 2
\end{axis}
\end{tikzpicture}
\end{center}
\caption{Pareto-frontier. All offers that are Pareto-optimal have been drawn here with a circle around them. }\label{fig:pareto_frontier}
\end{figure}

\begin{definition}\label{def:pareto_set}
For any negotiation domain $\dom$, its \textbf{Pareto set} $\OffPareto$ is the set of all Pareto-optimal offers. The \textbf{Pareto frontier} is the set of all utility vectors of the Pareto-optimal offers.
\end{definition}
Note that the Pareto set is a subset of $\Off$, while the Pareto frontier is a subset of $\mathbb{R}^\numAgents$. See Figure~\ref{fig:pareto_frontier} for the visualization of a Pareto frontier.

\section{Competitiveness}
In some negotiation domains it is easier to find good offers that are acceptable to all agents than in other domains. For example, if the domain contains a single offer $\off^*$ that yields the maximum utility to all agents (i.e. $\off^* = \offMax_1 = \offMax_2$), then it is obvious that that specific offer should be the one that the agents agree upon. After all, no agent would benefit from switching to any other agreement. In that case the interests of all agents are aligned and therefore we say the domain has zero \textit{competitiveness} or \textit{opposition} (we will use these two terms interchangeably).

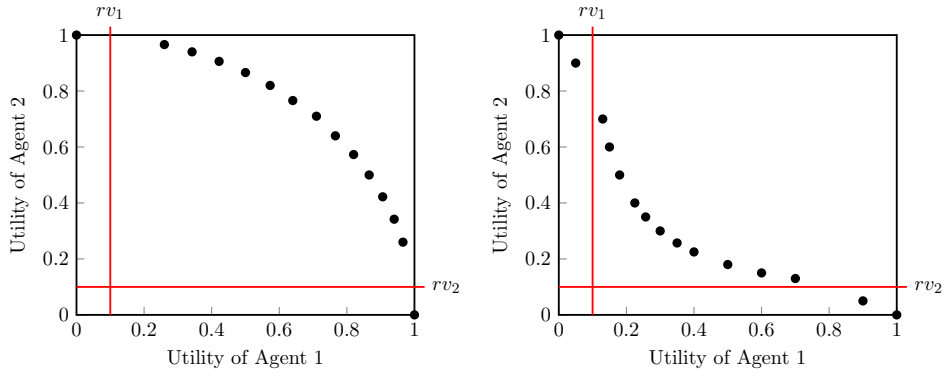
\begin{figure}
\tikzset{every picture/.style={line width=1pt}} %set default line width to 1pt        
\pgfplotsset{width=8cm}
\begin{subfigure}[h]{0.5\linewidth}
\begin{tikzpicture}[scale=0.7]
\begin{axis}[xmin=0, xmax=1,ymin=0,ymax=1,xlabel={Utility of Agent 1}, ylabel={Utility of Agent 2}, 
		clip=false, %ensures that you can draw outside the boundaries.
		xtick pos=left,
		ytick pos=left		
		]
	%%OFFERS:
    \filldraw [black] (0.0,1.0) circle [radius=2pt]
                      (0.26,0.966) circle [radius=2pt]   %15
                      (0.342,0.94) circle [radius=2pt]   %20
                      (0.422,0.906) circle [radius=2pt]  %25
                      (0.5,0.866) circle [radius=2pt]  %30
                     (0.573,0.82) circle [radius=2pt]	 %35
                     (0.64,0.766) circle [radius=2pt]  %40
                     (0.71,0.71) circle [radius=2pt]    %45
                     (0.766,0.64) circle [radius=2pt]   %40
                     (0.82,0.573) circle [radius=2pt]  %35
                      (0.866,0.5) circle [radius=2pt]  %30
                      (0.906,0.422) circle [radius=2pt]  %25
                      (0.94,0.342) circle [radius=2pt]   %20
                      (0.966,0.26) circle [radius=2pt]   %15
                     (1.0,0.0) circle [radius=2pt]
                     ;
    %% RESERVATION VALUES:
    \draw[red] (0,0.1) -- (1.03,0.1) node[anchor=west,color=black] {$\rv_2$}; %rv 1
    \draw[red] (0.1,0) -- (0.1,1.03) node[anchor=south,color=black] {$\rv_1$}; %rv 2
\end{axis}
\end{tikzpicture}
\end{subfigure}
\hfill
\begin{subfigure}[h]{0.5\linewidth}
\begin{tikzpicture}[scale=0.7]
\begin{axis}[xmin=0, xmax=1,ymin=0,ymax=1,xlabel={Utility of Agent 1}, ylabel={Utility of Agent 2}, 
		clip=false, %ensures that you can draw outside the boundaries.
		xtick pos=left,
		ytick pos=left		
		]
	%%OFFERS:
    \filldraw [black] (0,1.0) circle [radius=2pt]
    					(0.05,0.9) circle [radius=2pt]
    					(0.13,0.7) circle [radius=2pt]
    					(0.15,0.6) circle [radius=2pt]  
    					(0.18,0.5) circle [radius=2pt]  
                     (0.225,0.4) circle [radius=2pt] 
                     (0.257,0.35) circle [radius=2pt]
                     (0.3,0.3) circle [radius=2pt]
                     (0.35,0.257) circle [radius=2pt]
                     (0.4,0.225) circle [radius=2pt]
                     (0.5,0.18) circle [radius=2pt]
                     (0.6,0.15) circle [radius=2pt]
                     (0.7,0.13) circle [radius=2pt]
                     (0.9,0.05) circle [radius=2pt]
                     (1.0,0) circle [radius=2pt]
                     ;
    %% RESERVATION VALUES:
    \draw[red] (0,0.1) -- (1.03,0.1) node[anchor=west,color=black] {$\rv_2$}; %rv 1
    \draw[red] (0.1,0) -- (0.1,1.03) node[anchor=south,color=black] {$\rv_1$}; %rv 2
\end{axis}
\end{tikzpicture}
\end{subfigure}%
\caption{Left: a domain with low opposition. Right: a domain with high opposition.}\label{fig:opposition}
\end{figure}

On the other hand, in a split-the-pie domain there is high opposition, because the interests of the two agents are diametrically opposed. The better an offer is for one agent, the worse it is for the other. In fact, we can construct even more competitive domains where there is no good intermediate solution and every offer is really bad for at least one agent of the agents. See Figure~\ref{fig:opposition}.

In other words, the `competitiveness' or `opposition' of a domain measures how easy it is for all agents to receive high utility. Now, it would be nice to have a formula that allows us to quantify, for any given domain its competitiveness. It turns out, however, that many different such formulas have been proposed in the literature, so we will discuss a couple of them. For simplicity, we will assume the utility functions are normalized. Each of the expressions we discuss here is based on the idea that we first pick some `ideal' offer, and then measure the difference between the utility vector of that ideal offer and the `\textbf{utopian}' utility vector $(1,1,\dots ,1)$ that assigns the maximum utility to each agent. The higher this value, the higher the opposition of the domain.

% distance measure that measures the distance between the utility vector $(\util_1(\off), \util_2(\off), \dots, \util_n(\off))$ of any offer $\off$ and the `utopian' utility vector $(1,1,\dots ,1)$. The opposition is then defined as the minimum such distance over all offers $\off$ in $\Off$.

Perhaps the most commonly used definition of opposition is one based on the Euclidean distance~\cite{WilliamsANAC2012}. That is:
\begin{equation}\label{eq:opposition_euclid}
\opp(\dom) \quad := \quad \min_{\off\in \Off} \sqrt{\sum_{i=1}^\numAgents (1-\util_i(\off))^2}
\end{equation}
%The higher this value, the higher the opposition, and $\opp(\dom) = 0$ corresponds to the lowest possible opposition (this happens when there is an offer that yields a utility of 1 to all agents).

While this definition may initially seem intu\"itive, one could argue that it is not entirely satisfactory. For example, suppose that the minimum  Euclidean distance is attained for some offer with utility vector $(0.6,0.6)$. Now, it is easy to see that if we change the domain a bit, by replacing this offer with a new offer with utility vector $(0.6,0.65)$, then according to this Euclidean measure, the domain would become less competitive, even though we have only increased the utility of \textit{one} of the two agents. Moreover, we can even slightly decrease the utility of the other agent, to get $(0.59,0.65)$ and the Euclidean opposition measure would still indicate that this domain is less competitive than the original one. One could argue that this result is somewhat contrary to what you might expect from an accurate measure of opposition.

One alternative definition is the following~\cite{An2012}:
\begin{equation}\label{eq:opposition_min}
\opp(\dom) \quad := \quad  \min_{\off\in \Off}\ \ 1-\min_{i\in \{1,2,\dots,n\}}\util_i(\off)
\end{equation}
Here, the distance to the `utopian' outcome is defined as the difference between 1 and the utility obtained by the agent that receives lowest utility. The advantage of this measure is that to decrease the competitiveness of a domain, we need to increase the utility of \textit{all} agents.

Yet another definition~\cite{renting} also uses
the Euclidean distance, but defines the `ideal offer' as the one that  minimizes $| \util_1(\off) - \util_2(\off)|$ among all Pareto optimal offers. That is:
\begin{equation}\label{eq:opposition_euclid_kalai}
\opp(\dom) \quad := \quad  \sqrt{\sum_{i=1}^\numAgents (1-\util_i(\off^*))^2}
\end{equation}
where:
\begin{equation}
\off^* \quad := \quad  \min_{\off\in \OffPareto}\ \ | \util_1(\off) - \util_2(\off)|
\end{equation}

In the end, there is no obvious way to determine which of these measures is the `best'. I would say that this question mainly depends on the   purpose that you have in mind for which you want to measure opposition.

\section{The NegoSim Framework}\label{sec:simulation}

In order to implement negotiation algorithms and perform experiments on them, we need a software framework that allows us to run a simulation of a negotiation between agents. A commonly used framework for this is the NegMas platform~\cite{Mohammad2020NegMas}.

However, for this book we have implemented a very simple, toy-world negotiation simulator in Python. We call it NegoSim and it can be downloaded from the web page of this book: \\
%\url{https://gitlab.iiia.csic.es/davedejonge/negosimulator}
\url{https://www.iiia.csic.es/~davedejonge/intro_to_nego}

It does not rely on any libraries so you don't need to install anything, except of course Python itself, and any development environment that is suitable for Python. We will use this framework for various exercises throughout this book.

While the NegMas framework is much more extensive and developed for full-blown research purposes, NegoSim is specifically designed for absolute beginners. The fact that it only contains the bare minimum to get started makes it much easier to explore and understand. In fact, the core of NegoSim only consists of five files, and it should not take too much time for anyone to read all of the code and fully understand it. Specifically it consists of the following five files:
\begin{itemize}
\item \textbf{nego\_simulator.py} Contains the code to run a simulation of a single negotiation. When you run this code, it creates a negotiation domain and two agents which will then negotiate with each other over that domain. When this negotiation is over it prints the outcome of the negotiation (i.e. whether the agents made an agreement or not, plus the utility values obtained by the two respective agents). 
\item \textbf{domains.py} Contains the code that defines negotiation domains, and their components (issues, offers, utility functions and evaluation functions). Furthermore, it contains a function that constructs an example domain, and it contains functions to load and save negotiation domains from and to hard disk.
\item \textbf{agents.py} Contains the code of the `Agent' class, which should be the base class of any negotiating agent that you may implement in this framework. It also contains the implementation of an agent called `RandomAgent' that just makes random proposals and randomly accepts any offers with a probability of 1\%. 
\item \textbf{opponent\_utility\_models.py} Contains the base class for any opponent modeling algorithm that aims to learn the opponent's utility function (see Section \ref{sec:opp_util_modeling}). Furthermore, it contains the implementation of a `dummy' opponent model. This is a fake opponent modeling algorithm that you can use if you haven't implemented any real opponent modeling algorithms yet (it is `fake' in the sense that it uses the true utility function of the opponent, while the whole point of a `real' opponent modeling algorithm is that you don't have access to that utility function).
\item \textbf{opponent\_strategy\_models.py} Contains the base class for any opponent modeling algorithm that aims to learn the opponent's strategy (see Section \ref{sec:opp_strat_modeling}), plus the implementation of one very simple such algorithm based on linear extrapolation.
\end{itemize}

Of course, in order to run experiments you will need a lot more code. For example, you will need more negotiating agents and more domains, you will need code to run large tournaments, and you will need code to analyze the results of those tournaments. The idea, however, is that as you continue reading this book, you will implement that code yourself, by doing the exercises. We feel that in this way you get a much better understanding of how a negotiation framework works, in a step-by-step manner, rather than if we gave you the complete framework all at once.

However, if you don't want to do those exercises, you can skip most of them and find their solutions in the folder `Solutions to Exercises'. So, if you want to run experiments without implementing all the necessary code yourself, you can just copy-paste the required code from the solutions folder to the main folder.

\begin{exercise}\textbf{Explore the NegoSim framework.}
Download the the NegoSim framework, take a look at the five files mentioned above, and familiarize yourself with the code. In particular, make sure that you understand the source code of nego\_simulator.py and agents.py.

Next, run the file nego\_simulator.py. This will run a simulation of two RandomAgents that are negotiating over the example domain. 
\end{exercise}

The NegoSim framework allows you to define negotiation domains in two different ways: 
\begin{enumerate}
\item Directly in the code itself.
\item In JSON format.
\end{enumerate}
Currently, the framework comes with two example domains. One for each of these two methods. The example of the first method is given in the function get\_example\_domain() in the file domains.py. This method can be useful if you want to generate negotiation domains automatically.

For an example of the second method you can open the folder called `Domains'. The idea is that any negotiation domains that you create yourself can be stored in this folder. Each such domain should be stored in its own subfolder, which would then contain exactly one JSON file for each agent (so if it's a bilateral domain then the folder should contain exactly two JSON files), each defining the utility function of one of the agents. These files should all have the same structure, in the sense that they should each define the same domain name, the same number of issues, with the same issue names, and for each issue the two files should define the same options. The only things that can be different between the files are the numerical values (i.e. the issue weights, the evaluations, and the reservation values).

To obtain a NegotiationDomain object representing the domain defined in the folder named ``Cinema Date", you can call:
\begin{small}
\begin{lstlisting}[language=Python]
my_domain = load_domain_from_folder("Domains/Cinema Date")
\end{lstlisting}
\end{small}
This function is defined in the file domains.py.

Vice versa, any NegotiationDomain object can be converted to json format and stored to hard disk by calling:
\begin{small}
\begin{lstlisting}[language=Python]
save_domain_to_folder(my_domain, "Domains/My Domain", False)
\end{lstlisting}
\end{small}
This creates a new folder called `My Domain' inside the folder `Domains', and stores the domain there, in the form of two JSON files.

\begin{exercise}\label{ex:nego_domain}
\textbf{Create your own negotiation domain.} To do this, create a new subfolder inside the `Domains' folder, with the name of your domain. Then create two json files inside this subfolder with the same structure as the json files from the Cinema Date domain.
\newline

You can give your domain any number of issues you like and you can give each issue any number of options you like. Also, you can give any names you like to the issues and their options and to the domain itself. Just make sure that the two json files are identical, except for their numerical values.
\newline

Next, adapt the file nego\_simulator.py so that the two agents will negotiate over your new domain, and then run it.
\end{exercise}

\begin{exercise}\label{ex:random_domain_generator}
\textbf{Create a random domain generator.} Implement code that automatically generates an object of type `NegotiationDomain' (as defined in the file domains.py), with a randomly chosen number of issues, and with a random size for each issue, and with randomly chosen utility functions and reservation values.
\newline

Next, adapt the file nego\_simulator.py so that each time you run it,   it will randomly generate a new negotiation domain and then the two agents will negotiate over that domain.
\newline

To make it more challenging, you can add some constraints on the competitiveness of the domain. For example, you can add the constraint that for each offer $\off \in \Off$ the sum of the two agents' utilities $\util_1(\off) + \util_2(\off)$ must stay below a specified value.
\end{exercise}

\begin{exercise}\label{ex:visualizer}
\textbf{Implement a domain visualizer.} Implement a tool that, given an object of type `NegotiationDomain' (as defined in the file domains.py), generates a corresponding utility-space diagram, like in Figure~\ref{fig:util_space_diagram}. Use this tool to visualize some of the domains generated by the tool from Exercise~\ref{ex:random_domain_generator}.
\newline

Hint: a useful tool for generating such diagrams in Python is pyplot, which is part of the matplotlib library.
\end{exercise}

\section{Summary of Chapter}
\begin{itemize}
\item Negotiations between exactly two agents are called \textbf{bilateral} negotiations, while negotiations between more than two agents are called \textbf{multilateral} negotiations.
\item The most commonly used protocol for bilateral negotiation is the \textbf{alternating offers protocol}.
\item The set of all possible offers that the agents may propose to one another is called the \textbf{offer space} and denoted by $\Off$.
\item In some cases the offer space can be written as a Cartesian product of so-called \textbf{issues}:
\[\Off =  \issue{1}\times \issue{2} \times\dots \times \issue{\numIssues}\]
Each such issue may or may not be ordered.
\item The agents' preferences are typically defined by means of a \textbf{utility function} over the offer space. Whenever the agents agree on some offer $\off$, then each agent $\ag_i$ receives the utility value $\util_i(\off)$ for that offer according to his utility function $\util_i$.

\item Many authors study domains with \textbf{linear utility functions}. That is, utility functions that can be decomposed as a weighted sum of \textbf{evaluation functions}, with one evaluation function for each issue:
\[\util_i(\off) = \sum_{j=1}^\numIssues w_i^j \cdot \eval_i^j(\offComp{j})\]
where $\off = (\offComp{1}, \offComp{2}, \dots, \offComp{\numIssues})$
\item A negotiating agent needs to strike a balance between demanding high utility for itself, and offering high utility to its opponent, even if it is ultimately only interested in maximizing its own utility.
	
\item Each agent $\ag_i$ has a \textbf{reservation value} $\rv_i$, which is the utility it receives when the negotiations end without agreement.
\item A rational agent would never propose or accept any offer for which his utility is smaller than or equal to his reservation value.
\item Some authors assume that the value of an agreement decreases over time. The later the agents come to an agreement, the lower the utility they receive. This is typically modeled by means of \textbf{discount factors}. 
	
\item Authors that study automated negotiation from a theoretical point of view often assume that all agents have full knowledge of the negotiation domain. On the other hand, authors that focus more on algorithms and experiments typically assume that each agent $\ag_i$ only knows its own utility function $\util_i$ and reservation value $\rv_i$, but not the utility functions and reservation values of its opponents.
	
\item An offer $\off$ is called \textbf{individually rational} if for each agent $\ag_i$, its utility value $\util_i(\off)$ is strictly greater than its reservation value $\rv_i$.
\item We say an offer $\off$ \textbf{dominates} another offer $\off'$, if for each agent $\ag_i$ the utility it assigns to $\off$ is greater than or equal to the utility it assigns to $\off'$, and for at least one agent this inequality is strict.
\item We say an offer is \textbf{Pareto-optimal} if it is not dominated by any other offer in the offer space.
\item If all agents are perfectly rational, then they would never come to any agreement that is not individually rational.
\item Rational agents would always prefer Pareto-optimal agreements, but this is hard to guarantee in practice if the agents do not know each other's utility functions.
\item The \textbf{competitiveness} or \textbf{opposition} of a negotiation domain measures how close two agents could theoretically get to the \textbf{utopian} outcome in which each agent receives the maximum utility. There is no generally accepted equation to calculate competitiveness. Instead, different authors use different equations for this.
\end{itemize}

\chapter{Negotiation Strategies}\label{sec:negotiation_strategies}
We are now finally ready to discuss how we can actually implement a negotiation algorithm. This is probably the most important chapter of this book. We will describe several possible strategies and we will see that each of them has its own advantages and disadvantages.

The goal of this chapter is to discuss how we can develop our own agent, that will be able to negotiate with arbitrary unknown opponents. We will here always follow the convention that our agent is denoted as $\ag_1$, while its opponent is denoted as $\ag_2$.

It is important to understand that the only goal of our agent is always to maximize its own utility, so it does not care about other concepts such as fairness or social welfare, as explained in Section~\ref{sec:self_interested}, and we assume the same for the opponent.

There are many kinds of negotiation scenarios that we could consider, but in this chapter we will always make the following assumptions:
\begin{itemize}
\item Negotiations are bilateral (so our agent is negotiating with only one opponent).
\item Negotiations take place according to the alternating offers protocol (See Section \ref{sec:aop}).
\item Each of the two agents involved in the negotiation knows its \textit{own} utility function and its own reservation value, but neither of them knows the utility function or reservation value of the other.
\item The offer space $\Off$ is finite.
\item The agents have a finite deadline $\dead$ for the negotiations.
\item There is no maximum number of negotiation rounds (or equivalently, $\maxRounds = \infty$).
\item There are no discount factors (or equivalently, the discount factors are equal to 1).
\end{itemize}
On the other hand, we will not make any assumptions about whether the negotiation domain is a single-issue or multi-issue domain, nor about the type of utility functions the agents have (linear or non-linear).

We make these assumptions because they yield the simplest types of negotiation scenarios that are still interesting enough to allow us to discuss the most commonly used negotiation strategies. More advanced negotiation scenarios will be discussed later on in this book.

%\essential{Mention something, about the fact that agents would prefer to keep their utility functions private.}

\section{The BOA Model}

When implementing a negotiation algorithm, it is often useful to think of it as consisting of three separate components:
\begin{itemize}
\item A \textbf{Bidding strategy}: a strategy to determine when our agent will propose which offer to the opponent.
\item An \textbf{Opponent modeling algorithm}: an algorithm that allows our agent to approximately learn the opponent's utility function and/or its bidding strategy.
\item An \textbf{Acceptance strategy}: A strategy to determine which proposals received from the opponent should be accepted by our agent and which ones should be rejected. 
\end{itemize}
This model is known as the BOA model \cite{Baarslag2014BOA}. A typical BOA agent would be implemented as follows:
\begin{enumerate}
\item Receive an offer $\offRec$ proposed by the opponent.
\item Use the opponent modeling algorithm to update a model of the opponent's strategy and utility function, based on the received proposal.
\item Use the bidding strategy, in combination with the model of the opponent, to determine which counter offer $\offNext$ to propose next.
\item Use the acceptance strategy to determine whether or not to accept the received offer $\offRec$. If yes, then accept $\offRec$, if not, then propose $\offNext$. 
\end{enumerate}
An implementation in pseudo-code is displayed in Algorithm~\ref{alg:boa}. In the following sections we will present more specific strategies, but they all follow the same structure. One thing that may seem counter-intuitive, is that this algorithm first decides which offer to propose next, before it decides whether or not to accept the received offer. This is, because the decision whether or not the accept the received proposal often depends on which proposal you are going to make next (as we will see later in Section~\ref{sec:acceptance_strategies}).

\begin{algorithm}
\caption{BOA Agent for the Alternating Offers protocol. Generic implementation of a method that is called every time it is our agent's turn and determines whether the agent should accept the last proposal received from the opponent or reject it and, in case of rejection, which counter-offer to propose next.}\label{alg:boa}
\begin{algorithmic}[1]
	\Statex \textbf{Input:} 
	\Statex $\Off$ \Comment{The offer space.}
	\Statex $\util_1$ \Comment{The agent's own utility function.}
	\Statex $\rv_1$ \Comment{The agent's own reservation value.}	
	\Statex $\dead$ \Comment{The deadline.}
	\Statex $\om$ \Comment{A model of the opponent.}
	\Statex $t$ \Comment{The current time.}
    \Statex $\obsHist{1}$ \Comment{The observed negotiation history: a list containing all }
    \Statex \Commentt{proposals that have so far been proposed by both agents,}
    \Statex \Commentt{sorted in chronological order.}
    \Statex $\offRec$ \Comment{The offer last proposed by the opponent (if any).}
    \Statex \Commentt{Note that it is also contained in the history $\obsHist{1}$, }
     \Statex \Commentt{but for clarity we also display it here separately.}
    %\commentSlash From the history, extract the last offer proposed by the opponent.
    %\State $\offRec \leftarrow
    % \mi{getLastReceivedOffer}(\hist)$
     \Statex
    \comment{OPPONENT MODELING}
    \comment{First, update the opponent model:}
    		\State $\om \leftarrow updateOpponentModel(\Off, \dead, \om, t, \offRec)$
    		\Statex 
    		\comment{BIDDING STRATEGY}
    		\comment{Next, apply a bidding strategy to select the next offer to propose:}
    		\State $\offNext \leftarrow biddingStrategy(\Off, \util_1,\rv_1, \dead, \om, t, \obsHist{1})$ 
    		\Statex
    		\comment{ACCEPTANCE STRATEGY}
    		\comment{Then, determine whether or not to accept the opponent's last} 
    		\commentt{proposal. We store this decision in a boolean variable $\mi{acceptOffer}$:}
    		\State $\mi{acceptOffer} \leftarrow acceptanceStrategy(\Off, \util_1, \dead, \om, t, \offRec, \offNext)$
    		\Statex
    		\comment{RETURN SELECTED ACTION}
    		\comment{Finally, return the selected action (accept or propose):}
        \If{$\mi{acceptOffer}$}
        		%\State ACCEPT($\offRec$)
            \State \textsc{Return} ($\acc$, $\offRec$)
        \Else
        		%\State PROPOSE($\offNext$)
            \State \textsc{Return} ($\prop$, $\offNext$)
        \EndIf
\end{algorithmic}
\end{algorithm}

In the following section we will discuss various bidding strategies and present some example implementations in pseudo-code. These examples will also include various acceptance strategies, but we will not discuss them yet because we defer that discussion until  Section~\ref{sec:acceptance_strategies}. Furthermore, opponent modeling algorithms will be discussed in Chapter~\ref{sec:opponent_modeling}.

\section{Bidding Strategies}\label{sec:bidding_strategies}
In this section we will discuss the various negotiation strategies that have been studied in the literature. These strategies can be classified into the following three categories:
\begin{enumerate}
\item Time-based strategies.
\item Adaptive strategies.
\item Imitative strategies.
\end{enumerate}
We certainly do not claim that these are the only possible strategies, but they are the most commonly studied ones. In fact, in their seminal paper~\cite{Faratin1998} Faratin et al. also proposed a fourth type of strategy, known as a \textit{resource-based} strategy, but this type seems to have been given considerably less attention in the literature, so we will not discuss it in this book.

The basic idea behind all three types of strategy above is the same: our agent starts by proposing an offer that gives the highest possible utility to itself but, as time passes, our agent will propose offers that yield less and less utility to itself, which will hopefully make it more likely for the opponent to accept one of those offers. Every time an agent makes a new proposal that yields less utility to itself than any of its previous proposals, we say the agent is making a \textbf{concession}, or that the agent is \textbf{conceding}.

The big question is how to determine \textit{how much} to concede in every turn. On the one hand, our agent obviously should not concede too much, because its aim is to make a deal that gives itself the highest possible utility. An agent that concedes too much will only make deals that yield very little utility. But on the other hand, if our agent doesn't concede enough, there is the risk that it may not come to any agreement at all, which would often result in even less utility. Therefore, the key to a strong negotiation strategy is to make exactly the right trade-off between conceding enough, and not conceding too much. In the rest of this book we will refer to a strategy that concedes very little as a \textbf{\hard{} strategy}, while we will refer to a strategy that concedes very much as a \textbf{\soft{} strategy}. 

\subsection{Time-Based Strategies}
Time-based strategies are the simplest kind of negotiation strategy. A time-based strategy makes use of a function $\asp : \mathbb{R} \rightarrow \mathbb{R}$, known as the \textbf{aspiration function}, which would typically be strictly decreasing. This aspiration function controls the amount of concession the agent makes as a function of time. Specifically, the idea is that at any given time $t$ our agent $\ag_1$ will propose an offer $\off$ that concedes as much as possible, under the constraint that his utility value $\util_1(\off)$ must remain greater than, or equal to $\asp(t)$. 

Time-based agents can be either \hard{} or \soft{}, depending on the shape of the aspiration function. The faster $\asp$ decreases, the more \soft{} the agent will be. We will discuss this in more detail below.

\subsubsection{Choosing the Next Offer to Propose}

Given an aspiration function $\asp$, we need to implement a precise rule how to choose the next offer to propose $\offNext$ based on this function. One example would be to do it according to the following expression:
\begin{equation}\label{eq:time_based_max}
\offNext \quad = \quad \argmax_{\off\in \Off }\ \{\ \est{\util}_2(\off) \mid  \util_1(\off) \geq \asp(t) \ \land \ \off \not\in \Pro_{t}\}
\end{equation}
where $\est{\util}_2$ is an estimation that our agent $\ag_1$ makes of the opponent's utility function $\util_2$, by means of its opponent modeling algorithm.  The details about how such opponent modeling techniques work will be discussed in Chapter~\ref{sec:opponent_modeling}. For now, we will just see it as a `black box' that magically gives us an approximation of the opponent's utility function. Furthermore, $\Pro_t$ is the set off all offers that have already been proposed by $\ag_1$ before time $t$.
\begin{equation}\label{eq:Off_prop}
\Pro_{t} := \{ \off \in \Off \mid \exists t' \in [0,t] : (1,\prop, \off, t') \in \obsHist{1} \}
\end{equation}
In Equation~(\ref{eq:time_based_max}) we can clearly see how $\asp(t)$ controls the trade-off between demanding a high utility for yourself and conceding more utility to the opponent. On the one hand our agent is maximizing the opponent's estimated utility $\est{\util}_2$, but on the other hand this is restricted by the constraint that our agent should not propose any offer that yield less utility than $\asp(t)$.

The constraint $\off \not\in \Pro_{t}$ ensures that, if the best candidate has already been proposed, then instead of repeating that proposal, our agent will propose the second best candidate. After all, the opponent modeling algorithm may not be accurate, so even if $\est{\util}_2(\off)$ is greater than $\est{\util}_2(\off')$ it may happen that the opponent actually prefers $\off'$, so, if it has the chance, our agent should also try to propose $\off'$.

Of course, it may happen that there is no offer at all that satisfies the criteria, because all offers for which $\util_1(\off) \geq \asp(t)$ holds have already been proposed. In that case our agent can simply repeat the same proposal as in the last turn, or propose an arbitrary one that it has already proposed before.

%The rationale behind this is that, even though our agent ultimately only cares about its own utility, it also aims to propose offers with high utility to the opponent to maximize the probability that the opponent will be willing to accept the proposal.

The main disadvantage of Eq.~(\ref{eq:time_based_max}), however, is that it depends on having an accurate opponent modeling algorithm. Therefore, alternatively, one can instead use the following expression.
\begin{equation}\label{eq:time_based_min}
\offNext \quad = \quad \argmin_{\off\in \Off } \ \{\ \util_1(\off) \mid  \util_1(\off) \geq \asp(t) \ \land \ \off \not\in \Pro_{t}\}
\end{equation}
That is, it picks the offer with the \textit{lowest} utility value that is still greater than or equal to $\asp(t)$. There are two scenarios in which this alternative approach would make sense:
\begin{enumerate}
\item In domains where the utility functions of the two agents are strongly negatively correlated (that is, domains in which any offer that yields high utility to our agent, yields low utility to the opponent, and vice versa).
\item In domains with a very small offer space.
\end{enumerate}
An example of the first scenario is the case where a buyer and a seller negotiate the price of a car, or any other split-the-pie domain. In such cases, finding the offer that yields the highest utility to the opponent is (approximately) equivalent to finding the offer that yields the lowest utility to our agent. So, Eq.~(\ref{eq:time_based_min}) would yield approximately the same proposals as Eq.~(\ref{eq:time_based_max}), but without using any opponent modeling algorithm. Of course, the problem is that we have to \textit{know} that the utility functions are strongly correlated, so we need to have at least some prior knowledge about the opponent's utility function. 

In the second scenario Eq.~(\ref{eq:time_based_min}) may work, because there is enough time for our agent to propose \textit{all} the offers, one by one. For example, if it takes about 100 milliseconds for an agent to make a proposal, and the deadline is set to 1 minute, then there is  time to propose 6,000 different offers. So, if the offer space contains less than 6,000 different offers, then there is enough time for the two agents to propose all offers. In that case this approach may work even when there is no strong correlation between the utility functions, because it simply doesn't matter if our agent sometimes proposes offers that are bad for the opponent. If there is a better offer available, then our agent will simply propose that offer in any of the following turns. On the other hand, if the domain is too large (or the deadline too short), then this approach may fail because our agent cannot propose all offers, and therefore risks failing to propose those offers that are acceptable to the opponent.

\subsubsection{Choosing the Aspiration Function}
The aspiration function can be any monotonically decreasing function, but a good example would be the following \textit{exponential} function:
\begin{equation}\label{eq:asp_function}
\asp(t) \quad = \quad (\alpha-\targ) \cdot \frac{1- \gamma^{1-\frac{t}{\dead}}}{1 - \gamma} + \targ
\end{equation}
where $T$ is the deadline of the negotiations, and $\alpha$, $\targ$ and $\gamma$ are three parameters that can be freely chosen, but with $\alpha > \targ$ and $\gamma > 0$. We have plotted this expression in Figure \ref{fig:asp_function} for various different values of $\gamma$. An example implementation of a time-based agent that uses this aspiration function is displayed in Algorithm \ref{alg:time_based}.

An alternative aspiration function that is commonly used in the literature is the following  \textit{polynomial} one:
\begin{equation}\label{eq:asp_function_polynomial}
\asp(t) \quad = \quad (\alpha-\targ) \cdot (1- (\frac{t}{\dead})^{1/\gamma}) + \targ
\end{equation}
with the same constraints on the values of $\alpha$, $\targ$ and $\gamma$. We have plotted this expression in Figure~\ref{fig:asp_function_polynomial} (again, for various values of $\gamma$). There is no clear indication that either of these two expressions works better than the other.
%Note that the exponent is $1/\gamma$ rather than $\gamma$ to ensure that $\gamma$ has a similar interpretation as in Eq.~(\ref{eq:asp_function}).

\begin{figure}
\begin{center}
\begin{tikzpicture}
\begin{axis}[grid=both,
          xmin=0, xmax=1, ymin=0, ymax=1,
          xlabel={Time},
          ylabel={Aspiration Level},
          enlargelimits]
%%%%
\addplot[green,domain=0:1,samples=100]{(1 - pow(10,1-x))/(1-10)} node[pos=0.5,anchor=north east]{$\gamma=10$};
%%%%
\addplot[red,domain=0:1,samples=100]{(1 - x)} node[pos=0.5,anchor= south west]{$\gamma=1$};
%%%
\addplot[blue,domain=0:1,samples=100]{(1 - pow(0.01, 1-x))/(1-0.01)} node[pos=0.5,anchor= south west]{$\gamma=1/100$};

\end{axis}
\end{tikzpicture}
\end{center}
\caption{Exponential aspiration functions with $\alpha = 1$, $\targ=0$, $T=1$, and several different values for $\gamma$.}\label{fig:asp_function}
\end{figure}

\begin{figure}
\begin{center}
\begin{tikzpicture}
\begin{axis}[grid=both,
          xmin=0, xmax=1, ymin=0, ymax=1,
          xlabel={Time},
          ylabel={Aspiration Level},
          enlargelimits]
%%%%
\addplot[green,domain=0:1,samples=100]{1-pow(x,0.2)} node[pos=0.5,anchor=north east]{$\gamma=5$};
%%%%
\addplot[red,domain=0:1,samples=100]{1-pow(x,1)} node[pos=0.5,anchor= south west]{$\gamma=1$};
%%%%
\addplot[blue,domain=0:1,samples=100]{1-pow(x,5)} node[pos=0.5,anchor= south west]{$\gamma=1/5$};

\end{axis}
\end{tikzpicture}
\end{center}
\caption{Polynomial aspiration functions with $\alpha = 1$, $\targ=0$, $T=1$, and several different values for $\gamma$.}
\label{fig:asp_function_polynomial}
\end{figure}

\begin{algorithm}
\caption{Time-based bidding Strategy.}\label{alg:time_based}
\begin{algorithmic}[1]
    \Statex \textbf{Parameters:}  $\alpha$, $\targ$, $\gamma$
	\Statex \textbf{Input:} 
	\Statex $\Off$ \Comment{The offer space.}
	\Statex $\util_1$ \Comment{The agent's own utility function.}
	\Statex $\dead$ \Comment{The deadline.}
	\Statex $t$ \Comment{The current time.}
    \Statex $\obsHist{1}$ \Comment{The observed negotiation history.}
    \Statex $\offRec$ \Comment{The offer last proposed by the opponent (if any).}
     \Statex 
     %\Statex //OPPONENT MODELING
     \comment{OPPONENT MODELING}
    	\State $\om \leftarrow \mi{updateOpponentModel}(\Off, \dead, \om, t, \offRec)$
    	\State $\est{\util}_2 \leftarrow \mi{getEstimatedOpponentUtility}(\om)$
    	\Statex 
    	\comment{BIDDING STRATEGY}
    	\comment{Calculate the aspiration level:}
    \State $\asp \leftarrow (\alpha-\targ) \cdot \frac{1- \gamma^{1-t/\dead}}{1 - \gamma} + \targ$
    	\comment{Obtain the set of offers we have already proposed:} 
    \State $\Pro \leftarrow \mi{getOffersProposedByUs(\obsHist{1})}$
    \comment{Find the next offer to propose:}
    		\State $\offNext \leftarrow \argmax_{\off\in \Off} \{ \est{\util}_2(\off) \mid  \util_1(\off) \geq \asp \ \land \ \off \not \in \Pro\}$
    		 \Statex
    		 \comment{ACCEPTANCE STRATEGY}
    			\comment{Get the last proposal received from the opponent, and accept it if }
    		\commentt{it yields more utility to us than our aspiration level:}
    		\State $\mi{acceptOffer} \leftarrow u(\offRec) \geq \asp$
    		 \Statex
    		 \comment{RETURN SELECTED ACTION}
        \If{$\mi{acceptOffer}$}
            \State \textsc{Return} ($\acc$, $\offRec$) \quad \quad \Comment{accept the received offer}
        \Else
            \State \textsc{Return} ($\prop$, $\offNext$) \quad \quad \Comment{propose a new offer}
        \EndIf
\end{algorithmic}
\end{algorithm}

Let us now discuss how to interpret the parameters $\alpha$, $\targ$, and $\gamma$, and how to choose their values. This discussion applies equally to Equation (\ref{eq:asp_function}) and Equation (\ref{eq:asp_function_polynomial}). First, note that if $t=0$ then (for either of the two equations) we have $\asp(0) = \alpha$. Therefore, $\alpha$ represents the minimum utility our agent will demand for itself at the start of the negotiations. Similarly, if $t=\dead$ then we have $\asp(t) = \targ$. This means that $\targ$ represents the utility our agent will demand for itself at the end of the negotiations, when the deadline is near. We will call this the \textbf{target value}. A high target value represents a \hard{} strategy, while a lower target value represents a \soft{} strategy. Finally, the parameter $\gamma$ determines how quickly the agent concedes from $\alpha$ to $\targ$.

Typically, the value chosen for $\alpha$ is exactly the utility of the offer that the agent prefers most: $\alpha = \util_1(\offMax_1)$. After all, a typical negotiator would start with the proposal that yields the highest utility for itself. While it is certainly possible to start with a lower offer, there is typically not much reason to do so. So, the other two parameters are more important. 

Regarding the value for $\targ$, it should be obvious that it should always be greater than the agent's reservation value, because our agent should never propose any offers that yield less utility than that. One common choice is to set $\targ$ \textit{exactly} equal to the reservation value. The reasoning behind this is that making a deal that is just slightly above the reservation value is always better than making no deal at all, and thus one should be willing to concede all the way to the reservation value as the deadline gets close. 

While this reasoning may make sense if we focus only on one single negotiation in isolation, this choice is actually not optimal at all if we consider that our agent may be involved in many different negotiations and that our opponents may remember our agent's behavior from previous encounters and may be learning how to negotiate optimally against our agent.

The problem is this: if our agent always chooses $\targ = \rv_1$, then its opponents may anticipate this. That is, the opponents know that our agent will be conceding all the way to its reservation value and therefore they can exploit it by simply not conceding at all, or very little, and waiting until the very last moment before accepting any of our agent's proposals.

For example, consider a split-the-pie domain where the maximum utility is 1, and our reservation value is 0. If our agent plays a strategy with $\targ=0$ and the opponent chooses a strategy with $\targ = 0.99$ then all negotiations would end with an agreement that gives our agent a utility of $0.01$ and the opponent $0.99$ (assuming such an offer exists).

It is therefore often wiser to choose a higher target value (i.e. choose a more \hard{} strategy). This may sometimes cause the negotiations to fail, but in the long run that may actually be a good thing, because it sends a signal to our opponents that they will need to make concessions if they want to make an agreement with our agent. 

%In fact, even in a single negotiation this makes sense because it is fair to assume that the opponent is not going to choose a very high target value, and therefore you may not need to concede all the way to your reservation value in order to come to an agreement.\\

On the other hand, choosing the target value too high will not work  well either. It could work against a very \soft{} time-based agent (i.e. one with a low target value), but it will fail to come to an agreement if the opponent also chooses a high target value. For example, if both agents choose a target value of $0.99$ (when the maximum utility is 1), then they can only come to an agreement if there exists an offer that yields a utility of $0.99$ to both agents. It is rare to encounter a negotiation domain where this is the case.

\begin{figure}
\tikzset{every picture/.style={line width=1pt}} %set default line width to 1pt        
\pgfplotsset{width=8cm}
\begin{subfigure}[h]{0.5\linewidth}
\begin{tikzpicture}[scale=0.6]
\begin{axis}[xmin=0, xmax=1,ymin=0,ymax=1,xlabel={Utility of Agent 1}, ylabel={Utility of Agent 2}, 
		clip=false, %ensures that you can draw outside the boundaries.
		xtick pos=left,
		ytick pos=left		
		]
	%%OFFERS:
    \filldraw [black] 	(0.1,1) circle [radius=2pt]  
    						(0.1,0.6) circle [radius=2pt]  
    					  	(0.22,0.75) circle [radius=2pt]
    						(0.35,0.5) circle [radius=2pt]
    						(0.4,0.2) circle [radius=2pt]
                     (0.5,0.7) circle [radius=2pt]
                     (0.62,0.4) circle [radius=2pt]
                     (0.9,0.1) circle [radius=2pt]
                     (1.0,0.05) circle [radius=2pt];
    %% RESERVATION VALUES:
    \draw[red] (0,0.21) -- (1.03,0.21) node[anchor=west,color=black] {$\rv_2$}; %rv 2
    \draw[red] (0.15,0) -- (0.15,1.03) node[anchor=south,color=black] {$\rv_1$}; %rv 1
    
    %% ASPIRATION VALUES:
    \draw[blue] (0,0.95) -- (1.03,0.95) node[anchor=west,color=black] {$\asp_2(t)$}; %horizontal y=0.95
    \draw[blue] (0.9,0) -- (0.9,1.03) node[anchor=south,color=black] {$\asp_1(t)$}; %vertical x=0.9
\end{axis}
\end{tikzpicture}
\caption{Aspiration levels at $t/\dead=0.2$}
\end{subfigure}\hfill
\begin{subfigure}[h]{0.5\linewidth}
\begin{tikzpicture}[scale=0.6]
\begin{axis}[xmin=0, xmax=1,ymin=0,ymax=1,xlabel={Utility of Agent 1}, ylabel={Utility of Agent 2}, 
		clip=false, %ensures that you can draw outside the boundaries.
		xtick pos=left,
		ytick pos=left		
		]
	%%OFFERS:
    \filldraw [black] 	(0.1,1) circle [radius=2pt]  
    						(0.1,0.6) circle [radius=2pt]  
    					  	(0.22,0.75) circle [radius=2pt]
    						(0.35,0.5) circle [radius=2pt]
    						(0.4,0.2) circle [radius=2pt]
                     (0.5,0.7) circle [radius=2pt]
                     (0.62,0.4) circle [radius=2pt]
                     (0.9,0.1) circle [radius=2pt]
                     (1.0,0.05) circle [radius=2pt];
    %% RESERVATION VALUES:
    \draw[red] (0,0.21) -- (1.03,0.21) node[anchor=west,color=black] {$\rv_2$}; %rv 2
    \draw[red] (0.15,0) -- (0.15,1.03) node[anchor=south,color=black] {$\rv_1$}; %rv 1
    
    %% ASPIRATION VALUES:
    \draw[blue] (0,0.90) -- (1.03,0.90) node[anchor=west,color=black] {$\asp_2(t)$}; %horizontal y=0.90
    \draw[blue] (0.7,0) -- (0.7,1.03) node[anchor=south,color=black] {$\asp_1(t)$}; %vertical x=0.7
\end{axis}
\end{tikzpicture}
\caption{Aspiration levels at $t/\dead=0.4$}
\end{subfigure}

\medskip

%%%%%%%%%%%%%%%%%%%%  SECOND ROW:

\begin{subfigure}[h]{0.5\linewidth}
\begin{tikzpicture}[scale=0.6]
\begin{axis}[xmin=0, xmax=1,ymin=0,ymax=1,xlabel={Utility of Agent 1}, ylabel={Utility of Agent 2}, 
		clip=false, %ensures that you can draw outside the boundaries.
		xtick pos=left,
		ytick pos=left		
		]
	%%OFFERS:
    \filldraw [black] 	(0.1,1) circle [radius=2pt]  
    						(0.1,0.6) circle [radius=2pt]  
    					  	(0.22,0.75) circle [radius=2pt]
    						(0.35,0.5) circle [radius=2pt]
    						(0.4,0.2) circle [radius=2pt]
                     (0.5,0.7) circle [radius=2pt]
                     (0.62,0.4) circle [radius=2pt]
                     (0.9,0.1) circle [radius=2pt]
                     (1.0,0.05) circle [radius=2pt];
    %% RESERVATION VALUES:
    \draw[red] (0,0.21) -- (1.03,0.21) node[anchor=west,color=black] {$\rv_2$}; %rv 2
    \draw[red] (0.15,0) -- (0.15,1.03) node[anchor=south,color=black] {$\rv_1$}; %rv 1
    
    %% ASPIRATION VALUES:
    \draw[blue] (0,0.85) -- (1.03,0.85) node[anchor=west,color=black] {$\asp_2(t)$}; %horizontal y=0.85
    \draw[blue] (0.5,0) -- (0.5,1.03) node[anchor=south,color=black] {$\asp_1(t)$}; %vertical x=0.5
\end{axis}
\end{tikzpicture}
\caption{Aspiration levels at $t/\dead=0.6$}
\end{subfigure}\hfill
\begin{subfigure}[h]{0.5\linewidth}
\begin{tikzpicture}[scale=0.6]
\begin{axis}[xmin=0, xmax=1,ymin=0,ymax=1,xlabel={Utility of Agent 1}, ylabel={Utility of Agent 2}, 
		clip=false, %ensures that you can draw outside the boundaries.
		xtick pos=left,
		ytick pos=left		
		]
	%%OFFERS:
    \filldraw [black] 	(0.1,1) circle [radius=2pt]  
    						(0.1,0.6) circle [radius=2pt]  
    					  	(0.22,0.75) circle [radius=2pt]
    						(0.35,0.5) circle [radius=2pt]
    						(0.4,0.2) circle [radius=2pt]
                     (0.5,0.7) circle [radius=2pt]
                     (0.62,0.4) circle [radius=2pt]
                     (0.9,0.1) circle [radius=2pt]
                     (1.0,0.05) circle [radius=2pt];
    %% RESERVATION VALUES:
    \draw[red] (0,0.21) -- (1.03,0.21) node[anchor=west,color=black] {$\rv_2$}; %rv 2
    \draw[red] (0.15,0) -- (0.15,1.03) node[anchor=south,color=black] {$\rv_1$}; %rv 1
    
    %% ASPIRATION VALUES:
    \draw[blue] (0,0.72) -- (1.03,0.72) node[anchor=west,color=black] {$\asp_2(t)$}; %horizontal y=0.72
    \draw[blue] (0.2,0) -- (0.2,1.03) node[anchor=south west,color=black] {$\asp_1(t)$}; %vertical x=0.2
\end{axis}
\end{tikzpicture}
\caption{Aspiration levels at $t/\dead=0.8$}
\end{subfigure}
\caption{Negotiation between a \soft{} agent ($\ag_1$) and a \hard{} agent ($\ag_2$). Their aspiration levels are indicated with a vertical blue line and a horizontal blue line respectively. We see that the aspiration level of the \soft{} agent drops much further than the aspiration level of the \hard{} agent. 
The negotiations continue until they reach a point at which there is an offer that is acceptable to both agents. That is, when there is an offer for which its utility vector lies above the horizontal blue line, as well as to the right of the vertical blue line. In this example that happens at $t/\dead=0.8$. Note that the agreement yields more utility to the \hard{} agent than to the \soft{} agent.}\label{fig:asp_levels_example}
\end{figure}
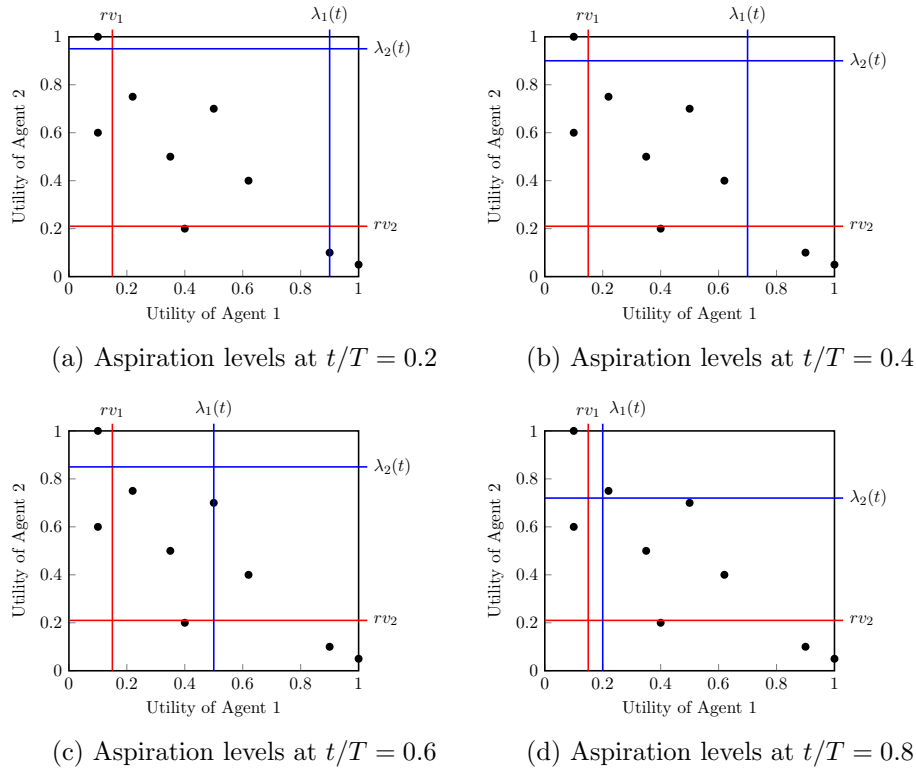

Figure \ref{fig:asp_levels_example} visualizes
the evolution of the aspiration levels of two time-based agents during a negotiation. The aspiration level of $\ag_1$ is indicated with a vertical blue line that over time moves from the right to the left, while the aspiration level of $\ag_2$ is indicated with a horizontal blue line that over time moves from the top to the bottom.
Note that in this example $\ag_1$ follows a  \soft{} strategy, while $\ag_2$ follows a \hard{} strategy. We see that they end up with an agreement that yields more utility to the \hard{} agent than to the \soft{} agent.

The parameter $\gamma$ is the \textbf{concession parameter}. It determines how fast our agent will concede towards its target value. If $\gamma$ is very small our agent will initially concede very slowly, as we can see in Figures \ref{fig:asp_function} and \ref{fig:asp_function_polynomial}, and only starts making large concessions towards the end of the negotiations. On the other hand, if $\gamma$ is very large, our agent will immediately start making large concessions. Note, however, that the question which values of $\gamma$ can be considered `very small' or `very large' depends heavily on which of the two expressions we are using. Finally, a value of $\gamma = 1$ represents an agent that concedes linearly.\footnote{Technically, the expression in Eq. \ref{eq:asp_function} is not defined for $\gamma=1$, but it can be shown that $\lim_{\gamma\rightarrow 1} f(t) = (\alpha-\targ) \cdot (1-t/\dead) + \beta$, which is a linear function of $t$.} 

In order to exploit the opponent as much as possible, our agent should make sure it concedes slower than the opponent. This suggest that we would always want a low value of $\gamma$. However, if we choose $\gamma$ too low, then our agent may start conceding so late, that by the time it finally makes a substantial concession there is no more time for the opponent to accept it. 

For example, suppose we choose an intermediate target value of $\beta = 0.5$, but our concession parameter is so low, that at 10 milliseconds before the deadline the aspiration value is still at $\asp(t) = 0.90$. While in theory the aspiration level will continue to decrease to $0.5$ in the last 10 milliseconds, this time might not be enough for our agent to actually exchange more proposals and come to an agreement. After all, every time our agent makes a proposal, it will take a small amount of time for that message to arrive at the opponent and then the opponent will still need some time to process it, and to send an `accept' message back.  This means that the optimal value of $\gamma$ largely depends on the speed at which the agents can send messages and at which they  are able to process them. In other words, it largely depends on practical considerations related to the infrastructure on which the agents are implemented. 

%It is therefore hard to determine a theoretically optimal value.

Furthermore, if we choose $\gamma$ very low, then our agent's aspiration level will remain very high for a long time, which means that for a long time there might not be any agreement possible. Then, when the deadline gets near, our agent will suddenly concede very fast towards its target value, meaning that the only possible agreement would be one close to the target value. But in that way we might miss out on any opportunities to obtain a better deal. Our agent would only be able to make a deal near its target value, or no deal at all. By choosing a somewhat higher value of $\gamma$ our agent has the time to propose several intermediate offers that yield utilities of, say, $0.8$, $0.7$ and $0.6$, which could be accepted by the opponent before our agent reaches its target level. 

Another reason why a low value of $\gamma$ might not be optimal is when there are discount factors (see Section~\ref{sec:discount_factors}), because in that case we would prefer our agent to come to an agreement as quickly as possible. Yet another example could be the case that the opponent is participating in multiple negotiations in parallel. For example, when a seller has one item to sell, and is negotiating with multiple potential buyers at the same time. In that case our agent, as a buyer, would also want to come to an agreement as soon as possible, before the seller sells the item to one of the other buyers.

Time-based strategies with a low value of $\gamma$, but with $\targ = \rv_1$ are also known as \textbf{Boulware strategies}.

Finally, it should be noted that Equations ~(\ref{eq:asp_function}) and (\ref{eq:asp_function_polynomial}) are sometimes adapted so that the agent reaches its target level already a bit \textit{before} the deadline, at a time $\dead'$ slightly less than $\dead$, which we will call the \textbf{target time}. After the target time, the aspiration level will just remain constant. So, Eq.~(\ref{eq:asp_function}) then becomes:
\begin{equation}\label{eq:asp_function_adapted}
\asp(t) =  
\begin{cases}
(\alpha-\targ) \cdot \frac{1- \gamma^{1-t/\dead'}}{1 - \gamma} + \targ & \text{if\ } t \in [0, \dead']\\
\targ & \text{if\ } t \in [\dead', \dead]
\end{cases}
\end{equation}
and Eq.~(\ref{eq:asp_function_polynomial}) becomes:
\begin{equation}\label{eq:asp_function_polynomial_adapted}
\asp(t) =  
\begin{cases}
(\alpha-\targ) \cdot (1- (\frac{t}{\dead})^{1/\gamma}) + \targ & \text{if\ } t \in [0, \dead']\\
\targ & \text{if\ } t \in [\dead', \dead]
\end{cases}
\end{equation}
This is to ensure that when our agent proposes its its ultimate offer (with utility equal to or very close to $\targ$), there is still enough time left for the opponent to accept that offer.

\begin{exercise}\label{ex:time_based}
\textbf{Time-based Agent.} Use the NegoSim framework (Section \ref{sec:simulation}) to implement an agent that applies a time-based negotiation strategy. Note that the framework already comes with the source code of a \mbox{RandomAgent}, so you can just copy its code and adapt it according to Algorithm \ref{alg:time_based}. \newline

Since we haven't discussed opponent modeling algorithms yet, you can use Equation (\ref{eq:time_based_min}) to determine the next offer, which doesn't require opponent modeling.\newline

Alternatively, you can use the DummyOpponentUtilityModel that comes with the framework. This is a fake opponent model that takes the opponent's real utility function as its input and returns a random approximation of that function. \newline

Run several negotiations between time-based agents and experiment with different parameter settings. Which values for the parameters $\alpha$, $\targ$ and $\gamma$ give the best results?
\end{exercise}

\subsection{Adaptive Strategies}\label{sec:adaptive_strategies}
We will now describe another type of strategy, known as an \textbf{adaptive strategy}. Adaptive strategies have probably received the most attention in the literature, and most agents that were successful in the various ANAC competition have been of this type.

In order to explain this type of strategy, let us first suppose that 
our opponent $\ag_2$ plays a time-based strategy with target value $\targ_2$. This means that, if we wait long enough, the opponent will be willing to propose or accept any offer $\off$ for which $\util_2(\off) \geq \targ_2$. Now, let $\off^*$ denote the offer that maximizes $\ag_1$'s own utility $\util_1$ among those offers. That is:
\begin{equation}\label{eq:optimal_target}
\off^* \ \ := \ \ \argmax_{\off \in \Off} \{\util_1(\off) \mid \util_2(\off) \geq \targ_2\}
\end{equation}
This means that $\ag_1$ cannot possibly receive any utility higher than $\util_1(\off^*)$. After all, by Eq.~(\ref{eq:optimal_target}) we know that for any offer $\off$ that yields a higher utility to $\ag_1$, we would have $\util_2(\off) < \targ_2$, and agent $\ag_2$ would never propose or accept any such offer, by definition of $\targ_2$. On the other hand, however, it also means that if we get close enough to the deadline, then $\ag_2$ will be willing to accept the offer $\off^*$ and therefore, \textit{ideally}, $\ag_1$ should not propose or accept any offers that yield less utility than $\util_1(\off^*)$. So, against this opponent, a theoretically optimal strategy for $\ag_1$ would be one that concedes no further than $\util_1(\off^*)$. For example, a time-based strategy with target value $\targ_1 = \util_1(\off^*)$.

Unfortunately, however, there are two problems with this idea. Firstly, $\ag_1$ typically does not know the target value $\targ_2$ of its opponent, and secondly $\ag_1$ typically also does not know the utility function $\util_2$ of its opponent. Therefore, $\ag_1$ cannot directly determine the ideal offer $\off^*$.

Instead, however, $\ag_1$ can try to infer it, using opponent modeling algorithms (which we will discuss in Chapter~\ref{sec:opponent_modeling}). The idea is then simple: every time our agent receives a proposal from the opponent, our agent uses it to update the opponent model to obtain a more accurate approximation of $\util_2$ and $\targ_2$, which it can then use to obtain a better prediction of $\off^*$. Then, our agent sets its target value equal to $\util_1(\off^*)$ (unless that is below our agent's reservation value, of course), and finally it uses this to determine its aspiration level at that moment, according to some formula such as Eq.~(\ref{eq:asp_function}) or~(\ref{eq:asp_function_polynomial}). 

This approach is called an \textit{adaptive strategy}, because it tries to adapt to its opponent. Just like a time-based strategy it applies an aspiration level that decreases over time, but the difference is that the target value is not constant. Instead, it is updated every time we gain more information about the opponent's strategy and utility function.  

In theory, if we are 100\% sure that our opponent is using a time-based strategy, and we have an opponent modeling algorithm that can predict $\off^*$ with 100\% accuracy, then an adaptive strategy is the theoretically optimal strategy against that opponent (in game theory terminology: it is a \textit{best response}, see Chapter \ref{sec:game_theory}). After all, it concedes exactly enough to ensure a deal, but no further than that, so it always achieves the maximum amount of utility that can possibly be achieved against that opponent.

Of course, in practice we don't really have a 100\% accurate opponent modeling algorithm. But besides that, another problem with the reasoning above is that it assumes the opponent does not know anything about our agent. The problem, is that if the opponent can somehow anticipate that we are using a purely adaptive strategy, then he may be able to exploit this knowledge by choosing a very \hard{} strategy. For example, in a split-the-pie domain where both agents have a reservation value of 0, he could choose a target value of $\beta = 0.99$. If we then apply a purely adaptive strategy, then our agent would always come to an agreement for which it gets no more than 0.01 utility. 

Therefore, in practice, many adaptive strategies  have a `minimum target' $\targ^{min}$ and they make sure that their target $\targ$ is never lower than that. That is:
\[\targ \ \ = \ \ \max \{\ \ \util_1(\off^*) \ \ , \ \ \targ^{min} \ \ \}\]
This means that such strategies are more of a hybrid between a time-based strategy and a \textit{purely} adaptive strategy.

Furthermore, since our opponent modeling will probably not be 100\% accurate, we may need to add another term $\safe$ to our target utility $\util_1(\off^*)$, where $\safe > 0$ and where $\safe$ decreases as we gain more and more knowledge about the opponent from the offers it proposes to us. So we would get:
\[\targ \ \ = \ \ \max \{\ \ \util_1(\off^*) + \safe \ \ , \ \ \targ^{min} \ \ \}\]
This is to prevent that an inaccurate estimation at the beginning of the negotiations causes our agent to concede too much.

Yet another problem with adaptive strategies, is that they kind of assume the opponent is following a purely time-based strategy, which allows the adaptive strategy to predict the optimal target value. This, however, gets much more complicated if the opponent is also playing an adaptive strategy. In that case we have two agents that are each trying to adapt to the other.

A basic implementation of an adaptive strategy is displayed in Algorithm~\ref{alg:adaptive}.

\begin{exercise}\label{ex:adaptive}
\textbf{Adaptive Agent.} Use the NegoSim framework  to implement an agent that applies an adaptive negotiation strategy. Note that the framework already comes with the source code of a \mbox{RandomAgent}, so you can just copy its code and adapt it according to Algorithm \ref{alg:adaptive}. \newline

Since we haven't discussed opponent modeling algorithms yet, you can again use the DummyOpponentUtilityModel that comes with the framework (See Exercise~\ref{ex:time_based}) to estimate the opponent's utility function.\newline

Furthermore, to estimate the optimal target value $\targ^*$ you can use the SimpleOpponentStrategyModel that also comes with the NegoSimulator framework. This class implements a very naive linear extrapolation algorithm to predict how far the opponent will concede.\newline

Experiment with several parameter settings and run a number of negotiations between your adaptive agent and your time-based agent(s) from Exercise~\ref{ex:time_based}.
\end{exercise}

\begin{algorithm}
\caption{Adaptive Strategy.}\label{alg:adaptive}
\begin{algorithmic}[1]
    \Statex \textbf{Parameters:}  $\alpha$, $\targ^{min}$, $\gamma$
	\Statex \textbf{Input:} 
	\Statex $\Off$ \Comment{The offer space.}
	\Statex $\util_1$ \Comment{The agent's own utility function.}
	\Statex $\rv_1$ \Comment{The agent's own reservation value.}
	\Statex $\dead$ \Comment{The deadline.}
	\Statex $\om$ \Comment{A model of the opponent.}
	\Statex $t$ \Comment{The current time.}
    \Statex $\obsHist{1}$ \Comment{The observed negotiation history.}
    \Statex $\offRec$ \Comment{The offer last proposed by the opponent (if any).}
    \Statex 
    \comment{OPPONENT MODELING}
    \comment{Update the opponent model:}
    	\State $\om \leftarrow \mi{updateOpponentModel}(\Off, \dead, \om, t, \offRec)$
    	\State $\est{\util}_2 \leftarrow \mi{getEstimatedOpponentUtility}(\om)$
    	\comment{Use the opponent model to estimate the optimal target value:} 
    	\Statex $\est{\targ}^* \leftarrow \mi{estimateOptimalTarget}(\om)$
    	\Statex 
    	\comment{BIDDING STRATEGY}
    	\comment{Calculate the aspiration value:}
    	\State $\targ \leftarrow \max \{ \est{\targ}^* \ , \ \targ^{min}\}$
    \State $\asp \leftarrow (\alpha-\targ) \cdot \frac{1- \gamma^{1-t/\dead}}{1 - \gamma} + \targ$
    	\comment{Obtain the set of offers we have already proposed:}
    \State $\Pro \leftarrow \mi{getOffersProposedByUs(\obsHist{1})}$
    \comment{Find the next offer to propose:}
    		\State $\offNext \leftarrow \argmax_{\off\in \Off} \{ \est{\util}_2(\off) \mid  \util_1(\off) \geq \asp \ \land \ \off \not \in \Pro\}$
    		 \Statex
    		 \comment{ACCEPTANCE STRATEGY}
    		\comment{Get the last proposal received from the opponent, and accept it if} 
    		\commentt{it yields more utility to us than our aspiration level:}
    		\State $\mi{acceptOffer} \leftarrow u(\offRec) \geq \asp$
    		 \Statex
    		\comment{RETURN SELECTED ACTION}
        \If{$\mi{acceptOffer}$}
            \State \textsc{Return} ($\acc$, $\offRec$) 
        \Else
            \State \textsc{Return} ($\prop$, $\offNext$)
        \EndIf
\end{algorithmic}
\end{algorithm}

\subsection{Imitative Strategies}

Above, we have seen that if we know the opponent plays a time-based strategy, then the best response for our agent would be to play an adaptive strategy. On the other hand, if the opponent is playing an adaptive strategy, then the best choice for our agent would be to play a \hard{} time-based strategy which can exploit the opponent's adaptiveness. Now, the question is how to choose between these two strategies when we don't know what strategy our opponent will choose. 

If one agent plays a \hard{} time-based strategy and the other plays an adaptive strategy, then the time-based agent would typically receive a higher utility than the adaptive agent. Therefore, one might be inclined to argue that  choosing a \hard{} time-based strategy is better. But the problem is that the opponent could follow exactly the same reasoning, and therefore choose a \hard{} strategy as well. But then we end up with two agents each playing a \hard{} strategy, and in that case it is unlikely that the two agents will come to an agreement, since neither of the two would be willing to make any considerable concessions.

For this reason, some might reason that it is better to play an adaptive strategy. But then again, the opponent might reason in the same way and also choose an adaptive strategy. In that case we would miss out on the opportunity of exploiting him. Furthermore, if we always choose an adaptive strategy, then that could be exploited by the opponent, by always choosing a \hard{} strategy. In other words, choosing between a \hard{} strategy and an adaptive strategy is a bit of a chicken-and-egg problem. The problem is that each of these strategies work well against the other, but neither of them is optimal when the opponent picks the same strategy.

We have seen that one way out would be to choose a hybrid approach that applies an adaptive strategy with a minimum target $\targ^{min}$, but then we still need to answer the question how to choose the optimal value for $\targ^{min}$. Another approach would be to flip a coin and decide between the two strategies randomly. We will discuss this option in more depth in Chapter~\ref{sec:game_theory}.

In this section, however, we will discuss an entirely different type of strategy that is designed specifically to play well against itself. Such strategies are known as \textbf{imitative strategies} \cite{Faratin1998}. Rather than trying to \textit{adapt} to the opponent (play \hard{} when the opponent plays \soft{} and vice versa), imitative agents instead try to \textit{imitate} the opponent. That is, when the opponent is \hard{} then play \hard{} as well, and when the opponent is \soft{}, play \soft{} as well. The rationale behind this, is that if the opponent plays too \hard{}, then our agent can `punish' it by also playing \hard{}, and when the opponent plays \soft{}, then our agent rewards the opponent by playing \soft{} as well. 

Of course, this is all based on the assumption that the opponent does not play a rigid time-based strategy, but rather observes our agent's actions and is able to adapt itself to our agent's strategy.

We will discuss two kinds of imitative strategies, namely the \textit{Classic Tit-for-Tat} strategy and the \textit{MiCRO} strategy.

\subsubsection{Classic Tit-for-Tat}

In game theory, Tit-for-That (TFT) strategies are strategies in which a player imitates the moves of the other player. This strategy has been proven especially useful in the iterated prisoner's dilemma~\cite{Axelrod1981evolution}. 

%In the context of the IPD, the idea of the tit-for-tat strategy is to always play cooperative moves, as long as the opponent does the same. However, whenever the opponent plays a non-cooperative move, you retaliate by also playing a non-cooperative move.

In the context of negotiation, this would mean that whenever our opponent makes a large concession, our agent replies to this by also making a large concession, and whenever our opponent makes a small concession (or no concession at all), then our agent replies with a small concession as well (or no concession at all). 

Before we continue, recall that $\Pro_t$ denotes the set of offers that have been proposed by our agent $\ag_1$ up until time $t$ (see Eq.(\ref{eq:Off_prop})). Similarly, we define $\Rec_t$ to be the set of offers that have been \textit{received} by our agent $\ag_1$ up until time $t$. In other words, it is set of offers that have been proposed by the \textit{opponent} $\ag_2$ up until time $t$. Formally:
\begin{equation}\label{eq:Off_rec}
\Rec_{t} := \{ \off \in \Off \mid \exists t' \in [0,t] : (2,\prop, \off, t') \in \obsHist{1} \}
\end{equation}

Now, in order to give a concrete implementation of a classic Tit-for-Tat negotiation strategy, we need a function $\con_1$ that, given $\Pro_t$ returns a value $\con_1(\Pro_t) \in \mathbb{R}$ that measures how much agent $\ag_1$ has so far conceded. Furthermore, we need a function $\con_2$ that, given $\Rec_t$ returns a value $\con_2(\Rec_t) \in \mathbb{R}$ that measures the amount of concession made by $\ag_2$.
\[\con_1,\con_2 : 2^\Off \rightarrow \mathbb{R}\]

In general, for any agent, when we say it makes a large `concession', this can be interpreted in two ways: it can mean that it makes a proposal with high utility for the opponent, or it can mean that it makes a proposal with low utility for itself. In a single-issue negotiation where the agents' interests are strictly opposing, such as the bargaining over the price of a second-hand car, we don't have to worry about this distinction because any concession of the first type is automatically also one of the second type and vice versa.

However, in more complex negotiation scenarios, where not every offer is Pareto-optimal, and where the agents do not know each others' utility functions, these two concepts are different.

This means that for $\con_1$ there are two obvious choices. Namely, we could define it in terms of our agent's own utility, or in terms of our opponent's (estimated) utility:
\begin{eqnarray}
\con_1(\Pro_t) &:=& \max \ \{\util_1(\offMax_1) - \util_1(\off) \mid \off \in \Pro_t\}\label{eq:con_1_self}\\
\text{or:} \nonumber\\
\con_1(\Pro_t) &:=& \max \ \{\est{\util}_2(\off) - \est{\util}_2(\offMin_2) \mid \off \in \Pro_t\}\label{eq:con_1_opp}
\end{eqnarray}
where $\est{\util}_2$ is an estimation of the opponent's utility function $\util_2$, made by an opponent modeling algorithm and where $\offMax_1$ and $\offMin_2$ are defined by Equations (\ref{eq:offMax}) and (\ref{eq:offMin}).

In the first case, our `concession' corresponds to the lowest amount of utility our agent has so far asked for itself, while in the second case it corresponds to the highest amount of utility it has so far offered to the opponent.

Similarly, we can measure the opponent's concession using either our agent's own utility function, or the opponent's estimated utility function:
\begin{eqnarray}
\con_2(\Rec_t) &:=& \max \ \{\util_1(\off) - \util_1(\offMin_1) \mid \off \in \Rec_t\}\label{eq:con_2_self}\\
\text{or:} \nonumber\\
\con_2(\Rec_t) &:=& \max \ \{\est{\util}_2(\offMax_2) - \est{\util}_2(\off) \mid \off \in \Rec_t\}\label{eq:con_2_opp}
\end{eqnarray}
Here, in the first case, the opponent's `concession' corresponds to the highest amount of utility the opponent has so far offered to our agent, while in the second case it corresponds to the lowest amount of utility the opponent has so far asked for itself.

%\todo{first give a simple concrete example, next give the more generic implementation}
%
% Simple examples could be: $\sigma_1(\obsHist{1}) = \max \{1 - \util_1(\off) \mid \off \in \Pro\}$ and $\sigma_2(\obsHist{1}) = \max \{\util_1(\off) \mid \off \in \Rec\}$, where $\Pro$ denotes the set of all offers that our agent has so proposed in the history $\obsHist{1}$ and $\Rec$ denotes the set of all offers our agent has received from the opponent in the history $\obsHist{1}$. But we  will discuss other possible choices below. 
 
Note that here, $\con_2$ is a function used by \textit{our} agent $\ag_1$ to measure the opponent's concession. In other words, it exists in the `mind' of our agent $\ag_1$ and the opponent itself may actually use an entirely different function to measure its own concession (if it even uses a Tit-for-Tat strategy at all).

Whenever it is our agent's turn, its goal is to propose an offer $\off_{next}$ such that the total amount of concession that our agent has made so far remains slightly higher than our opponent's. We therefore define, for any offer $\off \in \Off$, its \textit{concession gain}:
\[\conGain_t(\off) \ \ := \ \ \con_1(\Pro_t \cup \{\off\}) \ \  - \ \  \con_2(\Rec_t) \]
which allows us to quantify, for any offer $\off$, the difference between our agent's concession after proposing $\off$ and the concession made by the opponent.

Finally, the Tit-for-Tat strategy chooses our agent's next offer to propose $\offNext$ by selecting it from a set of possible offers that satisfy some criterion regarding to the concession gain. Again, there is no unique way to do this, so we provide two examples:
\begin{eqnarray}
\hspace{-4mm} \offNext \hspace{-2mm} &=& \hspace{-2mm} \argmax_{\off} \ \{\ \util_1(\off) \ |\ \conGain_t(\off) > \tftThresh_{min} \ \land \ \util_1(\off) > \rv_1\}\label{eq:tft_selfish}\\
%%%%%%%%%%%%%%%%%%
\text{or:} \nonumber\\
%%%%%%%%%%%%%%%%%%
\hspace{-4mm} \offNext \hspace{-2mm} &=& \hspace{-2mm} \argmax_{\off} \ \{\ \est{\util}_2(\off) \ |\ \conGain_t(\off) \in (\tftThresh_{min} , \tftThresh_{max})\land \util_1(\off) > \rv_1\}\label{eq:tft_altru}
\end{eqnarray}
where $\tftThresh_{min}$ and $\tftThresh_{max}$ are a minimum and a maximum required concession gain, respectively. In the first case our agent would select the offer that maximizes its own utility, under the constraint that it should also concede enough to the opponent. In the second case, our agent would select an offer that maximizes the \textit{opponent's} estimated utility, but that requires we also limit ourselves to a maximum concession gain, to prevent our agent from conceding too much. 

In each of these expressions, $\tftThresh_{min}$ can be equal to 0, but $\conGain_t(\off)$ must remain strictly greater than $0$. This is, because otherwise if it happens that both agents have made exactly the same amount of concession, then neither of them will be willing to concede more, and they get stuck in a deadlock (if they both use the same strategy). Therefore, each of the two agents should always strive to concede slightly more than the other.

We have now seen that for a concrete implementation of Tit-for-Tat we need to make 3 choices: an expression for $\con_1$, an expression for $\con_2$, and a method to choose $\off_{next}$ (e.g. Eq. (\ref{eq:tft_selfish}) or  Eq. (\ref{eq:tft_altru})).

At first sight, we might be tempted to choose the expressions that only depend on our agent's own utility function (i.e. Eqs. (\ref{eq:con_1_self}) , (\ref{eq:con_2_self}) and (\ref{eq:tft_selfish})), so that we don't have to rely on any opponent modeling algorithms. However, it turns out that this doesn't work very well. To understand the problem, we should first note that in that case, if both agents in each turn make sufficiently small concessions, then the final agreement would always be an offer $\off$ that satisfies $\util_1(\off) \approx \frac{1}{2}\util_1(\offMax_1) + \frac{1}{2} \util_1(\offMin_1)$. This can be seen easily as follows. Suppose for simplicity that we have a normalized utility function (i.e. $\util_1(\offMin_1)=0$ and $\util_1(\offMax_1)=1$). Now, if the opponent $\ag_2$ makes an offer that yields a utility of 0.1 to our agent, then our agent $\ag_1$ would reply with an offer that yields a utility of 1-0.1=0.9 to itself. Next, if $\ag_2$ makes a proposal with utility of, say, 0.3 for $\ag_1$, then $\ag_1$ replies with an offer with utility 1-0.3=0.7. Then, if $\ag_2$ proposes an offer with utility 0.35, our agent $\ag_1$ will reply with an offer that yields 1-0.35=0.65, next, if $\ag_2$ proposes an offer with utility 0.55 then $\ag_2$ replies with an offer that yields 1-0.55 = 0.45. It is easy to see that, no matter which offers the opponent proposes, this always either converges to an agreement with utility 0.5 for $\ag_1$, or the two agents' proposals don't converge at all, which means there will be no agreement. 

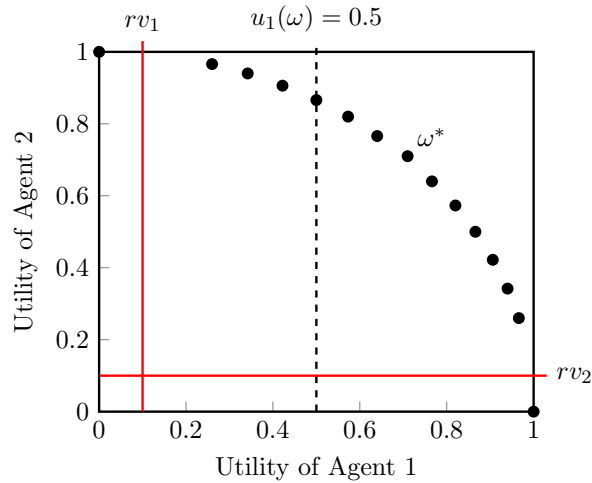
\begin{figure}
\tikzset{every picture/.style={line width=1pt}} %set default line width to 1pt        
\pgfplotsset{width=8cm}
\begin{center}
\begin{tikzpicture}[scale=0.9]
\begin{axis}[xmin=0, xmax=1,ymin=0,ymax=1,xlabel={Utility of Agent 1}, ylabel={Utility of Agent 2}, 
		clip=false, %ensures that you can draw outside the boundaries.
		xtick pos=left,
		ytick pos=left		
		]
	%%OFFERS:
    \filldraw [black] (0.0,1.0) circle [radius=2pt]
                      (0.26,0.966) circle [radius=2pt]   %15
                      (0.342,0.94) circle [radius=2pt]   %20
                      (0.422,0.906) circle [radius=2pt]  %25
                      (0.5,0.866) circle [radius=2pt]  %30
                     (0.573,0.82) circle [radius=2pt]	 %35
                     (0.64,0.766) circle [radius=2pt]  %40
                     (0.71,0.71) circle [radius=2pt] node[anchor=south west,color=black] {$\off^*$}   %45
                     (0.766,0.64) circle [radius=2pt]   %40
                     (0.82,0.573) circle [radius=2pt]  %35
                      (0.866,0.5) circle [radius=2pt]  %30
                      (0.906,0.422) circle [radius=2pt]  %25
                      (0.94,0.342) circle [radius=2pt]   %20
                      (0.966,0.26) circle [radius=2pt]   %15
                     (1.0,0.0) circle [radius=2pt]
                     ;
   \draw[black, dashed] (0.5,0) -- (0.5,1.03) node[anchor=south,color=black] {$\util_1(\off) = 0.5$}; 
    %% RESERVATION VALUES:
    \draw[red] (0,0.1) -- (1.03,0.1) node[anchor=west,color=black] {$\rv_2$}; %rv 2
    \draw[red] (0.1,0) -- (0.1,1.03) node[anchor=south,color=black] {$\rv_1$}; %rv 1
\end{axis}
\end{tikzpicture}
\caption{An example of a domain with low opposition. Here, the outcome with $\util_1(\off) = 0.5$ is highly unfair for $\ag_1$, since the opponent would receive $\util_2(\off) = 0.87$ for that same offer. Especially, since there exists a much fairer offer, here indicated as $\off^*$, for which both agents would receive 0.7.}\label{fig:tft_problem}
\end{center}
\end{figure}

Now, it happens that in many negotiation domains, if an offer yields 0.5 to one agent, then it yields much more utility to the other agent. This happens specifically in domains with low opposition, where there exist offers for which both agents receive a normalized utility greater than 0.5. This is illustrated in Figure \ref{fig:tft_problem}. In other words, our agent would receive much less utility than what it could potentially achieve with a better algorithm.

Moreover, if we already know that that this algorithm can only make agreements with a utility value of 0.5 for our agent, then we could just as well play a time-based strategy with target value of $\targ = 0.5$. This would at least give our agent the possibility of reaching agreements with higher utility.

So, what if we choose one of the other options? Well, if we choose the opponent's estimated utility $\est{\util}_2$ to calculate our own concession $\con_1$ as well as our opponent's concession $\con_2$, then we end up with essentially the same problem. In that case (assuming we have accurate opponent modeling algorithms), the only possible agreement the agents could make, would be one with $\util_2(\off) \approx 0.5$. While this may seem good, because such a solution would typically yield high utility for our own agent, the problem is that it would therefore be also less likely that the opponent would be willing to accept such a deal.

A better idea seems to be to use our own utility to measure our own concession and the opponent's utility to measure the opponent's concession, or vice versa. In either of these two cases the proposals would converge to some deal $\off$ with $\util_1(\off) \approx \util_2(\off)$, which would typically be better.

The problem with that, however, is that its success depends on the accuracy of our opponent modeling algorithms. If we cannot estimate $\util_2$ accurately, then our agent could be making concessions that are too large, yielding suboptimal agreements, or it could be making concessions that are too small, preventing the agents from coming to an agreement at all.

An alternative approach to reach good outcomes using TFT, is to use \textit{relative} concessions, instead of absolute ones~\cite{Baarslag2013TitForTat}. By this we mean that we first pick some ideal outcome $\off^*$, such as the max-sum solution, or the Nash bargaining solution (see Section \ref{sec:bargaining_solutions}) and then we measure concession relative to that ideal outcome:
\begin{eqnarray}
\con_1(\Pro_t) &=& \max \ \{\ \frac{\util_1(\offMax_1) - \util_1(\off)}{\util_1(\off_1^{max}) - \util_1(\off^*)} \mid \off \in \Pro_t \}\\
\con_2(\Rec_t) &=& \max \ \{\ \frac{\util_1(\off) - \util_1(\offMin_1)}{\util_1(\off^*) - \util_1(\offMin_1)} \mid \off \in \Rec_t \}
\end{eqnarray}
Note that this does require you to know which outcome $\off^*$ would be ideal, which would still depend on the opponent's utility function. However, it requires much less knowledge about $\util_2$ than if we used Eqs.~(\ref{eq:con_1_opp}) and (\ref{eq:con_2_opp}).

It may also be worth mentioning that in the  paper that originally proposed the TFT negotiation strategy \cite{Faratin1998}, the authors proposed a variant in which the agents' concessions were calculated only in terms of the the \textit{last few} proposals by each agent, rather than \textit{all} their proposals up to time $t$.

As explained before, the main idea of Tit-for-Tat is that it works well against itself. However, if the opponent uses a \hard{} time-based strategy, then Tit-for-Tat is likely to fail, because neither of the two agents will be making big concessions. If the opponent applies an adaptive strategy, or a \soft{} time-based strategy, Tit-for-Tat will likely come to an agreement, but it will not be able to exploit the opponent as much as a \hard{} strategy could have done.

Furthermore, even if we have a good opponent utility modeling algorithm, and the opponent is indeed using TFT as well, then the success of our agent also heavily relies on the accuracy of the \textit{opponent's} opponent modeling algorithms (i.e. the algorithm used by our opponent to estimate our utility function). After all, the opponent might \textit{intend} to make an offer that yields a lot of utility to our agent, but due to an inaccurate opponent model he might end up proposing one that actually yields very low utility to our agent, which would then respond with a counter-proposal that yields very low utility to the opponent. This would prevent them to reach an agreement, even though both agents have the intention to make large concessions.

\begin{exercise}\label{ex:tit_for_tat}
\textbf{Tit-for-Tat Agent.} Implement an agent that applies one of the various Tit-for-Tat strategies explained in this section.

Since we haven't discussed opponent modeling algorithms yet, you can use the DummyOpponentUtilityModel that comes with the NegoSim framework (See Exercise~\ref{ex:time_based}). 

Let your agent negotiate against the RandomAgent or against one of your agents from Exercises~\ref{ex:time_based} and \ref{ex:adaptive}, or  against a copy of itself.
\end{exercise}

\begin{algorithm}
\caption{A Classic Tit-for-Tat strategy.}\label{alg:tit_for_tat}
\begin{algorithmic}[1]
	\Statex \textbf{Parameters: $\tftThresh_{min}$} 
	\Statex \textbf{Input:} 
	\Statex $\Off$ \Comment{The offer space.}
	\Statex $\util_1$ \Comment{The agent's own utility function.}
	\Statex $\rv_1$ \Comment{The agent's own reservation value.}
	\Statex $\dead$ \Comment{The deadline.}
	\Statex $\om$ \Comment{A model of the opponent.}
	\Statex $t$ \Comment{The current time.}
    \Statex $\obsHist{1}$ \Comment{The observed negotiation history.}
        \Statex $\offRec$ \Comment{The offer last proposed by the opponent (if any).}
	\Statex    
    \comment{OPPONENT MODELING}
    	\State $\om \leftarrow \mi{updateOpponentModel}(\Off, \dead, \om, t, \offRec)$
    	\State $\est{\util}_2 \leftarrow \mi{getEstimatedOpponentUtility}(\om)$
    	\Statex 
    	\comment{BIDDING STRATEGY}
    		\comment{Get the next offer to propose according to Equation~ (\ref{eq:tft_selfish}).}
    		\comment{We split this calculation into two parts: }
    		\commentt{\quad 1) Get a set of candidate offers $\candidates$.}
    		\commentt{\quad 2) Find the offer that maximizes our utility.}
    \comment{Note that the calculation of $\conGain_t(\off)$ depends on the chosen}
    	\commentt{expressions for $\con_1$ and $\con_2$.}
 	\State $\candidates \leftarrow  \{\ \off \in \Off \ |\ \conGain_t(\off) > \tftThresh_{min} \ \land \ \util_1(\off) > \rv_1\}$
 		\If{$\candidates = \emptyset$}
 			\State $\offNext \leftarrow \dots$ \Comment{Use any alternative method to pick an offer here.}
 		\Else
   		 	\State $\offNext \leftarrow \argmax_{\off} \ \{\ \util_1(\off) \ |\ \off \in \candidates\}$
   		 \EndIf
    		  \Statex
    		 \comment{ACCEPTANCE STRATEGY}
    		\comment{Get the last proposal received from the opponent, and accept it if and}
    		\commentt{only if it is at least as good 
    		as the offer the agent is about the propose:}
    		\State $\mi{acceptOffer} \leftarrow u(\offRec) \geq u(\offNext)$
    		 \Statex
    		\comment{RETURN SELECTED ACTION}
        \If{$\mi{acceptOffer}$}
            \State \textsc{Return} ($\acc$, $\offRec$) 
        \Else
            \State \textsc{Return} ($\prop$, $\offNext$)
        \EndIf
\end{algorithmic}
\end{algorithm}

\subsubsection{The MiCRO Strategy}
We have seen above that classic TFT strategies depend heavily on the quality of the opponent modeling algorithms of both agents. However, recently a new kind of TFT strategy has been proposed based on the idea that our agent does not know anything about the opponent's utility function at all and moreover, that the opponent also does not know anything about \textit{our} agent's utility function~\cite{deJonge2022MiCRO}. This strategy is called MiCRO, which stands for \textit{Minimal Concession in Reply to new Offers}. Despite its simplicity and the fact that it does not require any form of opponent modeling, it has shown some remarkably good results.

MiCRO works as follows. Before the negotiations begin, our agent $\ag_1$ creates a list $(\off_1, \off_2, \dots, \off_K)$ containing all offers in the domain, sorted in order of decreasing utility for itself. That is, $u_1(\off_1) \geq u_1(\off_2) \geq \dots \geq u_1(\off_K)$. Then, when the negotiations start, our agent will first propose the offer with highest utility for itself. That is, $\off_1$, which is the first offer on the list. Then, in the following rounds, every time the opponent proposes a \textit{new} offer (i.e. an offer that it hasn't proposed before), our agent will respond by proposing the next offer on its list. So, it will first propose $\off_2$, then $\off_3$, then $\off_4$, etcetera. However, whenever the opponent $\ag_2$ proposes an offer it has already proposed before, $\ag_1$ will reply by also repeating an earlier proposal.

More precisely, whenever it is $\ag_1$'s turn to make a proposal, it counts how many \textit{different} offers it has so far proposed to the opponent (we denote this number by $m$), and how many different offers it has so far received from the opponent (we denote this number by $n$). That is, $m := |\Pro_t|$ and $n := |\Rec_t|$. Then, if $m \leq n$, our agent will propose $\off_{m+1}$. On the other hand, if $m>n$ then it will pick a random integer $r$ such that $1\leq r \leq m$ and propose $\off_r$.

%\footnote{Note that if the agents follow the AOP and one of them applies the MiCRO strategy, then we always have $|m-n| \leq 1$.}.

An implementation of the MiCRO strategy is given in Algorithm~\ref{alg:micro}.

The intuition behind MiCRO is that, like any other TFT algorithm, it tries to make a concession whenever the opponent makes a concession. However, since it assumes neither of the two agents know anything about the other agent's utility function, MiCRO does not care \textit{how large} the opponent's concessions are. After all, the size of the opponent's concession as perceived by our agent says nothing about the size of the concession the opponent \textit{intended} to make. The opponent might make a large concession in terms of its own utility $\util_2$, but this may result in a very small concession measured in our agent's own utility $\util_1$. For the same reason MiCRO never makes large concessions to its opponent. In fact, it always makes exactly the smallest possible concession: it just proposes the next offer on its list. Another difference between MiCRO and classic TFT is that MiCRO uses a different definition of `concession'. That is, even if the opponent's new proposal offers less utility to $\ag_1$ than the opponent's previous proposal, MiCRO still considers this a concession, as long as it is \textit{different} from any of the opponent's previous offers. After all, if the opponent makes offers in order of decreasing utility for itself, then every new proposal is indeed a concession from his point of view.

Note that MiCRO can indeed be seen as a TFT algorithm, with the following concession measures:
\[\con_1(\Pro_t) := |\Pro_t|\]
\[\con_2(\Rec_t) := |\Rec_t|\]
and that uses Eq.~(\ref{eq:tft_selfish}) to select the next offer to propose, with $\tftThresh_{min} = 0$.

At first sight, it may seem that MiCRO must be very slow in large negotiation domains, since it makes only minimal concessions. If a domain contains tens of thousands of offers, then you might expect it to take a long time before MiCRO has conceded enough for the opponent to be willing to accept any of MiCRO's proposals. However, in practice it turns out to be rather the opposite. When two MiCRO agents negotiate against each other they typically come to an agreement much faster than most other negotiation strategies. The reason for this, is that MiCRO does not spend any time updating any opponent modeling algorithms. In each turn it just performs a few very simple calculations and then proposes the next offer on its list, which makes it very fast.

However, perhaps the biggest advantage of MiCRO is that it is very simple to implement, since it does not require implementing any complicated opponent modeling algorithms, and that it does not require any parameters to be fine-tuned. This makes it especially ideal as a benchmark strategy for scientific experiments. After all, there is basically just one version of MiCRO, while almost any other negotiation strategy requires choosing some parameter values or choosing a particular opponent modeling algorithm. This makes it much harder to draw general conclusions about such strategies, or about agents that have been tested against such strategies.

Furthermore, what makes MiCRO particularly elegant, is that it makes a nearly optimal trade-off. On the one hand it is very \hard{} because it only makes minimal concessions and only keeps conceding as long as the opponent also keeps conceding. Yet, unlike \hard{} time-based agents, it typically still manages to come to agreement when negotiating against itself. This is because two MiCRO agents would always keep making concessions until sooner or later they reach an agreement. 

There are just two possible scenarios in which a negotiation between two MiCRO agents would fail to end with agreement. The first scenario is when one of the two agents has a very high reservation value so at some point it can't continue conceding because it has already reached its reservation value before the agents have reached an agreement. The second scenario is when the deadline is too short compared to the size of the domain, so there is no time to concede far enough to reach an agreement. However, as explained above, MiCRO is typically much faster than other strategies, so in this scenario many other strategies might also suffer to concede fast enough.

Apart from these two possible scenarios, the main disadvantage of \mbox{MiCRO} is that it will still fail to make an agreement against a \hard{} time-based agent that at some point refuses to make any further concessions.

\begin{exercise}\label{ex:micro}
\textbf{MiCRO.} Implement an agent based on the MiCRO strategy in the NegoSim framework and let it negotiate against the RandomAgent, or against any of the agents from the previous exercises, or against a copy of itself. 
\end{exercise}

In the literature, two important improvements to MiCRO have been proposed~\cite{deJonge2024theoreticalPropertiesMicro}. The first improvement is to include the technique of \textit{reproposing}, which is explained below in Section \ref{sec:reproposing}. The second improvement, known as \textit{MiCRO-Inspire}, does not only count the number of unique offers $m$ proposed by the agent itself and the number of unique offers $n$ proposed by the opponent, but also counts the number of unique offers $l$ that were first proposed by the opponent, and then later proposed by our agent, but rejected by the opponent. MiCRO-Inspire then proposes a new offer only if $m \leq n - l$.

The idea is that if some offer $\off$ was first proposed by the opponent, but then later rejected by that same opponent, then that offer should no longer be counted as a `concession'. So, we should subtract the number of such offers from the number $n$. This is especially useful as a defense mechanism against an opponent that purposefully makes proposals that are only intended to make MiCRO concede more, while the opponent doesn't really want them to be accepted (see~\cite{deJonge2024theoreticalPropertiesMicro} for more details).

\begin{algorithm}
\caption{The MiCRO strategy. Note that $\mathit{offers}[m]$ here corresponds to $\off_{m+1}$ in the text.}\label{alg:micro}
\begin{algorithmic}[1]

		\Statex \textbf{Input:} 
		\Statex $\mi{offers}$ \Comment{A list containing all possible offers, sorted in order }
		\Statex \Commentt{of decreasing utility.}
      % \Statex $\Rec$ \Comment{The set of all offers so far proposed by the opponent.}
		\Statex $\util_1$ \Comment{The agent's own utility function.}		
		\Statex $rv_1$ \Comment{The agent's own reservation value.}
    \Statex $\obsHist{1}$ \Comment{The observed negotiation history.}
    \Statex $\offRec$ \Comment{The offer last proposed by the opponent (if any).}
	\Statex   
	\State  $m \leftarrow \mathit{countUniqueOffersProposedByMe}(\obsHist{1})$ 
	\State  $n \leftarrow \mathit{countUniqueOffersProposedByOpponent}(\obsHist{1})$ 
	\Statex   
    \comment{If we have not proposed more unique offers than the opponent,}
    \commentt{and the next offer on our list is better than $\rv_1$, then we will}
    \commentt{propose a new offer.}
    \comment{We store this decision in a boolean variable $\mi{readyToConcede}$.}
    \State $\mi{readyToConcede} \ \ \leftarrow \ \  m \leq n$ \textbf{and} $\util_1(\mi{offers}[m]) > \rv_1$
    \Statex 
    	\comment{BIDDING STRATEGY}
    	\comment{If we are ready to concede then propose the next offer on the list.}
    \commentt{Otherwise, pick a random offer that we have already proposed before.}
    		\If{$\mi{readyToConcede}$}
    			\State $\offNext \leftarrow \mi{offers}[m]$ 
		\Else 
			\State $r \leftarrow \mi{getRandomInteger}(0,m)$ \hspace{5mm}\Comment{Pick a random integer $r$ with $0 \leq r < m$.}
			\State $\offNext \leftarrow \mi{offers}[r]$ 
		\EndIf	
		\Statex
		\comment{ACCEPTANCE STRATEGY}
			\comment{Determine the lowest utility we are willing to accept:}
		\If{$\mi{readyToConcede}$}
			\State $\asp \leftarrow \util_1(\mi{offers}[m])$ \quad \Comment{The utility of the offer we are about to propose next.}
		\Else
			\State $\asp \leftarrow \util_1(\mi{offers}[m-1])$ \quad  \Comment{The lowest utility among all offers we have already proposed.}
		\EndIf
		\State $\mi{acceptOffer} \ \ \leftarrow \ \ u(\offRec) \geq \asp$		
		\Statex
		\comment{RETURN SELECTED ACTION}
        \If{$\mi{acceptOffer}$}
            \State \textsc{Return} ($\acc$, $\offRec$) 
        \Else
            \State \textsc{Return} ($\prop$, $\offNext$)
        \EndIf
\end{algorithmic}
\end{algorithm}

\section{Acceptance Strategies}\label{sec:acceptance_strategies}
In the previous sections we have discussed a number of bidding strategies. In doing so, we also showed a number of different \textit{acceptance} strategies in the various examples (Algorithms \ref{alg:time_based}--\ref{alg:micro}). In this section we will discuss these acceptance strategies in a bit more detail.

In the following, let $\offNext$ denote the next offer to make, as decided by the bidding strategy, and let of $\offRec$ denote the last received offer.

Perhaps the most commonly used acceptance strategy in the literature is the $AC_{next}$ strategy that simply accepts $\offRec$ if and only if it is better than, or equal to $\offNext$:

\begin{definition}
The $AC_{next}$ acceptance strategy accepts if and only if:
\begin{equation}
\util_1(\offRec) \geq \util_1(\offNext)
\end{equation}
\end{definition}
At first sight, this makes perfect sense, because it simply let the bidding strategy do all the work to decide which offers our agent will consider acceptable. However, the problem with this strategy, is that it can lead to somewhat illogical decisions when the strategy is not purely monotonic. By `monotonic' we mean that the offers proposed by the agent keep always keep decreasing in terms of the utility for that agent. More precisely:
\begin{definition}
A bidding strategy for agent $\ag_i$ is \textbf{monotonic}, if for any negotiation history $\hist$, and any two proposals $(i,\prop, \off, t)\in \hist$, $(i,\prop, \off', t') \in \hist$ generated by that strategy  for which $t<t'$, we have $\util_i(\off) > \util_i(\off')$
\end{definition}
While each of the bidding strategies we discussed above \textit{in general} proposes offers in order of decreasing utility, it is certainly not the case that \textit{every} proposal is always followed by a proposal with lower utility. Therefore, none of these strategies are monotonic.

\begin{figure}
\tikzset{every picture/.style={line width=1pt}} %set default line width to 1pt        
\pgfplotsset{width=8cm}
\begin{center}
\begin{tikzpicture}[scale=1]
\begin{axis}[xmin=0.65, xmax=0.77,ymin=0,ymax=1,xlabel={Utility of Agent 1}, ylabel={\textit{Estimated} Utility of Agent 2}, 
		clip=false, %ensures that you can draw outside the boundaries.
		xtick pos=left,
		ytick pos=left		
		]
		
	%%OFFERS:
    \filldraw [black] 	

                     (0.7,0.3) circle [radius=2pt]
                     (0.715,0.4) circle [radius=2pt]
                     (0.73,0.2) circle [radius=2pt]
                     (0.68, 0.5)circle [radius=2pt]    
                      (0.75, 0.45)circle [radius=2pt]          
                  
                     (0.70, 0.8)circle [radius=2pt]
                 node[anchor=south,color=black] {$\off_2$}     
                     
                     (0.71, 0.6)circle [radius=2pt]
                node[anchor=south,color=black] {$\offRec$} 
                     
                     (0.73, 0.7)circle [radius=2pt]
                node[anchor=south,color=black] {$\off_3$} 
                     
                     (0.75, 0.8)circle [radius=2pt]
                 node[anchor=south,color=black] {$\off_1$} ;

    %% ASPIRATION VALUE:
    \draw[blue] (0.67,0) -- (0.67,1.03) node[anchor=south,color=black] {$\asp_1(t_3)$}; %vertical x=0.67
    \draw[blue] (0.69,0) -- (0.69,1.03) node[anchor=south west,color=black] {$\asp_1(t_2)$}; %vertical x=0.69
    \draw[blue] (0.74,0) -- (0.74,1.03) node[anchor=south,color=black] {$\asp_1(t_1)$}; %vertical x=0.74

\end{axis}
\end{tikzpicture}
\end{center}
\caption{The problem with $AC_{next}$. At $t_1$ agent 1 proposes $\off_1$, at $t_2$ agent 1 proposes $\off_2$, and at $t_3$ agent 1 has the choice between proposing $\off_3$ or accepting $\offRec$. According to $AC_{next}$, the agent should reject. However, this does not make sense, since he has already proposed $\off_2$ which is actually worse than $\offRec$.}\label{fig:ac_next_problem_example}
\end{figure}
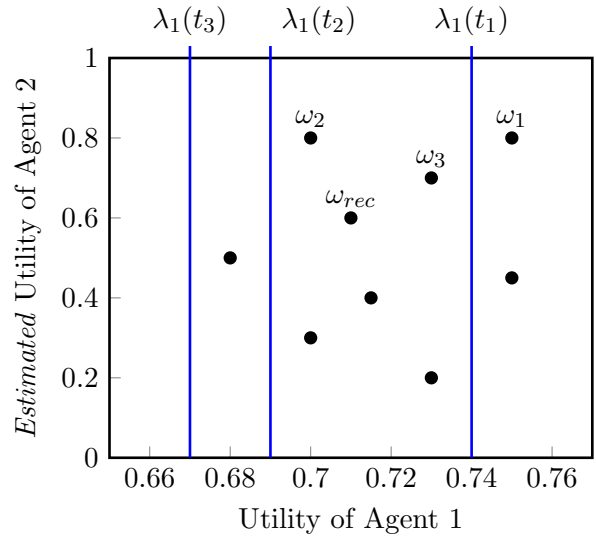

The problem with $AC_{next}$ and non-monotonic bidding strategies is illustrated in Figure~\ref{fig:ac_next_problem_example}. Before we explain the problem, we should first highlight a few important details about this figure. Firstly, note that the vertical axis does not represent $\ag_2$'s \textit{true} utility $\util_2$, but rather its \textit{estimated} utility $\est{\util}_2$, as estimated by agent 1's opponent modeling algorithm. Secondly, note that we have zoomed in a bit so that the horizontal axis shows only shows values between 0.65 and 0.77. Finally, note that we have drawn the aspiration levels of agent 1 in the diagram at three different times: $t_1$, $t_2$, and $t_3$, where $t_1 < t_2 < t_3$.

Now, let us suppose that our agent $\ag_1$ uses a time-based strategy, based on Equation~(\ref{eq:time_based_max}). Furthermore, suppose that at some time $t_1$ the aspiration level $\asp_1(t_1)$ of our agent is 0.74 and our agent proposes the offer $\off_1$ with utility $\util_1(\off_1) = 0.79$. Next, suppose that agent $\ag_2$ rejects this proposal, so after a small amount of time our agent gets to propose a new offer in the next turn, at time $t_2$. Meanwhile, our agent's aspiration level has dropped to, say, $\asp_1(t_2) = 0.69$. We see in the diagram that there are several offers with utility between 0.69 and  0.74 that can now be proposed but, according to Eq.~(\ref{eq:time_based_max}), our agent will propose the one with highest estimated opponent utility $\est{\util}_2$. This offer is denoted by $\off_2$ and we see that $\util_1(\off_2) = 0.7$. Again, suppose this offer is rejected and instead $\ag_2$ makes a counter-proposal, which is denoted $\offRec$ in the diagram, with utility $\util_1(\offRec) = 0.71$. Then, in the next turn, at time $t_3$, suppose the aspiration level has dropped to 0.67. Among all offers with $\util_1(\off) > 0.67$ that we have not proposed yet, the one with highest estimated opponent utility $\est{\util}_2$ is now $\off_3$, with utility $\util_1(\off_3) = 0.7$. So, the bidding strategy will select $\off_3$ to propose next.

Now, if our agent uses $AC_{next}$, it will compare $\offRec$ with $\off_3$. This means our agent will \textit{reject} $\offRec$, because $\off_3$ yields more utility. But this clearly does not make sense, because our agent has already proposed $\off_2$ which yielded less utility than $\offRec$. So, if our agent was willing to propose $\off_2$ with utility 0.7, then it should certainly be willing to accept $\offRec$ with utility 0.71. In fact, according to its aspiration level it should be willing to propose or accept any offer with utility higher than 0.67. 

Rejecting offer $\offRec$ only makes sense if our agent thinks it could obtain a better deal in the future, but if that's the case then our agent should have never proposed $\off_2$, and its aspiration level should not have dropped to 0.67.

The problem illustrated above can be resolved easily by using the aspiration level \textit{itself} to make the acceptance decision, rather than using the offer $\offNext$ that was chosen based on the aspiration level. Indeed, we used this acceptance strategy in Algorithms \ref{alg:time_based} and \ref{alg:adaptive}. We will denote this strategy by $AC_{asp}$.
\begin{definition}\label{def:ac_asp}
The $AC_{asp}$ acceptance strategy accepts if and only if:
\begin{equation}
\util_1(\offRec) \geq \asp(t)
\end{equation}
where $\asp$ is the aspiration function and $t$ is the time at which the decision is made.
\end{definition}
Of course, the problem with $AC_{asp}$ is that it only works if your bidding strategy indeed uses an aspiration function. For other bidding strategies, such as Tit-for-Tat or MiCRO, that do not make use of aspiration functions, there is another straightforward solution. Namely, to accept any offer that is better than the offer you are going to propose next, \textit{or} better than any of the offers you have already proposed before. We will denote this strategy by $AC_{low}$.
\begin{definition}
The $AC_{low}$ acceptance strategy accepts if and only if:
\begin{equation}
\util_1(\offRec) \geq \min \{\util_1(\off) \mid \off \in  \Pro_t \cup \{\offNext\}\}
\end{equation}
where $t$ is the time at which the decision is made and $\Pro_t$ denotes the set of offers so far proposed by our agent (as defined by Eq.~(\ref{eq:Off_prop})).
\end{definition}
Note that we used this acceptance strategy in our implementation of MiCRO in Algorithm \ref{alg:micro} (although this may not be immediately obvious from the notation). \later{To see this, note that...}

The strategies $AC_{next}$, $AC_{asp}$ and $AC_{low}$ are all based on the same principle: only accept an offer if you would also be willing to \textit{propose} that same offer yourself. While this principle makes sense, it may be somewhat too strict when the negotiations are close to the deadline. In that case it can be beneficial to even accept offers that are actually somewhat less valuable than those offers that you are willing to propose.

The idea is that near the deadline, proposing an offer is more risky than accepting an offer, because an acceptance yields a guaranteed amount of utility, while a proposal could be rejected by the opponent, so it brings along the risk that the negotiations may fail. The closer we get to the deadline, the more important this risk becomes.

Therefore, one could argue that when you decide to make a proposal, you should ask for a bit more utility than what you would be willing to accept, in order to offset the increased risk. This can modeled by a parametrized version of $AC_{next}$ \cite{Baarslag2013AcceptanceConditions}, which has two parameters $\alpha$ and $\beta$ and which is denoted by $AC_{next}(\alpha, \beta)$.
\begin{definition}Let $\alpha, \beta \in \mathbb{R}$ be two real numbers. Then the $AC_{next}(\alpha, \beta)$ acceptance strategy accepts if and only if:
\begin{equation}
\alpha \cdot \util_1(\offRec) + \beta \geq \util_1(\offNext)
\end{equation}
\end{definition}
Note that if $\alpha=1$ and $\beta=0$, then  $AC_{next}(\alpha, \beta)$ is just identical to $AC_{next}$. Typically, the values of $\alpha$ and $\beta$ would both be non-negative. While there is no mathematical reason why they could not be negative, there does not seem to be any obvious reason to ever consider such values. After all, it does not make a lot of sense to propose an offer with a utility of, say, $\util_1(\off) = 0.6$ if you are not willing to accept an offer with that same amount of utility, or better. The same generalization can also be applied to $AC_{asp}$ or $AC_{low}$. That is, we could define $AC_{asp}(\alpha, \beta)$ or $AC_{low}(\alpha, \beta)$ in an analogous manner. Of course, an obvious disadvantage of such parameterized strategies, is that it requires choosing the right values of $\alpha$ and $\beta$, which may be difficult.

Another reason why it could be advantageous for our agent to accept offers that yield less utility than the offers it is willing to propose, is that this would allow our agent to apply a very \hard{} bidding strategy, in order to entice the opponent to make large concessions, while at the same time it still allows our agent to come to an agreement in case the opponent is not willing to make such concessions. In other words, it allows our agent to pretend to be more \hard{} than what he really is.

\later{Here we could refer to Tim's alternative acceptance strategy.}
	
\begin{exercise}\label{ex:ac_low}
$\mathbf{AC_{low}}$. Adapt the implementation of your Tit-for-Tat agent from Exercise~\ref{ex:tit_for_tat} to apply the $AC_{low}$ acceptance strategy instead of $AC_{next}$.
\end{exercise}

\section{Reproposing}\label{sec:reproposing}

We will now discuss a simple technique that can be added on top of any of the previously described negotiation strategies, that can make them somewhat better. This approach was described, for example, in \cite{Williams2011GaussianProcesses}, \cite{deJonge2016gangster} and  \cite{deJonge2024theoreticalPropertiesMicro}.

Let us explain it with an example. Suppose that we have a negotiation domain with 10 possible offers: $\Off = \{\off_1, \off_2, \dots, \off_{10} \}$ and suppose that our agent's utility function is given by $\util_1(\off_j) = 0.1 j$. That is, $\util_1(\off_1) = 0.1$, $\util_1(\off_2) = 0.2$, etcetera, so our agent's most preferred offer is $\off_{10}$. Furthermore, suppose that our agent $\ag_1$ follows a time-based strategy with a linear aspiration function ($\gamma=1)$ and without opponent modeling, as given by Eq.~(\ref{eq:time_based_min}).

\begin{figure}
\begin{center}
\begin{tikzpicture}
\begin{axis}[grid=both,
          xmin=0, xmax=1, ymin=0, ymax=1,
          xlabel={Time},
          ylabel={Utility of Agent 1},
          enlargelimits]
%%%%
%\addplot[green,domain=0:1,samples=100]{(1 - pow(10,1-x))/(1-10)} node[pos=0.5,anchor=north east]{$\gamma=10$};
%%%%
\addplot[red,domain=0:1,samples=100]{(1 - x)} node[pos=0.8,anchor= south west]{$\asp_1$};
%%%
%\addplot[blue,domain=0:1,samples=100]{(1 - pow(0.01, 1-x))/(1-0.01)} node[pos=0.5,anchor= south west]{$\gamma=1/100$};
\filldraw [red] (0.0,1.0) circle [radius=2pt] node[anchor=south west]{$\off_{10}$};
\filldraw [black] (0.05,0.4) circle [radius=2pt]
node[anchor=south]{$\off_{4}$};
\filldraw [red] (0.1,0.9) circle [radius=2pt]
node[anchor=south west]{$\off_{9}$};
\filldraw [black] (0.15,0.6) circle [radius=2pt]
node[anchor=south]{$\off_{6}$};
\filldraw [red] (0.2,0.8) circle [radius=2pt]
node[anchor=south west]{$\off_{8}$};
\filldraw [black] (0.3,0.2) circle [radius=2pt]
node[anchor=south]{$\off_{2}$};
\draw [red] (0.5,0.5) circle [radius=2pt]
node[anchor=south west]{$\off_{5}$};
\end{axis}
\end{tikzpicture}
\end{center}
\caption{The benefit of reproposing. The red dots represent proposals made by $\ag_1$, the red line represents $\ag_1$'s aspiration level $\asp_1$ as a function of time, and the black dots represent proposals made by $\ag_2$. At time $t=0.5$, the bidding strategy of $\ag_1$ suggests to propose $\off_5$. However, it makes more sense for  $\ag_1$ to propose $\off_6$, which earlier was already proposed by $\ag_2$. Note that at that time $\ag_1$ cannot \textit{accept} $\off_6$, because the AOP only allows accepting the last received proposal, which was $\off_2$.}\label{fig:reproposing}
\end{figure}
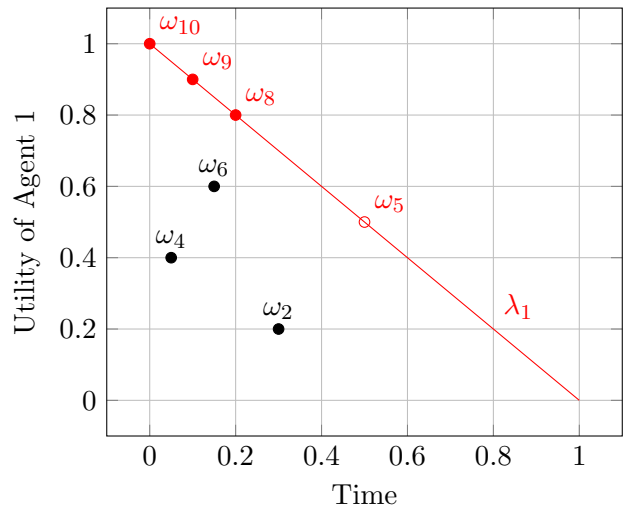

Now, suppose that, from the point of view of $\ag_1$,  the negotiations proceed as follows (see also Figure \ref{fig:reproposing}):

\begin{tabular}{lllll}
 & &  & & \\
1. & At $t=0.0$: & $\asp_1(t) = 1.0$ & $\ag_1$ proposes $\off_{10}$ & \ \\
2. & At $t=0.05$: & \ & \ & $\ag_2$ proposes $\off_{4}$\\
3. & At $t=0.10$: & $\asp_1(t) = 0.9$ & $\ag_1$ proposes $\off_{9}$ & \\
4. & At $t=0.15$: & \ & \ & $\ag_2$ proposes $\off_{6}$\\
5. & At $t=0.20$: & $\asp_1(t) = 0.8$ & $\ag_1$ proposes $\off_{8}$ & \\
6. & At $t=0.30$: & \ & \ & $\ag_2$ proposes $\off_{2}$\\
7. & At $t=0.50$: & $\asp_1(t) = 0.5$ & $\ag_1$ proposes ... & \\
 & &  & & \\
\end{tabular}

At time $t=0.50$, our agent's strategy prescribes that it should propose $\off_5$. Ideally, however, $\ag_1$ would like to accept $\off_6$ instead, because that would yield more utility. The problem is that the AOP does not allow that, because it only allows accepting the \textit{last} received offer, which is $\off_2$. Note that earlier our agent did not accept $\off_6$, because at the moment he received that offer, his aspiration level was still at $\asp_1(t) = 0.8$ which was greater than $\util_1(\off_6) = 0.6$. 

The solution, is to override the bidding strategy and propose $\off_6$ instead of $\off_5$. Since $\off_6$ was already proposed before by $\ag_2$, it is very likely that $\ag_2$ will now accept it, and therefore it should indeed be better for $\ag_1$ to propose $\off_6$, than to propose $\off_5$. We call this \textit{reproposing} because the agent is proposing an offer that was already proposed earlier by the opponent. Algorithm \ref{alg:reproposing} shows how this technique can be implemented on top of any generic agent.

%While there is no guarantee that $\ag_2$ will accept it, it is very likely that she will in fact do that, since she has already proposed $\off_6$ herself.

%Ideally, we would now like our agent to accept that offer, but the AOP does not allow that, since it only allows to accept the \textit{last} proposal. Instead, however, what we \textit{can} do is to propose $\off_6$ back to the opponent. 

%We call this \textit{reproposing} because the agent is proposing an offer that was already proposed earlier by the opponent.

%that at some point during the negotiations our agent has so far proposed $\off_{10}$, and $\off_9$. Next, the opponent $\ag_2$ proposes $\off_6$. After receiving this proposal $\ag_1$ calculates its aspiration level, and let's say that it is $\asp(t) = 0.8$. This means that $\ag_1$ currently does not yet consider $\off_6$ good enough to be accepted, because it would only yield $\util_1(\off_6) = 0.6$. So, instead $\ag_1$ proposes $\off_8$. Next, $\ag_2$ proposes $\off_2$, which our agent rejects, and now let us suppose that our agent's aspiration level has dropped to $\asp(t) = 0.5$. This is not low enough to accept the last received proposal $\off_2$, so $\ag_1$ rejects it. 

\begin{definition}
We say an agent $\ag_i$ \textbf{reproposes} an offer $\off$ if $\ag_i$ proposes it, while it was earlier already proposed by the other agent $\ag_j$ and $\ag_i$ itself has not yet proposed it since then.
\end{definition}

\begin{exercise}\label{ex:reproposing}
\textbf{Reproposing} Adapt the agents that you have implemented in the previous exercises to make them apply the reproposing technique, as described in Algorithm~\ref{alg:reproposing}.
\end{exercise}

\begin{algorithm}
\caption{Generic BOA Agent for the alternating offers protocol that applies reproposing.}\label{alg:reproposing}
\begin{algorithmic}[1]
	\Statex \textbf{Input:} 
	\Statex $\Off$ \Comment{The offer space.}
	\Statex $\util_1$ \Comment{The agent's own utility function.}
	\Statex $\rv_1$ \Comment{The agent's own reservation value.}
	\Statex $\dead$ \Comment{The deadline.}
	\Statex $\om$ \Comment{A model of the opponent.}
	\Statex $t$ \Comment{The current time.}
    \Statex $\obsHist{1}$ \Comment{The observed negotiation history.}
    \Statex $\offRec$ \Comment{The offer last proposed by the opponent (if any).}
    	\Statex
    	\comment{OPPONENT MODELING}
    		\State $\om \leftarrow \mi{updateOpponentModel}(\Off,  \dead, \om, t, \offRec)$
    		\Statex 
    		\comment{BIDDING STRATEGY}
    		\State $\offNext \leftarrow biddingStrategy(\Off, \util_1,\rv_1, \dead, \om, t, \obsHist{1})$ 
    		\Statex
    		\comment{CHECK IF WE CAN FIND A BETTER OFFER TO REPROPOSE}
    		\comment{From the negotiation history, extract the set of all offers that have}
    		\commentt{so far been proposed by this agent:}
    		\State $\Pro \leftarrow \mi{getProposedOffers(\obsHist{1})}$
    		\comment{From the negotiation history, extract the set of all offers that have}
    		\commentt{so far been proposed by the opponent:}
    		\State $\Rec \leftarrow \mi{getReceivedOffers(\obsHist{1})}$
    		\comment{See if we can find any offer that can be reproposed:}
    		\If{$\Rec \setminus \Pro \neq \emptyset$}
    			\State $\off_{rep} \leftarrow \argmax \{\util_1(\off) \mid \off \in \Rec \setminus \Pro\}$
    			\If{$\util_1(\off_{rep}) \geq \util_1(\offNext)$}
    				\State $\offNext \leftarrow \off_{rep}$
    			\EndIf
    		\EndIf
    		\Statex
    		\comment{ACCEPTANCE STRATEGY}
    		\State $\mi{acceptOffer} \leftarrow acceptanceStrategy(\Off, \util_1, \dead, \om, t, \obsHist{1},\offRec, \offNext)$
    		\Statex
    		\comment{RETURN SELECTED ACTION}
    		\comment{Finally, return the selected action (accept or propose):}
        \If{$\mi{acceptOffer}$}
            \State \textsc{Return} ($\acc$, $\offRec$) 
        \Else
            \State \textsc{Return} ($\prop$, $\offNext$)
        \EndIf
\end{algorithmic}
\end{algorithm}

\section{Summary of Chapter}
\begin{itemize}
\item We say an agent is \hard{} if it concedes very little to the opponent and keeps demanding high utility for itself. Such an agent will not make many agreements, but when it does, it will typically be a very profitable agreement.
\item We say an agent is \soft{} if it concedes a lot to the opponent and quickly lowers the amount of utility it demands for itself. Such an agent will come to an agreement more often, but those agreement will typically be less profitable.
\item A good negotiation strategy, therefore, is one that strikes the right balance between being \soft{} and being \hard{}.
	 
\item Negotiation algorithms are typically described in terms of three different components:
\begin{itemize}
	\item  A bidding strategy.
	\item  Opponent modeling strategies (an opponent strategy model and/or an opponent utility model).
	\item  An acceptance strategy.
\end{itemize}
This is known as the BOA model.
\item Bidding strategies determine when to propose which offer.
\item Opponent strategy models aim to estimate how far the opponent is willing to concede, based on the proposals he has so far made.
\item Opponent utility models aim to estimate the opponent's utility function, based on the proposals he has so far made.
\item Acceptance strategies determine when to accept which offer.
\item The literature describes three main types of bidding strategies:
	\begin{itemize}
	\item Time-based strategies.
	\item Adaptive strategies.
	\item Imitative strategies.
	\end{itemize}
\item For each of these types, the agent starts by proposing an offer with high utility for itself, but as the negotiations progress, the agent proposes offers with lower and lower utility for itself.
\item Time-based strategies do not adapt to their opponent, and only base their strategy on the amount of time that has passed.
\item They use a so-called aspiration function, which decreases over time, to determine, at any time $t$, the minimum utility it will demand.
\item To implement a time-based strategy, we need to make the following choices:
		\begin{itemize}
		\item An expression for the aspiration function (e.g. `polynomial' or `exponential')
		\item The parameters of the aspiration function:
		\begin{itemize}
			\item  The initial value $\alpha$.
			\item  The target value $\targ$.
			\item  The concession parameter $\gamma$.
		\end{itemize}
		\item How to select the next offer (e.g. maximizing estimated opponent utility as in Eq.~(\ref{eq:time_based_max}), or minimizing your own utility, as in Eq.~(\ref{eq:time_based_min})).
		\end{itemize}
\item Using Eq.~(\ref{eq:time_based_max}) is better, provided that you have an accurate opponent utility model.
\item However, Eq.~(\ref{eq:time_based_min}) has the advantage that you don't need any such opponent utility model at all.
\item A high value of $\targ$ creates a very \hard{} agent, while a low value of $\targ$ creates a very \soft{} agent.

\item Adaptive strategies aim to adapt perfectly to the opponent.
\item Their implementation is very similar to time-based strategies, except that their target value $\targ$ is not constant, but instead use an opponent strategy model in order to regularly adjust the target value during the negotiations.
\item Adaptive strategies are \soft{} when they negotiate against a \hard{} agent, and are \hard{} when they negotiate against a \soft{} opponent.
\item In theory, if they have access to a perfect opponent strategy model, then adaptive strategies are the optimal strategy against a time-based agent.
	
\item Imitative strategies aim to copy the behavior of the opponent, in order to entice the opponent to concede more.
\item We discussed two types of imitative strategies: 
	\begin{itemize}
		\item Classic Tit-for-Tat
		\item MiCRO
	\end{itemize}
\item A classic tit-for-tat agent tries to measure the amount of concession made by the opponent and then aims to concede an equal amount.
\item Classic tit-for-tat agents are \soft{} when they play against a \soft{} agent, and are \hard{} when they play against a \hard{} opponent.
\item To implement a classic tit-for-tat agent we need to make three decisions:
	\begin{itemize}
		\item How to measure the opponent's concession.
		\item How to measure our agent's own concession.
		\item How to select the next offer (e.g. using Eq.~(\ref{eq:tft_selfish}) or Eq.~(\ref{eq:tft_altru})).
	\end{itemize}
	
\item The MiCRO strategy is based on the assumption that our agent knows absolutely nothing about the opponent's utility function, and that the opponent knows nothing about our agent's utility function either.
\item Despite its simplicity, MiCRO has shown very strong performance against several top agents from various ANAC competitions.
\item MiCRO doesn't require any opponent modeling and does not require choosing any parameter values.
\item For these reasons, MiCRO is especially useful as a benchmark strategy.
	 
\item The $AC_{next}$ strategy is one of the most commonly used acceptance strategies.
\item However, there are two alternative acceptance strategies that improve over $AC_{next}$: 
	\begin{itemize}
	 \item $AC_{asp}$ can be used when the bidding strategy  uses an aspiration function.
	\item $AC_{low}$ can be used in other cases.
	\end{itemize}
\item Furthermore, there are some reasons to believe that agents may improve if they use an acceptance strategy that accepts offers with slightly lower utilities than what the acceptance strategies above would accept.
	 
\item Reproposing is a simple technique that can be used as an improvement on top of any of the bidding strategies mentioned before. The idea is that whenever 
	 the bidding strategy of our agent selects some offer $\off_{next}$, but there is another offer $\off$ with higher utility for our agent that was already proposed by the opponent before, then our agent should propose that offer $\off$ instead of $\off_{next}$.
\end{itemize}

\chapter{Opponent Modeling}\label{sec:opponent_modeling}

In this chapter we will discuss various techniques that have been proposed in the literature to model the opponent. Readers who are not interested in the details of such opponent modeling algorithms can safely skip this chapter, since the rest of this book does not depend on it.

We can distinguish between three types of opponent modeling:
\begin{enumerate}
\item Learning the opponent's utility function, during the negotiation.
\item Learning the opponent's strategy, during the negotiation.
\item Learning the opponent's strategy from earlier negotiations.
\end{enumerate}
We will discuss each of these types respectively in the following three sections.

Note that we do not discuss learning the opponent's utility function from earlier negotiations, because in most scenarios studied in the literature the utility function would change with every new negotiation, so this wouldn't make much sense.

\section{Learning the Opponent's Utility Function}\label{sec:opp_util_modeling}
In this section we will discuss several techniques that can be used by our agent to learn the opponent's utility function, based on the proposals that it receives from its opponent.

Specifically, we will discuss the following techniques:
\begin{enumerate}
\item Bayesian learning
\item Scalable Bayesian learning
\item Frequency Analysis
\end{enumerate}
We should note that each of these techniques assumes the negotiations take place over a multi-issue domain and that the opponent's utility function $\util_2$ is linear, so it is of the form of Eq.~(\ref{eq:linear_util_func}). Therefore, these techniques are not applicable to other types of negotiation domains.

\subsection{Bayesian Learning}

Bayesian learning \cite{Hindriks2008Bayesian} is one of the earliest and still most commonly used techniques in automated negotiation to learn the opponent's utility function. 

The idea is as follows. Suppose that we have some given set of possible utility functions $\Util$ and, based on the proposals $\pr_1, \pr_2, \dots, \pr_\numObs$ that our agent has so far received from its opponent, we want to calculate the probability, for each function $\util \in \Util$, that that function $\util$ is the actual utility function $\util_2$ of the opponent. That is, for each $\util \in \Util$ we want to calculate a probability $P(\util_2 = \util | \pr_1, \pr_2, \dots, \pr_\numObs)$.

\subsubsection{Bayesian Learning in General}\label{sec:bayesian_in_general}
Bayesian learning is a technique that is much older than automated negotiation and it has been used in many other applications. So, before we explain how it can be applied to automated negotiation, we will here first explain how it works in general.

The goal of Bayesian learning is, given a set of hypotheses $\Hypo$, a sequence of observations $\vec{\obs} = (\obs_1, \obs_2, \dots, \obs_k)$, and a \textit{prior probability} $P(\hypo)$ for each hypothesis $\hypo \in \Hypo$, to calculate the \textit{posterior probability} $P(\hypo | \vec{\obs})$ that the hypothesis $\hypo$ is true. Here, $P(\hypo)$ denotes the probability that we assign to hypothesis $\hypo$ \textit{before} making any observations, while $P(\hypo | \vec{\obs})$ represents the probability we assign to $\hypo$ \textit{after} making the observations $\obs_1, \obs_2, \dots, \obs_k$.

For example, suppose that somebody draws a card from a standard deck of 52 playing cards, without showing it to us. Then, for us, the prior probability that this card is the ace of spades would be $P(A\spadesuit) = \frac{1}{52}$. Next, suppose that this person tells us that the card is indeed a \textit{spades} card. Now, with this new information, the probability for us that it is the \textit{ace} of spades is suddenly four times higher: $P(A\spadesuit \mid \spadesuit) = \frac{1}{13}$.

In this example it was straightforward to calculate $P(\hypo |  \obs)$ directly. However, in practice, it often happens that it is much easier to calculate $P(\obs|\hypo)$ instead. In such cases we can use a theorem known as \textit{Bayes' rule} to express $P(\hypo |  \obs)$ in terms of $P(\obs|\hypo)$ and $P(\hypo)$.

It is important to understand that we always assume that there is exactly one hypothesis in $\Hypo$ that is true. Therefore, we always have:
\[\sum_{\hypo \in \Hypo} P(\hypo) = 1 \quad \text{and} \quad \sum_{\hypo \in \Hypo} P(\hypo | \vec{\obs} ) = 1 \]

To derive Bayes' rule, we start from the following identities, which are well-known from basic probability theory, and which hold for any arbitrary `events' $y$ and $o$:
\begin{equation}\label{eq:prob_1}
P(\hypo,\obs) \quad = \quad P(\hypo \mid \obs) \cdot P(\obs) \quad = \quad P(\obs \mid \hypo ) \cdot P(\hypo)
\end{equation}
\begin{equation}\label{eq:prob_2}
P(\obs) \quad = \quad \sum_{\hypo' \in \Hypo} P(\obs \mid \hypo') \cdot P(\hypo')
\end{equation}
From Equation (\ref{eq:prob_1}) we can directly derive:
\begin{eqnarray*}
P(\hypo \mid \obs)  &=& \frac{P(\obs \mid 
\hypo) \cdot P(\hypo)}{P(\obs)}\\
\end{eqnarray*}
and then using Equation (\ref{eq:prob_2}) we obtain Bayes' rule:
\begin{eqnarray*}
P(\hypo \mid \obs) &=& \frac{P(\obs \mid 
\hypo) \cdot P(\hypo)}{\sum_{\hypo'\in \Hypo} P(\obs \mid 
\hypo') \cdot P(\hypo')}
\end{eqnarray*}
Note that indeed, this rule allows us to express $P(\hypo | \obs)$ on the left-hand side in terms of $P(\obs | \hypo) $ and $P(\hypo)$ on the right-hand side.

If there are multiple observations $\obs_1, \obs_2, \dots, \obs_\numObs$, then this becomes:
\begin{equation}\label{eq:bayes}
P(\hypo \mid \obs_1, \obs_2, \dots,  \obs_\numObs) = \frac{P(\obs_1, \obs_2, \dots, \obs_\numObs \mid 
\hypo) \cdot P(\hypo)}{\sum_{\hypo'\in \Hypo} P(\obs_1, \obs_2, \dots, \obs_\numObs \mid 
\hypo') \cdot P(\hypo')}
\end{equation}
and if it holds that for any given hypothesis $\hypo$, the probabilities of observations $\obs_1, \obs_2, \dots, \obs_{\numObs}$, are all independent, then we can write this as:
\begin{equation}\label{eq:bayes_indep}
P(\hypo | \obs_1, \obs_2, \dots,  \obs_\numObs) = \frac{P(\obs_1 | \hypo) \cdot P(\obs_2|\hypo)\cdot \ \dots \ \cdot P(\obs_\numObs|\hypo) \cdot P(\hypo)}{\sum_{\hypo'\in \Hypo} P(\obs_1 | \hypo') \cdot P(\obs_2|\hypo')\cdot \ \dots \ \cdot P(\obs_\numObs|\hypo') \cdot P(\hypo')}
\end{equation}

Now, suppose that we have already calculated, for each hypothesis $\hypo \in\Hypo$, the probability $P(\hypo | \obs_1, \obs_2, \dots,  \obs_\numObs)$,  which we will denote as $P(\hypo | \vec{\obs})$. Next, suppose we make a new observation $\obs_{\numObs+1}$. We now want to update the probability of each hypothesis, taking into account this new observation. That is, for all $\hypo \in \Hypo$ we now want to calculate $P(\hypo | \vec{\obs}, \obs_{\numObs+1})$, given $P(\hypo | \vec{\obs})$.

To do this, first note that the denominator of Eq.~(\ref{eq:bayes_indep}) is just a normalization constant that ensures that the sum of all probabilities equals~1, which is the same for every hypothesis $\hypo \in \Hypo$. Ignoring this constant for a moment, we can define the \textit{unnormalized} probability $\tilde{P}(\hypo|\vec{\obs})$ as:
\begin{equation}\label{eq:p_tilde}
\tilde{P}(\hypo \mid \vec{\obs}) \ \ := \ \  P(\hypo) \cdot P(\obs_1 \mid \hypo) \cdot P(\obs_2 \mid \hypo) \cdot \ \dots \ \cdot P(\obs_\numObs \mid \hypo) 
\end{equation}
which is just the numerator of the right-hand side of Eq.~(\ref{eq:bayes_indep}).

We now see that to update this unnormalized probability  after a new observation $\obs_{\numObs+1}$ we just need to multiply it with $P(\obs_{\numObs+1}\mid \hypo)$. That is:
\begin{equation}\label{eq:bayesian_update}
\tilde{P}(\hypo \mid \vec{\obs}, \obs_{\numObs+1}) \ \  = \ \  \tilde{P}(\hypo \mid \vec{\obs}) \cdot P(\obs_{\numObs+1}\mid \hypo)
\end{equation}
Then, after we have done this for every possible hypothesis $\hypo \in \Hypo$ we can calculate the true probabilities $P(\hypo \mid \vec{\obs}, \obs_{\numObs+1})$ by normalizing:
\begin{equation}\label{eq:bayesian_normalizing}
P(\hypo \mid \vec{\obs}, \obs_{\numObs+1}) \ \ = \ \  \frac{\tilde{P}(\hypo \mid \vec{\obs}, \obs_{\numObs+1})}{\sum_{\hypo'\in \Hypo}\tilde{P}(\hypo' \mid \vec{\obs}, \obs_{\numObs+1})}
\end{equation}

\subsubsection{Implementation}
We will here discuss how the calculations discussed above can be implemented.

First determine, for every $\hypo \in \Hypo$, the prior probability $P(\hypo)$. Since initially we haven't made any observations yet, $\vec{\obs}$ will be empty and thus by Eq.~(\ref{eq:p_tilde}) we have $\tilde{P}(\hypo \mid \vec{\obs}) = P(\hypo)$, for all  $\hypo \in \Hypo$.

Then, every time we make a new observation $\obs_{\numObs+1}$, we take the following steps:
\begin{enumerate}
\item For each $y \in Y$, calculate: 
\[\tilde{P}(\hypo \mid \vec{\obs}, \obs_{\numObs+1}) = \tilde{P}(\hypo \mid \vec{\obs}) \cdot P(\obs_{\numObs+1}\mid \hypo)\]
\item Calculate the sum:
\[S = \sum_{\hypo\in \Hypo}\tilde{P}(\hypo \mid \vec{\obs}, \obs_{\numObs+1})\]
\item For each $y \in Y$, calculate: 
\[P(\hypo \mid \vec{\obs}, \obs_{\numObs+1})  = \frac{1}{S}\cdot \tilde{P}(\hypo \mid \vec{\obs}, \obs_{\numObs+1})\]
\end{enumerate}
Note that this requires two lists of size $|\Hypo|$ each: one list to store all the values of $\tilde{P}(\hypo \mid \vec{\obs})$ and one to store the values of $P(\hypo \mid \vec{\obs})$. However, this can be done a bit more efficiently. To see how, first note that we can modify the implementation as follows.

Every time we make a new observation $\obs_{\numObs+1}$, we take the following steps:
\begin{enumerate}
\item Pick an arbitrary number $C_{\numObs+1}$.
\item For each $\hypo \in \Hypo$, calculate: 
\[\tilde{P}(\hypo \mid \vec{\obs}, \obs_{\numObs+1}) =  \tilde{P}(\hypo \mid \vec{\obs}) \cdot P(\obs_{\numObs+1} \mid \hypo) \cdot C_{\numObs+1} \]
\item Calculate the sum:
\[S = \sum_{\hypo\in \Hypo}\tilde{P}(\hypo \mid \vec{\obs}, \obs_{\numObs+1})\]
\item For each $\hypo \in \Hypo$, calculate: 
\[P(\hypo \mid \vec{\obs}, \obs_{\numObs+1})  = \frac{1}{S}\cdot \tilde{P}(\hypo \mid \vec{\obs}, \obs_{\numObs+1})\]
\end{enumerate}
Note that the fact that in Step 2 each $\tilde{P}(\hypo \mid \vec{\obs}, \obs_{\numObs+1})$ is multiplied by a constant $C_{\numObs+1}$ does not affect the correctness of the calculations, because it means the sum $S$ in Step 3 will also be multiplied by the same constant, which means that in step 4 this constant will cancel out against itself.

Furthermore, note that every time we make a new observation we can choose a different value for this constant, and that instead of Eq.~(\ref{eq:p_tilde}), we are now calculating the unnormalized probability $\tilde{P}(\hypo \mid \vec{\obs})$ as:
\begin{equation}\label{eq:p_tilde_alt}
\tilde{P}(\hypo | \vec{\obs}) = P(\hypo) \cdot C_1\cdot P(\obs_1 | \hypo) \cdot C_2 \cdot P(\obs_2|\hypo) \cdot \ \dots \ \cdot C_{\numObs}\cdot P(\obs_\numObs|\hypo) 
\end{equation}
This means that if we choose each $C_{\numObs+1}$ as follows:
\begin{equation}\label{eq:bayesian_constant}
C_{\numObs+1} = \frac{1}{\prod_{i=1}^\numObs C_\numObs} \cdot \frac{1}{\sum_{\hypo'\in\Hypo}P(\hypo' \mid \vec{\obs})}
\end{equation}
then, by combining Eq.~(\ref{eq:p_tilde_alt}) and Eq.~(\ref{eq:bayesian_constant}) with Eq.~(\ref{eq:bayes}), we see that for every $\hypo\in \Hypo$ we now have:
\[C_{\numObs+1} \cdot \tilde{P}(\hypo \mid \vec{\obs}) = P(\hypo \mid \vec{\obs})\]

Knowing this, we can simplify our implementation, since it is now equivalent to the following:
\begin{enumerate}
\item For each $\hypo \in \Hypo$, calculate: 
\begin{equation}\label{eq:bayesian_update_alt}
\tilde{P}(\hypo \mid \vec{\obs}, \obs_{\numObs+1}) = P(\hypo \mid \vec{\obs}) \cdot P(\obs_{\numObs+1}\mid \hypo)
\end{equation}
\item Calculate the sum:
\[S = \sum_{\hypo\in \Hypo}\tilde{P}(\hypo \mid \vec{\obs}, \obs_{\numObs+1})\]
\item For each $\hypo \in \Hypo$, calculate: 
\[P(\hypo \mid \vec{\obs}, \obs_{\numObs+1})  = \frac{1}{S}\cdot \tilde{P}(\hypo \mid \vec{\obs}, \obs_{\numObs+1})\]
\end{enumerate}
While this looks very similar to our original implementation, the difference is that step 1 now involves $P(\hypo \mid \vec{\obs})$, rather than $\tilde{P}(\hypo \mid \vec{\obs})$. The great advantage of this, is that we now only need one list of size $|\Hypo|$. In Step~1 we can use this list to store the values of $\tilde{P}(\hypo \mid \vec{\obs}, \obs_{\numObs+1})$ and then in Step~3 we can simply overwrite it to store the values of $P(\hypo \mid \vec{\obs}, \obs_{\numObs+1})$. In our initial implementation this was not possible, because we needed to keep the values of $\tilde{P}(\hypo \mid \vec{\obs}, \obs_{\numObs+1})$ for the next iteration. Also note that we do not actually need to calculate the constants $C_{\numObs+1}$, since this last implementation does not use them. We only mentioned these constants and Eq.~(\ref{eq:bayesian_constant}) to show the correctness of the last implementation.

\subsubsection{Bayesian Learning for Automated Negotiation}
We will now explain how Bayesian Learning can be applied in automated negotiation to learn the utility function of the opponent.

In general, to apply Bayesian learning, we need the following ingredients:
\begin{itemize}
\item A set of possible observations $\Obs$.
\item A set of hypotheses $Y$.
\item For any hypothesis $\hypo\in \Hypo$, a prior probability $P(\hypo)$.
\item A formula that allows us to calculate, for any hypothesis $\hypo \in \Hypo$, and any observation $\obs \in \Obs$, the probability $P(\obs \mid \hypo)$.
\end{itemize}

In the context of automated negotiation, the observations that our agent makes are the proposals that it receives from the opponent. Recall that such a proposal $\pr$ is defined as a tuple of the form $(2,\prop, \off, t)$ for some offer $\off$ and some time $t$. So we have:
\[\Obs = \{(2,\prop, \off, t) \mid \off \in \Off, t\in [0,\dead]\}\]

The set of hypotheses would be some set of possible utility functions $\Util$ for the opponent. To stress that each hypothesis is now a utility function, we will from now on use the symbol $\Util$ to denote the set of hypotheses instead of $\Hypo$. We will discuss how to choose this set of  utility functions below in Section~\ref{sec:choosing_utilty_hyps}.

%So, Eq.(\ref{eq:bayes}) becomes:
%\begin{equation}\label{eq:bayes_nego}
%P(\util \mid \hist) = \frac{P(\hist \mid 
%\util) \cdot P(\util)}{\sum_{\util'\in \Util} P(\hist \mid 
%\util') \cdot P(\util')}
%\end{equation}
%Of course, the history $\hist$ does not only contain the opponent's proposals, but also our agent's own proposals, but that does not make any difference for the validity of this equation.

For the prior probabilities, the simplest approach is to assign all hypotheses the same prior probability. That is: $P(\util) = \frac{1}{|\Util|}$. However, depending on the domain of application, you could also choose different prior probabilities that take into account some background knowledge you may have about that specific application.

Finally, we need to determine how to calculate $P(\pr |
\util)$ for any arbitrary proposal $\pr \in \Obs$ and utility function $\util \in \Util$. That is, we have to make an assumption about which proposals the opponent would make, if he had utility function $\util$. In other words, we have to make some assumptions about his strategy. In order to do this, the authors of \cite{Hindriks2008Bayesian} modeled the opponent's strategy as a linear time-based strategy. So, at any time $t$ they \textit{expect} the opponent to propose an offer $\off$ with normalized utility $\util_2(\off)=1-c\cdot \frac{t}{\dead}$, where $c$ is some constant between 0 and 1. However, since this is of course not guaranteed to be exactly true, they assumed the opponent's \textit{actual} proposal at any time $t$ was drawn from the following probability distribution function:
\begin{equation}\label{eq:p_o_given_u}
P((2,\prop,\off,t) \mid \util) \ \ = \ \ \mc{N}(\util(\off) \mid 1-c\cdot\frac{t}{\dead} \ ,\ \sigma) 
\end{equation}
where the notation $\mc{N}(r | \mu, \sigma)$ represents the probability of drawing the number $r$ from a Gaussian probability distribution with mean $\mu$ and standard deviation $\sigma$. 

With this equation the Bayesian opponent model can be implemented straightforwardly using Equations (\ref{eq:bayesian_update_alt}) and (\ref{eq:bayesian_normalizing}). An example implementation is given in Algorithm~\ref{alg:bayes}.

Then, whenever our agent needs to have an estimation $\est{\util}_2(\off)$ of the opponent's utility for some offer $\off$, it can be calculated by taking the expectation value over all hypothetical utility functions $\util \in \Util$:
\begin{equation}
\est{\util}_2(\off) = \sum_{\util \in \Util} P(\util|\vec{\pr}) \cdot \util(\off)
\end{equation}
where $\vec{\pr}$ is the list of all proposals our agent has so far received from the opponent.

\begin{algorithm}
\caption{Opponent modeling algorithm based on Bayesian learning}\label{alg:bayes}
\begin{algorithmic}[1]
	\Statex \textbf{Parameters:} 
	\Statex $\sigma$ \Comment{Standard deviation of the Gaussian distribution.}
	\Statex $c$ \Comment{Concession speed of hypothesized opponent strategy.}
	\Statex $\Util$ \Comment{A set of hypothetical utility functions for the opponent.}
	\Statex \textbf{Input:} 
	%\Statex $\Off$ \Comment{The offer space.}
	\Statex $\dead$ \Comment{The deadline.}
	\Statex $t$ \Comment{The current time.}
    \Statex $\offRec$ \Comment{The last received offer.}
    \Statex \textit{probs}  \Comment{A map that maps each $\util \in \Util$ to the probability}
    \Statex \Commentt{value $P(\util \mid \pr_1, \pr_2, \dots, \pr_\numObs)$ as calculated in the}
    \Statex \Commentt{previous call to this algorithm.}
    \Statex
    \comment{Ensure that we initially assign the same probability to each}
    \commentt{hypothesis:}
    \If{this is our first turn}
    		\For{$\util \in \Util$}
    			\State $probs[\util] \leftarrow \frac{1}{|\Util|}$
    		\EndFor
    \EndIf
    \Statex
    \comment{Update all the values in $probs$, given the newly received offer $\offRec$}
    \commentt{and simultaneously calculate the sum of all these values:}
    \State $sum \leftarrow 0$
    \For{$\util \in \Util$}
		\State $probs[\util] \leftarrow probs[\util]\cdot \mc{N}(\util(\offRec) \mid 1-c\cdot \frac{t}{\dead} \ ,\  \sigma)$
		\State $sum \leftarrow sum + probs[\util]$
	\EndFor
	 \Statex
	\comment{Ensure that all probabilities are normalized:}
	 \For{$\util \in \Util$}
		\State $probs[\util] \leftarrow probs[\util] / sum$
	\EndFor
	\Statex
	\State \Return $probs$
\end{algorithmic}
\end{algorithm}

%In order to define a set of hypotheses, the authors of \cite{Hindriks2008Bayesian} made the following assumptions:
%\begin{itemize}
%\item The negotiation domain is a multi-issue domain with linear utility functions.
%\item Each issue is an \textit{ordered} set.
%\item For each issue $I_j$, the opponent's utility function $\util_2^j$ over that issue has a triangular shape. That is, the opponent has one preferred item and for all earlier items the utility function increases linearly, and for all later items it decreases linearly.
%\end{itemize}

\subsubsection{Choosing the Utility Hypotheses}\label{sec:choosing_utilty_hyps}
We now know how to apply Bayesian learning for some given set of hypothetical utility functions $\Util$. However, we still need to discuss how to choose this set.

To do this, let us first assume that the negotiation domain is a multi-issue domain with $\numIssues$ issues and that we know that the opponent's utility function $\util_2$ is linear, so it can be expressed in the form of Eq.~(\ref{eq:linear_util_func}). Therefore, it can be described in terms of its weights $w_2^1, w_2^2, \dots w_2^\numIssues$ and its evaluation functions $\eval_2^1, \eval_2^2, \dots \eval_2^\numIssues$.

To simplify the notation a bit, in the rest of this section we will suppress the subscript 2 and just write $w^j$ instead of $w_2^j$ and $\eval^j$ instead of $\eval_2^j$, since we are exclusively talking about the \textit{opponent}'s utility anyway.

Furthermore, we will use the notation $\offComp{{j,l}}$ to denote the $l$-th option for issue $\issue{j}$. For example, if $\issue{1}$ represents a movie to choose:
\[\issue{1} = \{\mathit{The\ Godfather}, \mathit{Casablanca}, \mathit{The\ Big\ Lebowski}\}\]
Then we have:
\[\offComp{{1,1}} = \mathit{The\ Godfather} \quad \offComp{{1,2}}= \mathit{Casablanca}\quad \offComp{{1,3}} = \mathit{The\ Big\ Lebowski}\]
%\begin{eqnarray*}
%\offComp{{1,1}} &=& \mathit{The\ Godfather}\\
%\offComp{{1,2}} &=& \mathit{Casablanca}\\
%\offComp{{1,3}} &=& \mathit{The\ Big\ Lebowski}\\
%\end{eqnarray*}
In addition, if $\eval^j$ is the evaluation function of agent $\ag_2$ for issue $\issue{j}$ then we use the notation $\eval^{j,l}$ as a shorthand for the value it assigns to option $\offComp{{j,l}}$. That is:
\[\eval^{j,l} \ \ := \ \ \eval_2^{j}(\offComp{{j,l}})\]

So, to fully specify a linear utility function, we need to specify the value of each weight $w^j$ and each $\eval^{j,l}$. This means that if the domain has $\numIssues$ issues and each issue has $\numOptions$ options, then we need to specify $\numIssues + \numIssues \cdot \numOptions$ parameters. For example, if $\numIssues = 4$ and $\numOptions=3$, then we could have the following parameters:
\begin{eqnarray*}
\ & & w^1=0.3,\quad  w^2=0.5,\quad w^3 = 0.1,\quad w^4 = 0.1 \\
& & \eval^{1,1} = 0.0,\quad  \eval^{2,1} = 0.3 ,\quad  \eval^{3,1} = 0.3,\quad  \eval^{4,1} = 1.0\\
& & \eval^{1,2} = 0.4,\quad  \eval^{2,2} = 0.7 ,\quad  \eval^{3,2} = 0.0,\quad  \eval^{4,2} = 1.0\\
& & \eval^{1,3} = 1.0,\quad  \eval^{2,3} = 0.9 ,\quad  \eval^{3,3} = 0.0,\quad  \eval^{4,3} = 0.2 \\
\end{eqnarray*}

Now, one way to select a finite set of hypothetical utility functions, is to restrict each of these parameters to only have values in some finite domain, such as the set $\{0, \ \ 0.1, \ \ 0.2, \ \ \dots,  \ \  0.9, \ \  1.0\}$. Since this set has 11 possible values, this gives us a total of $11^{\numIssues + \numIssues \cdot \numOptions}$ possible utility functions. Unfortunately, however, this is an astronomically large number, even for small domains with only $\numIssues = 3$ and $\numOptions=4$. This is a problem because, as can be seen in Algorithm~\ref{alg:bayes}, we need to loop over all elements of $\Util$, which is clearly unfeasible for such a large set.

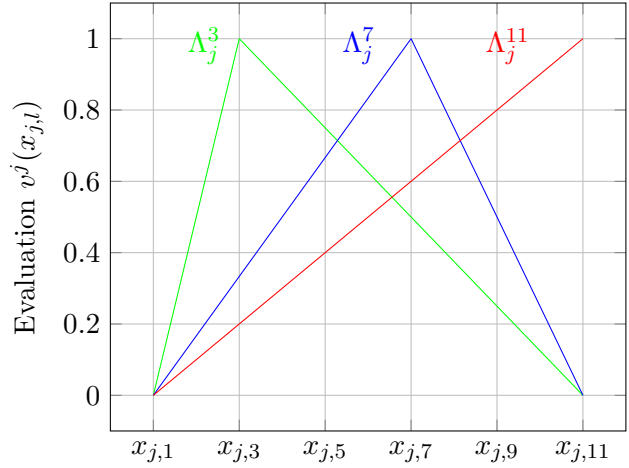
\begin{figure}
\begin{center}
\begin{tikzpicture}
\begin{axis}[grid=both,
          xmin=0, xmax=10, ymin=0, ymax=1,
          xticklabels={$ $, $\offComp{{j,1}}$, $\offComp{{j,3}}$, $\offComp{{j,5}}$,$\offComp{{j,7}}$,$\offComp{{j,9}}$,$\offComp{{j,11}}$},
          ylabel={Evaluation $\eval^{j}(\offComp{{j,l}})$},
          enlargelimits]
%%%%
\addplot[green,domain=0:2,samples=100]{x/2} node[pos=0.9,anchor=south east]{$\tri_j^3$};
\addplot[green,domain=2:10,samples=100]{(10-x)/(10-2)};
%%%%
\addplot[blue,domain=0:6,samples=100]{x/6} node[pos=0.9,anchor=south east]{$\tri_j^7$};
\addplot[blue,domain=6:10,samples=100]{(10-x)/(10-6)};
%%%%
\addplot[red,domain=0:10,samples=100]{x/10} 
node[pos=0.9,anchor= south east]{$\tri_j^{11}$};
%%%
%%%\addplot[blue,domain=0:10,samples=100]{(1 - pow(0.01, 1-x))/%%%(1-0.01)} node[pos=0.5,anchor= south west]{$\gamma=1/100$};
\end{axis}
\end{tikzpicture}
\end{center}
\caption{Some examples of triangular evaluation functions for an issue $I_j$ with 11 options.}\label{fig:triang_functions}
\end{figure}

The authors of \cite{Hindriks2008Bayesian} therefore made some simplifying assumption to decrease this number. For example, they assumed that all issues are \textit{ordered} sets, and that the evaluation functions are \textit{triangular}. That is, if $\offComp{{j,n}}$ denotes $\ag_2$'s most preferred option of issue  $I_j$, then they assume the evaluation function $\eval^j$ first increases linearly from 0 to 1 until the option $\offComp{{j,n}}$ is reached, after which it decreases linearly from 1 to 0. Figure~\ref{fig:triang_functions} displays a few examples of such functions. Formally, for any issue $I_j$ with size $\numOptions_j := |I_j|$ and any integer $n$ with $1\leq n \leq \numOptions_j$, the triangular function $\tri_j^n$ is defined as:
\begin{equation}\label{eq:triangular}
\tri_j^n(\offComp{{j,l}}) = \begin{cases}
\frac{l-1}{n-1} & \text{if\ } l< n \\
1 & \text{if\ } l = n \\ 
%%% the case l=n is really necessary, because the first case is not defined for n=1
\frac{\numOptions_j-(l-1)}{\numOptions_j-(n-1)} & \text{if\ } l>n \\
\end{cases}
\end{equation}
%This equation doesn't work for $n=s_j$, but that doesn't matter because in that case l cannot be greater than s.

This assumption of triangular evaluation functions greatly reduces the size of the set $\Util$ because now, to specify a single evaluation function $\eval^j$, we only need to specify the most preferred option $\offComp{{j,n}}\in \issue{j}$, 
rather than specifying a number $\eval^{j,l}$ for every single option $\offComp{{j,l}}\in \issue{j}$. This reduces the number of possible evaluation functions for $\issue{j}$ from $11^{\numOptions_j}$ to just $\numOptions_j$. And therefore it reduces the total number of utility functions to $11^{\numIssues} \cdot \numOptions^\numIssues$ (if all issues have the same size $\numOptions$).

\later{Discuss reducing the number of weights as well.}

With these reductions the set $\Util$ becomes small enough to apply Bayesian learning in practice to small domains with just a few issues. However, since the set $\Util$ still grows exponentially with the number of issues, this approach is still not feasible for scenarios with many issues. For this reason, luckily, the authors of~\cite{Hindriks2008Bayesian} also proposed a more scalable version of Bayesian opponent modeling, which we will discuss next.

\begin{exercise}\label{ex:bayesian_learning}
\textbf{Bayesian Learning.} Implement the Bayesian learning algorithm discussed above. Next, run some negotiations with your agents from Exercises~\ref{ex:time_based}, \ref{ex:adaptive}, and \ref{ex:tit_for_tat}, but using this new opponent modeling algorithm, instead of the DummyOpponentUtilityModel.
\end{exercise}

\subsection{Scalable Bayesian Learning}
Before we explain the scalable version of Bayesian learning \textit{for automated negotiation}, let us first take a step back and focus again on the general case.

Let us assume we have some set of hypotheses $\Hypo$ and that each hypothesis $\hypo \in \Hypo$ can be decomposed into a number of sub-hypotheses: $\hypo = (\hypo_1 , \hypo_2, \dots, \hypo_\numIssues)$, so the hypothesis space can be decomposed as the Cartesian product of a number of sub-hypothesis spaces: $\Hypo = \Hypo^1 \times \Hypo^2 \times \dots \times \Hypo^\numIssues$. For example, the hypothesis that a given playing card is the ace of spades can be written as $\hypo = (A, \spadesuit )$.  

Now, the probability $P(\hypo \mid \vec{\obs})$ can be written as:
\[P(\hypo \mid \vec{\obs}) = \prod_{j=1}^\numIssues P(\hypo_j \mid \vec{\obs})\]
and the Bayesian update rule (\ref{eq:bayesian_update_alt}) can be applied to each sub-hypothesis separately:
\begin{equation}\label{eq:bayesian_update_partial}
\tilde{P}(\hypo_j \mid \vec{\obs}, \obs_{k+1}) = P(\hypo_j \mid \vec{\obs}) \cdot P(\obs_{k+1} \mid \hypo_j)
\end{equation}
The question, now, is how to calculate $P(\obs_{k+1} \mid \hypo_j)$. After all, we typically need the full hypothesis $\hypo$ to be able to calculate the probability of some observation. 

%For this we can use the following:
%\[P(\obs_{k+1} \mid \hypo_j) = \sum_{\hypo \in \Hypo[\hypo_j]} P(\obs_{k+1} \mid \hypo) \]
%where $\Hypo[\hypo_j]$ is the set of all hypotheses for which the $j$-th component equals $\hypo_j$.

Before answering that question, let us first return to the topic of automated negotiation. In the previous section we have seen that each hypothesis $\hypo$ corresponds to a utility function $\util$, which is defined by a number of parameters: for each issue $I_j$ a weight $w^j$ and an evaluation function $\eval^j$. This means that the hypothesis space can be written as:
\[\Hypo \ \ = \ \ \Hypo_w^1 \times  \Hypo_w^2 \times \dots \times \Hypo_w^\numIssues \ \times \ \Hypo_\eval^1 \times  \Hypo_\eval^2 \times \dots \times \Hypo_\eval^\numIssues\]
where each $\Hypo_w^j$ is a set of possible values for weight $w^j$, and each $\Hypo_\eval^j$ is a set of possible evaluation functions defined over issue $\issue{j}$.

For example, if we assume that each weight must be an integer multiple of 0.1 and must be between 0 and 1, then we have:
\[\Hypo_w^1 =  \Hypo_w^2 = \dots = \Hypo_w^\numIssues = \{0, 0.1, 0.2, \dots, 0.9, 1.0\}\]
Furthermore, if we assume that each evaluation function must be a triangular function (see Eq.~(\ref{eq:triangular})), then for each $\Hypo_\eval^j$ we have:
\[\Hypo_\eval^j = \{\tri_j^1, \tri_j^2, \dots, \tri_j^{\numOptions_j} \}\]
where $\numOptions_j$ is the size of issue $\issue{j}$.

%If we assume that each $\Hypo_w^j$ and each $\Hypo_\eval^j$ is sorted, then we can use the notation $\weighthyp{j}{l}$ to denote the $l$-th element of $\Hypo_w^j$, and $\evalhyp{j}{l}$ to denote the $l$-th element of $\Hypo_\eval^j$.

%\essential{This notation is conflicting with notation in the previous section!}

%Furthermore, we will use $w^j$ to denote a variable that can take any value from the set $\Hypo_w^j$ and $\eval^j$ as a variable that can be any evaluation function from $\Hypo_\eval^j$.

So a hypothesis $\hypo$ is now a tuple $(w^1, w^2,\dots w^\numIssues, \eval^1, \eval^2, \dots \eval^\numIssues)$, where each $w^j$ is a value from the set of weight hypotheses $\Hypo_w^j$ and each $\eval^j$ is an evaluation function from the set of evaluation hypotheses $\Hypo_\eval^j$. Furthermore, each such hypothesis $\hypo$ corresponds to a utility function $\util_\hypo$:
\[\util_\hypo(\off) \ \ := \ \ \sum_{j=1}^\numIssues w^j\cdot \eval^j(\off)\]
Recall from Sec.~\ref{sec:linear_util_functions} that we may abuse notation by writing $\eval^j(\off)$ when we actually mean $\eval^j(\offComp{j})$, where $\offComp{j}$ is the $j$-th component of $\off$. 

For a given hypothesis $\hypo$ and a given sequence of received proposals $\vec{\pr}$ we can now express the posterior probability as:
\[
P(\hypo \mid \vec{\pr}) \ \ = \ \ \prod_{j=1}^\numIssues P(w^j \mid \vec{\pr})\cdot \prod_{j=1}^\numIssues  P(\eval^j \mid \vec{\pr})
\]
and each probability $P(w^j \mid \vec{\pr})$ and $P(\eval^j \mid \vec{\pr})$ can be updated separately. For example, for each weight $w^j$ the update rule (\ref{eq:bayesian_update_alt}) now becomes:
\begin{equation}\label{eq:update_weight_scalable}
\tilde{P}(w^j \mid \vec{\pr}, \pr_{k+1}) \ \ = \ \ P(w^j \mid \vec{\pr})\cdot P(\pr_{k+1} \mid w^j)
\end{equation}
and similarly, for the evaluation functions $\eval^j$:
\begin{equation}\label{eq:update_eval_scalable}
\tilde{P}(\eval^j \mid \vec{\pr}, \pr_{k+1}) \ \ = \ \ P(\eval^j \mid \vec{\pr})\cdot P(\pr_{k+1} \mid \eval^j)
\end{equation}
Note that these two equations are just special cases of  Eq.~(\ref{eq:bayesian_update_partial}), specific to automated negotiation. So, our original question how to calculate \mbox{$P(\obs_{k+1} \mid \hypo_j)$} can now be reformulated as the question how to calculate $P(\pr_{k+1} \mid w^j)$ and $P(\pr_{k+1} \mid \eval^j)$.

For example, suppose that we want to calculate $P(\pr_{k+1} \mid w^1=0.1)$. If we already know the correct values of all other weights, and the correct evaluation functions of all issues, then we can just use the utility function defined by all those weights and evaluation functions, plus the specific value of $w_1=0.1$. Let us denote this function by $\util_{[w^1=0.1]}$ and then define ${P(\pr_{k+1} \mid w^1=0.1)}$ to be equal to ${P(\pr_{k+1} \mid \util_{[w^1=0.1]})}$. Next, we can calculate  ${P(\pr_{k+1} \mid w^1)}$ in the same way for all other hypothetical values of $w^1$.

Of course, in reality, we do not know the correct values of all other weights, nor the expressions of all evaluation functions. However, what we can do instead, is take their \textit{expectation values}, based on the current probabilities that we have assigned to them.

To make this precise, we first define for each issue $\issue{j}$ its \textit{expected} weight $\expt{w}^j$ and its \textit{expected} evaluation function $\expt{\eval}^{j}$ as follows:
\begin{equation}\label{eq:expected_weight}
\expt{w}^j := \sum_{w^{j} \in \Hypo_w^j} w^{j} \cdot P(w^{j} \mid \vec{\pi})\end{equation}
\begin{equation}\label{eq:expected_eval}
\expt{\eval}^{j}(\off) := \sum_{\eval^{j} \in \Hypo_\eval^j} \eval^{j}(\off) \cdot P(\eval^{j} \mid \vec{\pi})
\end{equation}
which in turn can be used to define the expected utility function $\expt{\util}$:
\begin{equation}\label{eq:expected_util}
\expt{\util}(\off) := \sum_{j=1}^\numIssues \expt{w}^j \cdot \expt{\eval}^{j}(\off)
\end{equation}
The idea is then, that for every weight hypothesis we recalculate this expected utility, but with the expected weight replaced by that specific weight hypothesis.

That is, for any issue $\issue{j}$ and weight-hypothesis $w^{j} \in \Hypo_w^j$ we can calculate a function $\expt{\util}_{[w^j]}$ as follows:
\begin{equation}\label{eq:expected_util_w_j}
\expt{\util}_{[w^j]}(\off) \quad := \quad \sum_{\substack{k=1 \\ k\neq j}}^\numIssues \expt{w}^k \cdot \expt{\eval}^{k}(\off) + w^j \cdot \expt{\eval}^{j}(\off)
\end{equation}
which is the utility value calculated by taking, for each issue $\issue{k}$, the \textit{expectation} value of the weight $w^k$, and the expectation value of $\eval^k(\off)$, except for issue $\issue{j}$, for which we use the hypothesized weight $w^{j}$.

Then, the idea is that for any $w^j\in \Hypo_w^j$ we can calculate $P(\pr_{k+1} \mid w^j)$  as in Eq.~(\ref{eq:p_o_given_u}). but with the variable $\util$ replaced by $\expt{\util}_{[w^{j}]}$. However, there is still a problem with this. That is, that Eq.~(\ref{eq:p_o_given_u}) assumes the utility function is normalized, because it assumes the opponent's aspiration level decreases from 1 to 0. Unfortunately, however, even if every hypothetical utility function is normalized, the expected utility may no longer be normalized. For example, if we have two triangular functions $\tri_j^3$ and $\tri_j^5$, which each assume a maximum of 1, then the linear combination $0.5 \cdot \tri_j^3 + 0.5 \cdot \tri_j^5$ will have a maximum that is lower than 1.

So, we first need to calculate:
\[\expt{\util}_{[w^j]}^{max} \ \ := \ \ \max \ \{\expt{\util}_{[w^j]}(\off) \mid \off \in \Off\}\]
\[\expt{\util}_{[w^j]}^{min} \ \ := \ \ \min \ \{\expt{\util}_{[w^j]}(\off) \mid \off \in \Off\}\]
and then with this we can calculate the opponent's \textit{normalized} estimated utility:
\[\hat{\util}_{[w^j]}(\off) \ \ := \ \  \frac{\expt{\util}_{[w^j]}(\off) - \expt{\util}_{[w^j]}^{min}}{\expt{\util}_{[w^j]}^{max} - \expt{\util}_{[w^j]}^{min}}\]
Now, we can finally calculate $P(\pr_{k+1} \mid w^j)$ as follows:
\begin{equation}\label{eq:p_o_given_u_scalable}
P((2,\prop,\off,t) \mid w^j) \quad := \quad \mc{N}(\hat{\util}_{[w^j]}(\off) \mid 1-c\cdot\frac{t}{\dead} \ ,\  \sigma) 
\end{equation}

%Similarly, the probabilities for the evaluation functions can be updated according to:
%\begin{equation}\label{eq:update_eval_scalable}
%\tilde{P}(v^j|\vec{\pr}, \pr_{k+1}) \quad = \quad P(v^j|\vec{\pr})\cdot P(\pr_{k+1}|v^j)
%\end{equation}
%with:

Calculating the probabilities for the evaluation functions goes in essentially the same way. That is, we define:
\begin{equation}\label{eq:expected_util_v_j}
\expt{\util}_{[\eval^{j}]}(\off) \quad := \quad \sum_{\substack{k=1 \\ k\neq j}}^\numIssues \expt{w}^k \cdot \expt{\eval}^{k}(\off) + \expt{w}^{j} \cdot \eval^j(\off)
\end{equation}
and with this we can calculate the quantities $\expt{\util}_{[\eval^j]}^{max}$, $\expt{\util}_{[\eval^j]}^{min}$, and $\hat{\util}_{[\eval^j]}(\off)$ in a similar way as above, and then we can calculate $P(\pr_{k+1} \mid \eval^j)$ as:
\begin{equation}
 P((2,\prop,\off,t)\mid \eval^j) \quad := \quad \mc{N}(\hat{\util}_{[\eval^j]}(\off) \mid 1-c\cdot\frac{t}{\dead} \ ,\ \sigma) 
\end{equation}
See Algorithm~\ref{alg:bayes_scalable} for an implementation.

It should be noted, however, that these equations are just approximations. They are based on the assumption that the current expected utility function $\expt{\util}$ is already a good approximation to the opponent's true utility function $\util_2$.

\begin{algorithm}
\small
\caption{Opponent modeling algorithm based on Scalable Bayesian learning. This function is called every time a new proposal is received, in order to update our agent's model of the opponent's utility function.}\label{alg:bayes_scalable}
\begin{algorithmic}[1]
	\Statex \textbf{Parameters:} 
	\Statex $\sigma$ \Comment{Standard deviation of the Gaussian distribution.}
	\Statex $c$ \Comment{Concession speed of hypothesized opponent strategy.}
	\Statex \textbf{Input:} 
	%\Statex $\Off$ \Comment{The offer space.}
	\Statex $\dead$ \Comment{The deadline.}
	\Statex $t$ \Comment{The current time.}
    %\Statex $\hist$ \Comment{The history.}
    \Statex $\offRec$ \Comment{The last received offer.}
    %\Statex $\Hypo_w^1, \dots, \Hypo_w^\numIssues$ \Comment{For each issue $\issue{j}$ a set of possible values for  weight $w^j$.}
    \Statex \textit{w\_hyps}  \Comment{A double array that contains for each issue $\issue{j}$ a list of possible} 
    \Statex \Commentt{weights. So, $w\_hyps[j]$ is a single array that represents $\Hypo_w^j$.}
   %\Statex \Commentt{possible weight $w^{j,l}$ a probability value $P(w^{j,l}|\vec{\pr})$.}
    \Statex \textit{w\_probs}  \Comment{A double array that contains for each issue $\issue{j}$ and each } 
   \Statex \Commentt{possible weight $w^{j} \in \Hypo_w^j$ a probability value $P(w^{j}|\vec{\pr})$.}
    %\Statex $\Hypo_\eval^1, \dots, \Hypo_\eval^\numIssues$ \Comment{For each issue $\issue{j}$ a set of possible evaluation functions $\Hypo_\eval^j$.} 
    \Statex \textit{e\_hyps}  \Comment{A double array that contains for each issue $\issue{j}$ a list of possible} 
    \Statex \Commentt{evaluation functions. So, $e\_hyps[j]$ is a single array that}
    \Statex \Commentt{represents $\Hypo_\eval^j$.}
    \Statex \textit{e\_probs} \Comment{A double array that contains for each issue $\issue{j}$ and each}
    \Statex \Commentt{possible evaluation function $\eval^{j}\in \Hypo_\eval^j$ a probability value $P(\eval^{j}|\vec{\pr})$.}
    \Statex
    \comment{Calculate the values of $\expt{w}^j$ and $\expt{\eval}^j(\offRec)$ according to Eqs.~(\ref{eq:expected_weight}) and (\ref{eq:expected_eval}):}
 	\For{each issue $\issue{j}$ of the domain}
 		%\Statex
 		\State $\expt{w}^j \leftarrow \sum_{l=1}^{|\Hypo_w^j|} w\_hyps[j][l] \cdot w\_probs[j][l]$
 		\State $\expt{v}^j \leftarrow \sum_{l=1}^{|\Hypo_v^j|} e\_hyps[j][l](\offRec) \cdot e\_probs[j][l]$
% 		 \For{each evaluation hypothesis $\eval^{j,l} \in \Hypo_\eval^j$}
% 			\State $\est{v}^j \leftarrow \est{v}^j + \eval^{j,l}(\off_{rec}) \cdot eval\_probs[j][l]$
% 		\EndFor
 		%\Statex
 	\EndFor
 	\Statex
    	\For{each issue $\issue{j}$ of the domain}
    		\Statex
    		\For{$l \in \{0, 1, \dots, |\Hypo_w^j|-1\}$}

    				\State $\expt{\util}_{[w^{j}]} \leftarrow \sum_{k=1, k\neq j}^\numIssues \expt{w}^k \cdot \expt{\eval}^{k} + w\_hyps[j][l] \cdot \expt{\eval}^{j}$  \quad \quad \quad $\triangleright$ {\color{CommentColor} Eq.~(\ref{eq:expected_util_w_j})}
    				\State $\hat{\util}_{[w^{j}]} \leftarrow normalize\_util(\expt{\util}_{[w^{j}]})$
    				\State $w\_probs[j][l] \leftarrow w\_probs[j][l] \cdot \mc{N}(\hat{\util}_{[w^{j}]} \mid 1-c\cdot \frac{t}{\dead}, \sigma)$ \quad \quad $\triangleright$ {\color{CommentColor} Eq.~(\ref{eq:update_weight_scalable})}
    		\EndFor
    		\State $normalize\_probs(weight\_probs[j])$
    		\Statex
    		%%%
    		%%%
    		\For{$l \in \{0, 1, \dots, |\Hypo_\eval^j|-1\}$}
    			\State $\expt{\util}_{[\eval^{j}]} \leftarrow \sum_{k=1, k\neq j}^\numIssues \expt{w}^k \cdot \expt{\eval}^{k} + \expt{w}^{j} \cdot e\_hyps[j][l](\offRec)$ \quad \quad $\triangleright$ {\color{CommentColor} Eq.~(\ref{eq:expected_util_v_j})} 
    			\State $\hat{\util}_{[\eval^{j}]} \leftarrow normalize\_util(\expt{\util}_{[\eval^{j}]})$
    			\State $e\_probs[j][l] \leftarrow e\_probs[j][l] \cdot \mc{N}(\expt{\util}_{[\eval^{j}]} \mid 1-c\cdot \frac{t}{\dead}, \sigma)$ \quad \quad $\triangleright$ {\color{CommentColor} Eq.~(\ref{eq:update_eval_scalable})}
    		\EndFor
    		\State $normalize\_probs(eval\_probs[j])$
    		\Statex
%    			\For{each evaluation hypothesis $\eval^{j,l} \in \Hypo_\eval^j$}
%    				\State $\est{\util}_{[\eval^{j,l}]} \leftarrow \sum_{i=1, i\neq j}^\numIssues \est{w}^i \cdot \hat{\eval}^{i} + \est{w}^j \cdot  \eval^{j,l}(\offRec)$ 
%    				\State $eval\_probs[j][l] \leftarrow$
%    			   	\Statex \hspace{3cm} $\eval\_probs[j][l] \cdot \mc{N}(1-c\cdot \frac{t}{\dead}, \sigma)(\est{\util}_{[\eval^{j,l}]})$
%    			\EndFor
%    			\State $normalize(eval\_probs[j])$
    	\EndFor    
	\State \Return $(weight\_probs, eval\_probs)$
\end{algorithmic}
\end{algorithm}
\normalsize

While scalable Bayesian learning largely solves the problem of scalability, the main disadvantage 
is that we still need to make a lot of assumptions. For example, we still need to assume that the opponent's utility function is linear, that the issues are ordered, and that the opponent has triangular evaluation functions. Furthermore, it depends on the chosen model of the opponent's bidding strategy and on the chosen standard deviation $\sigma$ for the Gaussian distribution.

\begin{exercise}\label{ex:scalable_bayesian_learning}
\textbf{Scalable Bayesian Learning.} Implement the scalable Bayesian learning algorithm discussed in this section. Next, run some negotiations with some of the agents that you implemented for previous exercises, but now with this new opponent modeling algorithm, instead of the dummy opponent model or the regular Bayesian learning algorithm from Exercise~\ref{ex:bayesian_learning}.
\end{exercise}

\subsection{Frequency Analysis}
In this section we will discuss a simpler alternative to Bayesian learning, called \textit{frequency analysis}, which is based on the idea that the opponent's evaluation functions and weights can be estimated from the frequency with which the opponent proposes the respective options for each issue. While this method is perhaps not as elegant or sophisticated as Bayesian learning, it turns out that in practice it often performs equally well, or even better~\cite{Baarslag2013OpponentModels}.

The basic idea of frequency analysis is that for any issue $\issue{j}$ and any option $\offComp{{j,l}} \in \issue{j}$ of that issue, the value $\eval_2^j(\offComp{{j,l}})$ that the opponent assigns to it can be estimated from the number of times that the opponent makes proposals containing that option.

For example, in the scenario that Alice and Bob are negotiating about a visit to the cinema, if Alice keeps making proposals that include the movie \textit{The Godfather}, then that is a clear indication that Alice probably likes that movie very much. 

Furthermore, to estimate the opponent's weights $w_2^j$, the idea is that if
the opponent proposes many different options for the same issue $\issue{j}$, then this is an indication that that issue is probably not very important to the opponent, so the weight $w_2^j$ should have a low value.

For example, if Alice first proposes to see the movie at 18:00, but then proposes to see it at 20:00, and then proposes to see it at 22:00, then apparently she does not really care much about the time at which the movie starts.

As usual, there are many ways how these ideas can be implemented. As an example, we here present the implementation by van Galen Last~\cite{vanGalenLast2012AgentSmith}.\footnote{The cited paper itself actually does not explain this opponent modeling algorithm, but it can be found in the source code of their agent, which can be found at \url{https://tracinsy.ewi.tudelft.nl/pubtrac/Genius/browser/src/main/java/agents/anac/y2010/AgentSmith}}

Let $\numRec$ denote the total number of proposals made by the opponent:
\[\numRec := |\{(i,\actype, \off, t) \in \hist \mid i=2 \ \land \ \actype = \prop\}|\]
and let $\offComp{{j,l}}$ denote the $l$-th option for issue $\issue{j}$. Furthermore, let $\freq_\hist(\offComp{{j,l}})$ denote the number of times that the opponent has proposed an offer that contained $\offComp{{j,l}}$:
\begin{equation}\label{eq:def_freq_hist_1}
\freq_\hist(\offComp{{j,l}}) := |\{(i,\actype, \off, t) \in \hist \mid i=2 \ \land \ \actype = \prop \ \land \ \offComp{{j,l}} \in \off\}|
\end{equation}
Then, each value $\eval_2^j(\offComp{{j,l}})$ can be estimated  as the number of times the option $\offComp{{j,l}}$ has been proposed by the opponent, divided by the total number of proposals made by the opponent:
\[\est{\eval}_2^j(\offComp{{j,l}}) \quad = \quad \frac{\freq_\hist(\offComp{{j,l}})}{\numRec}\]
and each weight $w_2^j$ can be estimated as:
\[\est{w}_2^j \quad = \quad \frac{\max \ \ \{\freq_\hist(\offComp{{j,l}}) \mid \offComp{{j,l}} \in \issue{j}\}}{k}\]

Note that this approach in general will not yield a normalized utility function, so you may optionally still want to apply some normalization to these weights and evaluation functions.

Another thing to note, is that if the opponent proposes the same offer $\off$ multiple times, then for each of these proposals, the frequency $\freq_\hist(\offComp{{j,l}})$ of the options in that offer is increased. This may yield wrong estimations, because the fact that an offer is repeated many times, does not necessarily mean that it yields very high utility for the opponent. For example, when the MiCRO algorithm repeats on offer, it chooses that offer randomly. Furthermore, in general, whenever an agent can't find any good new offers to propose, it may simply choose to repeat the \textit{last} offer it already proposed, rather than its \textit{favorite} offer. Therefore, one could argue that, when calculating the frequency $\freq_\hist(\offComp{{j,l}})$, it would be better to ignore repeated offers and define it as follows:
\begin{equation}\label{eq:def_freq_hist_2}
\freq_\hist(\offComp{{j,l}}) \ \ := \ \ |\{\off \in \Off \mid \exists t: (2, \prop, \off, t) \in \hist \ \land \ \offComp{{j,l}} \in \off\}|
\end{equation}
which yields an alternative version of Frequency Analysis.
Note that this definition involves the size of a set of \textit{offers}, rather than the size of a set of \textit{proposals}, which means that indeed, an offer that is proposed twice is only taken into account once.

\begin{exercise}\label{ex:freq_analysis}
\textbf{Frequency Analysis.} Implement the frequency analysis algorithm discussed in this section, based either on Eq.~(\ref{eq:def_freq_hist_1}) or on Eq.~(\ref{eq:def_freq_hist_2}). Next, run some negotiations with your agents from Exercises~\ref{ex:time_based}, \ref{ex:adaptive} and \ref{ex:tit_for_tat}, but using this new opponent modeling algorithm.
\end{exercise}

\section{Learning the Opponent's Strategy}\label{sec:opp_strat_modeling}
In this section we will discuss how to model the opponent's bidding strategy, based on the proposals he makes during the negotiations. More precisely, given the set of proposals that our agent received from the opponent until time some time $t$, we aim to predict which offers the opponent will propose later on, between time $t$ and the deadline. 

The ability to make such predictions is essential for the implementation of an adaptive negotiation strategy, as explained in Section \ref{sec:adaptive_strategies}.

To formalize this, let 
\[\pr_1 = (2, \prop, \off_1, t_1), \ \ \pr_2 = (2, \prop, \off_2, t_2), \ \  \dots, \ \ \pr_\numObs = (2, \prop, \off_\numObs, t_\numObs)\]
denote the sequence of proposals that our agent has received from its opponent and let $z_1, z_2, \dots, z_\numObs$ denote their corresponding utility values, for \textit{our} agent. That is:
\[z_j \ \ := \ \  \util_1(\off_j)\]
Then our goal is to implement an algorithm that can take as its input the sequence 
\[(z_1, t_1), \ \ (z_2, t_2), \ \ \dots, \ \  (z_\numObs, t_\numObs)\]
plus some arbitrary time $t_{\numObs+1}$ in the future,
and that outputs a prediction for the corresponding utility value $z_{\numObs+1}$.

Of course, in general it is unlikely that we can make such a prediction perfectly, so rather than outputting the actual value $z_{\numObs+1}$, a typical opponent modeling algorithm would instead output a probability distribution $P(z_{\numObs+1})$  over all the possible values of $z_{\numObs+1}$.

Many different techniques to do this have been proposed in the literature. For example, Agent~K~\cite{agentK}, the winner of ANAC 2010, used an extrapolation algorithm  based on the average and standard deviation of the values of $z_i$. Other agents used non-linear regression  (IAMhaggler~\cite{IAMhaggler}), or wavelet decomposition and cubic smoothing splines (OMAC~\cite{OMAC}). Here, however, we will only focus on the technique of Gaussian Processes (IAMHaggler2011~\cite{Williams2011GaussianProcesses}).

\subsection{Gaussian Processes}
Due to the technical nature of this topic we cannot discuss Gaussian processes in detail, so we will only give a global idea of how this technique works.  For a more detailed discussion we refer to \cite{williams2006gaussian} or \cite{Bishop2007}.

The idea behind Gaussian processes is that we assume that at any given time the probability that the opponent will propose an offer $\off$ with utility $u_1(\off) = z$ is given by a Gaussian distribution:
\[P(z)\ \ = \ \ \mc{N}(z \mid \mu, \sigma) \ \ =  \ \ \frac{1}{\sqrt{2\pi}\sigma}e^{-\frac{(z-\mu)^2}{2\sigma^2}}\]
Now, in order to be able to use this for our purposes, we first need to determine an expression for the probability that the opponent proposes a certain \textit{sequence} of offers with utility values $z_1, z_2, \dots, z_\numObs$ respectively.

If we could assume that each offer is drawn \textit{independently} from the same normal distribution, then this would be easy, as we could simply multiply the probabilities. This would yield the following expression:
\[P(z_1, z_2, \dots, z_\numObs) \ \ = \ \ \frac{1}{(2\pi)^{\numObs/2}}\cdot \frac{1}{\sigma^\numObs}\cdot e^{-\frac{(z_1-\mu)^2 + (z_2-\mu)^2 + \dots + (z_\numObs-\mu)^2  }{2\sigma^2}}\]
which can be rewritten using vector-notation:
\begin{equation}\label{eq:gaussian_indep}
P(\vec{z}) \ \ = \ \  \frac{1}{(2\pi)^{\numObs/2}}\cdot \frac{1}{\sigma^\numObs}\cdot e^{-\frac{1}{2\sigma^2}(\vec{z}-\vec{\mu})^T  \mathbf{I} (\vec{z}-\vec{\mu})}
\end{equation}
where $\mathbf{I}$ is the $\numObs \times \numObs$ identity matrix and $\vec{\mu} = (\mu, \mu, \dots, \mu)^T$ is the $\numObs$-dimensional column vector containing just $k$ copies of the number $\mu$.

However, the offers proposed by the opponent are typically not independent. After all, it is fair to assume that the opponent is following some negotiation strategy that concedes over time with respect to his utility $\util_2$ and that this utility function is at least to some extent correlated with our own utility $\util_1$. 

For example, in the extreme case that the opponent follows a strictly monotonic bidding strategy and that the negotiation domain is a split-the-pie domain, then our agent would perceive the offers it receives from the opponent as strictly increasing over time, i.e. $z_1 \leq z_2 \leq \dots \leq z_k$. So, their values are clearly not independent.

Of course, in practice many negotiation scenarios will not be split-the-pie domains in which the utility functions are that strongly correlated. Nevertheless, it is still reasonable to assume that there will at least be some correlation. In fact, we have to make this assumption, because if there is no correlation between the two utility functions at all, then there would be no way for our agent to make any predictions based on the received proposals. After all, the utility values of the received proposals would just appear as a completely random sequence with no pattern whatsoever. 

We will therefore assume that, \textit{in general}, two consecutive proposals $\pr_i$ and $\pr_{i+1}$ will often have similar values: $z_i \approx z_{i+1}$. To state this more formally, we will assume that the closer two proposals $\pr_i$ and $\pr_{j}$ are to each other in time, the stronger the correlation between the corresponding random variables $z_i$ and $z_j$.

Whenever a sequence of Gaussian random variables is not independent, we can model their joint distribution by replacing the identity matrix in Eq.~(\ref{eq:gaussian_indep}) with some other matrix $\covmat$ (which has to be symmetric and positive semi-definite) so that the expression for the joint probability becomes:
\begin{equation}\label{eq:gaussian_covariance}
P(\vec{z}) = \frac{1}{(2\pi)^{\numObs/2}}\cdot \frac{1}{|\covmat|^{1/2}}\cdot e^{-\frac{1}{2}(\vec{z}-\vec{\mu})^T \covmat^{-1} (\vec{z}-\vec{\mu})}
\end{equation}
where $|\covmat|$ is the determinant of $\covmat$.

%This matrix can be interpreted in two ways. Firstly, it can be seen as a coordinate transformation, that maps the vector $\vec{z}-\vec{\mu}$ to \essential{some alternative vector}. This can  be seen in Figure~\ref{fig:gaussian_covariance}.

\begin{figure}
\tikzset{every picture/.style={line width=1pt}} %set default line width to 1pt        
\pgfplotsset{width=8cm}
\begin{subfigure}[h]{0.45\linewidth}
\begin{tikzpicture}[scale=0.6]
\begin{axis}[xmin=0, xmax=1,ymin=0,ymax=1,xlabel={\huge $z_1$}, ylabel={\huge $z_2$}, 
		xtick pos=left,
		ytick pos=left		
		]:
	\draw [black] (0.5,0.5) circle [radius=0.1];
	\draw [black] (0.5,0.5) circle [radius=0.2];
	\draw [black] (0.5,0.5) circle [radius=0.3];
	\draw [black] (0.5,0.5) circle [radius=0.4];
	%\draw [black] (0.5,0.5) circle [radius=0.5];
	\draw[red] (0,0.5)--(1,0.5);
\end{axis}
\end{tikzpicture}
\caption{Countour plot of a multi-variate Gaussian distribution with identity matrix.}\label{fig:gaussian_identity}
\end{subfigure}\hfill
\begin{subfigure}[h]{0.45\linewidth}
\begin{tikzpicture}[scale=0.6]
\begin{axis}[xmin=0, xmax=1,ymin=0,ymax=1,xlabel={\huge $z_1$}, ylabel={\huge $z_2$}, 
		xtick pos=left,
		ytick pos=left		
		]
	\begin{scope}[shift={(0.5,0.5)}, rotate=30]
		\draw[black] (0,0) ellipse [x radius=0.1, y radius=0.05];
		\draw[black] (0,0) ellipse [x radius=0.2, y radius=0.1];
		\draw[black] (0,0) ellipse [x radius=0.3, y radius=0.15];
		\draw[black] (0,0) ellipse [x radius=0.4, y radius=0.20];
		\draw[black] (0,0) ellipse [x radius=0.5, y radius=0.25];
	\end{scope}
	\draw[red] (0,0.12)--(1,0.88);
\end{axis}
\end{tikzpicture}
\caption{Countour plot of  multi-variate Gaussian distribution with alternative covariance matrix.}\label{fig:gaussian_covariance}
\end{subfigure}\hfill
\caption{Multi-variate Gaussian distributions.}\label{fig:multi_variate_gaussian}
\end{figure}
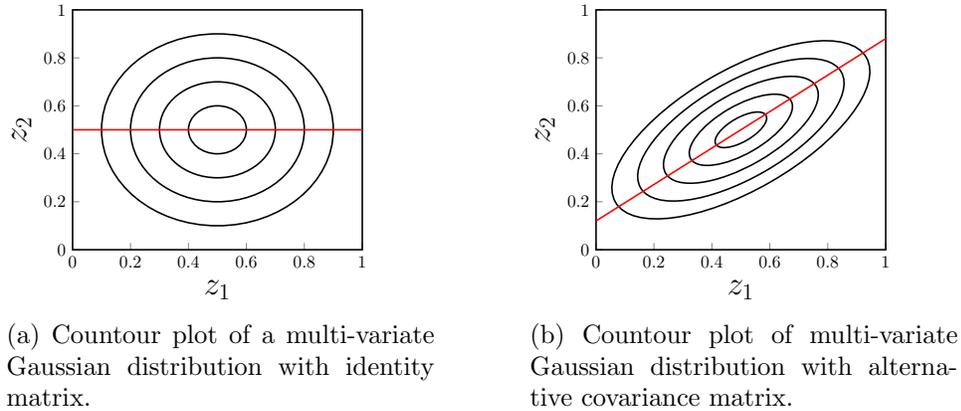

The fact that this matrix indeed introduces a dependency between each pair of variables $z_i$ and $z_j$ can be seen clearly from Figure~\ref{fig:multi_variate_gaussian}. In this figure we have drawn two contour plots for a Gaussian distribution over just two variables $z_1$ and $z_2$.  Figure \ref{fig:gaussian_identity} shows the case where $\covmat$ is just the identity matrix, so this corresponds to Eq.~(\ref{eq:gaussian_indep}). We see that for any arbitrary value of $z_1$, the probability distribution for $z_2$ is maximized at the same value $z_2 = 0.5$ (indicated with a red line). Similarly, for any value of $z_2$ the probability distribution for $z_1$ is maximized at the same value $z_1 = 0.5$. In other words, the probability distribution for $z_2$ does not depend on $z_1$ and vice versa.

On the other hand, in Figure~\ref{fig:gaussian_covariance}, where we have drawn the contour plot of a Gaussian distribution with an alternative matrix $\covmat$ we see that as $z_1$ increases, the value of $z_2$ with maximum probability also increases (again, indicated with a red line). That is, the larger the value of $z_1$, the greater the expectation value of $z_2$.

%it can also be seen as a way to introduce correlation between each pair of variables $z_i$ and $z_j$. Indeed this can also bee seen from Figure~\ref{fig:gaussian_covariance} 

Furthermore, it can be shown that if we use  Eq.~(\ref{eq:gaussian_covariance}) to calculate the \textit{covariance}  $\mathbb{E}\Big((z_i - \mu)\cdot (z_j-\mu)\Big)$ between any two variables $z_i$ and $z_j$, then the result will be exactly the element $K_{i,j}$ of the  matrix $\covmat$. For this reason, $\covmat$ is called the \textit{covariance} matrix. From this it follows immediately that if $\covmat$ is the identity matrix, then there is no covariance among any two different variables $z_i$ and $z_j$, which means that they are indeed independent.

The question now, is how to choose the correct matrix $\covmat$. For this, we use a so-called \textit{kernel} function. A kernel function is a function $\kernel : \mathbb{R}^2 \rightarrow \mathbb{R}$ that represents how the correlation between any two variables $z_i$ and $z_j$ depends on the times $t_i$ and $t_j$. That is, we set:
\begin{equation}\label{eq:kernel}
K_{i,j} := \kernel(t_i, t_j)
\end{equation}
where $K_{i,j}$ is an entry of the matrix $\covmat$, representing the covariance between variables $z_i$ and $z_j$, and $t_i$ and $t_j$ are the times of the proposals $\pr_i$ and $\pr_j$. 

Of course, we have now only replaced our original question \textit{``How do we select the correct covariance matrix?"} by a new question: \textit{``How do we select the correct kernel function?"}.

We will not go into the details of how to select the best such kernel function. We will just mention that it should be consistent with our requirement that the smaller the difference between $t_i$ and $t_j$, the more the two variables $z_i$ and $z_j$ should be correlated. So, this should be reflected in the kernel function: the smaller $|t_i-t_j|$, the greater $\kernel(t_i, t_j)$. Furthermore, let us mention that Williams et al.~\cite{Williams2011GaussianProcesses} used a so-called \textit{Matérn} kernel.

Once we have determined the covariance matrix, we know the expression for $P(\vec{z})$. The next step, is to use this to calculate an expression for $P(z_{\numObs+1} \mid z_1, z_2, \dots, z_{\numObs})$. This is indeed the expression that we are looking for, because it calculates the probability of some future value $z_{\numObs+1}$, given the observed sequence $z_1, z_2, \dots, z_{\numObs}$. 

The expression for $P(z_{\numObs+1} \mid z_1, z_2, \dots, z_{\numObs})$ can be obtained directly from the expression for $P(\vec{z})$ using straightforward, but somewhat tedious, algebra. We will not go into the details of this calculation here, but the key point is that $P(z_{\numObs+1} \mid z_1, z_2, \dots, z_{\numObs})$ will again be a Gaussian distribution. Therefore, this distribution is determined by just two parameters $\mu$ and $\sigma$, representing the mean and standard deviation.

Note that, technically, the probability $P(z_{\numObs+1} \mid z_1, z_2, \dots, z_{\numObs})$ also depends on the times $t_1, t_2, \dots, t_\numObs$, of the received proposals, as well as on the chosen future time $t_{\numObs+1}$, because they determine the covariance matrix $\covmat$, through the kernel function $\kernel$, as in Eq.~(\ref{eq:kernel}). We may therefore write this probability more correctly as $P(z_{\numObs+1} \mid \pr_1, \pr_2, \dots, \pr_{\numObs}, t_{\numObs+1})$

Finally, let us mention that  instead of using \textit{all} received proposals from the opponent as their input, Williams et al.~\cite{Williams2011GaussianProcesses} divided time into a number of time-windows and only used the proposal with highest utility from each time window. This has the advantage that it reduces noise in the data, and it also reduces the size of the input data, which in turn reduces the required computation time.

\subsection{Choosing the Optimal Target Value for an Adaptive Negotiation Strategy}
The typical use case for Gaussian processes, is to determine an optimal target value $\targ^*$ for an adaptive negotiation strategy. Let us here explain in more detail how that can be done.

In order to do this, we first have to select a time point $t_{k+1}$ which is close to the deadline $\dead$. This will allow us to predict the utility value of the last offer that the opponent will propose to us. The output of our Gaussian process algorithm will then consist of the two parameters $\mu$ and $\sigma$, which are the mean and the standard deviation of the Gaussian probability distribution that represents the probability that the opponent will propose an offer $\off_{\numObs+1}$ at time $t_{\numObs+1}$ with utility $z_{\numObs+1}$:
\[P(z_{\numObs+1} \mid \pr_1, \pr_2, \dots \pr_\numObs, t_{\numObs+1}) \ \ = \ \ \mc{N}(z_{\numObs+1} \mid \mu, \sigma)\]

Now, let us suppose for a moment that we know the exact value $z_{k+1}$ of the offer $\off_{\numObs+1}$ that the opponent will propose at time $t_{k+1}$, and furthermore that we have a good approximation $\est{\util}_2$ of the opponent's utility function, so we can ensure that our own proposals are Pareto-optimal. In that case we can assume that the opponent will accept any Pareto-optimal offer $\off$ for which $\util_1(\off) < z_{k+1}$. After all, if, \textit{for our agent}, the offer $\off$ is worse than the offer $\off_{\numObs+1}$ that the opponent would propose, then by Pareto-optimality, \textit{for the opponent}, the offer $\off$ would be \textit{better} than the offer $\off_{\numObs+1}$ that he would propose. So, it is fair to assume that the opponent would be willing to accept $\off$.

%The target value is calculated by calculating, for each possible target value, the expected utility, based on the output of the Gaussian Process. Note that this output is a Gaussian distribution representing the probability that the opponent will propose or accept an offer with some given utility value for us.

%Suppose that we use the Gaussian process to predict the utility that the opponent will propose to us at time $t_{k+1}$ which is close to the deadline $\dead$. The output of the Gaussian process will be two numbers $\mu$ and $\sigma$ representing the mean and the standard deviation of Gaussian distribution. This means that the probability that the opponent will propose an offer $\off$ with utility $\util_1(\off) = z_{k+1}$ at time $t_{k+1}$ is given by:
%\[P(z_{k+1}) = \mc{N}(z_{k+1} \mid \mu, \sigma)\]
%Assuming the opponent will accept any offer $\off$ if and only if $\util_1(\off) < z_{k+1}$, 

Of course, in reality we only have a \textit{probability distribution} for $z_{k+1}$, so we can calculate, for any offer $\off$ with utility $\util_1(\off) = z$ the \textit{probability} that the opponent will accept it, by integrating over all values of $z_{k+1}$ that are greater than $z$. That is:
\[P_\acc(z) = \int_z^{\infty} P(z_{k+1} \mid \pr_1, \pr_2, \dots \pr_\numObs)\ dz_{k+1}\]
where $P_\acc(z)$ denotes the probability that $\ag_2$ would accept an offer $\off$ with utility $u_1(\off) = z$.

Let us now make the pessimistic assumption that if our target value is $\targ$, then we will indeed need to concede all the way to that value and we will not be able to get any agreement with higher utility than that. Therefore, our expected utility would be given by $\targ \cdot P_\acc(\targ)$. That is, the utility $\targ$ in case of agreement, multiplied by the probability that the opponent will indeed accept such an agreement. We can now determine our optimal target value $\targ^*$ as follows:
\[\targ^* = \argmax_\targ \ \targ \cdot P_\acc(\targ)\]

%\[\mathbb{E}(\util_1 \mid \targ) = \beta \cdot P_\acc(\beta)\]
%%\[\mathbb{E}(\util_1 \mid \targ) = \int_\targ^{\util_1^{max}} z \cdot P_\acc(z) \  dz \]
%So the optimal target value $\targ^*$ can be calculated as:
%\[\targ^* = \argmax_\targ \ \mathbb{E}(\util_1 \mid \targ)\]

\section{Learning the Opponent's Strategy from Previous Negotiation Sessions}
COMING SOON!

\newpage
\section{Summary of Chapter}
\begin{itemize}
\item We can distinguish between three types of opponent modeling:
\begin{enumerate}
\item Learning the opponent's utility function, during the negotiation.
\item Learning the opponent's strategy, during the negotiation.
\item Learning the opponent's strategy from earlier negotiations.
\end{enumerate}
\item To learn the opponent's utility function, we have discussed three techniques:
\begin{enumerate}
\item Bayesian Learning
\item Scalable Bayesian Learning
\item Frequency analysis
\end{enumerate}
\item Bayesian Learning assumes the opponent follows a time-based strategy, and considers a finite set of hypotheses regarding the utility function of the opponent.
\item Initially, it assigns a uniform probability distribution to the set of hypothesis, but then uses Bayes' rule to update these probabilities, based on the proposals it receives from the opponent.
\item Unfortunately, the number of hypotheses grows exponentially in the number of issues and the number of options per issue. This makes the basic version of Bayesian Learning only practical on very small domains.
\item Scalable Bayesian Learning solves this, by decomposing the hypotheses into smaller sub-hypotheses, for which the probabilities are updated separately.
\item This makes Scalable Bayesian Learning applicable to somewhat larger domains than the basic Bayesian Learning approach.
\item However, Scalable Bayesian Learning still suffers from a number of other disadvantages:
\begin{enumerate}
\item We need to assume that the opponent's utility function is linear.
\item We need to assume that that the issues are ordered.
\item We need to assume that that the opponent has triangular evaluation functions.
\end{enumerate}
\item Furthermore, it depends on the chosen model of the opponent's bidding strategy and on the chosen standard deviation $\sigma$ for the Gaussian distribution.
\item Frequency Analysis is a much simpler approach that often performs just as well as (scalable) Bayesian Learning, or even better.
\item It counts, for each issue and each option, how often that option appears in the proposals received from the opponent.
\item It then assumes that the evaluation of that option is proportional to this frequency.
\item Furthermore, the opponent's weights $w_j$ are estimated based on how many different options are proposed for issue $\issue{j}$. The more variety in options, the lower the weight for that issue.
\item To learn the opponent's strategy (or, more precisely: to learn how far the opponent will be willing to concede), several approaches have been proposed in the literature, such as linear extrapolation, Gaussian process, and wavelet decomposition and cubic smoothing splines.
\end{itemize}

\chapter{Game Theory}\label{sec:game_theory}

In Chapter \ref{sec:negotiation_strategies} we discussed various negotiation strategies. The big question now, is which one is the ``best". Unfortunately, it turns out that there is no definitive answer to this question. Nevertheless, we may still want to investigate how close we can get to such an answer, and for that it is absolutely essential to have a basic understanding of the topic of game theory.

Game theory, as the name indicates, deals with the analysis of games. However, it should be understood that the notion of a `game' here is much more general than what one would normally consider a game in daily life. Specifically, \textit{game theory applies to any scenario that involves multiple agents whose goals are at least partially conflicting, and in which the outcome for each agent also depends on the the actions taken by the other agents}. In particular, this means it applies to automated negotiation.

Game theory is a very large subject and it would go much too far to go into an in-depth discussion in this book. Therefore, we will here only explain the most basic concepts that are relevant for the rest of this book. For a more in-depth study of game theory I recommend the book `\textit{A Course in Game Theory}' by Osborne and Rubinstein \cite{osborne1994course}.

Readers who are not interested in theory may want to skip this chapter, since most of it will not come back in the rest of this book.  The only exception is that the theory of normal-form games, discussed in Section~\ref{sec:normal_form_games}, as well as the notions of a symmetric game and symmetric equilbria discussed in Section~\ref{sec:symmetric_equilibria} will come back later in Section~\ref{sec:egta}.

\section{Cooperative vs. Non-Cooperative Game Theory}

In general, in game theory it is assumed that there are two or more agents, that each agent can perform certain actions, and that each agent chooses its actions so as to maximize its own individual utility function. Furthermore, it is assumed that for each agent, its utility function does not only depend on the agent's own actions, but also on the actions of the other agents.

We can distinguish between two main branches of game theory, namely \textit{cooperative} game theory, and \textit{non-cooperative} game theory. The difference is that in the case of cooperative game theory it is assumed that the agents are able to coordinate their actions, which may allow them to achieve outcomes that are mutually beneficial. In non-cooperative game theory, on the other hand, it is assumed that each agent chooses its actions in an entirely individual way, without any form of explicit coordination with the other agents.

Another way to see it, is to say that non-cooperative game theory purely focuses on the question which actions each agent will take, while cooperative game theory assumes that there is a kind of `communication layer' superimposed on top of the game, which allows the agents to coordinate or negotiate the actions they will take.

It should be understood however, that even in the case of cooperative game theory, each agent is still assumed to have its own individual utility function and that each agent is still assumed to be purely self-interested. In other words, an agent is only willing to cooperate with the other agents if that yields an individual benefit to that agent. Therefore, cooperative game theory should not be confused with \textit{distributed optimization} in which all agents share the same goal or utility function and are all programmed to work together.

We can summarize the differences as follows.
\begin{itemize}
\item \textbf{Distributed Optimization}:
\begin{itemize}
\item All agents have the same goals.
\item The agents work together to achieve their common goals.
\item \textbf{Example}: A swarm of fire-fighting drones that aim to extinguish a bush fire.
\end{itemize}
\item \textbf{Cooperative Game Theory}:
\begin{itemize}
\item Each agent has its own individual goals, which may  conflict with the goals of the other agents.
\item Agents may work together, but they only do so if that benefits them individually.
\item \textbf{Example}: Political parties that form coalitions to create a government.
\end{itemize}
\item \textbf{Non-Cooperative Game Theory:}
\begin{itemize}
\item Each agent has its own individual goals, which may  conflict with the goals of the other agents.
\item No cooperation or coordination between the agents at all. Each agent chooses its actions purely individually.
\item \textbf{Example}: a game of chess.
\end{itemize}
\end{itemize}

%\begin{table}[h]
%\begin{tabular}{|l|l|}
%\hline
%\textbf{Distributed Optimization} & All agents have the same goal or utility function.\\
%  & The agents work together to achieve their common goal.\\
%  & Example: A swarm of fire-fighting drones that aim to extinguish a bush fire.\\
%\hline
%\textbf{Cooperative Game Theory} & bla \\
%\hline
%\textbf{Non-cooperative Game Theory} & bla \\
%\hline
%\end{tabular}
%\end{table}

Automated negotiation is clearly related to cooperative game theory, since indeed it considers agents that are aiming to find a joint solution, but only if that increases their own individual utilities. In fact, one could see automated negotiation as a sub-field of cooperative game theory, although in practice the literature usually treats them as two distinct fields. A main difference, is that in the field of cooperative game theory one typically assumes that all agents have full knowledge of each others' utility functions, while in automated negotiation we usually assume the agents only have limited or no knowledge about their opponents' utility functions. Furthermore, in automated negotiation we mainly focus on the \textit{process} of how the agents agree on some final outcome (i.e. the negotiation), while in most work on cooperative game theory this process is entirely abstracted away and one only focuses on the \textit{outcome} of such negotiations.

%In particular, the field of cooperative game theory one typically assumes all agents have full knowledge of each others' utility functions and researchers mainly focus on finding mathematical solutions that satisfy certain given axioms. On the other hand, in automated negotiation we often assume the agents only have limited or no knowledge about their respective utility functions and---more importantly---one  focuses more on the \textit{process} of how the agents agree on some final outcome. In most work on cooperative game theory, on the other hand, this process is entirely abstracted away and one only focuses on the \textit{outcome} of the negotiation.

Given the close relationship between automated negotiation and cooperative game theory, it may come as a surprise that in this section and in the rest of this book we are actually more interested in \textit{non-cooperative} game theory, rather than in cooperative game theory. The reason for this, is that in order to determine which negotiation strategies are best, we need to model the process of negotiation \textit{itself} as a game. This contrasts with cooperative game-theory, in which negotiation is considered as a process that is superimposed \textit{on top of} a game. So, if we model negotiation itself as a game, it would be a non-cooperative one.

Within the field of non-cooperative game theory, we can further distinguish between two main types of games:
\begin{enumerate}
\item Normal-form games
\item Extensive-form games
\end{enumerate}
Normal-form games are games in which all players simultaneously choose exactly one action and then the game is over. Probably the most well-known example of a normal-form game is `Paper-Scissors-Rock'. 
Extensive-form games, on the other hand, are the more common type of games that take place over multiple rounds. Examples are chess, go, and poker. We will first discuss normal-form games in Sections~\ref{sec:normal_form_games} and \ref{sec:equilibrium_selection}, and then we will discuss extensive-form games in Section~\ref{sec:turn_taking_games}. Finally, in Sections~\ref{sec:nego_as_a_game} and \ref{sec:bargaining_solutions}
we will apply our knowledge of game theory to the topic of automated negotiation.

\section{Normal-Form Games}\label{sec:normal_form_games}
Formally, normal-form games are defined as follows.
\begin{definition}
Let $\numAgents$ be a positive integer. Then, an $\numAgents$-player \textbf{normal-form game} consists of:
\begin{itemize}
\item For each $i \in \{1, 2, \dots, \numAgents\}$ a set of \textbf{actions} $\Ac_i$ (sometimes also referred to as \textbf{moves}).
\item For each $i \in \{1, 2, \dots, \numAgents\}$ a \textbf{utility function} $\util_i$ that maps the Cartesian product of all action sets to the set of real numbers:
\[\util_i \quad : \quad \Ac_1 \times \Ac_2 \times \dots \times \Ac_\numAgents \rightarrow \mathbb{R}\] 
\end{itemize}
\end{definition}
Note that in game theory the agents are typically referred to as `players'. So, we will refer to each set $\Ac_i$ as the set of ``actions of player $i$" and to each utility function $\util_i$ as the ``utility function of player $i$". Furthermore, we may use the notation $\ag_i$ to refer to player $i$. In the rest of this section we will mainly focus on 2-player games. 

A tuple of actions, consisting of one action for each player is called an \textbf{action profile}. In other words, an action profile is an element of the set $\Ac_1 \times \Ac_2 \times \dots \times \Ac_\numAgents$.

Note that for each player, its utility function depends on the actions chosen by \textit{all} players. For example, in the case of Papers-Scissors-Rock (with two players), each player has the same action set $\Ac_1 = \Ac_2 = \{paper, \mi{scissors}, rock\}$. The utility function $\util_1$ for player 1 could be given by:
\begin{alignat*}{3}
\util_1(paper, paper) &= 1, &\quad \util_1(paper, \mi{scissors}) &= 0, &\quad \util_1(paper, rock) &= 2\\
\util_1(\mi{scissors}, paper) &= 2, &\quad \util_1(\mi{scissors}, \mi{scissors}) &= 1, &\quad \util_1(\mi{scissors}, rock) &= 0\\
\util_1(rock, paper) &= 0, &\quad \util_1(rock, \mi{scissors}) &= 2, &\quad \util_1(rock, rock) &= 1
\end{alignat*}
That is, player 1 receives 2 utility `points' if she wins, 0 utility points if she loses, and 1 utility point in case of a draw. Similarly, the utility function for player 2 can then be defined as $\util_2(\ac_1, \ac_2) = 2-\util_1(\ac_1, \ac_2)$, for any pair of actions $(\ac_1, \ac_2) \in \Ac_1 \times \Ac_2$.

Two-player normal-form games are typically represented using so-called \textit{pay-off matrices}. That is, a matrix for which each row corresponds to an action $\ac_1 \in \Ac_1$, and each column corresponds to an action $\ac_2 \in \Ac_2$, so it's an $|\Ac_1| \times |\Ac_2|$ matrix. Each cell of the matrix therefore corresponds to a pair of actions $\ac_1, \ac_2$ and it contains the corresponding pair of utility values $(\util_1(\ac_1, \ac_2) \ , \ \util_2(\ac_1, \ac_2))$ for the two players. For example, the payoff matrix of Paper-Scissors-Rock is displayed in Table~\ref{tab:paper_scissors_rock}.
\begin{table}[h]
\begin{center}
\begin{tabular}{l|c|c|c|}
\ & Paper & Scissors & Rock \\
\hline
Paper & (1 , 1) & (0 , 2) & (2 , 0)\\
\hline
Scissors & (2 , 0) & (1 , 1) & (0 , 2)\\
\hline
Rock & (0 , 2) & (2 , 0) & (1 , 1)\\
\end{tabular}
\caption{Payoff-matrix of the game Paper-Scissors-Rock}\label{tab:paper_scissors_rock}
\end{center}
\end{table}

In this book we will always follow the convention that player 1 is  the `\textit{row player}' and that player 2 is the `\textit{column player}'. That is, the rows of the matrix correspond to the actions of player 1, and the columns correspond to the actions of player 2.

For any action profile $(\ac_1, \ac_2)$ we may use the notation $\vec{\util}(\ac_1, \ac_2)$ to denote the utility vector of that profile. That is:
\[\vec{\util}(\ac_1, \ac_2) \ \ := \ \ (\ \util_1(\ac_1, \ac_2) \ ,\ \util_2(\ac_1, \ac_2)\ )\]

\subsection{Zero-sum Games}
Note that in the game of 2-player Paper-Scissors-Rock, no matter what actions the players choose, the sum of their respective utilities ($\util_1 + \util_2$) will always be 2. In other words, the agents' objectives are diametrically opposed. The higher the utility for player $\ag_1$, the lower the utility for player $\ag_2$ and vice versa. Such games are also known as \textbf{constant-sum games} or, more commonly, \textbf{zero-sum games}. This last name comes from the fact that we can add any arbitrary constant to the utility function of either player, without affecting the essence of the game (as per the principle of `Invariance under Linear Transformations', see Def.~\ref{def:invariance_lin_trans}). Therefore, any constant-sum game can be transformed into an equivalent game for which the sum of the players' utility values is always exactly zero. Games in which the sum of the players' utility values is not always the same are called \textbf{non-zero-sum games} or \textbf{general-sum games}.

Many board games such as chess, checkers, or go, can indeed be seen as zero-sum games because they either end with one player as the winner and the other as the loser, or in a draw. So, we can assign 2 points to the winner, 0 points to the loser, and 1 point to each player in case of a draw. Conversely, for any 2-player zero-sum game we can say that the player that achieved the highest utility is the `winner' and the other player the `loser', or that the game ended in a draw if both players achieved the same utility.

However, it is important to understand that when we study \textit{non}-zero-sum games there is not always a clear winner or loser. For example, one could encounter a game that has one action profile for which both players achieve the maximum utility, while it also has one action profile for which both players achieve the minimum utility. Therefore, in such games we cannot say that the goal is to \textit{win} the game. Instead, \textit{the goal for each player is purely to maximize its own utility value}. Especially, we should stress that in non-zero-sum games \textit{it is \underline{not} the goal of the players to `beat' the opponent, or to achieve more utility than the opponent.}

For example, if one action profile leads to a utility of 10 for player 1 and a utility of 5 for player 2, while another action profile yields a utility of 100 for player 1 and a utility of 200 for player 2, then player 1 prefers the second action profile, because it yields more utility. In particular, player 1 does \textit{not} care about the fact that with the second action profile player 2 achieves more utility than player 1. 

\subsection{Simultaneous Moves}
As we mentioned above, in a normal-form game the players choose their actions simultaneously. What we mean by this, is that each player has to choose his or her action without knowing which actions the other players are choosing. It does \textit{not} mean that the players \textit{literally} have to choose their actions at exactly the same moment. Instead, we can imagine, for example, that each player first secretly writes down his action on a piece of paper and only once all players have written down their chosen actions, those actions are revealed. While in this way the players do not literally choose their actions at exactly the same moment, the point is that each player has to make his choice without knowing the choices of the other players, which, for all intents and purposes, is the same as the situation that all agents really do choose their actions at exactly the same time.

\subsection{Pure Nash Equilibria}\label{sec:pure_nash_equilibria}
Naturally, the main question any player in any game wants to answer, is the question which action is the best action to choose. In order to study this question we will focus on 2-player games and we will assume that each player has full knowledge of the other player's utility function.

If we knew which action the opponent was choosing, then this question would be easy to answer, because then our best action would simply be the one that maximizes our utility, given the opponent's action. We call this the \textit{best response} to the opponent's action.

\begin{definition}
Let $\nfgame$ be some 2-player normal-form game and let $\ac_1 \in \Ac_1$ be any action for player $1$. Then, we say that an action $\ac_2 \in \Ac_2$ for player $2$ is a \textbf{best response} to $\ac_1$ if the following holds:
\[\forall \ac \in \Ac_2: \ \  \util_2(\ac_1, \ac) \ \ \leq \ \  \util_2(\ac_1, \ac_2)\]
Analogously, an action $\ac_1 \in \Ac_1$ for player $1$ is a \textbf{best response} to some action $\ac_2 \in \Ac_2$ for player 2, if the following holds:
\[\forall \ac \in \Ac_1: \ \  \util_1(\ac, \ac_2) \ \ \leq \ \  \util_1(\ac_1, \ac_2)\] 
\end{definition}
%\begin{definition}
%Let $\nfgame$ be some two-player normal-form game and let $i \in \{1, 2\}$ and $j \in \{1, 2\}$ with $i\neq j$. Furthermore, let $\ac_i \in \Ac_i$ be any action for player $i$. Then, we say that an action $\ac_j \in \Ac_j$ for player $j$ is a \textbf{best response} to $\ac_i$ if the following holds:
%\[\forall \ac_j' \in \Ac_j \ \ : \ \  \util_j(\ac_i, \ac_j') \ \ \leq \ \  \util_j(\ac_i, \ac_j)\]
%\end{definition}
In other words, for any action $\ac_i$ of player $i$, a `best response' for player~$j$ is an action that yields highest utility to player~$j$, when player~$i$ chooses action $\ac_i$.

For example, in the game of `Paper-Scissors-Rock', if player 1 chooses the action `scissors' then the best response for player 2 is to choose `rock'.

Note that the best response may not be unique, because multiple actions may yield the same utility. Therefore, in general, for any action $\ac_i$ there is a \textit{set} of actions which are all best responses. We will denote this set by $\br_j(\ac_i)$. That is:
\begin{eqnarray*}
\br_1(\ac_2) \quad &:=& \quad \argmax_{\ac} \quad \{\util_1(\ac, \ac_2) \mid \ac \in \Ac_1\}\\
\br_2(\ac_1) \quad &:=& \quad \argmax_{\ac} \quad \{\util_2(\ac_1, \ac) \mid \ac \in \Ac_2\}
\end{eqnarray*}
So, the phrase ``$\ac_j$ \textit{is a best response to} $\ac_i$" can be formally denoted as $\ac_j \in \br_j(\ac_i)$.

%In other words, $\br(\ac_i) \in \Ac_j$ satisfies:
%\[\forall \ac_j \in \Ac_j \ \ : \ \  \util(ac_i, \ac_j) \ \ \leq \ \  \util(ac_i, \br(ac_i))\]
Of course, the problem is that, in principle, we do not know the opponent's action. However, to solve this, we can assume that the opponent is rational, which may allow us to \textit{reason} about what action the opponent would choose.

In the following we will follow our usual convention that we are implementing agent $\ag_1$ and therefore that $\ag_2$ is our opponent.

The idea is as follows. Before the game starts, we first choose some arbitrary action $\ac_1 \in \Ac_1$. We then assume that, if there is indeed a good reason for us to pick that action, then the opponent would be able to follow that reasoning and therefore would be able to conclude that we are picking $\ac_1$. But that means that if the opponent is rational she would now choose an action $\ac_2$ that is a best response against our action (i.e.  $\ac_2 \in \br_2(\ac_1)$). Now, assuming that the opponent will indeed choose that action, we can change our mind (before the game starts), and instead pick a new action $\ac_1'$ that is a best response to \textit{that} action $\ac_2$. That is, we choose $\ac_1' \in \br_1(\ac_2)$. Now, again, we can make the assumption that the opponent is able to reason in the same way as us, and therefore is able to anticipate our change of mind, which allows her to also change her mind, and pick a best response to our new choice. That is, we now assume the opponent will actually choose some action $\ac_2'\in \br_2(\ac_1')$. If we keep reasoning like this, then either of the following two things can happen: 
\begin{enumerate}
\item The two players keep changing their actions infinitely often.
\item At some point they reach an equilibrium were neither of the two players changes their mind anymore, because they have chosen two actions that are best responses \textit{to each other}.
\end{enumerate}
In the second case, we say the players have reached a \textit{Nash equilibrium}. More precisely, we say the two players have reached a \textit{pure} Nash equilibrium. There also exists a different kind of equilibrium that is called a \textit{mixed} Nash equilibrium, but we will discuss that later on.

It is important to understand that this process of players changing their actions until they reach equilibrium, only describes the \textit{thought process} of the players \textit{before the game has started}. In other words, it only takes place \textit{in their minds}. After all, once the game starts, the player reveal their moves simultaneously, and after that they cannot change their moves anymore.

Formally, a pure Nash equilibrium is a pair of actions, such that each of the two actions is a best response to the other one.
\begin{definition}
Let $(\ac_1, \ac_2) \in \Ac_1 \times \Ac_2$ be any pair of actions of a two-player normal-form game. We say it is a \textbf{pure Nash equilibrium} iff:
\[\ac_1 \in \br_1(\ac_2) \quad \text{and} \quad \ac_2 \in \br_2(\ac_1)\]
\end{definition}

The following observation states the importance of Nash equilibria.
\begin{observation}
If a normal-form game has exactly one Nash equilibrium, then the action profile chosen by two optimal players would be exactly that Nash equilibrium. 
\end{observation}
To see this, assume the opposite. Suppose that they choose an action profile $(\ac_1, \ac_2)$, that is not a Nash equilibrium. In particular, let us assume that $\ac_1$ is not a best response to $\ac_2$. That means that player $1$ could have achieved more utility if he had chosen a different action $\ac_1'$ that \textit{is} a best response to $\ac_2$ (i.e. $\ac_1' \in \br(\ac_2)$). So, by choosing $\ac_1$ player $1$ did not make an optimal choice, which contradicts the assumption that they were playing optimally.

Now, imagine that before they play the game, all players have decided which action they each will play. However, suppose that right before they reveal their chosen actions, one player changes his mind and switches to another action,  \textit{while the other player keeps her decisions unchanged}. We then say the first player is making a \textbf{unilateral deviation}. With this terminology the notion of a pure Nash equilibrium can be defined alternatively as: ``\textit{a strategy profile such that no agent can increase his utility by making a unilateral deviation}''.

Unfortunately, not all games have a pure Nash equilibrium. One example is the Paper-Scissors-Rock game. If we apply our reasoning above to this game, it is easy to see that we keep looping forever. For example, if we initially choose `paper', then our opponent will choose the best response, which is `scissors'. Then, we can change our mind and choose the best response against `scissors', which is `rock'. Next, the opponent will change to the best response against `rock' which is `paper', etcetera.  Clearly, this will continue forever.

An example of a game that does have a pure Nash equilibrium, is the well-known Prisoner's dilemma, which we will discuss next.

\subsection{The Prisoner's Dilemma}\label{sec:prisoners_dilemma}
The prisoner's dilemma is probably the most commonly used example in game theory, because it shows the counter-intuitive result that when every player plays optimally from his own individual point of view, the final outcome may actually turn out to be very bad for each individual player.

The prisoner's dilemma is typically explained as follows: two prisoners are each being questioned separately by the police. They each have two options: to confess that they committed a crime, or to deny that they did it. If they both confess then they both have to stay in prison for 8 years. If they both deny, then they both only have to stay in prison for 2 years. However, if one of them denies and the other confesses, then the one who confessed will be released from prison immediately and be free, while the other one will have to stay in prison for 10 years. 

We should stress that we are discussing this game in the context of non-cooperative game theory, so the prisoners are not able to communicate and each of them has to make his decision in complete isolation from the other.

This game can be displayed as the following payoff matrix.
\begin{center}
\begin{tabular}{l|c|c|}
\ & Deny & Confess \\
\hline
Deny & (8 , 8) & (0 , 10) \\
\hline
Confess & (10 , 0) & (2 , 2) \\
\end{tabular}
\end{center}
Note that the utilities here are given as $10-x$, where $x$ is the number of years they stay in prison. So, if a prisoner is released immediately he will get a utility of 10. The payoff vector $(8,8)$ represents that they both go to prison for 2 years, while the payoff vector $(2,2)$ represents that they both go to prison for 8 years. This is because we follow the standard convention that the matrix displays \textit{utility} values, which the players are aiming to \textit{maximize}.

Now, the question is what the optimal strategy for each of the two prisoners would be. Most people who see this game for the first time would argue that the best strategy is to play `deny', because if both players choose that action, they will both get a low punishment. However, perhaps surprisingly, we will see that the optimal strategy is actually to play `confess'.

To see this, let us first imagine that player 1 is choosing to play `deny'. What is now the best response for player 2? We see from the matrix that if player 2 chooses `deny' as well, then she receives a utility of 8 (2 years in prison), while if she chooses `confess' she receives a utility of 10 (immediate freedom). So, `confess' is the best response.
\[\br_2(deny) = \{\mi{confess}\}\]
Next, suppose that player 1 chooses to play `confess'. We now see that if player 2 chooses `deny' she will get a utility of 0 (i.e. 10 years in prison), while if she chooses `confess' she will get a utility of 2 (i.e. 8 years in prison). Again, we see that `confess' is the best option.
\[\br_2(\mi{confess}) = \{\mi{confess}\}\]
In other words: \textit{No matter what player 1 chooses, player 2 is always better off if she chooses `confess'}. Vice versa, the same holds for player 1. Player~1 is always better off by playing `confess', no matter what player 2 chooses. We therefore see that the action profile $(\mi{confess}, \mi{confess})$ is the unique pure Nash equilibrium of this game. From this we conclude that if both players are perfectly rational, they would each choose to play `confess' and therefore they would each go to prison for 8 years. 

The conclusion of our analysis may seem highly counter-intuitive, because if they cooperated they could have ensured to go to prison for only 2 years. The problem with that cooperative solution, however, is that even if the players could somehow make an agreement to each play `deny', then, \textit{by assumption of non-cooperative game theory}, still neither of the two players could be forced to keep their promise. So, if you agree with your opponent to play `deny', then the best thing you can do is to break your promise and play `confess' anyway. Formally speaking, we say that players cannot \textit{commit} to their actions in advance.

The reason this outcome seems so counter-intuitive, is that in real life most situations we encounter do not follow the strict rules of non-cooperative game theory. For example:
\begin{itemize}
	\item In real life people are social:
	\begin{itemize}
		\item The prisoners could be friends or family that prefer to help each other rather than to make purely selfish choices.
		\item People are hardwired to often be helpful and friendly, even to strangers.
	\end{itemize}
	\item In real life, people may fear repercussions if they betray others.
	\item In real life, people \textit{can} commit to their actions:
		\begin{itemize}
		\item They can sign legally binding contracts.
		\item They may feel obliged to keep their promises as a matter of honor.
		\end{itemize}
\end{itemize}
On the other hand, in non-cooperative game theory we assume:
\begin{itemize}
\item that the players are \textit{only} interested in maximizing their own individual utility functions,
\item that each game is played in complete isolation, so actions in the current game do not have repercussions in later games,
\item that players cannot commit in advance to their actions.
\end{itemize}

\later{Maybe we should discuss here why it is still important to study non-cooperative game theory.}

Note that indeed, as per the definition of a Nash equilibrium, neither of the two players can increase their utility by making a \textit{unilateral} deviation. On the other hand, in the prisoner's dilemma it \textit{is} possible for the players to increase their utility if they \textit{both} switch from `confess' to `deny'. In other words, if they make a \textit{bilateral} deviation. However, the definition of a Nash equilibrium does not take such bilateral deviations into consideration. The reason for this, again, is that we are talking about \textit{non-cooperative} game theory, which, by definition, assumes the players cannot coordinate their actions. So, whenever a player switches to a different action, he has to assume that this will not affect the opponent, and thus that the opponent's action remains unchanged.

\subsection{Multiple Pure Nash Equilibria}\label{sec:multiple_nash_eq}
As discussed above, some games, such as Paper-Scissors-Rock, do not have any pure Nash equilibrium. Other games, on the other hand, have the problem that they actually have \textit{multiple} pure Nash equilibria.

A simple example is the game known as `Battle of the Sexes'. It can be explained as follows. The two players are a married couple and they want to go out. They each can choose between two options: to go to a football match or to go to a ballet performance. While the husband prefers to see the football match, the wife prefers to go to the ballet performance. However, for both, the most important thing is that they go together. That is, they each prefer to choose the same activity, rather than that they each choose a different activity. This can be summarized in the following payoff matrix:

\begin{center}
\begin{tabular}{l|c|c|}
\ & Football & Ballet \\
\hline
Football & (2 , 1) & (0 , 0) \\
\hline
Ballet & (0 , 0) & (1 , 2) \\
\end{tabular}
\end{center}

Note that no matter what the wife chooses, the best response for the husband is to choose the same option, and similarly, no matter what the husband chooses, the best response for the wife is also to choose the same option:
\[\forall i \in \{1,2\}: \ \ \br_i(\mi{football}) = \{\mi{football}\} \ \ \text{and} \ \ \br_i(ballet) = \{ballet\}\]
This means that there are two pure Nash equilibria:
\[(\mi{football}, \mi{football}) \quad \text{and} \quad (\mi{ballet}, \mi{ballet})\]

\subsection{Mixed Nash Equilibria}\label{sec:mixed_nash_equilibria}
We have seen that the Paper-Scissors-Rock game does not have any pure Nash equilibria.  No matter which of the three actions we choose, if the opponent can anticipate our action, then she can choose the best response to that action, and we lose. So, how then do we determine our optimal strategy? The answer is simple: by making sure that the opponent cannot anticipate our action. Specifically, we can do that by picking an action randomly. We call this a \textit{mixed strategy}.

\begin{definition}
Let $\Ac_i$ be the set of actions of player $i$. Then, a  \textbf{mixed strategy} for player $i$ is a probability distribution over the set $\Ac_i$. That is, a function $\ms : \Ac_i \rightarrow \mathbb{R}$ such that $\sum_{\ac_i \in \Ac_i} \ms(\ac_i) = 1$ and $\forall \ac_i \in \Ac_i: \ms(\ac_i) \geq 0$. We will denote the set of all mixed strategies of player $i$ by $\MS_i$.
\end{definition}
The interpretation is that the player selects each action $\ac_i$ with probability $\ms(\ac_i)$. Note that even if the game only has a finite number of actions, each player has an infinite number of possible mixed strategies.

Whenever a player does not choose his action randomly, but instead just  chooses one specific action deterministically, then this is also known as a \textbf{pure strategy}. Of course, one can say that a pure strategy is actually just a special case of a mixed strategy, for which there is exactly one action $\ac_i$ with $\ms(\ac_i) = 1$ and therefore $\ms(\ac_i') = 0$ for all other actions $\ac_i'\in \Ac_i$.

A tuple $\vec{\ms} = (\ms_1, \ms_2, \dots, \ms_\numAgents)$ consisting of one mixed strategy for each player is called a \textbf{strategy profile}.

Previously, we defined the utility function of a player as a function that assigns a utility value to every possible action profile. This can now be extended to profiles of mixed strategies, by defining it as the  \textit{expected} utility over all pure action profiles. That is, for games with two players:
\[\util_i(\ms_1, \ms_2) \ \ := \ \ \sum_{\ac_1 \in \Ac_1} \sum_{\ac_2 \in \Ac_2} \ms_1(\ac_1) \cdot \ms_2(\ac_2) \cdot \util_i(\ac_1, \ac_2)\]
We may use the notation $\vec{\util}(\ms_1, \ms_2)$ or $\vec{\util}(\vec{\ms})$ to denote the utility vector of $\vec{\ms}$:
\[\vec{\util}(\vec{\ms}) \ \ := \ \ (\ \util_1(\vec{\ms})\ ,\ \util_2(\vec{\ms})\ )\]

For example, in the game of Paper-Scissors-Rock, suppose that player $\ag_1$ chooses a mixed strategy $\ms_1$ in which he plays `paper' with a probability of 40\% and `scissors' with a probability of 60\%, and  suppose that player $\ag_2$ chooses a mixed strategy $\ms_2$ in which she plays `scissors' with a probability of 20\% and `rock' with a probability of 80\%, then, the expected utility of player~1 will be:
\begin{eqnarray*}
\util_1(\ms_1, \ms_2) &=& 0.4\cdot 0.2 \cdot \util_1(paper,\mi{scissors}) + 0.6\cdot 0.2 \cdot \util_1(\mi{scissors},\mi{scissors}) + \\
& & 0.4\cdot 0.8 \cdot \util_1(paper,rock) + 0.6\cdot 0.8 \cdot \util_1(\mi{scissors},rock)\\
&=& 0.4\cdot 0.2 \cdot 0 + 0.6\cdot 0.2 \cdot 1 +  0.4\cdot 0.8 \cdot 2 + 0.6\cdot 0.8 \cdot 0\\
%&=& 0.12 \cdot 1 +  0.32 \cdot 2 \\
&=& 0.76 \\
\end{eqnarray*}
while for player $\ag_2$ it will be:
\begin{eqnarray*}
\util_2(\ms_1, \ms_2) &=& 0.4\cdot 0.2 \cdot \util_2(paper,\mi{scissors}) + 0.6\cdot 0.2 \cdot \util_2(\mi{scissors},\mi{scissors}) + \\
& & 0.4\cdot 0.8 \cdot \util_2(paper,rock) + 0.6\cdot 0.8 \cdot \util_2(\mi{scissors},rock)\\
&=& 0.4\cdot 0.2 \cdot 2 + 0.6\cdot 0.2 \cdot 1 +  0.4\cdot 0.8 \cdot 0 + 0.6\cdot 0.8 \cdot 2\\
%&=& 0.08\cdot 2 + 0.12 \cdot 1 +  0.32 \cdot 0 + 0.48\cdot 2 \\
%&=& 0.16 + 0.12 + 0.96 \\
&=& 1.24 \\
\end{eqnarray*}
This, in turn, allows us to extend the definition of `best response' to mixed strategies.

\begin{definition}
Let $\nfgame$ be some two-player normal-form game and let $\ms_1 \in \MS_1$ be a mixed strategy for player $1$. Then, we say that a mixed strategy $\ms_2 \in \MS_2$ for player $2$ is a \textbf{best response} to $\ms_1$ if the following holds:
\[\forall \ms \in \MS_2: \ \  \util_2(\ms_1, \ms) \ \ \leq \ \  \util_2(\ms_1, \ms_2)\]
Analogously, a mixed strategy $\ms_1 \in \MS_1$ for player $1$ is a \textbf{best response} to some mixed strategy $\ms_2 \in \MS_2$ for player $2$, if the following holds:
\[\forall \ms \in \MS_1: \ \  \util_1(\ms, \ms_2) \ \ \leq \ \  \util_1(\ms_1, \ms_2)\]
\end{definition}
%\begin{definition}
%Let $\nfgame$ be some two-player normal-form game and let $i \in \{1, 2\}$ and $j \in \{1, 2\}$ with $i\neq j$. Furthermore, let $\ms_i$ be a mixed strategy for player $i$. Then, we say that a mixed strategy $\ms_j $ for player $j$ is a \textbf{best response} to $\ms_i$ if the following holds:
%\[\forall \ms_j' \in \MS_j  \ \ : \ \  \expt{\util}_j(\ms_i, \ms_j') \ \ \leq \ \  \expt{\util}_j(\ms_i, \ms_j)\]
%\end{definition}
As before, we use the notation $\br_j(\ms_i)$ to denote the set of best responses to a mixed strategy $\ms_i$.
\begin{eqnarray*}
\br_1(\ms_2) \quad &:=& \quad \argmax_{\ms_1} \quad \{\util_1(\ms_1, \ms_2) \mid \ms_1 \in \MS_1\}\\
\br_2(\ms_1) \quad &:=& \quad \argmax_{\ms_2} \quad \{\util_2(\ms_1, \ms_2) \mid \ms_2 \in \MS_2\}
\end{eqnarray*}

\later{Introduce special notation for the set of \textit{mixed} best responses?}

Finally, we can now also generalize the concept of a pure Nash equilibrium to mixed strategies.
\begin{definition}
Let $(\ms_1, \ms_2)$ be any pair of mixed strategies of a two-player normal-form game. We say it is a \textbf{mixed Nash equilibrium} if:
\[\ms_1 \in \br_1(\ms_2) \quad \text{and} \quad \ms_2 \in \br_2(\ms_1)\]
\end{definition}
It can be shown that every pure Nash equilibrium is also a mixed Nash equilibrium (if we consider a pure strategy to be a special case of a mixed strategy). To prove this, one must show that if a player cannot deviate to a better action, he also cannot deviate to a better mixed strategy. It is not hard to see that this is indeed true, so we will leave this as an exercise to the reader. We refer to \cite{osborne1994course} for more details.

%\begin{definition}
%For any mixed strategy its \textbf{support} is the set of actions for which the probability is non-zero:
%\[sup(\ms) = \{\ac_i \in \Ac_i \mid \ms(\ac_i) > 0\}\]
%\end{definition}

While we have seen that not every game has a pure Nash equilibrium, it turns out that every finite 2-player normal-form game does have at least one mixed Nash equilibrium. A proof of this theorem can be found in \cite{osborne1994course}.
\begin{theorem}\label{thm:existence_nash_eq}
Every finite 2-player normal-form game has at least one mixed Nash equilibrium.
\end{theorem}

It is relatively straightforward to determine the pure Nash equilibria of a normal-form game. All it amounts to is to determine for each action of either player which actions are best responses. This can be seen directly from the pay-off matrix. Determining the \textit{mixed} Nash equilibria, on the other hand, is a  computationally hard problem that you would typically not do manually. Instead there are various algorithms for this task, such as the Lemke-Howson algorithm \cite{LemkeHowson1964}. A commonly used software package that implements such algorithms is the Gambit library \cite{gambit}.

\section{The Equilibrium Selection Problem}\label{sec:equilibrium_selection}
As mentioned above, our aim is to determine, for any given normal-form game, what the optimal strategy would be for each of the players. So far, we have only partially answered this question. Namely, we now know that the players should be playing a Nash equilibrium (pure or mixed). Furthermore, we know from Theorem~\ref{thm:existence_nash_eq} that such a Nash equilibrium always exists. However, that still leaves us with the question \textit{which} Nash equilibrium to choose if the game has \textit{multiple} Nash equilibria.  This problem is known as the \textit{equilibrium selection problem} (ESP). This is especially important for us because, as we will see later on (in Sec.~\ref{sec:nash_equilibria_nego}), in automated negotiation there are typically indeed many Nash equilibria.

While many solutions to the ESP have been proposed in the literature, 
none of them is widely accepted as being fully satisfactory for general normal-form games. However, there are a number of solutions to this problem that are applicable to special cases. We will here discuss some of them. But before that, we will first discuss some apparent solutions to the ESP that might seem to make sense at first, but that, upon closer inspection, actually turn out not to be satisfactory.

\subsection{Wrong Solutions to the Equilibrium Selection Problem}\label{sec:wrong_solutions}

A naive solution to the ESP, would be to assume that a player could simply flip a coin to choose one of the equilibria at random and then play his strategy from that equilibrium. However, we will see that this solution typically does not work.

Suppose that a certain 2-player game has exactly two Nash equilibria: $(\ms_1, \ms_2)$ and $(\ms_1', \ms_2')$. Now, suppose that player~1 flips a coin to choose the first equilibrium with probability $P$ and the second equilibrium with probability $1-P$. The problem, is that this means that essentially, player~1 is playing neither $\ms_1$ nor $\ms_1'$, but in fact is playing an entirely different mixed strategy, namely: $P \cdot \ms_1 + (1-P)\cdot \ms_1'$. And since we assumed there were only two Nash equilibria, this means that player 1 is in fact not playing any equilibrium strategy \textit{at all}. He's playing a different mixed strategy that may not be a best response to the opponent's strategy. Therefore, if player~2 could reason that player 1 is playing that strategy, then player 2 could play a best response against it, which may yield a much better outcome for player~2 (and a much worse outcome for player 1) than if they had played either of the Nash equilibria. Furthermore, it would mean that player~1 could improve by deviating to a different strategy and therefore that it is currently not playing an optimal strategy.

Another idea could be that player 1 chooses a Nash equilibrium based on some entirely different criterion that is not related to his utility function at all. For example, for each of his potential strategies $\ms_1$ and $\ms_1'$, he could look at the \textit{name} of the action that receives the highest probability, and then select the strategy for which this name comes earliest in alphabetical order. However, this solution suffers from essentially the same problem. Since the choice of player~1 is not based on his utility function, player 2 cannot reason which strategy player~1 would choose, and therefore instead has to \textit{guess} it. Therefore, player~2 would reason that there is a 50\% chance that player~1 chooses strategy $\ms_1$ ad a 50\% chance that player~1 chooses strategy $\ms_1'$. This means that the optimal strategy for player~2 would be to pick the best response against $0.5 \cdot \ms_1 + 0.5\cdot \ms_1'$. Again, this would typically mean that the players end up playing an entirely different strategy profile, which is not a Nash equilibrium.

\subsection{Factorizable Sets}\label{sec:factorizable_sets}

%In the previous few sections we have discussed various techniques to choose the optimal Nash equilibrium, but unfortunately none of these methods is guaranteed to yield a unique solution. In other words, each of these methods may only yield us a subset of possible `candidate' equilibria $\cand$. For example, in Section \ref{sec:pareto_optimal_nash} this was the set of all Nash equilibria that are Pareto-optimal among Nash equilibria. In Section \ref{sec:symmetric_equilibria} this was the set of all symmetric Nash equilibria that aren't dominated by any other symmetric equilibrium, and in Section \ref{sec:AoRE} this was the set of all Nash equilibria that maximize the utility-sum \later{and minimize absolute utility difference}.
%
%In the ideal case, the set of candidates just contains one element, in which case we have a unique solution to the equilibrium selection problem. But if it contains multiple equilibria, then we still need a way to choose among them and, 

The first valid solution to the ESP that we will discuss, applies only to the very special case that all Nash equilibria have exactly the same utility vector and, moreover, the set of all Nash equilibria happens to be \textit{factorizable} or \textit{semi-factorizable}.

\begin{definition}
Let $\nfgame$ be a 2-player normal-form game and let $\cand \subseteq \MS_1 \times \MS_2$ be a set of strategy profiles of $\nfgame$. We say that $\cand$ is  \textbf{factorizable}, if for each player $\ag_i$ there is a set of strategies $\sfs_i \subseteq \MS_i$ such that $\cand$ is the Cartesian product of those two sets: $\cand = \sfs_1 \times \sfs_2$.
\end{definition}

For example, suppose the set $\cand$ consists of the following four strategy profiles:
\[\cand \ \ = \ \ \{ \ (\ms_1, \ms_2)\ ,\ (\ms_1', \ms_2)\ ,\ (\ms_1, \ms_2')\ ,\ (\ms_1', \ms_2')\ \}\]
then $\cand$ is indeed factorizable, because it can be written as:
\[\cand = \{\ms_1, \ms_1'\} \times \{\ms_2, \ms_2'\}\]
That is, every combination of $\ms_1$ or $\ms_1'$ with either $\ms_2$ or $\ms_2'$ is contained in $\cand$. On the other hand, if we have 
\[\cand = \{ \ (\ms_1, \ms_2)\ ,\  (\ms_1', \ms_2')\ \}\]
then $\cand$ is not factorizable.

Note that any set of strategy profiles $\cand$ that is a \textit{singleton} set (i.e. it contains exactly one strategy profile), is factorizable.

Now, remember from the previous section that, in general, players cannot just select a Nash equilibrium at random, because if they do try to do that, then the resulting strategy profile may actually turn out not to be a Nash equilibrium. However, if it happens that the players are completely indifferent between all Nash equilibria of the game and on top of that the set of all Nash equilibria happens to be factorizable, then that argument no longer holds and the players can safely choose any random Nash equilibrium.

%there is an exception to this rule. Namely,  , and  the set of all Nash equilibria 
% 
% is the Cartesian product of two sets of strategies for the two respective players.
%
%
%The point is that, if the set of all Nash equilibria of a game is factorizable as $S_1 \times S_2$ then each player $\ag_i$ can just select a random strategy from $S_i$, and the resulting strategy profile will be guaranteed to be a Nash equilibrium. 

\begin{observation}
If, for some 2-player normal form game $\nfgame$ the set of all Nash equilibria $\mi{NE}$ happens to satisfy the following two conditions:
\begin{enumerate}
\item $\mi{NE}$ is factorizable.
\item Every element of $\mi{NE}$ has the same utility vector:
\[\forall \vec{\ms} , \vec{\ms}'\in \mi{NE}: \quad \vec{\util}(\vec{\ms}) = \vec{\util}(\vec{\ms}')\]
%\[\forall (\ms_1, \ms_2) , (\ms_1', \ms_2') \in \mi{NE}: \quad \vec{\util}(\ms_1, \ms_2) = \vec{\util}(\ms_1', \ms_2')\]
\end{enumerate}
then each agent can simply pick a random Nash equilibrium and play his strategy according to that equilibrium.
\end{observation}

Note that the second condition is necessary to ensure both agents are completely indifferent between all possible Nash equilibria. After all, if an agent did prefer some equilibria over some other equilibria, then it would not be optimal to just select any arbitrary equilibrium at random.

Next, we will see that this observation can be improved, with the following definition.
\begin{definition}
Let $\nfgame$ be a 2-player normal-form game and let $\cand \subseteq \MS_1 \times \MS_2$ be a set of strategy profiles of $\nfgame$. Furthermore, let us define $\fs_i$ to be the set of all strategies of agent $\ag_i$ that appear in any of those strategy profiles:
\[\fs_1 := \{\ms_1 \in \MS_1 \mid \exists \ms_2 \in \MS_2: (\ms_1, \ms_2) \in \cand\}\]
\[\fs_2 := \{\ms_2 \in \MS_2 \mid \exists \ms_1 \in \MS_1: (\ms_1, \ms_2) \in \cand\}\]
We say that $\cand$ is  \textbf{semi-factorizable}, if for each player $\ag_i$ we can find a set $\sfs_i \subseteq \MS_i$ such that the following conditions both hold:
\begin{enumerate}
\item $\sfs_1 \times \fs_2 \subseteq \cand$
\item $\fs_1 \times \sfs_2 \subseteq \cand$
\end{enumerate}
we will refer to the sets $\sfs_1$ and $\sfs_2$ as \textbf{safe sets}, and their elements as \textbf{safe strategies}.
\end{definition}
Now, if we assume that the two agents will each choose a strategy profile from $\cand$ and then play their own strategy from that profile, then the first condition guarantees that if $\ag_1$ chooses a strategy from $\sfs_1$, then the two strategies chosen by the two players will definitely be in $\cand$. The second condition says the same, but for $\ag_2$. In other words, each player $\ag_i$ can choose any arbitrary safe strategy without having to worry about the choice of the opponent.

Note that each factorizable set is also semi-factorizable (with $\sfs_i = \fs_i$). An example of a set that is semi-factorizable but not factorizable, is the following: 
\[\cand \ \ = \ \ \{ \ (\ms_1, \ms_2)\ ,\ (\ms_1', \ms_2)\ ,\ (\ms_1, \ms_2')\ \}\]
with $\sfs_1 = \{\ms_1\}$ and $\sfs_2 = \{\ms_2\}$. Indeed, we have:
\[\sfs_1 \times \fs_2 \ \ = \ \ \{\ms_1\} \times \{\ms_2, \ms_2'\} \ \ =  \ \ \{ \ (\ms_1, \ms_2)\ ,\ (\ms_1, \ms_2')\ \} \ \ \subseteq \ \ \cand\]
\[\fs_1 \times \sfs_2 \ \ = \ \  \{\ms_1, \ms_1'\} \times  \{\ms_2\} \ \ = \ \   \{ \ (\ms_1, \ms_2)\ ,\ (\ms_1', \ms_2)\ \} \ \ \subseteq \ \ \cand\]

\begin{observation}
If, for some 2-player normal form game $\nfgame$ the set of all Nash equilibria $\mi{NE}$ happens to satisfy the following two conditions:
\begin{enumerate}
\item $\mi{NE}$ is semi-factorizable.
\item Every element of $\mi{NE}$ has the same utility vector:
\[\forall \vec{\ms} , \vec{\ms}'\in \mi{NE}: \quad \vec{\util}(\vec{\ms}) = \vec{\util}(\vec{\ms}')\]
%\[\forall (\ms_1, \ms_2) , (\ms_1', \ms_2') \in \mi{NE}: \quad \vec{\util}(\ms_1, \ms_2) = \vec{\util}(\ms_1', \ms_2')\]
\end{enumerate}
then each agent can pick a random strategy from its safe set $\sfs_i$.
\end{observation}

While this exact situation may not happen very often, this solution is still very important because we will later see that the same ideas can be used as a refinement on the other solutions that we will discuss next.

\subsection{Pareto-Optimality among Nash Equilibria}\label{sec:pareto_optimal_nash}
Perhaps the most obvious way to partially resolve the equilibrium selection problem, is to argue that players would never choose a Nash equilibrium that is dominated by some other Nash equilibrium.

In Section \ref{sec:pareto_optimal_offers} we gave the definition of `domination' and `Pareto-optimality' for offers. The same concepts can also be defined for strategy profiles.
%\begin{definition}
%We say that a strategy profile $(\ms_1, \ms_2)$ \textbf{dominates} another strategy profile $(\ms_1', \ms_2')$ if:
%\[\forall i\in \{1, 2\} \ : \ \util_i(\ms_1, \ms_2) \geq \util_i(\ms_1', \ms_2')\]
%and there is at least one player for which this inequality is strict:
%\[\exists i\in \{1, 2\} \ : \ \util_i(\ms_1, \ms_2) > \util_i(\ms_1', \ms_2')\]
%We say a strategy profile $\ms'$ \textbf{is dominated} by $\ms$, if $\ms$ dominates $\ms'$. A strategy profile $(\ms_1, \ms_2)$ is \textbf{Pareto optimal} if it is not dominated by any other strategy profile.
%\end{definition}
\begin{definition}
We say that a strategy profile $\vec{\ms}$ \textbf{dominates} another strategy profile $\vec{\ms}'$ if:
\[\forall i\in \{1, 2\} \ : \ \util_i(\vec{\ms}) \geq \util_i(\vec{\ms}')\]
and there is at least one player for which this inequality is strict:
\[\exists i\in \{1, 2\} \ : \ \util_i(\vec{\ms}) > \util_i(\vec{\ms}')\]
We say a strategy profile $\vec{\ms}'$ \textbf{is dominated} by $\vec{\ms}$, if $\vec{\ms}$ dominates $\vec{\ms}'$. A strategy profile $\vec{\ms}$ is \textbf{Pareto optimal} if it is not dominated by any other strategy profile.
\end{definition}
Clearly, if a game has two Nash equilibria and one of them yields a utility of 10 to each player, while the other one yields a utility of 20 to each player, then both players would choose the second one.

%Therefore, we can assume that if the game has multiple Nash equilibria then the players would choose one that is not dominated by any other Nash equilibrium.

We therefore argue that in a game with multiple Nash equilibria, the players would only consider choosing those that are Pareto-optimal \textit{among the Nash equilbria}.
\begin{definition}
We say a Nash equilibrium $\vec{\ms}$ is \textbf{Pareto-optimal among Nash equilibria}, if it is not dominated by any other Nash equilibrium.
\end{definition}

Note that we make a distinction between a Nash equilibrium being `\textit{Pareto-optimal}' and being `\textit{Pareto-optimal among Nash equilibria}'. The first concept means that it is not dominated by any other \textit{action profile}. The second concept is  much weaker because it only says that it is not dominated by any other \textit{Nash equilibrium}. 

For example, in the prisoner's dilemma, the Nash equilibrium $(\mi{confess}, \mi{confess})$ is dominated by the action profile $(deny, deny)$. Therefore, $(\mi{confess}, \mi{confess})$ is not Pareto-optimal. However, $(deny, deny)$ is not a Nash equilibrium. So, while $(\mi{confess}, \mi{confess})$ is dominated by some other action profile, it is not dominated by any other Nash equilibrium (after all, it is the \textit{only} Nash equilibrium) and therefore we can say that it is Pareto-optimal \textit{among Nash equilbria}.

Unfortunately, however, this solution still does not completely solve the equilibrium selection problem, because it is perfectly possible for a game to have multiple Nash equilibria that are Pareto-optimal among Nash equilibria.

\later{mention something about combining this with factorizability}

\subsection{Symmetric Games and Symmetric Equilibria}\label{sec:symmetric_equilibria}
There is another way to (partially) solve the equilibrium selection problem, but it only applies to so-called \textit{symmetric} games.

A symmetric game is a game for which it does not matter which player you are, because the game looks exactly the same from the point of view of either player. The game of Paper-Scissors-Rock and the prisoner's dilemma are both examples of symmetric games. In each of these games it clearly does not matter whether you are `player 1' or `player 2', because those are just labels. If you switch the players' roles, nothing changes. 

To keep things simple we will here give a definition of the concept of a `symmetric game' that is actually somewhat too strict, but easier to understand than the full definition.
\begin{definition}\label{def:symmeteric_game_1}
Let $\nfgame$ be a 2-player normal-form game. We say it is a \textbf{symmetric game} if $\Ac_1 = \Ac_2$, and for any $(\ac_1, \ac_2) \in \Ac_1 \times \Ac_2$ we have:
\begin{equation}\label{eq:symmetric_game_1}
\util_1(\ac_1, \ac_2) \ \ = \ \ \util_2(\ac_2, \ac_1)
\end{equation}
\end{definition}

It is easy to see that Paper-Scissors-Rock satisfies this definition. For example, suppose that Alice is player 1 and she plays `scissors', while Bob is player 2 and he plays `rock'. Then Alice loses so she receives 0 points. That is, we have: $\util_1(\mi{scissors}, rock) = 0$. Now, imagine that the roles are switched, but that the players still play exactly the same actions. That is, Bob is now player 1, but he still plays `rock' and Alice is now player 2, but she still plays `scissors'. Clearly, Bob still wins the game and Alice still receives 0 points. However, because we have switched their `roles', this is now formalized as: $\util_2(rock, \mi{scissors}) = 0$. Indeed, we see that it doesn't matter who is `player 1' and who is `player 2' and that we have $\util_1(\mi{scissors}, rock) = \util_2(rock, \mi{scissors})$, which is indeed an instance of Eq.~(\ref{eq:symmetric_game_1}).

As we mentioned, Def.~\ref{def:symmeteric_game_1} is actually too strict in the sense that it requires the two action sets $\Ac_1$ and $\Ac_2$ to be \textit{exactly} equal. This means that if we just change the \textit{names} of the actions of one of the two players, then the game will trivially fail Definition~\ref{def:symmeteric_game_1}. For example, suppose we said that player 1 still has the actions $\Ac_1 = \{paper, \mi{scissors}, rock\}$, but that player~2 now has the actions $\Ac_2 = \{parrot, \mi{sizzlers}, rack\}$. The payoff matrix stays exactly the same as in Table~\ref{tab:paper_scissors_rock}, but the columns are now labeled with these new actions, while the rows are still labeled with the original actions. Since we now have $\Ac_1 \neq \Ac_2$, this game would---according to Def.~\ref{def:symmeteric_game_1}---no longer be symmetric. Of course, this should not be the case, because the names of the actions should not matter, so there is clearly something wrong with the definition. A similar problem can occur if we multiply the utility function of one of the two players by a fixed constant. Anyway, we will not go into the details of a proper definition of `symmetric game'. The given definition suffices for our purposes.

%
%The following definition, which is more general thanDef~\ref{def:symmeteric_game_1} is therefore the correct one: 
%\begin{definition}\label{def:symmeteric_game_2}
%Let $G$ be a 2-player normal-form game. We say it is a \textbf{symmetric game} if there exists a bijection $f$ from $\Ac_1$ to $\Ac_2$, and for any $(\ac_1, \ac_2) \in \Ac_1 \times \Ac_2$ we have 
%\begin{equation}\label{eq:symmetric_game_2}
%\util_1(\ac_1, \ac_2) \ \ = \ \ \util_2(\ac_2, \ac_1)
%\end{equation}
%\end{definition}
%\later{Actually, this definition is not strong enough either, because we can apply a linear transformation to the utility function of one of the players}.

Note that to specify the payoff matrix of a symmetric game, it is sufficient to only provide the utilities of the \textit{row}-player. After all, if for some action profile $(\ac_1, \ac_2)$, you want to know the corresponding utility value $\util_2(\ac_1, \ac_2)$ of the column player, then you can just look for $\util_1(\ac_2, \ac_1)$ in the table. See Table~\ref{tab:psr_row_player_only}.

\begin{table}[h]
\begin{center}
\begin{tabular}{l|c|c|c|}
\ & Paper & Scissors & Rock \\
\hline
Paper & 1 & 0 & 2\\
\hline
Scissors & 2 & 1 & 0\\
\hline
Rock & 0 & 2 & 1\\
\end{tabular}
\caption{Payoff-matrix for the game Paper-Scissors-Rock, with only the utilities for the row-player. Given the knowledge that it is a symmetric game, it is not necessary to explicitly display the utility values of the column player. For example, if you want to know the utility of the column player for the profile $(paper , \mi{scissors})$, then you can just look up the utility of the row player for the profile $(\mi{scissors}, paper)$, which we can see is 2.}\label{tab:psr_row_player_only}
\end{center}
\end{table}

While a negotiation is typically not a symmetric game (because the two agents typically have different utility functions), the topic of symmetric games is still very important for the study of automated negotiation, as we will see later on in this book when we discuss the evaluation of negotiation strategies using `empirical game-theoretic analysis'.

We can now define the notion of a symmetric Nash equilibrium (for symmetric games).
\begin{definition} 
Let $G$ be a symmetric 2-player normal-form game. We say a strategy profile $(\ms_1, \ms_2)$ for this game is a \textbf{symmetric Nash equilibrium} if it is a Nash equilibrium, and it satisfies $\ms_1 = \ms_2$.
\end{definition} 

The following theorem is proven in \cite{cheng2004symmetricGames}.
\begin{theorem}
Any finite symmetric game has a symmetric Nash equilibrium.
\end{theorem}

We now claim that in a symmetric game, if the players play optimally, they would choose a symmetric Nash equilibrium.

The idea behind this, is that if the game is perfectly symmetrical, and the players are perfectly rational, then, whenever player 1 reasons that some mixed strategy $\ms$ is the optimal strategy, player 2 would come to exactly the same conclusion, and thus they would always choose the same mixed strategy. Therefore, the only Nash equilibria they could possibly end up choosing, are the symmetric ones.

%Note that since every symmetric equilibrium is a Nash equilibrium, but not vice versa, this solution concept is indeed a refinement. 

However, it can still happen that a symmetric game has multiple symmetric equilibria. In that case, we can apply the Pareto-optimality criterion from Section \ref{sec:pareto_optimal_nash} to make a choice among the symmetric equilibria. Note that for any symmetric equilibrium $(\ms, \ms)$ in a symmetric game, the two players will always receive the same utility: $\util_1(\ms, \ms) = \util_2(\ms, \ms)$. Therefore, if we have two symmetric equilibria, with different utility vectors, then one will dominate the other. For example, if one symmetric equilibrium yields utility vector $(20,20)$ and another one yields utility vector $(10,10)$, then the first one dominates the second one.

%Therefore, in a symmetric game there is always exactly one optimal symmetric Nash equilibrium, except if it happens that there are two or more non-dominated Nash equilibria with exactly the same utility vectors.

Now, a valid question would be what happens if this solution conflicts with the solution we discussed in Section \ref{sec:pareto_optimal_nash}. That is, what happens if a game is symmetric, but every symmetric equilibrium is dominated by a non-symmetric Nash equilibrium. For example, suppose we have a symmetric Nash equilibrium with utility vector $(10, 10)$ and a non-symmetric Nash equilibrium with utility vector $(20, 15)$. On the one hand, our discussion in Section \ref{sec:pareto_optimal_nash} told us that the players should choose the Pareto-optimal one, but on the other hand, we have just discussed in this section that the players should choose the symmetric one.

We argue that in this situation the players would typically choose the symmetric Nash equilibrium. To see this, note that because the game is symmetric, we know that there must also exist a third Nash equilibrium, with utility vector $(15, 20)$. This means that whenever player 1 reasons that he should choose the equilibrium with outcome $(20, 15)$, by the symmetry of the game, player~2 would reason that she should choose the third equilibrium, with outcome $(15, 20)$. Therefore, just as in Section \ref{sec:wrong_solutions} they would end up playing an entirely different strategy profile that typically  wouldn't be a Nash equilibrium.\footnote{We're using the word `typically' here, because there may exist games in which the resulting strategy profile would actually still be a Nash equilibrium.} So, in the end, any Nash equilibrium they could actually end up playing, would have to be a symmetric one.

%%%%%%%%%%%%%%%%%%%%%%%%%%%%%%%%%%%%%
%%%%%%%%%%%%%%%%%%%% Suppose we have: (a, b) ~ (10,15) and (c, d) ~(15,20) Now, if player 1 chooses a and player 2 chooses d Then we have the following possibilities:
% 1. (a, d)  is not a Nash equilibrium
% 2. (a, d)  is a symmetric equilibrium (i.e. a = d)
% 3. (a, d)  is a non-symmetric equilibrium, but with symmetric utility vector.
% 4. (a, d)  is a non-symmetric equilibrium, with a non-symmetric utility vector. e.g. (14, 21)
%
% In the first two cases, our conclusion that the players should always pick a symmetric equilibrium, still hold.
%
% In the third case...
%
% In the fourth case, note that (d,a) would also be a Nash equilibrium, with utility vector (21, 14)...

% IDEA: perhaps we should weaken our statement, to say that the players would always select an equilibrium with symmetric utility vector. 

% Actually, that idea is not correct. For example, say we have a game with two actions a, and b for each player and the following utils:
% \vec{\util}(a,b) = \vec{\util}(b,a) = (10, 20)
% \vec{\util}(a,a) = \vec{\util}(b,b) = (18, 18).
% in that case the players would each end up playing 0.5a + 0.5b
% which yields an expected util vector of (19, 19).

%%%%%%%%%%%%%%%%%%%%%%%%%%%%%%%%%%%%%
%%%%%%%%%%%%%%%%%%%%%%%%%%%%%%%%%%%%%

This, however, still does not solve the equilibrium selection \textit{completely}, even for symmetric games, because it may still happen that some symmetric game has multiple symmetric equilibria with exactly the same utility vector.  If that set happens to be semi-factorizable, then we can just pick any random safe strategy. If it is not semi-factorizable, then we have to discard that entire set and pick the next best option, etcetera. This solution to the equilibrium selection problem for symmetric games is displayed in Algorithm~\ref{alg:opt_strategy_symm}.

\begin{algorithm}
\caption{Algorithm that chooses the optimal strategy for either of the two players of any symmetric 2-player game $\nfgame$.}\label{alg:opt_strategy_symm}
\begin{algorithmic}[1]
\Statex \textbf{Input:}
	\Statex $\nfgame$ \Comment{The game to play (must be a symmetric game).}
	%\Statex $i$ \Comment{The role our agent will be playing in this game }
	%\Statex \Commentt{(i.e. 1 if it is the row-player and 2 if it is the column-player.)}
	\Statex
	\comment{Determine the set of all Nash equilibria $\mi{NE}$ of $\nfgame$:}
	\State $\mi{NE} \leftarrow \mi{getNashEquilibria}(\nfgame)$ 
	\Statex
    \comment{Determine the set of \textit{symmetric} Nash equilibria $\mi{SNE}$:} 
    \State $\mi{SNE} \leftarrow \{(\ms_1, \ms_2) \in \mi{NE} \mid \ms_1 = \ms_2 \}$
    \Statex
    \While{$\mi{SNE} \neq \emptyset$}
    \comment{From this, extract the subset of symmetric Nash equilibria for}
     \commentt{which the utility of the players is maximal.}
    \commentt{It doesn't matter if we use $\util_1$ or $\util_2$ for this because for any symmetric}
    \commentt{game and any mixed strategy $\ms$ we have $\util_1(\ms, \ms) = \util_2(\ms, \ms)$ anyway.}
    \State $\mi{MAX} \leftarrow \argmax_\ms \{\util_1(\ms, \ms) \mid (\ms , \ms) \in \mi{SNE}\} $
    \Statex
    \comment{If this set is semi-factorizable (which includes the case that it contains only one element) then we can determine a safe set and pick any random strategy from it.}
    \If{$\mi{MAX}$ is semi-factorizable}
    \State $\sfs \leftarrow$ getSafeSet($\mi{MAX}$)
    \State $\ms\leftarrow $ getRandomElement($\sfs$)
	\State \Return $\ms$
    \EndIf
    \Statex
    \comment{If not, then we have to discard this set of equilibria, and keep looking for the next best set.}
    \State $\mi{SNE} \leftarrow \mi{SNE} \setminus \mi{MAX}$
    \EndWhile
    \Statex
    \comment{If this approach does not yield any solution, then return the empty set.}
\State \Return $\emptyset$
\end{algorithmic}
\end{algorithm}

\subsection{The Assumption of Role-Equifrequency}\label{sec:AoRE}
We will now discuss our last solution to the ESP, which applies not only to symmetric games, but to normal-form games in general. However, instead we do need to make an alternative assumption, called \textit{the Assumption of Role-Equifrequency} (AoRE) which we will explain below. In a nutshell, this solution says that whenever the AoRE holds, one should pick the Nash equilibrium that maximizes the sum of the utilities of the two players \cite{deJonge2023bargainingSolution}.

The main idea behind this solution is that in most realistic situations, instead of trying to find an optimal \textit{strategy} for a single game, one actually tries to implement an optimal \textit{algorithm} to  select a strategy, that can be applied \textit{multiple times} to one or more different games. After all, you are not going to delete the algorithm directly after playing one game and then implement an entirely new algorithm for the next game (especially if that next game is identical to the previous one). 

This means we are essentially solving a kind of `meta-game' where the possible actions are the possible algorithms that we can implement, and the utility function we are trying to optimize is the agent's expected utility, averaged over some set of 2-player games $\mc{G}$ that we expect our algorithm is going to be playing. Furthermore, if we assume the that our agent is going to play each of the two `roles' of each game equally often (as `row-player' and as `column-player'), then this meta-game is symmetric (even if the games in $\mc{G}$ themselves are not symmetric), so we can solve it using the approach of Section \ref{sec:symmetric_equilibria}.

In order to formalize this precisely, we first need to make a clear distinction between two concepts that have until now mostly treated as equal, namely the concept of a `\textit{player}' and the concept of an `\textit{agent}'. 

In our terminology, an `agent' is an entity, such as a computer program, an application, a robot, or even a human being, that is capable of playing a game. For example, if we are talking about the game of chess, then examples of agents are Deep Blue (the first chess program to ever beat the human world champion of chess), Stockfish (one of the strongest chess engines in the world, at the time of writing), or Magnus Carlsen (the highest ranking human chess player in the world, at the time of writing). On the other hand, when we use the term `player' we are referring to the \textit{role} that the agent is playing in the game. For example, in the game of chess there are exactly two `roles' or `players', namely, \textit{black} and \textit{white}.

Many text books and papers on game theory do not make such a clear distinction between `players' and `agents', because they typically just study a single instance of a single game, so there are only two agents present in that context and each agent plays exactly one of the two roles. In those cases there is a one-to-one correspondence between the two agents and the two players, and one can use the term \textit{`agent 1'} to refer to \textit{`the agent that plays the role of player 1'}. 

However, in our current context the distinction between players and agents is crucial because we will assume that the same agent will be playing the same game multiple times, sometimes in the role of `player 1' and sometimes in the role of `player 2'. To avoid confusion, in this section we will refer to our own agent as $\ag_A$ and to its opponent as $\ag_B$. So, this means that in some games agent $\ag_A$ will play the role of player 1 (a.k.a. the `row player') while $\ag_B$ plays the role of player 2 (a.k.a. the `column player'), and in other games (or other instances of the same game) the roles will be reversed and $\ag_A$ will play the role of player 2 while $\ag_B$ plays the role of player 1.

\begin{definition}
Let $\mc{G}$ be a set of 2-player normal-form games. Then, a \textbf{role-frequency function} $\ff : \mc{G} \times \{1,2\} \rightarrow \mathbb{R}^+$ is a function that assigns to each game $\nfgame \in \mc{G}$ and each player index $i\in \{1,2\}$ a non-negative real number  $\ff(\nfgame,i)$. 
\end{definition}
The number $\ff(\nfgame,i)$ represents the relative frequency with which our agent $\ag_A$ is going to be (or is expected to be) playing game $\nfgame$ in the role of player~$i$. For example, if $\nfgame$ is the `Battle of the Sexes' game (Sec. \ref{sec:multiple_nash_eq}) and if $\ff(\nfgame,1) = 2 \cdot \ff(\nfgame,2)$, then this means that we expect our agent $\ag_A$ to be playing this game in the role of the `husband' twice as often than as in the role of the `wife'.

There are several ways to interpret $\ff$ in a more precise manner. For example $\ff(\nfgame, i)$ could literally be the number of times the agent is going to play the game $\nfgame$ as player $i$, or it can be just the \textit{probability} that it will play the game $\nfgame$ as player $i$. The precise interpretation does not matter, however, for the rest of this section. 

For any game $\nfgame$, let $\ms_A(\nfgame,i)$ denote the mixed strategy selected by our agent $\ag_A$ when playing game $\nfgame$ in the role of player $i$, and similarly, let $\ms_B(\nfgame,i)$ denote the mixed strategy selected by agent $\ag_B$ when playing game $\nfgame$ in the role of player $i$. Then, for any set of 
2-player normal-form games $\mc{\nfgame}$ and any role-frequency function $\ff$ for this set, we can define `meta-utility' functions $\mc{U}_A$ and $\mc{U}_B$  as follows:
\begin{eqnarray*}
\mc{U}_A(\ag_A, \ag_B) &=& \sum_{\nfgame \in \mc{\nfgame}} \ff(\nfgame,1) \cdot \util^\nfgame_1(\ms_A(\nfgame,1), \ms_B(\nfgame,2)) \quad + \\ 
\ & & \quad \quad \quad \ff(\nfgame,2) \cdot \util^\nfgame_2(\ms_B(\nfgame,1), \ms_A(\nfgame,2)) \\
\ & & \ \\
\mc{U}_B(\ag_A, \ag_B) &=& \sum_{\nfgame \in \mc{\nfgame}} \ff(\nfgame,1) \cdot \util^\nfgame_2(\ms_A(\nfgame,1), \ms_B(\nfgame,2)) \quad + \\ 
\ & & \quad \quad \quad \ff(\nfgame,2) \cdot \util^\nfgame_1(\ms_B(\nfgame,1), \ms_A(\nfgame,2)) 
\end{eqnarray*}
where $\util^\nfgame_i$ is the utility function of player $i$ in game $\nfgame$. 

Our goal is to implement an agent $\ag_A$ that maximizes the meta-utility $\mc{U}_A$, which, however, depends on the opponent $\ag_B$ that aims to maximize $\mc{U}_B$. Note that the only difference between these two expressions is that the places of $\util_1$ and $\util_2$ have been switched. In particular, $\mc{U}_B$ is still described in terms of the role-frequency function $\ff$ of agent $\ag_A$.

We can see this model as if our agent is playing a \textit{tournament}, defined by $\mc{\nfgame}$ and $\ff$, rather than just a single game, and its final score in the tournament is given by $\mc{U}_A  $, which is an aggregation of the utilities it achieved in the individual games of the tournament.

It is important to understand that even though we model our agent's opponent as a single agent that we denote as $\ag_B$, our agent doesn't literally have to be playing against the same opponent in every game. It is perfectly possible that in reality our agent is playing against an entirely different opponent in each game. However, that doesn't change the fact that we can simply model this collection of opponents as a single agent $\ag_B$, which just happens to behave differently in each game. For this reason we also assume that each game is played completely independently from any previous or future games. That is, the agents do not have a memory of what happened in previous games, and do not try to influence each others' actions in future games. After all, it wouldn't make sense to do so, because we assume the agents don't know whether they are playing against the same opponents or against different opponents in every game.

\begin{definition}\label{def:AoRE}
Let $\mc{\nfgame}$ be a set of 2-player normal-form games and $\ff$ a role-frequency function for this set, then we say the \textbf{Assumption of Role-equifrequency} (AoRE) holds if and only if for all $\nfgame \in \mc{\nfgame}$ we have $\ff(\nfgame,1) = \ff(\nfgame,2)$
\end{definition}
In other words, the AoRE holds if for each game in $\mc{\nfgame}$ we expect our agent to play either of the two roles of that game with the same frequency or with the same probability.

The idea is that the tournament defined by $\mc{\nfgame}$ and $\ff$ can be seen as a giant normal-form game for which the set of actions consists of the set of all possible agents that we can implement and the utility functions are the functions $\mc{U}_A$ and $\mc{U}_B$. Furthermore, the AoRE ensures that this game is symmetrical---even if the games $\nfgame$ inside $\mc{\nfgame}$ are not symmetrical---which means we can apply the solution discussed in Section \ref{sec:symmetric_equilibria} to determine the optimal Nash equilibrium.

\begin{definition}
Let $\mc{\nfgame}$ be a set of 2-player games and $\ff$ a role-frequency function for this set, then we define the corresponding \textbf{meta-game} to be a 2-player normal-form game such that:
\begin{itemize}
\item For each player, its set of actions is the set of all possible algorithms that can take as their input a description of any 2-player normal-form game $\nfgame \in \mc{\nfgame}$ and a number $i\in\{1,2\}$ and that then output a mixed strategy $\ms$ for player $i$ in that game.
\item The utility functions are the functions $\mc{U}_A$ and $\mc{U}_B$ as defined above.
\end{itemize}
\end{definition}

At first it may seem like a very complicated problem to even find any Nash equilibrium at all for this meta-game, since there are infinitely many different algorithms that one could implement, so it's a game with infinitely many actions. However, it turns out that one can show that a pair of agents $\ag_A, \ag_B$ forms a Nash equilibrium of the meta-game, if and only if for every game $\nfgame \in \mc{\nfgame}$ those agents select a Nash equilibrium of $\nfgame$~\cite{deJonge2023bargainingSolution}. Furthermore, writing an algorithm that finds a Nash equilibrium for any arbitrary finite 2-player normal-form game $\nfgame$ is not overly complicated. For example, the well-known Lemke-Howson algorithm~\cite{LemkeHowson1964} does exactly that.

Now, since the AoRE ensures the meta-game is symmetrical, we are only interested in the \textit{symmetric} equilibria of the meta-game. Specifically, this means we are interested in implementing a \textit{single} agent $\ag$ such that, for any game $\nfgame \in \mc{G}$ the two strategies $\ms(\nfgame, 1)$ and $\ms(\nfgame,2)$ it would select for the two respective roles, together always form a Nash equilibrium of $\nfgame$. Furthermore, we have to make sure that our agent picks these strategies in such a way that the agent $\ag$, when seen as an action of the meta-game, satisfies the solution concept of Section~\ref{sec:symmetric_equilibria}. 

Using this model, it can be shown that under the AoRE, for any 2-player normal-form game $\nfgame$ the optimal Nash equilibrium is, in general, the one that maximizes the sum of the utilities of the two players. If there is more than one such Nash equilibrium then, as was argued in~\cite{deJonge2023bargainingSolution}, one should choose, among those equilibria, one that minimizes the absolute utility difference $|\util_1(\vec{\ms}) - \util_2(\vec{\ms})|$. If we then still have multiple possible equilibria, then we need to check if that set of equilibria is semi-factorizable. If yes, then we can randomly choose any of our safe strategies. If not, then we have to discard this set and repeat the process. This procedure is displayed in Algorithm \ref{alg:opt_strategy_AoRE}. Note that this algorithm is a small improvement with respect to the one presented in~\cite{deJonge2023bargainingSolution}, because it uses the solution discussed in Section~\ref{sec:factorizable_sets} as a refinement.

%\begin{theorem}
%Let $\mc{G}$ be a set of 2-player normal-form games and $\ff$ a role-frequency function for this set that satisfies the AoRE. Then, an optimal agent would select, for any game $\nfgame \in \mc{G}$ a Nash equilibrium $\vec{\ms}^*$ that maximizes the sum of the utilities of the two players:
%\[\vec{\ms}^* \quad \in \quad \argmax_{\vec{\ms}}\ \{\util_1(\vec{\ms}) + \util_2(\vec{\ms}) \mid \vec{\ms} \in \mathit{NE}\}\]
%where $\mathit{NE}$ is the set of all Nash equilibria of $\nfgame$.
%\end{theorem} 

To get an intuitive idea of why one should choose the Nash equilibrium that maximizes the utility-sum, imagine that $\mc{G}$ consists of just one game $\nfgame$, which our agent will be playing just once. Furthermore, suppose the AoRE holds, which means that there is a 50\% chance that our agent will be playing this game as player 1, and a 50\% chance that it will be playing it as player 2. Then, this means that for any Nash equilibrium $\vec{\ms}$ the expected utility for our agent under that equilibrium would be $0.5 \cdot \util_1(\vec{\ms}) + 0.5 \cdot \util_2(\vec{\ms})$. So, in order for our agent to play optimally, he should choose the Nash equilibrium that maximizes this quantity.

\begin{algorithm}
\caption{Algorithm that chooses the optimal strategy for player $i$ for any 2-player normal-form game $\nfgame$ (not necessarily symmetrical). This is based on the assumption that for any game $\nfgame$ this algorithm will be called equally often with $i=1$ as with $i=2$.}\label{alg:opt_strategy_AoRE}
\begin{algorithmic}[1]
\Statex \textbf{Input:}
	\Statex $\nfgame$ \Comment{The game to play.}
	\Statex $i$ \Comment{The role our agent will be playing in this game }
	\Statex \Commentt{(i.e. 1 if it is the row-player and 2 if it is the column-player.)}
	\Statex
	\comment{Determine the set of all Nash equilibria $\mi{NE}$ of $\nfgame$.}
	\State $\mi{NE} \leftarrow \mi{getNashEquilibria}(\nfgame)$ 
	\Statex
%    \comment{Determine the set of \textit{degenerate} Nash equilibria $\mi{DNE}$. }
%    \commentt{That is, those Nash equilibria $\vec{\ms}$ for which there exists another Nash}
%    \commentt{equilibrium $\vec{\ms}$' such that the utility  vector of $\vec{\ms}$' is the `reflection'}
%    \commentt{ of the utility vector of $\vec{\ms}$:}
%    \State $\mi{DNE} \leftarrow \{\ \vec{\ms} \in \mi{NE} \ \mid \ \exists \vec{\ms}' \ : \ \ (\util_1(\vec{\ms}),\util_2(\vec{\ms})) = (\util_2(\vec{\ms}),\util_1(\vec{\ms})) \ \}$
%    \Statex
%    \comment{Determine the \textit{non-degenerate} equilibria $\mi{NDNE}$:}
%\State $\mi{NDNE}\leftarrow \mi{NE} \setminus \mi{DNE}$   
\While{$\mi{NE} \neq \emptyset$}
    \comment{Among the candidate Nash equilibria, find the ones that}
    \commentt{maximize the sum of the two players' utility values:}
\State $\mi{MaxSum} \leftarrow \argmax_{\vec{\ms}} \{\util_1(\ms) + \util_2(\ms) \mid  \vec{\ms} \in \mi{NE} \}$
\Statex
\comment{Among those Nash equilibria, find the ones that minimizes the}
\commentt{absolute difference between the utility values of the two players:}
\State $\mi{MinAbsDiff} \leftarrow \argmin_{\vec{\ms}} \{ |\util_1(\ms) - \util_2(\ms)| \mid  \vec{\ms} \in \mi{MaxSum} \}$
\Statex
\comment{If this set is semi-factorizable (which includes the case that it contains only one element) and all its elements have the same utility vector, then pick a random safe strategy and return it.}
\If{$\mi{MinAbsDiff}$ is semi-factorizable \textbf{and} all equilibria in $\mi{MinAbsDiff}$ have the same utility vector}
\State $\sfs_i \leftarrow $ getSafeSet($\mi{MinAbsDiff}$, $i$)
\State $\ms \leftarrow $ getRandomElement($\sfs_i$)
\State \Return $\ms$
\Else
\State $\mi{NE}\leftarrow \mi{NE} \setminus \mi{MinAbsDiff}$
\EndIf
\EndWhile
    \comment{If this did not yield a solution, return the empty set.}
\State \Return $\emptyset$
\end{algorithmic}
\end{algorithm}

It is important to note that the games $\nfgame$ in $\mc{\nfgame}$ do not need to be symmetric. However, if any game $\nfgame$ in $\mc{\nfgame}$ does happen to be symmetric, then for that game, Algorithm \ref{alg:opt_strategy_AoRE} is equivalent to Algorithm~\ref{alg:opt_strategy_symm}. So there is no conflict between the current solution and the solution from Section~\ref{sec:symmetric_equilibria}.

%Now,at first sight, it might look as if something is wrong with this solution, because it seems that it does not obey the principle of \textit{Invariance under Linear Transformations} (see Def.~\ref{def:invariance_lin_trans}). For example, suppose we have a game with two Nash equilibria $\vec{\ms}$ and $\vec{\ms}'$, and suppose they have the following utility-vectors:
%\[\vec{\util}(\vec{\ms}) \quad = \quad (19\ ,\ 10)\]
%\[\vec{\util}(\vec{\ms}^{\ '}) \quad = \quad (10\ ,\ 20)\]
%%
%%\[\util_1(\vec{\ms}) = 19 \quad \util_2(\vec{\ms}) = 10\]
%%\[\util_1(\vec{\ms}^{\ '}) = 10 \quad \util_2(\vec{\ms}^{\ '}) = 20\]
%Then we see that $\vec{\ms}^{\ '}$ has a utility-sum of $10+20=30$, which is greater than the utility sum of $\vec{\ms}$, which is $19+10=29$. So our agent should choose $\vec{\ms}^{\ '}$.
%
%However if we now multiply the utility values of player 1 by a factor of 2, then we get:
%\[\vec{\util}(\vec{\ms}) \quad = \quad (38\ ,\ 10)\]
%\[\vec{\util}(\vec{\ms}^{\ '}) \quad = \quad (20\ ,\ 20)\]
%%\[\util_1'(\vec{\ms}) = 38 \quad \util_2(\vec{\ms}) = 10\]
%%\[\util_1'(\vec{\ms}^{\ '}) = 20 \quad \util_2(\vec{\ms}^{\ '}) = 20\]
%Now suddenly, $\vec{\ms}$ has the greater utility sum ($38+10=48$ vs. $20+20=40$).

Now, at first sight, it might look as if something is wrong with this solution, because it seems to disobey the principle of \textit{Invariance under Linear Transformations} (see Def.~\ref{def:invariance_lin_trans}). For example, suppose we have a game with two Nash equilibria $(\ms_1, \ms_2)$ and $(\ms_1', \ms_2')$, and suppose they have the following utility-vectors:
\[\vec{\util}(\ms_1, \ms_2) \quad = \quad (10\ ,\ 20)\]
\[\vec{\util}(\ms_1', \ms_2') \quad = \quad (19\ ,\ 10)\]
%
%\[\util_1(\vec{\ms}) = 19 \quad \util_2(\vec{\ms}) = 10\]
%\[\util_1(\vec{\ms}^{\ '}) = 10 \quad \util_2(\vec{\ms}^{\ '}) = 20\]
Then we see that $(\ms_1, \ms_2)$ has a utility-sum of $10+20=30$, which is greater than the utility sum of $(\ms_1', \ms_2')$, which is $19+10=29$. So, our agent should choose $(\ms_1, \ms_2)$.

However, if we multiply the utility values of player 1 by a factor of 2, then we get:
\[\vec{\util}(\ms_1, \ms_2) \quad = \quad (20\ ,\ 10)\]
\[\vec{\util}(\ms_1', \ms_2') \quad = \quad (38\ ,\ 20)\]
Now suddenly, $(\ms_1', \ms_2')$ has the greater utility-sum ($38+20=58$ vs. $20+10=30$).

The solution to this paradox lies in the important distinction between \textit{agents} and \textit{players} that we mentioned before. The point is, that the principle of Invariance under Linear Transformations applies to \textit{agents}, while in this example we have erroneously applied it to a \textit{player}. What we mean is that each \textit{agent} is allowed to freely apply a linear transformation to his utilities. However, since we are assuming the AoRE, each agent is going to be playing in both roles of the game and therefore he will have to apply the same linear transformation to \textit{both} utility functions. Importantly,  this does not impede the other agent to apply a totally different linear transformation to the utility functions. Let us illustrate this with another example. 

Suppose Alice and Bob are going to play a very simple game against each other, in which they can win a monetary prize in euros. For example, suppose this game has the following pay-off matrix:

\begin{center}
\begin{tabular}{c|c|c}
\ & $L$ & $R$ \\
\hline
$T$ & (\euro\ 40 , \euro\ 30) & (\euro\ 0 , \euro\ 0) \\
\hline
$B$ & (\euro\ 0 , \euro\ 0) & (\euro\ 10 , \euro\ 20) \\
\end{tabular}
\end{center}

Furthermore, suppose they are going to play this game twice. In the first game Alice will be the row-player and Bob will be the column player, and in the second game their roles are swapped, so the AoRE holds. Note that this game has two Nash equilibria: $(T,L)$ and $(B,R)$, but the first one is clearly the better one, for both players. 

Now, suppose that Alice is American, so she would convert her prize money to dollars. Let's say that 1 euro equals 1.10 dollar. Then, from her point of view the payoff matrix would look like:
\begin{center}
\begin{tabular}{c|c|c}
\ & $L$ & $R$ \\
\hline
$T$ & (\$ 44 , \$ 33) & (\$ 0 , \$ 0) \\
\hline
$B$ & (\$ 0 , \$ 0) & (\$ 11 , \$ 22) \\
\end{tabular}
\end{center}

In other words, she has applied a linear transformation of the form ${\util' = 1.1 \cdot \util}$, but since she is going to play the game in both roles, she applies it to both utility functions. Of course, it is clear that this does not make any difference to the question which equilibrium has the highest utility-sum. The equilibrium $(T,L)$ is still the best for both players.

Furthermore, let us suppose that Bob is British, so he will convert his prize money to British pounds, and let's say that 1 euro equals 0.90 pound. Then, from his point of view the payoff matrix would look like:
\begin{center}
\begin{tabular}{c|c|c}
\ & L & R \\
\hline
T & (\pounds\ 36 , \pounds\ 27) & (\pounds\ 0 , \pounds\ 0) \\
\hline
B & (\pounds\ 0 , \pounds\ 0) & (\pounds\ 9 , \pounds\ 18) \\
\end{tabular}
\end{center}
Again, this does not make any difference to the question which equilibrium is better.

We see that the two \textit{agents} are each still perfectly allowed to  apply a different linear transformation, and that this does not affect the outcome. Therefore, the solution concept discussed in this section is  perfectly compatible with the principle of Invariance under Linear Transformations.

\begin{observation}
Under the AoRE, the solution that maximizes the sum of the players' utilities still obeys the principle of Invariance under Linear Transformations (as long as the transformations are applied to \underline{agents}, rather than \underline{players}).
\end{observation}

As a final remark, we should stress that the solution that we discussed in this section can be applied even if we don't know exactly which set of games $\mc{\nfgame}$ our agent is going to play. Moreover, we also don't even need to know the role-frequency function $\ff$. \textit{The only thing we really need to know, is that it is reasonable to assume that our agent is going to play each role of each game (approximately) equally often.} For example, if we are implementing a chess-bot, then it seems perfectly reasonable to assume that this bot will play as `black' approximately equally often as as `white'.

\section{Turn-taking Games}\label{sec:turn_taking_games}

As explained above, a normal-form game is a game in which each player makes just one move, and then the game is over. However, most games we play in real life are not over after just one action. Typical games like chess or poker involve multiple rounds. Such games are called \textit{extensive-form games}. To keep things simple we here only focus on one specific type of extensive-form game in which in each turn only one player makes a move. Such games are called \textit{turn-taking games}. Again, games like chess and poker fall into this category. On the other hand, the game of \textit{Diplomacy} does not fall into this category because in that game in each round the players choose their moves simultaneously.

\subsection{Tuples}
Before we can formally define the notion of a turn-taking game, we first need to introduce some other mathematical concepts.

First, for any set $X$ and any integer $n$, let $X^n$ denote the $n$-fold Cartesian product of $X$ with itself. That is, $X^1 := X$, \quad $X^2 := X \times X$, \quad $X^3 := X \times X \times X$, etcetera. 

Then, let $X^\kleene$ denote the set of all finite \textbf{tuples} over $X$. That is:
\[X^\kleene \ \ := \ \ \bigcup_{n\in \mathbb{N}} X^n \ \ = \ \ X^0 \cup X^1 \cup X^2 \cup X^3 \cup \dots \]
In particular, note that $X^\kleene$ also includes $X^0$, which is  just the singleton set containing only the empty tuple~$()$. In the rest of this book we will use the symbol $\emptyTuple$ to denote the empty tuple.

For example, if $X = \{a,b,c\}$, then some examples of tuples over $X$ are $(b)$,  $(a,a)$, $(a, b, c)$, and $(b, c, b, a, a, b, a)$. Note that tuples can have arbitrary length (as long as they are \textit{finite}), that a tuple may contain the same element multiple times, and that the elements may appear in any arbitrary order. Also note that the order of the elements matters. That is, $(a, b, c)$ is considered a different tuple than $(c, b, a)$.

We use the symbol $\circ$ to denote the \textbf{concatenation} of two tuples. For example $(a,b,c) \circ (d,e) = (a,b,c,d,e)$.

\begin{definition}
For any tuple $x\in X^\kleene$ its \textbf{length} $n$, denoted $|x|=n$, is defined as the integer $n$ for which $x\in X^n$. 
\end{definition}
For example, the tuple $(a,b,c)$ has length 3.

For any tuple $x \in X^\kleene$ of length $n$ and any integer $m$ with $m\leq n$, we will use the notation $x[m]$ to denote the $m$-th element of $x$. For example, if $x = (a, b, c)$ then $x[1] = a$, $x[2] = b$ and $x[3] = c$.

\begin{definition}
Let $x,y \in X^\kleene$ be two tuples with $|x| < |y|$. Then we say that $x$ is a \textbf{prefix} of $y$ if there exists some tuple $z$ such that $x \circ z = y$.
\end{definition}
%
%\begin{definition}
%Let $x,y \in X^*$ be two tuples with $|x| < |y|$. Then we say that $x$ is a \textbf{prefix} of $y$ if, for every integer $m$ with $m \leq |x|$ we have $x[m] = y[m]$. 
%\end{definition}
In other words, if $x$ is a tuple of length $n$ (i.e. $|x|=n$), and $x$ is a prefix of $y$, then that means that $x$ consists of exactly the first $n$ elements of $y$. For example, the tuple $x = (a, b, c)$ is a prefix of the tuple $y = (a,b,c,d,e)$, because we have $(a, b, c) \circ (d,e) = (a,b,c,d,e)$. In particular, note that the empty tuple is a prefix of every tuple in $X^\kleene$.

\begin{definition}
Let $Y$ be a set of tuples over some set $X$. That is $Y \subseteq X^\kleene$. Then we say that $Y$ is \textbf{prefix closed}, if for any $y\in Y$ and any prefix $y'$ of $y$ we also have $y'\in Y$.
\end{definition}
For example, the set $Y = \{\emptyTuple, (a), (b), (a,b), (a, b, c, d)\}$ is \textit{not} prefix closed, because the tuple $(a,b,c)$ is a prefix of $(a, b, c, d)$ but $(a,b,c)$ is not contained in $Y$, while $(a,b,c,d)$ \textit{is} contained in $Y$.

On the other hand, the set $Y' = \{\emptyTuple, (a), (b), (a,b), (a,b,c), (a, b, c, d)\}$ \textit{is} prefix closed. To verify this, we just need to check for any tuple $y\in Y'$, except the empty tuple, that if we remove the last element of $y$, then the resulting tuple $y'$ is also contained in $Y$.

\begin{definition}
Let $Y$ be a set of tuples over some set $X$. That is $Y \subseteq X^\kleene$. We say a tuple $y \in Y$ is \textbf{non-terminal} in $Y$ if there exists another tuple $y' \in Y$ such that $y$ is a prefix of $y'$. On the other hand, if there is no such tuple $y'$ then we say that $y$ is \textbf{terminal}. The set of all terminal tuples in $Y$ is denoted as $Y^\term$.
\end{definition}
For example, again let $Y = \{\emptyTuple, (a), (b), (a,b), (a, b, c, d)\}$. The tuple $(a)$ is non-terminal in $Y$, because it is a prefix of $(a,b)$. Similarly, $(a,b)$ is non-terminal because it is a prefix of $(a, b, c, d)$. On the other hand, $(b)$ is terminal, because there is no other tuple in $Y$ that starts with $b$. Similarly, $(a,b,c,d)$ is also terminal. The empty tuple $\emptyTuple = ()$ is of course always non-terminal, except in the case that it is the only tuple in the entire set.

\subsection{Tree Diagrams}
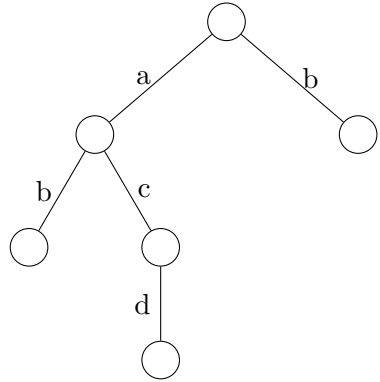
\begin{figure}
\begin{center}
\tikzstyle{treenode}=[circle, draw, minimum size=0.5cm]
\begin{tikzpicture}[
  level distance=1.5cm,
  level 1/.style={sibling distance=3.5cm},
  level 2/.style={sibling distance=1.75cm},
  %every node/.style={circle, draw, minimum size=0.5cm}
]

\node [treenode]{} % Root
  child {
  	node [treenode]{} 
    child {node [treenode]{} edge from parent node [left]{b}}
    child {
    		node [treenode]{} 
    	    child {node [treenode]{} edge from parent node [left]{d}}	
    		edge from parent node [right]{c}
    	}
    edge from parent node [left]{a}
  }
  child {
    node [treenode]{}
    edge from parent node [right]{b}
  };

\end{tikzpicture}
\caption{Example of a tree corresponding to a set of tuples $\{\emptyTuple, (a), (b), (a,b), (a,c), (a, c, d)\}$. The root corresponds to the empty tuple $\emptyTuple$. The two children of the root correspond to the tuples $(a)$ and $(b)$ respectively. The two `grand children' of the root correspond to the tuples $(a,b)$ and $(a,c)$ respectively. Finally, the last node corresponds to the tuple $(a,c,d)$.}\label{fig:example_tree}
\end{center}
\end{figure}

Any finite set of tuples that is prefix closed can be visually displayed as a tree. The easiest way to see this, is to simply look at Figure~\ref{fig:example_tree}, which displays the tree corresponding to  the set of tuples $\{\emptyTuple, (a), (b), (a,b), (a,c), (a, c, d)\}$.

Formally, a \textbf{tree} is a connected acyclic graph, for which one of the nodes is marked as the \textbf{root}. The \textbf{depth} of a node is the length of the unique path from the root to that node. For any node $\node$ with depth $d$, its \textbf{children} are those neigbors of $\node$ that have depth $d+1$. Furthermore, its \textbf{parent} is its unique neighbor with depth $d-1$. A \textbf{leaf node} is a node that does not have any children.

Formally, for any set $X$ and any prefix closed set of tuples $Y\subset X^\kleene$, the tree-diagram of $Y$ is a tree such that the following holds:
\begin{itemize}
\item There is a one-to-one correspondence between $Y$ and the set of nodes of the tree. We will use the notation $y(\node)$ to denote the tuple corresponding to node $\node$.
\item The root node corresponds to the empty tuple.
\item For any pair of nodes $\node$ and $\node'$ such that $\node'$ is a child of $\node$, there exists an element of $X$ such that $y(\node')$ can be obtained from $y(\node)$ by concatenating it with a single element of $X$:
\[\exists x \in X \ :\ \ y(\node') = y(\node) \circ (x)\]
and the edge $(\node, \node')$ is labeled with $x$.
\end{itemize}

%Now, if $X$ is some set any $Y$ is a set of tuples over $X$ that is prefix closed, then we can draw a tree such that there is a one-to-one correspondence between $Y$ and the set of nodes of the tree, such that the root corresponds to the empty tuple, and such that every node of depth $d$ corresponds to a tuple of length $d$. We will use the notation $y(\node)$ to denote the tuple corresponding to node $\node$ and similarly $\node(y)$ to denote the node corresonding to tuple $y$.

%Specifically, for any node $\node$ and any child $\node'$ of $\node$ we have that $y(\node')$ can be obtained from $y(\node)$ by concatenating it with a single element of $X$:
%\[\exists x \in X \ \ :\ \ y(\node') = y(\node) \circ (x)\]

%Note that in this way we can label every edge between $\node$ and $\node'$ is labeled with that element $x$.

%\later{I think this can be explained better.}

\subsection{Definition of a Turn-taking Game}
We are now ready to define the notion of a turn-taking game. However, before we give the formal definition, let us first explain it informally, using the game of Tic-Tac-Toe as an example.

The game of Tic-Tac-Toe is a turn-taking game, which means that in each turn one of the players chooses an action to play. So, in order to define the rules of this game, we first need to specify the set of actions that the players can choose from. In Tic-Tac-Toe choosing an action consists in marking a symbol \textbf{X} or \textbf{O} in a $3\times 3$ grid. We can formalize such an action as a tuple $(r,c,s)$ where $r \in \{1,2,3\}$ is the row in which the symbol is marked, $c\in\{1,2,3\}$ is the column, and $s \in \{\textbf{X}, \textbf{O}\}$ is the symbol itself. For example, when a player puts the symbol \textbf{X} in the center of the grid, this action is denoted by $(2,2,\textbf{X})$. So, we have a set of \textit{actions} $\Ac = \{1,2,3\} \times \{1,2,3\} \times \{\textbf{X}, \textbf{O}\}$.

Every time a player makes a move, the state of the game changes. Therefore, any state of the game can be identified with the sequence of  of actions that have already been played. In other words, the set of all possible states of the game is a subset of $\Ac^\kleene$.

For example, suppose that in the first turn player 1 plays $(2,2,\textbf{X})$. Then, in the second turn player 2 plays $(1,1,\textbf{O})$, and then, in the third turn player 1 plays $(1,2,\textbf{X})$. At that point, the state of the game is the tuple:  
\[\Big((2,2,\textbf{X}) , (1,1,\textbf{O}) , (1,2,\textbf{X})\Big)\]
Of course, not every action in $\Ac$ is legal in every state of the game. For example, after the first player has played $(2,2,\textbf{X})$, the second player is not allowed to play $(2,2,\textbf{O})$, because the cell $(2,2)$ is already filled. Therefore, the set of \textit{legal} sequences of actions is only a \textit{subset} of $\Ac^\kleene$. We will refer to such legal sequences as \textit{action histories} and we will denote the set of all such histories by $\Hist$.

In particular, note that $\Hist$ must be prefix closed. After all, the state $\Big((2,2,\textbf{X}) , (1,1,\textbf{O}) , (1,2,\textbf{X})\Big)$ can only be reached if the previous state was $\Big((2,2,\textbf{X}) , (1,1,\textbf{O}) \Big)$. In other words, if $\Big((2,2,\textbf{X}) , (1,1,\textbf{O}) , (1,2,\textbf{X})\Big)$ is legal, then $\Big((2,2,\textbf{X}) , (1,1,\textbf{O}) \Big)$ must also be legal.

Furthermore, to fully define the game of Tic-Tac-Toe, we have to specify the goals of the respective players. This can be formalized by defining a utility function for each player. For example, we can assign a value of 2 to the winner, a value of 0 to the loser, and in case of a draw we can assign a utility value of 1 to each of the players. Of course, the notion of a winner or loser is only defined at the end of the game. Therefore, the utility functions are defined over the set of all \textit{terminal} histories.

Finally, we have to specify which player can can choose an action when. We call the player who's turn it is, the \textit{active player}. Formally, we need a function that maps each non-terminal history to the index of the active player:
\[\act \ \ :\ \ \Hist \setminus \Hist^\term \rightarrow \{1, 2, \dots, \numAgents\}\]
where $\numAgents$ is the number of players.

In Tic-Tac-Toe, just as in most other turn-taking games, the active player simply alternates each turn. So, in each odd turn, player 1 is the active player and in each even turn player 2 is the active player. That is: ${\act(\hist) = |\hist| \pmod{2} + 1}$

Whenever the current state of the game is a non-terminal history $\hist$ and it's the turn of player $i$, then this player can choose any action $\ac \in \Ac$ such that the concatenation $\hist \circ (\ac)$ is legal (i.e. $\hist \circ (\ac) \in \Hist$). Such an action certainly exists, because we assumed $\hist$ was non-terminal. On the other hand, if $\hist$ is terminal, then, by definition, the game is over and the utility function determines the outcome of the game.

In summary, a turn-taking game is formally defined as follows.
\begin{definition}
A \textbf{turn-taking} game for $\numAgents$ players, consists of the following components:
\begin{itemize}
\item A set $\Ac$, which we call the set of \textbf{actions}.
\item A set $\Hist$, called the set of all legal \textbf{action histories}, which is a subset of the set of all finite tuples over $\Ac$ (i.e. $\Hist \subseteq \Ac^\kleene$), such that $\Hist$ is prefix closed.
\item A function $\act$ called the \textbf{active player map}, that maps each non-terminal history $\hist \in \Hist \setminus \Hist^\term$ to the index of the player whose turn it is:
\[\act \ \  : \ \ \Hist \setminus \Hist^\term \rightarrow \{1, 2, \dots, \numAgents\}\]
\item For each $i\in \{1, 2, \dots, \numAgents\}$ a \textbf{utility function}  $\util_i$ that assigns a utility value for player $i$ to each terminal history in $\Hist$:
\[\util_i \ \  : \ \  \Hist^\term \rightarrow \mathbb{R}\]
\end{itemize}
\end{definition}

\subsection{Game Trees}
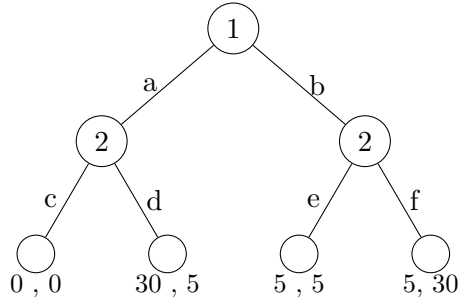
\begin{figure}
\begin{center}
\tikzstyle{treenode}=[circle, draw, minimum size=0.5cm]
\begin{tikzpicture}[
  level distance=1.5cm,
  level 1/.style={sibling distance=3.5cm},
  level 2/.style={sibling distance=1.75cm},
  %every node/.style={circle, draw, minimum size=0.5cm}
]

\node [treenode]{1} % Root
  child {
  	node [treenode]{2} 
    child {
    		node [treenode]{} 
    		node[below=4pt] {\small 0 , 0} 
    		edge from parent node [left]{c}}
    child {
    		node [treenode]{} 
    		node[below=4pt] {\small 30 , 5} 
    		edge from parent node [right]{d}}
    edge from parent node [left]{a}
  }
  child {
    node [treenode]{2}
    child {
    		node [treenode]{}
    		node[below=4pt] {\small 5 , 5} 
    		edge from parent node [left]{e}}
    child {
    		node [treenode]{}
    		node[below=4pt] {\small 5, 30} 
    		edge from parent node [right]{f}}
    edge from parent node [right]{b}
  };

\end{tikzpicture}
\caption{A game tree that visualizes a very simple 2-player turn-taking game that only lasts for two rounds. Each edge is labeled with an action from the game, and therefore each node corresponds to the history consisting of all actions along the path from the root to that node. Furthermore, each non-terminal node is labeled with the index of the active player after that history and each terminal node is labeled with the utility values of the two respective players.}\label{fig:non_credible_threat}
%\caption{A simple game that illustrates the concept of a `non-credible' threats and why the concept of a Nash equilibrium in unsatisfactory in extensive-form games.}\label{fig:non_credible_threat}
\end{center}
\end{figure}

Since a turn-taking game is essentially a set of tuples that is prefix closed, together with utility functions and an active player function, we can visually display it as a tree. See for example the game displayed in Figure~\ref{fig:non_credible_threat}.

Note that in this case the nodes corresponding to the non-terminal histories are labeled with the index of the active player, and that the nodes corresponding to the terminal histories are labeled with the utility values of the respective players. We will call such diagrams \textbf{game trees}.

In the game of Figure~\ref{fig:non_credible_threat}, each player has just one turn. In the first turn, player 1 can choose between actions $a$ and $b$. If player 1 chooses $a$ then in next player 2 can choose between actions $c$ and $d$. Otherwise, if player 1 chooses to play $b$, then next player 2 can choose between actions $e$ and $f$.

If the two players choose $a$ and $c$ respectively, then each of them will receive a utility of 0. On the other hand, if they choose actions $b$ and $f$ respectively, then player 1 will receive a utility of 5, while player 2 will receive a utility of 30.

\subsection{Strategies}
We will now define the notion of a `strategy' for a turn-taking game.

Let $\Hist_i$ denote the set of all non-terminal histories after which player $i$ is the active player. That is:
\[\Hist_i \ \ := \ \ \{\hist \in \Hist \setminus \Hist^{\term} \mid \act(\hist) = i\}\]
%Note that this induces a partition of $\Hist$:
%\[\Hist = \Hist_1 \cup \Hist_2 \cup \dots \cup \Hist_\numAgents \quad \text{with} \quad i\neq j \ \ \rightarrow \ \ \Hist_i \cap \Hist_j = \emptyset \]

Furthermore, for any non-terminal history $\hist$, let $\Ac_\hist$ denote the set of \textit{legal actions} that the active player is allowed to choose after history $\hist$. More formally, it is the set of actions that yield a legal history when concatenated with $\hist$.
\[\Ac_\hist \ \ := \ \ \{\ac \in \Ac \mid \hist \circ (\ac) \in \Hist\}\]

\begin{definition}
For any turn-taking game, a \textbf{strategy} $\strat$ for player $i$ is a map that assigns to each history $\hist$ after which $\ag_i$ is the active player, a legal action for $\ag_i$. 
\[\strat \ \ : \ \ \Hist_i \rightarrow \Ac \quad \text{such that} \quad \forall \hist \in \Hist_i : \ \ \strat(\hist) \in \Ac_{\hist}\]
\end{definition}
In line with our earlier definitions, we refer to a tuple of strategies $(\strat_1, \strat_2, \dots, \strat_\numAgents)$, one for each agent, as a \textbf{strategy profile}.

In the example game of Figure~\ref{fig:non_credible_threat}, there is only one history after which player 1 is the active player, namely the empty history (i.e. at the beginning of the game). Therefore his strategy is entirely determined by the action he chooses at the start of the game. Since he can choose between two actions, $a$ and $b$, he also only has a total of two strategies, defined by $\strat(\emptyTuple) = a$ and $\strat(\emptyTuple) = b$, respectively.

On the other hand, for player 2 there are two possible histories after which she needs to choose an action. Namely after the history $(a)$ and after the history $(b)$. So, to choose a strategy, she has to make two choices: what to do after history $(a)$ and what to do after history $(b)$. For each of these two histories she has two actions to choose from, so she has $2^2 = 4$ possible strategies:
\begin{enumerate}
\item $\strat(a) = c$, $\strat(b) = e$
\item $\strat(a) = c$, $\strat(b) = f$
\item $\strat(a) = d$, $\strat(b) = e$
\item $\strat(a) = d$, $\strat(b) = f$
\end{enumerate}
In general, if there are $m$ histories after which it is your turn, and after each of these histories you have exactly $n$ possible actions, then you have $n^m$ possible strategies (although in general the number of legal actions may be different after each history).
%Since in this game each action occurs in only one place in the tree, we can denote the strategies of player 2 simply by listing the actions she chooses. So, here strategies can be denoted as $ce$, $cf$, $de$ and $df$ respectively.

Note that once every player has chosen a strategy, and each player follows his chosen strategy throughout the game, then the evolution of the game is completely fixed, so the terminal state in which the game will end will be fixed.

For example, in the case of Tic-Tac-Toe, in the first round player 1 will play the action given by $\strat_1(\emptyTuple)$. Let's say he chooses the center square, so we have: $\strat_1(\emptyTuple) = (2,2,\textbf{X})$. Next, player 2 plays the action given by $\strat_2\Big((2,2,\textbf{X})\Big)$. Let's say that this is $\strat_2\Big((2,2,\textbf{X})\Big) = (1,1,\textbf{O})$. This continues until a terminal history is reached.

\begin{enumerate}
\item Player 1 chooses action $\strat_1(\emptyTuple) = (2,2,\textbf{X})$. 
\item Player 2 chooses action $\strat_2((2,2,\textbf{X})) = (1,1,\textbf{O})$.
\item Player 1 chooses action $\strat_1((2,2,\textbf{X}), (1,1,\textbf{O}) ) = (1,2,\textbf{X})$.
\item Player 2 chooses action $\strat_2((2,2,\textbf{X}), (1,1,\textbf{O}), (1,2,\textbf{X})) = (3,2,\textbf{O})$.
\item etcetera...
\end{enumerate}

Let $\vec{\strat} = (\strat_1, \strat_2, \dots, \strat_\numAgents)$ be a strategy profile. Then we use the notation $\hist_{\vec{\strat}}$ to denote the unique terminal history generated by this strategy profile.
Formally, it is defined as the unique terminal history that satisfies:
\[\hist_{\vec{\strat}} \quad := \quad \Big( \ac_1 \ \ , \ \ \ac_2\ \ , \ \ \ac_3 \ \ , \ \ \dots \ \ , \ \ \ac_k \Big)\]
where:
\[\ac_j \quad := \quad
\begin{cases}
 \strat_1(\emptyTuple) &  \text{if\ } j=1\\
\strat_i(\ac_1 \ ,\ \ac_2 \ ,\ \dots \ ,\ \ac_{j-1}) & \text{if\ } j>1 
\end{cases}
\]
\[\text{with\ } i \ \ := \ \ \act(\ac_1 , \ac_2 , \dots , \ac_{j-1})\]

Furthermore, we may use the notation $\util_i(\strat_1, \strat_2, \dots, \strat_n)$ or $\util_i(\vec{\strat})$  as a shorthand for $\util_i(\hist_{\vec{\strat}})$.

\subsection{Non-credible Threats}
Just like in our section about normal-form games, the main question we aim to answer is how to find the optimal strategy for each player. We will first explore a naive potential solution to this problem, which will turn out to be wrong. 

The idea behind this wrong solution is as follows: for any given turn-taking game $\ttgame$ we can consider the set of all possible strategies for player $i$, which we will denote by $\Strat_i$. As mentioned before, if each player chooses a strategy, this will uniquely determine a terminal history $\hist_{\vec{\strat}}$, and therefore it will uniquely determine a tuple of utility values $(\util_1(\hist_{\vec{\strat}}) , \util_2(\hist_{\vec{\strat}}), \dots, \util_\numAgents(\hist_{\vec{\strat}}))$, which we may denote as $(\util_1(\vec{\strat}) , \util_2(\vec{\strat}), \dots, \util_\numAgents(\vec{\strat}))$. We can then define the notion of a pure Nash equilibrium for a turn-taking game in an analagous manner as for normal-form games.

\begin{definition}
Let $\ttgame$ denote a two-player turn-taking game and let $\strat_1$ and $\strat_2$ denote two strategies for player $1$ and player 2, respectively. Then, we say that $\strat_1$ is a best response against $\strat_2$ if:
\[\forall \strat \in \Strat_1: \ \  \util_1(\strat, \strat_2) \leq \util_1(\strat_1, \strat_2)\]
and similarly, we say that $\strat_2$ is a best response against $\strat_1$ if:
\[\forall \strat \in \Strat_2: \ \ \util_2(\strat_1, \strat) \leq \util_2(\strat_1, \strat_2).\]
We say a pair of strategies $\strat_1, \strat_2$ is a \textbf{pure Nash equilibrium} of the turn-taking game $\ttgame$, if $\strat_1$ is a best response against $\strat_2$ and $\strat_2$ is a best response against $\strat_1$.
\end{definition}

Another way to look at this, is to say that each turn-taking game corresponds to a normal-form game. That is, given the $\numAgents$-player turn-taking game $\ttgame$ we can define a corresponding $\numAgents$-player normal-form game $\nfgame$ as follows:
\begin{itemize}
\item For each $i \in \{1,2, \dots, \numAgents\}$ the set of \textit{actions} $\Ac_i^\nfgame$ of player $i$ in $\nfgame$ is exactly the set of \textit{strategies} $\Strat_i$ for player $i$ in $\ttgame$. That is:
\[\Ac_i^\nfgame \quad := \quad \Strat_i\]
\item For each $i \in \{1,2, \dots, \numAgents\}$ the utility function $\util_i^\nfgame$ of player $i$ in $\nfgame$ is defined as 
\[\util_i^\nfgame(\strat_1, \strat_2, \dots, \strat_\numAgents) \quad := \quad \util_i(\hist_{(\strat_1, \strat_2, \dots, \strat_\numAgents)})\]
where the utility functions $\util_i$ on the right-hand side are the utility functions of $\ttgame$.
\end{itemize}
It should now be clear that the pure Nash equilibria of the turn-taking game $\ttgame$ conincide exactly with the corresponding pure Nash equilibria of the normal-form game $\nfgame$. 

In principle, we could now also define a \textit{mixed} Nash equilibrum of $\ttgame$ to be exactly mixed Nash equilibrum of $\nfgame$. However, there is no reason to consider such mixed equilibria, because, as we   recall from Section \ref{sec:mixed_nash_equilibria}, the purpose of  a mixed strategy is to be unpredictable to your opponent. But that doesn't work in a turn-taking game, because in each turn, the active player already knows what action the opponent has chosen in the previous turn, 
anyway.

Now that we have re-interpreted the turn-taking game $\ttgame$ as a normal-form game $\nfgame$, one might think that the optimal solution for each player is to choose a strategy $\strat_i$ such that the strategy profile $(\strat_1, \strat_2, \dots, \strat_\numAgents)$ forms a Nash equilibrium of $\nfgame$. However, we will show that this solution is not satisfactory. This is demonstrated with the game displayed in Figure~\ref{fig:non_credible_threat}. As explained above, in this game player 1 has two possible strategies, corresponding to the actions $a$ and $b$, and player 2 has four possible strategies, which we will here denote as $ce$, $\mi{cf}$, $de$ and $df$ respectively. So, we can model this game as a $2\times 4$ normal-form game, of which the payoff matrix is displayed in Table~\ref{tab:non_credible_threat}.

\begin{table}
\begin{center}
\begin{tabular}{l|c|c|c|c|}
\ & $ce$ & $\mi{cf}$ & $de$ & $df$ \\
\hline
$a$ & (0 , 0) & (0 , 0) & (30 , 5) & (30 , 5)\\
\hline
$b$ & (5 , 5) & (5 , 30) & (5 , 5) & (5 , 30)\\
\end{tabular}
\caption{The pay-off matrix corresponding tot the game of Figure~\ref{fig:non_credible_threat}}\label{tab:non_credible_threat}
\end{center}
\end{table}

Note that this game has three pure Nash equilibria:
\begin{enumerate}
\item Actions: $(a\ ,\ de)$ \quad utilities: $(30\ ,\  5)$
\item Actions: $(a\ ,\ df)$ \quad utilities: $(30\ ,\  5)$
\item Actions: $(b\ ,\ \mi{cf})$ \quad utilities: $( 5\ ,\ 30)$
\end{enumerate}
We will argue that the third Nash equilibrium is, in a certain sense, unrealistic. To see that it is a Nash equilibrium indeed, first note that in this strategy profile player 2 receives the maximum utility she can possibly achieve, so indeed she cannot benefit from any deviation. Furthermore, note that if player 1 were to deviate to action $a$, then the resulting action profile would be $(a, \mi{cf})$, which means that player 1 would play action $a$, followed by player 2 playing action $c$. The resulting utility vector would then be $(0\ ,\ 0)$, so player 1 does not benefit from any deviation either. 

However, this all depends on the assumption that player 1 indeed makes a \textit{unilateral} deviation. The problem, is that if player 1 would indeed switch to action $a$, then it would be highly unlikely that player 2 would still stick with strategy $\mi{cf}$. After all, playing $c$ after $a$ is essentially a form of `suicide' by player 2. In principle, player 2 could play $d$ and obtain 5 points, but instead she plays $c$ yielding 0 points to herself.

This problem occurs because, as explained in Section~\ref{sec:prisoners_dilemma}, the definition of a Nash equilibrium only takes \textit{unilateral} deviations into consideration. This makes sense if the game was truly a normal-form game in which each player has to fully commit to its own strategy without observing the actions of the opponent. But in this case we are playing a turn-taking game. This means that if player~1 deviates to strategy $a$, then player 2 will \textit{observe} that player 1 plays action $a$, which means that player 2 now has the possibility to also change her strategy, based on that observation. Indeed, if she is rational, she would also deviate and choose action $d$ instead of action $c$. Therefore, in turn-taking games it is not enough to only consider unilateral deviations, and thus the concept of a Nash equilibrium is too weak.

We say that the third Nash-equilbrium in our example is based on a so-called \textit{non-credible threat}. It is as if player 2 is saying to player 1: ``\textit{If you play action $a$ then I will play action $c$ and you will end up with 0 utility. Therefore, you'd better play action $b$}". This threat is not credible, because playing action $c$ does not only hurt player~1, but also player 2 herself. Therefore, player~1 could simply ignore this threat and play action $a$ anyway, knowing that player 2 is rational and therefore would not follow through with her threat but play action $d$ instead.

From this, we conclude that the concept of a Nash equilibrium is not satisfactory for turn-taking games, because some Nash equilibria may be based on non-credible threats. Therefore, we need a refined solution concept that only considers those Nash equilibria that do not involve such non-credible threats.

\subsection{Subgame Perfect Equilibria}
We will now discuss an alternative solution concept, known as the the `\textit{subgame perfect equilibrium}', which is widely regarded as the `correct' solution concept for turn-taking games.

To explain this concept, we first need to define the notion of a subgame. Informally, for any turn-taking game $\ttgame$ and any given non-terminal history $\hist$ of that game, the subgame of $\ttgame$ at $\hist$ is exactly the same as $\ttgame$, except that it doesn't start from the same initial state as $\ttgame$, but rather it starts from some non-empty history $\hist$ of $\ttgame$. In other words, it is as if we start somewhere in the middle of the game.

For example, let $\ttgame$ be the game of Tic-Tac-Toe, and let $\hist$ be the history given by:
\[{\hist = \Big((2,2,\textbf{X}), (1,1,\textbf{O}), (1,2,\textbf{X})\Big)}\]
Then the subgame of $\ttgame$ at $\hist$ follows the same rules as ordinary Tic-Tac-Toe, except that the game does not start from an empty grid, but rather starts from the state:
\begin{center}
\begin{tabular}{c|c|c}
\textbf{O} & \textbf{X} & \phantom{O} \\
\hline
\ & \textbf{X} & \ \\
\hline
\ & \  & \ \\
\end{tabular}
\end{center}
This can be formalized as follows.
\begin{definition}
Let $\ttgame$ be a turn-taking game and let $\Hist$ denote the set of histories of that game. Furthermore, let $\hist \in \Hist \setminus \Hist^{\term}$ be any non-terminal history of $G$. Then the \textbf{subgame} of $\ttgame$ at $\hist$ is a turn-taking game, denoted $\ttgame_\hist$, such that its histories (denoted $\Hist_\hist$) are exactly those histories in $\Hist$ that have $\hist$ as a prefix.
\[\Hist_\hist = \{\hist' \in \Hist \mid \hist \text{\ is\ a\ prefix\ of\ } \hist'\}\]
The active player function and the utility functions of $\ttgame_\hist$ are just the same as those of $\ttgame$, but restricted to the set $\Hist_\hist$.
\end{definition}

\later{Include a figure here to give an example}

Note that any strategy for the game $\ttgame$ can naturally be interpreted as a strategy for the game $\ttgame_\hist$ as well, simply by restricting it to the histories $\Hist_\hist$ of $\ttgame_\hist$.

\begin{definition}\label{def:spe}
Let $\ttgame$ be an $\numAgents$-player turn-taking game and $(\strat_1, \strat_2, \dots, \strat_\numAgents)$ a strategy profile for this game. We say that this strategy profile is a \textbf{subgame-perfect equilibrium} if it is a Nash equilibrium on all subgames of $\ttgame$.
\end{definition}

The proof of the following theorem can be found in~\cite{osborne1994course}.
\begin{theorem}\label{thm:spe_existence}
Every finite turn-taking game has a subgame perfect equilibrium.
\end{theorem}

Let us now try to find the subgame-perfect equilibria of our example game from Figure~\ref{fig:non_credible_threat}. First note that  that Definition~\ref{def:spe} implies that every subgame-perfect equilibrium of a turn-taking game $\ttgame$ is also a Nash equilibrium of $\ttgame$. After all, by definition it has to be a Nash equilibrium on \textit{all} subgames of $\ttgame$, which includes $\ttgame$ itself. Since we already know the Nash equilibria of $\ttgame$, namely $(a, de)$, $(a, df)$ and $(b, \mi{cf})$, we can restrict our attention to those three strategy profiles.

Next, let us look at the subgame $\ttgame_{(a)}$ defined by the history $(a)$. In this subgame there is only one player, namely player 2, who can choose between actions $c$ and $d$. Action $c$ will yield a utility of 0 to player 2 and action $d$ will yield her a utility of 5, so she would choose action $d$. Therefore, the strategy profile $(b, \mi{cf})$, is not a Nash equilibrium on the subgame $\ttgame_{(a)}$, since it prescribes that player 2 would choose action $c$ instead of $d$.

Finally, let us look at the subgame $\ttgame_{(b)}$ defined by the history $(b)$. Again, in this subgame player 2 is the only player, and this time she can choose between actions $e$ and $f$. Action $e$ will yield a utility of 5 to player 2 and action $f$ will yield her a utility of 30, so she would choose action $f$. Therefore, the strategy profile $(a, \mi{de})$, is not a Nash equilibrium on the subgame $\ttgame_{(b)}$, since it prescribes that player 2 would choose action $e$ instead of $f$.

In conclusion, we see that the strategy profile $(a, df)$ is the only subgame-perfect equilibrium of our example game, because indeed it forms a Nash equilibrium on all three subgames  of $\ttgame$ (that is, $\ttgame_{(a)}$, $\ttgame_{(b)}$, and $\ttgame$ itself).

\subsection{Non-deterministic Turn-taking Games}\label{sec:non_deterministic_games}
Games like chess or Tic-Tac-Toe are completely deterministic. However, many other games, such as backgammon or poker involve randomness because players need to throw dice or shuffle cards.

A common way to formally model non-deterministic games is to introduce an extra player to the game, which is often called `nature'. The idea is that, unlike the other players, nature does not have a utility function and always selects its actions randomly. For example, whenever a 6-sided die is thrown, we say it is nature's turn and that nature will randomly choose an action $\ac \in \{1,2,3,4,5,6\}$.

\begin{definition}
A \textbf{non-deterministic turn-taking} game for $\numAgents$ players, consists of the following components:
\begin{itemize}
\item A set $\Ac$, which we call the set of \textbf{actions}.
\item A set $\Hist$, called the set of all \textbf{histories}, which is a subset of the set of all finite tuples over $\Ac$ (i.e. $\Hist \subseteq \Ac^\kleene$), such that $\Hist$ is prefix closed.
\item A function $\act$ called the \textbf{active player map}, that maps each non-terminal history $\hist \in \Hist \setminus \Hist^\term$ to the index of the player whose turn it is, or to $\nat$, representing `nature':
\[\act \ \  : \ \ \Hist \setminus \Hist^\term \rightarrow \{\nat, 1, 2, \dots, \numAgents\}\]
\item For each $i\in \{1, 2, \dots, \numAgents\}$ a \textbf{utility function}  $\util_i$ that assigns a utility value for player $i$ to each terminal history in $\Hist$:
\[\util_i \ \  : \ \  \Hist^\term \rightarrow \mathbb{R}\]
\item For each history $\hist$ such that $\act(\hist)=\nat$, a probability distribution $P_h$ over the set $\Ac_\hist$ of legal actions after $\hist$.
\end{itemize}

\end{definition}
Note that we still refer to this game as an $n$-player game, even though it technically has $\numAgents + 1$ players, including nature. This is of course because we don't want to count `nature' as a real player.

In an $n$-player non-deterministic turn-taking game, it no longer holds that any $n$-tuple of strategies $\vec{\strat}$ yields a unique terminal history, because the terminal history now also depends on the random choices made by nature. Instead, however, each $\numAgents$-tuple of strategies $\vec{\strat}$ leads to a probability distribution $P(\hist \mid \vec{\strat})$ over the set of all terminal histories $\hist \in \Hist^\term$. This means that, for any player $i$ and any strategy profile $\vec{\strat}$, we can only calculate an \textit{expected} utility $\expt{\util}_i(\vec{\strat})$:
\[\expt{\util}_i(\vec{\strat}) \ \ := \ \ \sum_{\hist \in \Hist^\term} P(\hist \mid \vec{\strat}) \cdot \util_i(\hist)\]

In order to define the notion of an `optimal' strategy, we can now follow the same procedure as for deterministic turn-taking games, except that we need to define everything in terms of the \textit{expected} utility functions. That is, a non-deterministic turn-taking game $\ttgame$ corresponds to a normal-form game $\nfgame$, where the actions of $\nfgame$ are exactly the strategies of $\ttgame$ and the utility functions of $\nfgame$ are exactly the \textit{expected} utility functions of $\ttgame$. Then, the pure Nash equilibria of $\ttgame$ are defined as the pure Nash equilibria of $\nfgame$ and a subgame perfect equilibrium of $\ttgame$ is defined as a strategy profile that forms a Nash equilibrium on every subgame of $\ttgame$.

%
%However, for this book we find it convenient to model indeterminism in a slightly different way.
%
%\essential{Define non-deterministic extensive-form game}

\subsection{Turn-taking Games with Imperfect Information}
Another property that many games satisfy, but that we haven't discussed yet, is the property of \textit{imperfect information}. This means that during the game the players do not have full knowledge of the state of the game, or of the actions played by the other players. Typical examples of such games are card games, such as poker, where each player can only see his own cards but not the cards in the hands of the other players.

To model the notion of a turn-taking game with imperfect information, we assume that whenever a player plays an action, this action is not seen by the other players. Instead, every player receives a signal that may or may not reveal some (limited) information about which action was played. For example, imagine the players are playing a card game, and imagine that player~1 discards one of his cards, say, his ace of spades. So, while player~1 is playing the action $(\mi{discard}, Ace, \spadesuit)$, the other players will only observe the signal $(\mi{discard})$. From this signal, the other players will understand that player~1 discarded a card, but they will not be able to tell \textit{which} card player~1 was discarding.

In order to formalize this, we will assume that the game has a predefined set of possible \textbf{observations} (or `signals') $\Obs$ and that each player has a so-called \textbf{observation function} $\obsfunc_i : \Hist \rightarrow \Obs^\kleene$ that maps each legal action history to a sequence of observations for that player. 

For example, suppose the current state of some game is given by a history $(\ac_1, \ac_2, \ac_3)$, but player 1 has received the following sequence of observations: $\obsfunc_1(\ac_1, \ac_2, \ac_3) = (\obs_1, \obs_2, \obs_3)$. Then, after player 2 plays action $\ac_4$, player 1 will receive some observation $\obs_4$, so we have $\obsfunc_1(\ac_1, \ac_2, \ac_3, \ac_4) = (\obs_1, \obs_2, \obs_3, \ac_4)$. Typically, $\obsfunc_1$ would be a non-invertible function, so just from the observations $(\obs_1, \obs_2, \obs_3, \obs_4)$ the player would not be able to deduce the actual actions $(\ac_1, \ac_2, \ac_3, \ac_4)$ that have been played. In other words, at any point during the game, a player will, in general, not be aware of the history of actions that have so far been played, but instead will only be aware of the sequence of observations he or she has so far received. Also note that each player has its own individual observation function, so each player may receive different observations.

%For some games we may want to model the case that a player does not receive any observation at all, when another player has made a move. That is, the player will not even be aware that the other player has made his move until some third player has also made his move. We model this with the special symbol $\silent$, which is a special observation that is in fact not observed at all.

\begin{definition}
Let $\Hist$ be some set of action histories and $O$ be some set of observations, then an \textbf{observation function} $\obsfunc_i\ :\ \Hist \rightarrow \Obs^\kleene$ is a function that maps every possible history to a tuple of observations, such that for any pair of histories $\hist, \hist' \in \Hist$ where $\hist$ is a prefix of $\hist'$, we also have that $\obsfunc_i(\hist)$ is a prefix of $\obsfunc_i(\hist')$.
\end{definition}
We will refer to $\obsfunc_i(\hist)$ as the \textbf{observed history} of agent $i$ and we may sometimes use the notation $\obsHist{i}$ as a shorthand for $\obsfunc_i(\hist)$.

\later{Maybe give a bit more explanation about why $\obsfunc_i$ should preserve prefixes.}

Note that this definition allows for the possibility that a player sometimes may not receive any observation at all, when another player plays an action. For example, we could have something like: $\obsfunc_1(\ac_1, \ac_2, \ac_3) = \obsfunc_1(\ac_1, \ac_2, \ac_3, \ac_4)$. This means that when player 2 plays action $\ac_4$, player 1 will not even be aware that player 2 played any action at all.

%\begin{definition}
%Let $\hist$ be some action history of a turn-taking game: $\hist = (\ac_1, \ac_2, \ac_3, \dots, \ac_k)$ and let $\obsfunc_i$ be an observation function for player $i$. Then, the \textbf{observed history} $\hist^{obs}_i \in \Obs^\kleene$ of player $i$, corresponding to the full history $\hist$, is a tuple of observations
%
%
% with the same length as $\hist$: 
%\[\hist^{obs}_1 = (\obs_1, \obs_2, \obs_3, \dots, \obs_k)\]
%where the observations $\obs_j$ satisfy:
%\[\forall j\in \{1, 2,\dots, k\}: \ \ \obs_j = \obsfunc_i(\ac_1, \ac_2 \dots, \ac_j)\]
%\end{definition}

With these definitions we can now formally define the notion of a turn-taking game with imperfect information.
\begin{definition}
A \textbf{turn-taking game with imperfect information} (for $\numAgents$ players) is a turn-taking game together with a set of possible \textbf{observations} $\Obs$ and for each player $\ag_i$ an \textbf{observation function} $\obsfunc_i : \Hist \rightarrow \Obs^\kleene$.
Furthermore, apart from the active-player function, $\act$, each player $\ag_i$ also has its own individual active-player function $\act_i : \Obs^\kleene \rightarrow \{1, 2, \dots, \numAgents, ?\}$ which must satisfy:
\[\forall \hist \in \Hist \ \forall i  \in \{1, 2, \dots, \numAgents\}:\quad \act_i(\obsfunc_i(\hist)) = i \quad \text{if and only if} \quad  \act(\hist) = i\]
%\[\forall \hist \in \Hist \ \forall i  \in \{1, 2, \dots, \numAgents\}: \quad \act_i(\hist^{obs}_i) = \act(\hist) \]
\end{definition}
The last constraint in this definition ensures that, even though the players do not have full information about the current state of the game, each player is still able to correctly determine  whether or not it is his turn to make a move, based only on his own observations. Technically, we should also include similar constraints to ensure the players always have full knowledge of their legal actions and their utility functions. However, we will skip that to avoid overcomplicating things.

Furthermore, note that we have included the symbol `?' in the codomain of the functions $\act_i$. This symbol represents the case that player $i$ does not know whose turn it is.

Now, a strategy for a turn-taking game with imperfect information can be defined as a function that maps observation histories to actions.
\begin{definition}
Let $\ttgame$ be a turn-taking game with imperfect information. Furthermore, let $\Obs_i$ denote the set of all possible observed histories after which it is player $i$'s turn:
\[\Obs_i := \{\vec{\obs} \in \Obs^\kleene \mid \act_i(\vec{\obs}) = i\}\]
Then, a \textbf{strategy} for player $i$ is a map that assigns to each observed history $\vec{\obs}$ after which $\ag_i$ is the active player, a legal action for $\ag_i$. 
\[\strat \ \ : \ \ \Obs_i \rightarrow \Ac \quad \text{such that} \quad \forall \hist \in \Hist_i : \ \ \strat(\obsfunc_i(\hist)) \in \Ac_{\hist}\]
\end{definition}
This definition implies that a player can only choose his actions based on the observations that he has seen, rather than on the actual actions that have been played. This represents the fact that in general the player doesn't know exactly which actions have been played, and that the `observations' are indeed the only thing the player observes.

Of course, in most games a player would at least be able to fully observe his \textit{own} actions. This means the observation made by the active player would typically simply be the action itself.
%\[\act(\ac_1, \ac_2, \dots, \ac_k) = i \quad \rightarrow \quad  \obsfunc_i(\ac_1, \ac_2, \dots, \ac_k) = \ac_k\]
%Nevertheless, for the sake of generality, we allow the possibility that even the active player himself may not be able to see his own action, and thus receives an observation different from his own action.

Furthermore, note that a turn-taking game with \textit{perfect} information (i.e. a game such as chess or go where all the players do have a full view of all the players' actions), can be seen as a special case of a game with imperfect information, where each observed history is just the full history itself:
\[\forall \hist \in \Hist \ \forall i  \in \{1, 2, \dots, \numAgents\}: \quad \obsfunc_i(\hist) \ = \ \hist\]
%\[\forall i \ \forall (\ac_1, \ac_2 \dots, \ac_k) \in \Hist: \quad \obsfunc_i(\ac_1, \ac_2 \dots, \ac_k) \ \ = \ \ \ac_k\]

The question how to determine the optimal strategy profile for games with imperfect information is, however, a lot more difficult to answer than for ordinary turn-taking games. We will just comment that the commonly accepted solution concept for such games is known as the \textit{sequential equilibrium}, without going into detail about how it is defined. For more information about this topic we refer to \cite{osborne1994course}.

\subsection{Turn-taking Games with Incomplete Information}
In game theory, the notion of \textit{incomplete} information is  similar to, but not exactly the same as, the notion of \textit{imperfect} information.

As we discussed in the previous section, \textit{imperfect} information refers to the lack of knowledge about the \textit{current state} of the game. In other words, at any moment \textit{during} the game the players may not know exactly at which node of the game tree they are currently situated.

On the other hand, a game with \textbf{incomplete} information, is a game for which the players do not even have full information about the \textit{structure} of the game itself. For example, the players may not know each others' utility functions or may not know exactly which actions are available to the other players. Note that this lack of information already exists even \textit{before} the game has started.

So, in a game of \textit{complete}, but \textit{imperfect} information, the players initially have full knowledge of the structure of the game, but after the game has started some of the actions played by some of the players may be invisible to other players, yielding uncertainty about the current state. Most card games, such as poker, are examples of such games. 

On the other hand, in a game of \textit{incomplete} information, the players already lack knowledge of some aspects of the game, even before the game has started. Of course, it is also possible that a game has both incomplete \textit{and} imperfect information.

In automated negotiation one often assumes that the negotiating agents do not know each others' utility functions. So, under that assumption, automated negotiation is indeed an example of a game with incomplete information.

\section{Automated Negotiation as a Game}\label{sec:nego_as_a_game}
Now that we have discussed the basic principles of game theory, we can finally come back to the topic of automated negotiation, and discuss in what sense it is a game.

The basic idea is simple: each negotiating agent is a player of the game and the actions they can play are exactly the negotiation actions as defined in Definition \ref{def:action}. However, since each action is followed by a small unpredictable delay due to network latency, it is a non-deterministic game and since this delay itself cannot be observed, it is also a game of imperfect information. Furthermore, in case we assume the agents do not no each others' utility functions, then it is also a game of incomplete information.

So, in this section we will formally define, for any negotiation domain $\dom$, a corresponding non-deterministic turn-taking game with imperfect information for 2 players, denoted $\ttgame_\dom$. Note that essentially we are just repeating the definition of a bilateral negotiation under the alternating offers protocol that we already gave in Chapter~\ref{sec:basic_negotiations}, but this time we are using game-theoretical terminology.

\subsection{Actions}\label{sec:nego_game_actions}
The actions of the two players in the game $\ttgame_\dom$ are exactly the negotiation actions as defined in Def. \ref{def:action}. We use the notation $\Ac_i^\dom$ to refer to the set of negotiation actions for player~$i$. That is:
\[\Ac_1^\dom := \{1\} \times \{\prop, \acc\} \times \Off \times \mathbb{R}^+\]
\[\Ac_2^\dom := \{2\} \times \{\prop, \acc\} \times \Off \times \mathbb{R}^+\]

However, since it is a non-deterministic game, we also need an extra player called `nature', as explained in Section \ref{sec:non_deterministic_games}. Every time after one  of the two real players has submitted a negotiation action, it is nature's turn to ``choose" a random delay for the message to arrive at the other agent. This delay can be any positive real number, so the set of actions for nature is the set of positive real numbers $\mathbb{R}^+$.

So, in total, the set of actions $\Ac$ of the game $\ttgame_\dom$ is:
\[\Ac \quad = \quad \Ac_1^\dom \cup A_2^\dom \cup \mathbb{R}^+\]

\subsection{The Active Player Map}\label{sec:nego_game_active_player}
Since we are modeling the alternating offers protocol, the agents' turn to make a proposal will alternate between players 1 and 2. However, since each negotiation action is followed by a random `delay', every turn in which one of the two players chooses an action has to be followed by a turn for nature to choose the delay. Therefore, the game has the following turn-taking structure:
\ \\

\noindent Player 1, nature, player 2, nature, player 1, nature, player 2, nature, etc... 
\ \\

Formally, we can define this as follows:
\[\act(\hist) = 
\begin{cases}
1 & \text{if\ } |\hist|(\text{mod}\ 4) = 0\\
2 & \text{if\ } |\hist|(\text{mod}\ 4) = 2\\
\nat & \text{otherwise (i.e.\ } |\hist| \text{\ is\ odd).} \\
\end{cases}
\]
where $\hist$ is any tuple over the set of actions $\Ac$, i.e. $\hist \in \Ac^\kleene$.

Note that we here follow the convention that it is always player 1 that starts the negotiation (unlike in some of the previous sections in which we followed the convention that player 1 is `our' agent).

\subsection{The Set of Legal Histories}\label{sec:nego_game_legal_histories}
The set of legal histories of $\Gamma_\dom$ is exactly the set of negotiation histories as defined by Definitions \ref{def:history} and \ref{def:aop}.

We can define it recursively. That is, let $\hist'$ be any legal history, then we can define the criteria that an action $\ac \in \Ac$ would need to satisfy in order for the history $\hist := \hist' \circ (\ac)$ to be legal as well. Then, given that the empty history $\emptyTuple$ is 
legal, we can construct all other legal histories.

Suppose the current state of the game is given by some history $\hist'$. If, in this state, it is player $1$'s turn (i.e. $\act(\hist')=1$), then she can either propose an offer or accept an offer. That is, she can play an action of the form $(1,\prop, \off, t)$ or $(1,\acc, \off, t)$. In other words, she can choose an action $\ac$ from the set $\Ac_1^\dom$. And analogously for player 2. On the other hand, when it is nature's turn (i.e. $\act(\hist')=\nat$), nature can select any positive number $\ac \in \mathbb{R}^+$.

Formally, this means that an action $\ac$ is only legal in state $\hist'$ if the following conditions hold:
\begin{itemize}
\item if $\act(\hist') = 1$ then $\ac \in \Ac_1^\dom$
\item if $\act(\hist') = 2$ then $\ac \in \Ac_2^\dom$
\item if $\act(\hist') = \nat$ then $\ac \in \mathbb{R}^+$
\end{itemize}
Furthermore, there are a number of other constraints that must be satisfied as well.

%\later{Note that $\hist' = (\dots, (i_j, \actype_j, \off_j, t_j), \dots )$ }

Specifically, in order for the numbers $t_j$ and $\delay_j$ to be interpretable as \textit{times} we have to impose the condition that, for any index $j$ the number $t_{j+1}$ must be larger than $t_j + \delay_j$. That is, if $(i_k, \actype_k, \off_k, t_k)$ and $\delay_k$ are the last two actions of the history $\hist'$, and $\ac = (i_{k+1}, \actype_{k+1}, \off_{k+1}, t_{k+1})$, then we must have:
\[t_k + \delay_k < t_{k+1}\]

In addition, recall that the definition of the AOP specifies that an agent can only accept the \textit{last} offer proposed by the other agent. That is, we must have:
\[\text{if} \quad \actype_{k+1} = \acc \quad \text{then} \quad \off_{k} = \off_{k+1}\]

%\essential{finish this}

Finally, the history $\hist'$ is terminal (meaning that there is no action $\ac$ such that $\hist' \circ (\ac)$ is legal), if and only if
its length is an even number (i.e. $|\hist'|(\text{mod}\ 2) = 0$) and 
at least one of the following holds:
\begin{itemize}
\item $t_k + \delay_k \geq \dead$
\item $k = \maxRounds$
\item $\actype_k = \acc$
\end{itemize}

The condition that the length has to be an even number, means that the negotiations have finished only after `nature' has made its move, which means that the last propose- or accept-message must have arrived at its recipient.

\subsection{The Observation Functions}\label{sec:nego_game_observation_functions}
Suppose we have the following history:
\[\hist = \Big( (1, \prop, \off_1, t_1), \delay_1, (2, \prop, \off_2, t_2), \delay_2, (1, \prop, \off_3, t_3), \delay_3, (2, \prop, \off_4, t_4), \delay_4, \dots \Big)\]

As explained in Section \ref{sec:aop}, whenever player 1 proposes an offer, he will only be aware of the time $t$ at which he proposed it, but he will not know how much time $\delay$ it takes for that message to arrive at player~2, and therefore he will not know the time $t + \delay$ at which player~2 receives it. Similarly, player~2 will not be able to observe the time $t$ at which the message was sent, nor the delay $\delay$, but will only observe the time $t + \delay$ at which she receives the message.

Therefore, the observed history for player~1 looks as follows:
\[\obsfunc_1(\hist) = \Big( (1, \prop, \off_1, t_1), (2, \prop, \off_2, t_2 + \delay_2), (1, \prop, \off_3, t_3), (2, \prop, \off_4, t_4 + \delay_4), \dots \Big)\]
and for player~2:
\[\obsfunc_2(\hist) = \Big( (1, \prop, \off_1, t_1 + \delay_1), (2, \prop, \off_2, t_2), (1, \prop, \off_3, t_3 + \delay_3), (2, \prop, \off_4, t_4), \dots \Big)\]

That is, the set of observations of $\Gamma_\dom$ is just the set of negotiation actions:
\[\Obs = \Ac_1^\dom \cup \Ac_2^\dom \]

Formally, let $\obs_j^i$ denote the $j$-th observation received by player $i$, so we have:
\[\obsfunc_i(\hist) = (\obs_1^i, \obs_2^i, \obs_3^i, \dots,\obs_k^i)\]
Then, if $(i_j, \actype_j, \off_j, t_j)$ denotes the $j$-th negotiation action of $\hist$, each $\obs_j^i$ must satisfy:
\[
\obs_j^i = 
\begin{cases}
(i_j, \actype_j, \off_j, t_j) & \text{if\ } i=i_j \\
(i_j, \actype_j, \off_j, t_j+\delay_j) & \text{if\ } i\neq i_j \\
\end{cases}
\]

\subsection{The Individual Active-Player functions}\label{sec:nego_game_indiv_active_player_functions}
Recall that for a game of imperfect information, besides the active player function $\act$, we also need to define an \textit{individual} active-player function $\act_i$, representing each player's \textit{knowledge} about whose turn it is.

Note that when player 1 proposes an offer, then directly after this action, he knows that it is now the turn of `nature', until the message has arrived, after which it will be player 2's turn. However, since player 1 cannot observe the duration of the delay, he will not know when exactly it stops being nature's turn and when it starts being player 2's turn. In other words, player 1 will typically not know whose turn it is, until it is his own turn. And the same holds of course for player 2.

So, if $\obsHist{i}$ denotes the observed history of player $i$ (i.e. $\obsHist{i} := \obsfunc_i(\hist)$), then:
\[
\act_i(\obsHist{i}) = 
 \begin{cases}
 i & \text{if\ } \act(\obsHist{i}) = i \\
 ? & \text{otherwise}
 \end{cases}
\]

\subsection{The Utility Functions}\label{sec:nego_game_util_functions}
The utility functions of the game $\ttgame_\dom$ are defined in terms of the utility functions of the negotiation domain $\dom$. However, the utility functions of the game are defined over the set of terminal histories.

If the negotiation ended with an acceptance that arrived before the deadline, then each player receives their respective utility value $\util_i(\off_k)$ corresponding to the accepted offer $\off_k$. Otherwise, each player $i$ receives his reservation value $\rv_i$.

Formally, let $\hist$ be a terminal history, and let $(i_k, \actype_k, \off_k, t_k)$ denote the last negotiation action of $\hist$. Then:
\[\util_i(\hist) = 
 \begin{cases}
 \util_i(\off_k) & \text{if\ } \actype_k = \acc \text{\ and\ } t_k + \delay_k < \dead.\\
 \rv_i & \text{otherwise} \\
 \end{cases}
\]
where the $\util_i$ on the left-hand side is a utility function of the game $\ttgame_\dom$ and the $\util_i$ on the right-hand side is a utility function of the negotiation domain $\dom$. Furthermore $\rv_i$ is the reservation value of player $i$ of negotiation domain $\dom$.

\subsection{Formal Definition}
We can now put all this together into the formal definition of the  game $\ttgame_\dom$.

\begin{definition}
Let $\dom$ be a bilateral negotiation domain with offer space $\Off$. Then a negotiation over this domain, according to the alternating offers protocol, with deadline $\dead$ and maximum number of rounds $\maxRounds$, can be modeled as a non-deterministic turn-taking game with imperfect information $\ttgame_\dom$, defined as follows:
\begin{itemize}
\item The set of actions $\Ac$ of the game $\ttgame_\dom$ is defined as in Section \ref{sec:nego_game_actions}.
\item The active player map $\act$ of $\ttgame_\dom$ is defined as in Section \ref{sec:nego_game_active_player}.
\item The set of legal histories $\Hist$ of $\ttgame_\dom$ is defined as in Section \ref{sec:nego_game_legal_histories}.
\item The observation functions $\obsfunc_i$ of $\ttgame_\dom$ are defined as in Section \ref{sec:nego_game_observation_functions}.
\item The individual active-player functions $\act_i$ of $\ttgame_\dom$ are defined as in Section \ref{sec:nego_game_indiv_active_player_functions}
\item The utility functions $\util_i$ of $\ttgame_\dom$ are defined as in Section \ref{sec:nego_game_util_functions}.
\end{itemize}
\end{definition}

Now that we have formalized negotiation using the terminology of game-theory, we would like to apply techniques from game theory to determine the optimal negotiation strategy. Unfortunately, however, this turns out extremely difficult for several reasons. 

The first reason, is that most techniques from game theory assume that the players have full information about each others' utility functions. An assumption that often does not hold in automated negotiation. 

A second reason, is that we had to model negotiation as a turn-taking game with \textit{imperfect} information. This means that to find the optimal strategy profile, we would need to determine the sequential equilibria of $\ttgame_\dom$, which is known to be an extremely hard problem to solve, even for very simple games. We therefore have to lower our expectations, and ignore the fact that the agents are not able to observe the delay times. If we pretend that they do know this information, then we can treat the game as if it was as a turn-taking game with perfect information, so we can try to determine its subgame-perfect equilibria.

A third reason, is that even if we assume that the agents could somehow observe the delays of \textit{past} messages and therefore treat the game as a turn-taking game with \textit{perfect} information, it would still be very difficult to find its subgame-perfect equilibria. This is because in order to play optimally, the agents would also have to be able to deal with the randomness of the delays of \textit{future} messages. That is, the agents would still have to deal with the fact that it is a \textit{non-deterministic} game. While in general there are techniques to deal with this, the problem is that in our case the random choices $\delay_j$ of nature can take an infinite number of possible values, which makes it hard to apply any well-known techniques.

For these reasons, the best result we can expect to obtain here, is to find the ordinary Nash equilibria of $\ttgame_\dom$, when regarded as a game of perfect information, and under the assumption that we know the utility functions and reservation values of both players.

%However, we can ignore that limitation and simply pretend that the negotiators do have full knowledge about each other's utility functions and reservation values.

While the assumption of full knowledge of both agents' utility functions and reservation values may be unrealistic in many real-life negotiation scenarios, it does allow us to determine a theoretical upper bound to what an agent could achieve in the ideal case that it had a perfect opponent modeling algorithm. In other words, it can be used in a laboratory setting to compare a real negotiation algorithm with a theoretically optimal one.

%Furthermore, if $(i,\eta, \off, t) \in \hist$ and $(j,\eta, \off', t') \in \hist$ with 

\subsection{Nash Equilibria of a Negotiation}\label{sec:nash_equilibria_nego}
In this section we will show that the game $\ttgame_\dom$ typically has many pure Nash equilibria.

%Of course, since $\ttgame_\dom$ is a turn-taking game, the real question we would like to answer, is the question which pair of negotiation strategies would yield a subgame-perfect equilibrium. Unfortunately, however, this turns out to be an extremely difficult question to answer. This difficulty arises from the fact that each proposal takes an  unknown delay time $\delay$ to arrive at the opponent. The fact that this time is unpredictable, plus the fact that this random variable can take an infinite number of possible values, makes it hard to apply any well-known techniques to determine the subgame-perfect equilibria.

%Instead, we will therefore only try to find the Nash equilibria. 

As explained above, ideally, we would like to find the subgame-perfect equilibria, or even the sequential equilibria, of the game $\ttgame_\dom$. However, since this is very hard, we will instead just try to determine its ordinary Nash equilibria. We could then hope to find that this game has only one Nash equilibrium, which would then automatically also have to be its subgame-perfect equilibrium. Unfortunately, however, it turns out that this is typically not the case. In fact, the following theorem shows that the game $\ttgame_\dom$ typically has \textit{many} Nash equilibria: at least one for every offer that is Pareto-optimal and individually rational. An important consequence of this, is that if we want to determine an optimal negotiation strategy, we would need to apply some of the techniques discussed in Section~\ref{sec:equilibrium_selection}.

\begin{theorem}\label{thm:nash_equilibria_nego}
Let $\dom$ be a bilateral negotiation domain with a finite offer space $\Off$ and let $T$ be the deadline for the negotiations. If $T$ is sufficiently large then for every offer $\off \in \Off$ that is Pareto-optimal and individually rational, there exists a pair of negotiation strategies $(\strat_1, \strat_2)$ that forms a Nash equilibrium and that leads to $\off$ as the final agreement (or another offer $\hat{\off}$ with exactly the same utility values).
\end{theorem}
\begin{proof}
Let $\off$ be any arbitrary Pareto-optimal and individually rational offer. Given $\off$, let $\strat_1$ be a time-based strategy based on Eq.~(\ref{eq:time_based_max}) or Eq.~(\ref{eq:time_based_min}) and with an aspiration function defined by Eq.~(\ref{eq:asp_function_adapted})  with target value $\targ_1 = \util_1(\off)$, in combination with the $AC_{asp}$ acceptance strategy (Def.~\ref{def:ac_asp}). Similarly, let $\strat_2$ be a time-based strategy, defined by the same equations, and with target value $\targ_2 = \util_2(\off)$. 

Now, to prove the theorem, we will prove the following three claims one by one:
\begin{enumerate}
\item If these two strategies come to an agreement, then it must either be the offer $\off$, or some other offer $\hat{\off}$ with exactly the same utility vector.
\item These two strategies will indeed come to an agreement.
\item Neither of the two agents can improve by making a unilateral deviation.
\end{enumerate}
To prove the first claim, note that for any arbitrary offer $\off'$ one of the following must hold:
\begin{enumerate}
\item $\off'$ dominates $\off$.
\item $\util_1(\off') < \util_1(\off)$
\item $\util_2(\off') < \util_2(\off)$
\item $\util_1(\off') = \util_1(\off)$ and  $\util_2(\off') = \util_2(\off)$.
\end{enumerate}
However, the first case is impossible, because we assumed that $\off$ was Pareto-optimal. In the second case we would have that $\util_1(\off') < \targ_1$ which means, by definition of $\targ_1$, that $\ag_1$ would never propose or accept $\off'$. Similarly, in the third case we would have that $\util_2(\off') < \targ_2$ which means that $\ag_2$ would never propose or accept $\off'$. This means that in the second or third case, $\off'$ could not be the final agreement of the negotiations. Therefore, the only case in which $\off'$ could be the final agreement is the fourth case, which is what we wanted to prove.

Now, to prove the second claim, note that if $T$ is large enough, then sooner or later either of the two agents will have proposed all other offers that are better for him than $\off$, so that agent will eventually propose $\off$. Furthermore, sooner or later the other agent's aspiration level $\asp_i(t)$ will become equal to her utility value $\util_i(\off)$  and thus she will eventually accept the offer $\off$.

Finally, to prove the third claim, we will show that agent $\ag_2$ cannot deviate unilaterally to a better strategy (we should also show the same for $\ag_1$, but that goes analogously).  To do this, note that if $\ag_2$ does deviate to any alternative strategy $\strat_2'$, then this must yield one of the following outcomes:
\begin{enumerate}
\item The negotiations end without agreement.
\item The negotiations end with the same agreement $\off$.
\item The negotiations end with a different agreement $\off'$ such that ${\util_2(\off') \leq \util_2(\off)}$. \item The negotiations end with a different agreement $\off'$ such that ${\util_2(\off') > \util_2(\off)}$. 
\end{enumerate}
In the first case, the deviation did not improve the outcome for agent $\ag_2$, because she ends up with her reservation value $\rv_2$. Note that we assumed that $\off$ was individually rational, and therefore we have $\rv_2 < \util_2(\off)$, so indeed she would have been better off if she didn't deviate.

In the second case the deviation did not improve the outcome for $\ag_2$ either, because the outcome is the same as for the original strategy profile.

In the third case, again, the deviation did not improve her outcome, because agent $\ag_2$ ends up with less or equal utility than in the original situation.

In the fourth case agent $\ag_2$ does improve, but we will show that this case cannot happen. The reason for this, is that we assumed that $\off$ was Pareto-optimal. This means that if $\util_2(\off') > \util_2(\off)$, we must necessarily have  $\util_1(\off') < \util_1(\off)$, otherwise $\off'$ would dominate $\off$ and therefore $\off$ would not be Pareto-optimal. However, since we assumed that $\ag_1$ applies a time-based strategy with target value $\targ_1 = \util_1(\off)$, we know that $\ag_1$ would never accept or propose any offer with utility lower than $\util_1(\off)$, so in particular she would never propose or accept $\off'$, which means that $\off'$ could never become an agreement.

We have therefore proved that $\ag_2$ cannot make a unilateral deviation that increases her utility. The fact that this also holds for $\ag_1$ can be proved in exactly the same way.
\end{proof}

\subsection{Non-credible Threats in a Negotiation}

Now that we have determined the Nash equilibria of a negotiation, the question we will investigate next, is whether or not any of them are based on non-credible threats. It turns out that indeed, such non-credible threats do appear when either of the agents blindly follows a Nash equilibrium.

Imagine we are very close the the deadline and agent 1 has proposed some Pareto-optimal and individually rational offer $\off$. Furthermore, suppose that for the remainder of the negotiations, agent 1 has chosen the following strategy: ``reject any counter offer from agent 2 that is worse for me than $\off$, and do not make any further concessions, no matter what". Now, it is easy to see that for agent 2 the best response to this strategy would be to accept the offer $\off$. However, let us assume that  agent 2 does not play this best response (that is, player 2 `deviates') and instead makes a counter offer $\off'$ with slightly less utility for agent 1. Furthermore, suppose that there is not enough time left for agent 1 to propose any new offer. So, agent 1 can only accept $\off'$ or accept that the negotiations will fail. Assuming $\off'$ is also individually rational, it would be sub-optimal for agent 1 to stick to his strategy (which would cause the negotiations to fail) because he would be better off by accepting $\off'$. Therefore, he would be forced to also deviate, which means that his original strategy was indeed based on a non-credible threat.

\section{Bargaining Solutions}\label{sec:bargaining_solutions}

Now that we have developed a basic understanding of game theory and modeled negotiations as a game, we can start investigating how game theory can help us finding an optimal negotiation strategy. This question is also known as \textbf{the bargaining problem}. 

Despite the fact that this is a very old research topic, dating back to the 1950's, however, there is still no definitive general solution to this problem. However, just as for the equilibrium selection problem, many different solutions have been proposed that are each applicable to different special cases. Such solutions are often referred to as \textbf{bargaining solutions}. We will here discuss just a few of them.

Note that these bargaining solutions do not actually find any optimal negotiation strategies \textit{directly}. Instead, they merely determine the offer that two `optimal' negotiation strategies would agree upon (depending on the definition of `optimal', which differs for each such bargaining solution). However, once we have found such an offer $\off^*$, we know that our agent shouldn't be proposing or accepting anything worse than that, so we can implement a strategy that concedes towards, but no further than, that offer. For example, it could be a time-based strategy that uses the value $\util_1(\off^*)$ as its target value. We then know from the proof of Theorem~\ref{thm:nash_equilibria_nego} that a pair of such strategies would form a Nash equilibrium.

%One of the main impediments to finding a game-theoretically optimal negotiation strategy, is that most practically usable solutions proposed by game theory assume that each player has full knowledge of the opponent's utility function. This assumption typically does not hold in automated negotiation. 

We should point out, however, that each of these bargaining solutions is based on the assumption that we know the utility functions of \textit{both} agents. This means we typically cannot apply them directly in practice, because a negotiating agent typically only knows his own utility function, but not his opponent's. Nevertheless, these bargaining solutions can still be very useful, because they can serve as a theoretical upper bound to what a real negotiation algorithm could achieve. That is, we can evaluate a real negotiation algorithm (that does not have knowledge of its opponent's utility function) by comparing it with the theoretically optimal solution that could be achieved if one does have knowledge of the opponent's utility function. For example, it has been shown \cite{deJonge2024theoreticalPropertiesMicro} that if two MiCRO agents negotiate against each other, then they often come remarkably close to to the \textit{Nash Bargaining Solution} as well as the \textit{Max-Sum solution} (we will discuss these solutions below).

Another important simplification made by most bargaining solutions, is that they model the the negotiations as a normal-form game, even though we have seen that negotiation should technically be seen as an extensive-form game.

\subsection{The Nash Bargaining Solution}\label{sec:nbs}
The solution to the bargaining problem that is by far the best-known, is the so-called \textit{Nash Bargaining Solution} (NBS), which was invented by John Nash in 1950~\cite{Nash1950} (the same person that also invented the concept of the Nash equilibrium).

%The Nash Bargaining Solution applies to bilateral negotiations in which a number of conditions hold.

So far, we have always considered negotiation domains in this book that have \textit{finite} offer spaces. However, the NBS applies to negotiation domains with an \textit{infinite} number of possible agreements. In particular, Nash assumed that the offer space formed a subset of an $\numIssues$-dimensional vector space $\mathbb{R}^\numIssues$, for some positive integer $\numIssues$. That is, $\Off \subseteq \mathbb{R}^\numIssues$. This means that we can now talk about `linear combinations' of offers. That is, given two offers $\off$ and $\off'$ and two real numbers $a$ and $b$, the linear combination ${a \cdot \off + b \cdot \off'}$ is also a vector, which may or may not lie inside the agreement space. Moreover, it means we can calculate the Euclidean distance $d(\off, \off')$ between those two offers. More specifically, Nash worked under the assumption that the offer space is a \textit{closed}, \textit{bounded} and \textit{convex} subset of $\mathbb{R}^\numIssues$. We will now explain these terms.

\begin{definition}
Let $S$ be some subset of $\mathbb{R}^\numIssues$, i.e. $S \subseteq \mathbb{R}^\numIssues$. Then, we say that $S$ is \textbf{bounded} iff there exists a number $M$ such that for any two vectors $v, w \in S$ their distance is smaller than $M$:
\[\exists M\in \mathbb{R} : \quad \forall v,w\in S : \quad d(v,w) \leq M\]
where $d(v,w)$ denotes the Euclidean distance between $v$ and $w$.
\end{definition}
In other words, there is a maximum distance $M$ between any two points in $S$.

\begin{definition}
Let $S$ be some subset of $\mathbb{R}^\numIssues$, i.e. $S \subseteq \mathbb{R}^\numIssues$. Then, we say that $S$ is \textbf{closed} iff for any convergent sequence of vectors in $S$, the limit of that sequence is also contained in $S$. 
\end{definition}
For example, the following set is \textit{not} closed:
\[S \quad = \quad  \{(x,y) \in \mathbb{R}^2 \mid x^2 + y^2 < 1\}\]
because the following sequence:
\[(0,0) \quad , \quad (0, \frac{1}{2}) \quad , \quad (0, \frac{3}{4}) \quad , \quad (0, \frac{7}{8}) \quad , \quad (0, \frac{15}{16}) \quad , \quad \dots \]
converges to the point $(0,1)$, but that point does not lie inside $S$. On the other hand, the following set \textit{is} closed:
\[S' \quad = \quad  \{(x,y) \in \mathbb{R}^2 \mid x^2 + y^2 \leq 1\}\]

Roughly speaking, one could say that a set is closed if the ``border'' of that set is also part of the set itself.

\begin{definition}
A subset $S$ of $\mathbb{R}^\numIssues$ is called \textbf{compact} if it is both bounded and closed.
\end{definition}

\begin{definition}
Let $S$ be some subset of $\mathbb{R}^\numIssues$, i.e. $S \subseteq \mathbb{R}^\numIssues$. Then, we say that $S$ is \textbf{convex} iff for any two vectors $v, w \in S$, and any real number $x \in [0,1]$, it holds that the vector $x\cdot v + (1-x)\cdot w$ is also contained in $S$.
\end{definition}
Intuitively, the term `convex' means that if you take any two points $v$ and $w$ in $S$, then the line between $v$ and $w$ must be completely contained inside $S$. See Figure \ref{fig:convex} for a visualization of this concept. Note that a convex set must necessarily contain \textit{infinitely} many elements (because there are infinitely many real numbers $x$ in the interval $[0,1]$).

\begin{figure}
\begin{center}
\begin{tikzpicture}[dot/.style={circle,inner sep=1pt,fill,label={#1},name=#1},
  extended line/.style={shorten >=-#1,shorten <=-#1},
  extended line/.default=1cm]

% Draw the convex set (a circle with radius 2).
\draw[fill=gray!30] (-5,2) circle (2);

% Draw two points insisde the convex set and a line between them.
\draw[fill=gray!256] (-4,3) circle (0.05);
\draw[fill=gray!256] (-4,1) circle (0.05);
\draw (-4,3) to (-4,1);
 
% Draw the non-convex set.
\draw[fill=gray!30] (0,0) to[out=180, in=-90] (-2,2) to[out=90, in=180] (0,4) to[out=0, in=90] (2,3) to[out=-90,in=90] (0,2) to[out=-90, in=90] (2,1) to[out=-90, in=0] (0,0);

% Draw two points insisde the non-convex set, and the line between them.
\draw[fill=gray!256] (1,3) circle (0.05);
\draw[fill=gray!256] (1,1) circle (0.05);
\draw (1,3) to (1,1);
\end{tikzpicture}
\caption{Left: a convex set. For any pair of points inside this set, the line between them will also lie completely inside the set. Right: a non-convex set. We can find two points that are both inside the set, but the line between them is not completely contained inside the set.}\label{fig:convex}
\end{center}
\end{figure}

The offer spaces that Nash discusses (compact and convex) are very different from the offer spaces we have so far discussed in this book. An example of a compact and convex negotiation domain, is the scenario that one agent aims to sells a certain commodity, such as oil or gold, to another agent, and they negotiate the quantity to sell and the price, which are both constrained to some finite domain. For example, Alice has 1000 liters of crude oil for sale, and Bob has a budget of \$800 that he can spend on oil. In that case the offer space is given by the set $\Off = [0, 1000] \times [0 , 800]$, which is indeed a compact and convex\footnote{Technically, this is not entirely true, because the price can only be specified as a finite number of dollar cents. However, this is such a small step-size that we can ignore this and pretend that money is a continuous commodity.} subset of $\mathbb{R}^2$.

Furthermore, Nash himself argued that even if there are really only a finite number of possible offers, then we could still consider the offer space to be compact and convex, if we also allow the agents to propose and accept \textit{lotteries} over offers. For example, they could agree that they will flip a coin. If the coin comes up `heads' then they will execute solution $\off$, while if the coin comes up `tails' then they will execute solution $\off'$. If we furthermore allow the agents to agree that the coin can be biased with any probability $P$ for heads or tails, then the set of lotteries is indeed compact and convex.

\begin{opinion}
Personally, I have always found the possibility of negotiating over lotteries rather far-fetched. I just can't imagine many real-world situations in which two negotiators would jointly agree to flip a coin.
\end{opinion}

Apart from these assumptions about the agreement space, Nash also assumed that the two agents have full knowledge of each others' utility functions, that the offer space contains at least one individually rational offer, and that the two utility functions are `\textit{distributive}' over linear combinations of offers. 
That is:
\[\forall i\in \{1,2\}: \ \forall \off, \off' \in \Off: \ \forall a,b\in \mathbb{R}: \quad \util_i(a\off + b\off') = a \util_i(\off) + b\util_i(\off')\]
Note that this is a stronger assumption than merely assuming the functions are linear in the sense of Eq.~(\ref{eq:linear_util_func}), because this implies that every \textit{evaluation} function $\eval_i^j$ must also be a linear function itself (from $\mathbb{R}$ to $\mathbb{R}$).

In summary, the Nash Bargaining Solution applies to negotiations that satisfy the following conditions:
\begin{itemize}
\item C1: The negotiations are bilateral.
\item C2: The agreement space $\Off$ is a compact and convex subset of $\mathbb{R}^m$ (for some positive integer $m$).
\item C3: The agreement space contains at least one individually rational offer (see Def.~\ref{def:individually_rational}).
\item C4: The utility functions are distributive over the offers.
\item C5: The two agents have full knowledge of each others' utility functions.
\end{itemize}

For any negotiation domain $\dom$, let us define the \textbf{utility space} $\utilSpace_{\dom} \subseteq \mathbb{R}^2$ to be the set of all utility vectors. That is:
\[\utilSpace_{\dom}  := \{ \big(\util_1(\off), \util_2(\off) \big) \mid \off \in \Off\} \]
Note that if $\Off$ is convex, and the two utility functions are distributive, then $\utilSpace_\dom$ is also a convex set. 

\begin{definition}
Let $\dom$ be a negotiation domain. We say that $\dom$ is \textbf{symmetric} if for any $(x,y) \in \utilSpace_{\dom}$ we also have $(y,x) \in \utilSpace_{\dom}$.
\end{definition}

If $\dom$ and $\dom'$ are two negotiation domains, then we say that $\dom'$ is a linear transformation of $\dom$ if they both have the same offer space, and the utility functions and reservation values of $\dom'$ are just linear transformations of the utility functions and reservation values of $\dom$. More precisely:
\begin{definition}
Let $\dom = (\Off, \util_1, \util_2, \rv_1, \rv_2)$ and  $\dom' = (\Off', \util_1', \util_2', \rv_1', \rv_2')$ be two negotiation domains. We say that $\dom'$ is a \textbf{linear transformation} of $\dom$ iff
$\Off = \Off'$ and there exist four real numbers $a_1,b_1,a_2,b_2 \in \mathbb{R}$, with $a_1>0$ and $a_2>0$, such that all of the following conditions are satisfied:
\begin{eqnarray*}
\forall \off \in \Off: \quad \util_1'(\off) &=& a_1\cdot \util_1(\off) + b_1 \\
\rv_1' &=& a_1\cdot \rv_1 + b_1 \\
\forall \off \in \Off: \quad \util_2'(\off) &=& a_2 \cdot \util_2(\off) + b_2 \\
\rv_2' &=& a_2 \cdot \rv_2 + b_2
\end{eqnarray*}
\end{definition}

We say that a domain $\dom'$ is an \textit{extension} of another domain $\dom$, if $\dom$ can be obtained by removing some offers from the offer space of $\dom'$. More precisely:
\begin{definition}
Let $\dom = (\Off, \util_1, \util_2, \rv_1, \rv_2)$ and  $\dom' = (\Off', \util_1', \util_2', \rv_1', \rv_2')$ be two negotiation domains. We say that $\dom'$ is an \textbf{extension} of $\dom$ iff
$\Off \subseteq \Off'$ and the following conditions are all satisfied:
\begin{eqnarray*}
\forall \off \in \Off: \quad \util_1'(\off) &=& \util_1(\off) \\
\rv_1' &=& \rv_1\\
\forall \off \in \Off: \quad \util_2'(\off) &=& \util_2(\off) \\
\rv_2' &=& \rv_2\\
\end{eqnarray*}

\end{definition}

Nash argued that, for any negotiation domain $\dom$, if the conditions C1--C5 hold, then the agreement between two optimal negotiators would be an offer $\off_\dom^*$ that satisfies all of the following axioms:
\begin{itemize}
\item A1 \textit{Pareto-optimality}:\\
 The agreement $\off_{\dom}^*$ should be Pareto-optimal (see Def. \ref{def:pareto_optimal}).
\item A2 \textit{Symmetry}:\\
 If $\utilSpace_{\dom}$ is symmetric, then $\util_1(\off_{\dom}^*) = \util_2(\off_{\dom}^*)$
\item A3 \textit{Invariance under linear transformations}: \\ If $\dom'$ is a linear transformation of $\dom$, then $\off_{\dom}^* = \off_{\dom'}^*$.
\item A4 \textit{Independence of irrelevant alternatives}: \\ If $\dom'$ is an extension of $\dom$, then either $\off_{\dom'}^* = \off_{\dom}^*$, or $\off_{\dom'}^*  \in \Off' \setminus \Off$.
\end{itemize}

The first two of these axioms speak for themselves. The third axiom follows directly from the principle of von Neumann-Morgenstern utilities, which we discussed in Section \ref{sec:neumann_morgenstern}. That is, since the application of a linear transformation to either of the two utility functions of $\dom$ does not essentially change the domain, we can say that $\dom'$ is essentially the same as $\dom$, and therefore the outcome of the negotiations should be the same.

The fourth axiom is probably less obvious, and is easiest to explain for the case that $\Off'$ contains just one offer more than $\Off$. That is, suppose that in domain $\dom$, two optimal negotiators would agree on offer $\off_{\dom}^*  \in \Off$. Now, suppose that we add a new offer $\off'$ to the offer space, so we have $\Off' = \Off \cup \{\off'\}$, and
we repeat the negotiations. Then, it may or may not happen that the two negotiators will now agree on that new offer $\off'$. However, in the case that this does not happen, then this new offer can be considered `irrelevant' to the negotiations and should not affect the outcome. Therefore, in that case the agents should come to exactly the same agreement~$\off_{\dom}^*$ as in the original negotiations. After all, it would be strange if the addition of some new offer $\off'$ would lead the negotiators to accept a \textit{different} offer $\off''$. This is exactly what axiom A4 says, except that it allows for any arbitrary number of `irrelevant' new offers.

Nash proved the following theorem \cite{Nash1950}:
\begin{theorem}
Under conditions C1, C2, C3, C4 and C5, any offer $\off_\dom^*$ that satisfies axioms A1, A2, A3, and A4, satisfies:
\[\off_\dom^* \quad \in \quad \argmax_{\off \in \Off} \ \ \big\{\ (\util_1(\off)-\rv_1)\cdot(\util_2(\off)-\rv_2)\ \big\}\]
Furthermore, all such offers have exactly the same utility vector.
\end{theorem}
In other words, Nash argued that under the given conditions two optimal negotiators would agree on the offer that maximizes the product of the two agents' utility values (minus reservation values). This implies that an optimal negotiation strategy, according to Nash, would be one that aims to achieve that offer as the final agreement. For example, this could be a time-based strategy with $\targ_i = \util_i(\off_D^*)$.

\begin{definition}\label{def:nbs}
Let $\dom$ be a bilateral negotiation domain. Then, for any offer $\off \in \Off$ of this domain, the product $(\util_1(\off)-\rv_1) \cdot (\util_2(\off)-\rv_2)$ is called the \textbf{Nash product}. Furthermore, the offer that maximizes the Nash product is called the \textbf{Nash Bargaining Solution}.
\end{definition}

The essence behind the NBS, is that it models negotiation as a normal-form game. To simplify it a bit, we can restrict the game to only allow the players to choose \textit{time-based} strategies. We know from Theorem~\ref{thm:nash_equilibria_nego} that such a game has many Nash equilibria.  Then, if $\dom$ is symmetric, this means that the game is a symmetric game, so we can use the solution of Section~\ref{sec:symmetric_equilibria} to pick an optimal Nash equilibrium, and it can be shown that this will indeed yield an agreement that maximizes the Nash product (there are many such pairs of time-based strategies, but this is a factorizable set). Furthermore, for those cases that the negotiation domain $\dom$ is not symmetric, Nash used his various conditions and axioms to show that there must exist some other domain $\dom'$ that \textit{is} symmetric and that should yield the same outcome as $\dom$.

Now, we should warn the reader that the NBS is sometimes discussed by authors as if it was the one and only correct solution to the bargaining problem. However, one should not forget that the NBS in principle only applies to those negotiations where the (rather strong) conditions C1--C5 hold, and even then it only applies under the (controversial) assumption that the axioms A1--A5 are indeed a `correct' description of an optimal solution. Authors sometimes tend to forget or ignore these restrictions.

Finally, we should warn that some authors erroneously define the NBS as the offer that directly maximizes the product of the utilities $\util_1(\off) \cdot \util_2(\off)$, without subtracting the reservation values. However, this is wrong, because it would not satisfy the axiom of Invariance under Linear Transformations. Unless, of course, the reservation values both happen to be zero.

\subsection{The Kalai-Smorodinsky Solution}
In 1975, two researchers by the names of Kalai and Smorodinsky~\cite{KalaiSmorodinsky} argued that there is a problem with the NBS, which they demonstrated with the following example.

Suppose we have two bilateral negotiation domains $\dom_1$ and $\dom_2$ that satisfy the conditions C1--C5 from Section \ref{sec:nbs}. Furthermore, suppose that $\dom_2$ is an extension of $\dom_1$. To be specific, suppose that their utility spaces are given as follows:
\begin{itemize}
\item $\utilSpace_1$ is the interior of the following four points:
\[\{ \ \ (0,0) \ \ ,\ \   (1,0) \ \ , \ \  (0,1) \ \  , \ \  (0.75, 0.75) \ \ \}\]
\item $\utilSpace_2$ is the interior of the following four points: 
\[\{ \ \ (0,0) \ \ ,\ \   (1,0) \ \ , \ \  (0,1) \ \  , \ \  (1, 0.7) \ \ \}\]
\end{itemize}
See Figure~\ref{fig:kalai_paradox} for an illustration. 

\begin{figure}
\begin{center}
\begin{tikzpicture}[scale=4]

% Draw the first utility space
\draw[fill=gray!50] (0,0) to (0,1) to (0.75,0.75) to (1,0) to (0,0);

\node at (0.5,0.5) {$\utilSpace_1$};

% Draw the second utility space
\draw[fill=gray!20] (0,1) to (1,0.7) to (1,0) to (0.75,0.75) to (0,1);

\node at (0.90,0.5) {$\utilSpace_2$};

% Draw horizontal axis label
\node[anchor=north] at (0.5,0) {Utility for agent 1};

% Draw vertical axis label
\node[anchor=east] at (0,0.5) {\rotatebox{90}{Utility for agent 2}};

\end{tikzpicture}
\end{center}
\caption{The dark grey area represents all $\utilSpace_1$, while the light grey area represents the difference between $\utilSpace_1$ and $\utilSpace_2$, i.e. $\utilSpace_2 \setminus \utilSpace_1$. The horizontal axis represents the utility of agent $\ag_1$ while the vertical axis represents the utility of agent $\ag_2$.}\label{fig:kalai_paradox}
\end{figure}

Now, note that for any utility value $x \in [0,1]$ there exists a $y\in [0,1]$ such that $(x,y) \in \utilSpace_2$ and such that for all $(x,y') \in \utilSpace_1$ we have $y'< y$. So, in words: no matter what utility value $\ag_1$ would receive from the optimal solution of domain $\dom_1$, there always exists some offer $\off$ in $\Off_2$ that gives the same utility to $\ag_1$, but that is strictly better for $\ag_2$. It therefore makes sense to expect that $\ag_2$ would at least receive the same utility from $\dom_2$ as from $\dom_1$, and possibly even better. 

However, if we follow the NBS, then we would conclude that the outcomes $\off_{\dom_1}^*$ and $\off_{\dom_2}^*$ for the two respective domains should be given by:
\begin{eqnarray*}
\vec{\util}(\off_{\dom_1}^*) &=&  (0.75\ ,\ 0.75)\\
\vec{\util}(\off_{\dom_2}^*) &=& (1\ ,\ 0.7)\\
\end{eqnarray*}
%\begin{eqnarray*}
%(\ \util_1(\off_{\dom_1}^*)\ ,\ \util_2(\off_{\dom_1}^*)\ ) &=&  (0.75\ ,\ 0.75)\\
%(\ \util_1(\off_{\dom_2}^*)\ ,\ \util_2(\off_{\dom_2}^*)\ ) &=& (1\ ,\ 0.7)\\
%\end{eqnarray*}
That is, even though we have extended domain $\dom_1$ with offers that are better for $\ag_2$, the final outcome of the negotiation would, according to the NBS, actually be \textit{worse} for $\ag_2$ (i.e. 0.7 instead of 0.75).

Kalai and Smordinsky argued that this result is highly unsatisfactory and that it is caused by the axiom of independence of irrelevant alternatives. They therefore proposed an alternative axiom, called \textit{The Axiom of Monotonicity}.

In order to define this axiom we must first introduce some more notation. For any negotiation domain $\dom$, let $\pf$ denote its pareto-frontier (see Def.~\ref{def:pareto_set}):
\[\pf \quad := \quad \{\ ( \util_1(\off) , \util_2(\off) ) \mid \off \in \OffPareto \ \}\]
where $\OffPareto$ is the Pareto-set of $\dom$. Furthermore, let $\kf_{\dom}$ be a function from the interval $[\rv_1, \utilMax_1]$ to $\mathbb{R}$, defined as follows:
\begin{equation}\label{eq:kalai_function}
\kf_{\dom}(x) = 
\begin{cases}
y & \text{if\ } \exists y: (x,y) \in \pf \\
\utilMax_2 & \text{otherwise}
\end{cases}
\end{equation}
This function is visualized in Figure~\ref{fig:kalai_function}. Finally, for any two given negotiation domains $\dom$ and $\dom'$, let $\utilMax_1$ and ${\utilMax_1}'$ denote the highest possible utility values that agent 1 can achieve in these two respective domains (as per Eq.~(\ref{eq:utilMax})).

\begin{figure}
\begin{center}
\begin{tikzpicture}[scale=4]
% Draw the utility space
\draw[fill=gray!30] (0.3,1) to[out=0, in=135] (0.8, 0.8) to[out=-45, in=90] (1,0.3) to[out=-90, in=0] (0.8,0.1) to[out=180, in=-90] (0.1,0.8) to[out=90, in=180] (0.3,1);

% Draw the axes
\draw (0,0) to (0,1);
\draw (0,0) to (1,0);

% Draw the function
\draw [line width=0.5mm, red ] (0,1) to (0.3,1) to[out=0, in=135] (0.8, 0.8) to[out=-45, in=90] (1,0.3);

% Draw the label.
\node at (0.6,0.6) {$\utilSpace_\dom$};

% Draw the reservation point.
\node[anchor=north east] at (0,0) {$(\rv_1, \rv_2)$};

% Draw horizontal axis label
\node[anchor=north] at (0.5,0) {Utility for agent 1};

% Draw vertical axis label
\node[anchor=east] at (0,0.5) {\rotatebox{90}{Utility for agent 2}};
\end{tikzpicture}
\end{center}
\caption{The red line indicates the function $\kf_{\dom}$ as defined by Eq.~(\ref{eq:kalai_function})}\label{fig:kalai_function}
\end{figure}

The Axiom of Monotonicity is then defined as follows.
\begin{itemize}
\item A5 \textit{The Axiom of Monotonicity}: If $\dom$ and $\dom'$ are two negotiation domains such that $\utilMax_1 = {\utilMax_1}'$
%\[\max_{\off \in \Off} \util_1(\off) \quad = \quad \max_{\off \in \Off'} \util_1(\off)\]
 and such that 
\[\forall x\in [\rv_1, \utilMax_1]: \quad \kf_\dom(x) \leq \kf_{\dom'}(x)\] 
 then we should have:
\[\util_2(\off_\dom^*) \quad \leq \quad \util_2(\off_{\dom'}^*).\]
\end{itemize}
Kalai and Smorodinsky themselves described this axiom informally as follows: \textit{``If, for every utility level that agent 1 may demand, the maximum feasible utility level that agent 2 can simultaneously reach is increased, then the utility level assigned to agent 2 according to the solution should also be increased."}

So, they argue that under the same conditions C1--C5 as Nash, the agreement made by two optimal negotiators should actually satisfy axioms A1, A2, A3 and A5 (instead of A4). 

They then prove the following theorem:
\begin{theorem}
Under conditions C1, C2, C3, C4 and C5, any offer $\off_\dom^*$ that satisfies axioms A1, A2, A3, and A5, is given by:
\[\off_\dom^* \quad = \quad \argmax_{\off \in \Off} \{\util_1(\off) \mid (\util_1(\off) , \util_2(\off)) \in L(\dom)\}\]
where $L(\dom)$ is the line from the point $(\rv_1 , \rv_2)$ to the `utopian' point $(\utilMax_1, \utilMax_2)$.
Furthermore, all such offers have exactly the same utility vector.
\end{theorem}
\begin{figure}
\begin{center}
\begin{tikzpicture}[scale=4]

% Draw the utility space
\draw[fill=gray!50] (0.3,1) to[out=0, in=135] (0.8, 0.8) to[out=-45, in=90] (1,0.3) to[out=-90, in=0] (0.8,0.1) to[out=180, in=-90] (0.1,0.8) to[out=90, in=180] (0.3,1);

% Draw bounding rectangle
\draw (0,0) to (0,1) to (1,1) to (1,0) to (0,0);

% Draw the label.
\node at (0.5,0.7) {$\utilSpace_\dom$};

% Draw the diagonal line
\draw (0,0) to (1,1);

% Draw a label for the reservation point
\node[anchor=north east] at (0,0) {$(\rv_1, \rv_2)$};

% Draw a label for the utopian point
\node[anchor=south west] at (1,1) {$(\utilMax_1, \utilMax_2)$};

% Draw the Kalai-Smorodinsky point.
\draw[fill=gray!256] (0.8,0.8) circle (0.025);

% Draw horizontal axis label
\node[anchor=north] at (0.5,0) {Utility for agent 1};

% Draw vertical axis label
\node[anchor=east] at (0,0.5) {\rotatebox{90}{Utility for agent 2}};

\end{tikzpicture}
\end{center}
\caption{The black dot represents utility vector of the Kalai-Smorodinsky solution. It lies, by definition, on the intersection of $\utilSpace_\dom$ and the diagonal between $(\rv_1, \rv_2)$ and $(\utilMax_1, \utilMax_2)$. Specifically, it is the point on this intersection that lies closest to $(\utilMax_1, \utilMax_2)$.}\label{fig:kalai}
\end{figure}
In other words, Kalai and Smorodinsky argued that, under the given conditions, two optimal negotiators would agree on the offer for which the utility vector lies on the intersection of the utility space and the diagonal between the reservation values and the `utopian' point $(\utilMax_1, \utilMax_2)$, and such that it is closest to that utopian point. This is known as the \textbf{Kalai-Smorodinsky Solution} (KSS). See Figure~\ref{fig:kalai} for an illustration of this concept.

\subsection{The Max-Sum Solution}
The NBS and the KSS are widely used as a reference to compare to what extent real negotiation algorithms are able to negotiate optimally. This is quite striking, however, given that both solutions technically only apply to negotiations over convex offer spaces, and with distributive utility functions, while many negotiation scenarios that are studied in the literature do not satisfy these conditions.

For this reason, in \cite{deJonge2023bargainingSolution} a new bargaining solution was proposed specifically for negotiations over \textit{finite} offer spaces, rather than convex ones. Just as for the NBS, the idea behind this solution is to model negotiations as a normal-form game, which, as discussed in Section \ref{sec:nash_equilibria_nego}, typically has many Nash equilibria. However, this time, to choose the optimal equilibrium we use the solution discussed in Section \ref{sec:AoRE}. This means that this bargaining solution only applies to those situations in which the Assumption of Role Equifrequency (AoRE) holds (Def.~\ref{def:AoRE}). That is, given any domain $\dom$, our agent assumes that it will be negotiating over that domain with each of the two utility functions equally often or with equal probability. Note that for most studies in the literature, this assumption indeed holds. See for example the various ANAC competitions~\cite{anac2010,anac2019}.

From this it follows that two optimal negotiators would (typically) agree on the offer that maximizes the \textit{sum} of the utilities of the two agents. We will therefore call this the \textbf{Max-Sum} solution.\footnote{In the literature this bargaining solution has also been called the \textit{maximum social welfare solution}, but this name may be a bit misleading, because it actually has nothing to do with maximizing social welfare.}

More precisely, he max-sum solution works as follows:
\begin{enumerate}
\item Select all offers that are Pareto-optimal and individually rational.
\item Among those offers, select the subset that maximizes the utility-sum $\util_1(\off) + \util_2(\off)$.
\item Among those offers, select the subset that minimizes the absolute utility difference $|\util_1(\off) - \util_2(\off)|$.
\item If these offers all have the same utility vector, then return any of those offers.
\item Otherwise, discard this set of offers and go back to step 2 with the remaining Pareto-optimal and individually rational offers.
\end{enumerate}

Note that this is essentially just Algorithm~\ref{alg:opt_strategy_AoRE}, where we identify the set of Nash equilibria with the set of Pareto-optimal and individually rational offers (as per Theorem~\ref{thm:nash_equilibria_nego}).

%\begin{definition}
%Let $\dom$ be a bilateral negotiation domain with finite offer space, and assume that the AoRE holds. Then, we can obtain an optimal pair of negotiation strategies by modeling the negotiations as a normal-form game and then applying the solution of Section~\ref{sec:AoRE} to find an optimal Nash equilibrium. This is known as the \textbf{Max-Sum} solution. 
%\end{definition}

The main advantage of this bargaining solution is that we can drop the assumption that the offer space has to be compact and convex, or that the utility functions are distributive. Furthermore, we no longer need the controversial axioms of Independence of Irrelevant Alternatives, or of Monotonicity. 

Regarding to the axiom of Invariance under Linear Transformations, however, we have to be a bit careful. Nash's axiom states that we can apply two different linear transformations to the two utility functions of the domain. However, as we also discussed in Section~\ref{sec:AoRE}, under the AoRE, this is actually too strong. That is, we have to make a careful distinction between `players' and `agents', and while it makes perfect sense to allow each \textit{agent} to apply any arbitrary linear transformation, we cannot say the same about \textit{players}.

Now, to demonstrate that, under the AoRE and on finite domains, the Max-Sum solution is indeed better than the NBS, we will now give an explicit example of such a negotiation domain and show that a strategy based on the Max-Sum solution indeed outperforms a strategy based on the NBS.

Suppose Alice and Bob are negotiating against each other over some negotiation domain $\dom$. We will assume that they will negotiate twice. Once with Alice having utility $\util_1$ and Bob having utility $\util_2$, and once with the utility functions assigned in the opposite way. Therefore, the AoRE holds. Furthermore, let us assume that Alice and Bob are each only considering two possible negotiation strategies, which we will call `sum-maximizer' (denoted $S$) and `product-maximizer' (denoted $P$). The sum-maximizer strategy only accepts the offer that maximizes the utility-sum or anything with higher utility, while the product-maximizer only accepts the offer that maximizes the Nash product, or anything with higher utility.

Now, let us assume that the negotiation domain $\dom$ only has two offers: $\off_s$ and $\off_p$. These offers have respective utility vectors $\vec{\util}(\off_s) = (5,1)$ and $\vec{\util}(\off_p) = (2,3)$. Furthermore, both reservation values are 0. Note that $\off_s$ has the highest utility-sum ($5+1=6$ vs. $2+3=5$), while $\off_p$ has the highest Nash product ($(5-0)\cdot (1-0)=5$ vs. $(2-0)\cdot (3-0) = 6$).

We now observe the following facts:
\begin{itemize}
\item If both agents choose strategy $S$, then the only possible agreement is $\off_s$.
\item If both agents choose strategy $P$, then the only possible agreement is $\off_p$.
\item If the agent with utility $\util_1$ chooses $S$ and the agent with utility $\util_2$ chooses $P$, then neither of the two offers can become an agreement (because $\ag_1$ demands a utility of at least $\util_1(\off_s) = 5$, while $\ag_2$ demands a utility of at least $\util_2(\off_p) = 3$ and there is no offer that satisfies both constraints).
\item If the agent with utility $\util_1$ chooses $P$ and the agent with utility $\util_2$ chooses $S$, then either of the two offers can become an agreement (because $\ag_1$ demands a utility of at least $\util_1(\off_p) = 2$, while $\ag_2$ demands a utility of at least $\util_2(\off_s) = 1$ and both offers satisfy both constraints). 

Whenever this happens we will assume there is a 50\% chance that the agreement will be $\off_s$ and a 50\% chance that the agreement will be $\off_p$.
\end{itemize}

%%%%%%%%%%%%%%%%%%%%%%%%%%%%%%%%%%%%%%%%%%%%%%%%%% 
%
% Note that this game has two pure Nash equilibria:
% (S,S) with utils (5 , 1)
% (P,P) with utils (2 , 3)
%
% and a mixed equilibrium:
% (4/7 S + 3/7 P  ,  1/2 S + 1/2 P) with utils (2.86 , 1.5)
%%%%%%%%%%%%%%%%%%%%%%%%%%%%%%%%%%%%%%%%%%%%%%%%%% 

From this, we conclude that, depending on which strategies Alice and Bob choose, we get the following:
\begin{itemize}
\item If Alice and Bob both choose strategy $S$, then in both negotiations the agreement will be $\off_s$, and thus they will each obtain a utility of 6 (in the first negotiation Alice will receive 5 and Bob will receive 1, and in the second negotiation vice versa).
\item If Alice and Bob both choose strategy $P$, then in both negotiations the agreement will be $\off_p$, and thus they will each obtain a utility of 5 (in the first negotiation Alice will receive 3 and Bob will receive 2, and in the second negotiation vice versa).
\item If Alice chooses strategy $S$ and Bob chooses $P$, then the first negotiation will end without agreement, so both will receive a utility of 0. The second negotiation will end with either of the two offers. Therefore, Alice will have an expected utility of $0.5\cdot 1 + 0.5\cdot 3 = 2$, while Bob will have an expected utility of $0.5\cdot 5 + 0.5\cdot 2 = 3.5$.
\item If Alice chooses strategy $P$ and Bob chooses $S$, then the situation is reversed, so now Alice has an expected utility of 3.5 and Bob has an expected utility of 2.
\end{itemize}

This can now be modeled as a normal-form game with the following (symmetric) payoff matrix:
\begin{center}
\begin{tabular}{c|c|c}
 & $S$ & $P$ \\
 \hline
 $S$ & $(6\ ,\ 6)$ & $(2\ ,\ 3.5)$ \\
  \hline
 $P$ & $(3.5\ ,\ 2)$  & $(5\ ,\ 5)$ \\
\end{tabular}
\end{center}
While this game has two Nash equilibria $(S,S)$ and $(P,P)$, it is clear that $(S,S)$ is the better choice.

\later{Rubinstein}

\later{The Balance Set}

\later{Discuss negotiations with fixed number of rounds?}
%\chapter{Game Theory}\label{sec:game_theory}

%\section{Normal-Form Games}
%COMING SOON!
%
%\section{Extensive-Form Games}
%COMING SOON!
%
%\section{Automated Negotiation as a Game}
%COMING SOON!

%
%\later{Discuss negotiations with fixed number of rounds?}

%
%
%\later{The Balance Set}

\chapter{Evaluation of Negotiation Algorithms}\label{sec:evaluation}
In Chapter \ref{sec:negotiation_strategies} we discussed various types of negotiation strategies. Then, in Chapter \ref{sec:game_theory} we discussed how one could try to use game theory to determine which strategy is the best. Unfortunately, however, the `bargaining solutions' we discussed there were based on two major simplifications. Firstly, we had to model negotiations as a normal-form game, rather than as an extensive-form game. Secondly, they all depended on having full knowledge of both the agents' utility functions. For these reasons, game theory is only of limited use when implementing an actual negotiation strategy.

Therefore, in practice, whenever we implement a negotiation strategy we have to resort to experimental methods to assess its strength. In this chapter we will discuss how to do that. In fact, we will describe three different methods to evaluate agents:
\begin{enumerate}
\item Tournament Evaluation
\item Empirical Game-Theoretical Analysis (EGTA)
\item Sequential Elimination Ranking
\end{enumerate}

Throughout this chapter we will assume we have some set of agents, denoted $\Ag = \{\ag_{\tx{1}}, \ag_{\tx{2}}, \dots, \ag_{\tx{\numAgents}}\}$, that we want to compare to each other and that are all developed to negotiate under some bilateral negotiation protocol.

Note that we here underline the indices of the agents. This is to make a clear distinction between the $\tx{i}$-th agent from our collection of agents $\Ag$, and the $i$-th agent in a given bilateral negotiation. More precisely, for any negotiation over some given bilateral domain $\dom$, the notation $\ag_1$ refers to the agent that aims to maximize the utility function $\util_1$ from that domain, and $\ag_2$ refers to the agent that aims to maximize utility $\util_2$. On the other hand $\ag_{\tx{1}}$ refers to the first agent from our collection of agents $\Ag$, and $\ag_{\tx{2}}$ refers to the second agent from that collection. 

For example, suppose that the domain $\dom$ represents a car sale, in which $\util_1$ is the utility function of the buyer and $\util_2$ is the  utility function of the seller. Then, the expression $\ag_{\tx{1}} = \ag_2$ would mean that the first agent from our set of agents $\Ag$ is acting as the `seller'  in the negotiation, while the expression $\ag_{\tx{4}} = \ag_1$ would mean that the fourth agent from our set of agents is acting as the buyer in this negotiation. 

In general, any underlined index always refers to the index of an agent within the set $\Ag$, while a regular index refers to the role the agent is playing within a given negotiation.

\section{Tournament Evaluation}

The most basic way to evaluate a negotiating agent, is to compare it with a number of other benchmark agents, by means of a tournament. That is, we first pick a number of well-known existing benchmark agents, plus a number of negotiation domains, and then let all agents (i.e. all benchmark agents plus our own agent) negotiate against each other, in every domain. We may repeat this several times, in order to obtain enough data to get statistically significant results, and then finally we calculate the average utility obtained by each agent, over all the negotiations it was involved in, and rank all agents based on that average utility. Hopefully, our agent then ends in first place in that tournament.

\subsection{Tournament Score}
Suppose we have an agent $\ag_{\tx{1}}$ and we want to compare it to a number of benchmark agents $\ag_{\tx{2}}, \ag_{\tx{3}}, \dots, \ag_{\tx{\numAgents}}$, all developed to negotiate under some given bilateral negotiation protocol $\prot$. Furthermore, suppose we have a number of different bilateral negotiation domains $\dom_1, \dom_2, \dots, \dom_\numDomains$. Since all domains are bilateral we can create $\numDomains \times \numAgents \times \numAgents$ different negotiation \textit{scenarios}.

\begin{definition}\label{def:scenario}
We define a bilateral negotiation \textbf{scenario} $\scen$ as a tuple $\scen = (\prot, \dom, \ag, \ag')$, where $\prot$ is a negotiation protocol, $\dom$ is a negotiation domain, and $\ag$ and $\ag'$ are two negotiating agents.
\end{definition}
A negotiation scenario represents a negotiation between agents $\ag$ and $\ag'$ over domain $\dom$, under negotiation protocol $\prot$. It is important to realize that if $\ag$ and $\ag'$ are two different agents, then $(\prot, \dom, \ag, \ag')$ and $(\prot, \dom, \ag', \ag)$ are two different scenarios. This is because in the first scenario $\ag$ has utility function $\util_1$ from domain $\dom$ and $\ag'$ has utility function $\util_2$ from that same domain. On the other hand, in the second scenario $\ag'$ has utility function $\util_1$ and $\ag$ has utility function $\util_2$.

For example, if $\dom$ represents a negotiation between a car seller and a buyer, then in the first scenario $\ag$ plays the role of the seller and $\ag'$ plays the role of the buyer, while in the second scenario the roles are reversed, so now $\ag'$ is the seller and $\ag$ is the buyer. 

Furthermore, it is also important to realize that we allow a scenario to be of the form $(\prot, \dom, \ag, \ag)$ where both roles are played by the same agent $\ag$. Of course, it doesn't make sense to literally have a \textit{single} agent negotiating against itself, but what we mean is that we may have two agents that are each applying exactly the same negotiation algorithm. Or, in other words, we may have two identical \textit{copies} of the same agent negotiating against each other. We should stress that in that case, even though the two agents are identical, they should still be seen as two separate agents that have opposing interests. That is, the first copy is purely interested in maximizing utility function $\util_1$, while the second copy is purely interested in maximizing utility function $\util_2$. The fact that both agents have identical implementations doesn't change that. There is no reason to assume that an agent would treat his opponent any differently from other opponents if that opponent happens to be an identical copy of himself.

For example, imagine again the case of a seller and a customer that are negotiating the price of a car. This time, assume that they are humans, but that they each download a negotiation algorithm from the Internet that will do the negotiations for them. It is then perfectly conceivable that they each happen to download exactly the same algorithm, so there will be two copies of the same agent negotiating against each other. The agent downloaded by the seller will then try to negotiate the highest possible price, while the agent downloaded by the buyer will try to negotiate the lowest possible price.

%\begin{definition}\label{def:encounter}
%We define a bilateral negotiation \textbf{encounter} $\enc$ as a tuple $\enc = (\prot, \dom, \ag, \ag', l)$, where $l\in \mathbb{N}$ is an integer that serves as an identifier to distinguish different negotiation sessions between the same agents in the same domain.
%\end{definition}

\begin{definition}\label{def:tournament}
A bilateral \textbf{negotiation tournament} is a tuple $(\prot, \Dom, \Ag, \mc{\numReps})$ where:
\begin{itemize}
\item $\prot$ is a bilateral negotiation protocol.
\item $\Dom = \{\dom_1, \dom_2, \dots, \dom_\numDomains\}$ is a set of bilateral negotiation domains.
\item $\Ag = \{\ag_{\tx{1}}, ag_{\tx{2}}, \dots, ag_{\tx{\numAgents}}\}$ is a set of agents.
\item $\mc{\numReps}\ :\ \Dom \times \Ag \times \Ag \rightarrow \mathbb{N}$ is a function that maps each possible negotiation scenario to a non-negative integer.
\end{itemize}
\end{definition}
The function $\mc{\numReps}$ indicates how often each negotiation scenario is repeated. We will use the notation $\numReps^{d,\tx{i},\tx{j}}$ as a shorthand for $\mc{\numReps}(\dom_d, \ag_{\tx{i}}, \ag_{\tx{j}})$. So,  $\numReps^{d,\tx{i},\tx{j}}$ is the number of times the agents $\ag_{\tx{i}}$ and $\ag_{\tx{j}}$ will be negotiating (with utility functions $\util_1$ and $\util_2$ respectively) over domain $\dom_d$. In general, we would initially choose the number of repetitions to be the same for every scenario. However, it may turn out that some scenarios yield a lot more variance in their outcomes than others, so for those scenarios we might afterwards choose to increase the number of repetitions to get more accurate data.

The goal of running the tournament, is to gather data. Specifically, for each repetition of any given scenario, we obtain two data points, namely the utility values obtained by the two respective agents in that negotiation. Let us use the notation $\util_{1}^{d,\tx{i},\tx{j},r}$ to denote the utility obtained by the \textit{first} agent in the $r$-th repetition of the scenario $(\prot, \dom_d, \ag_{\tx{i}}, \ag_{\tx{j}})$. So, it is the utility obtained by $\ag_{\tx{i}}$. Similarly, let $\util_{2}^{d,\tx{i},\tx{j},r}$ denote the the utility obtained by the \textit{second} agent in the $r$-th repetition of the scenario $(\prot, \dom_d, \ag_{\tx{i}}, \ag_{\tx{j}})$. So, it is the utility obtained by $\ag_{\tx{j}}$. For example, $\util_{1}^{7,\tx{3},\tx{4},9}$ represents the utility obtained by agent $\ag_{\tx{3}}$ in the ninth repetition of the scenario $(\prot, \dom_7, \ag_{\tx{3}}, \ag_{\tx{4}})$.

Specifically:
\[
\util_{l}^{d,\tx{i},\tx{j},r} := 
\begin{cases}
\util_l(\off) & \text{If\ the\ $r$-th\ repetition\ of\ scenario\ } (\prot, \dom_d, \ag_{\tx{i}}, \ag_{\tx{j}}) \\
 & \text{ended\ with\ agreement\ } \off. \\
\rv_l & \text{If\ the\ $r$-th\ repetition\ of\ scenario\ } (\prot, \dom_d, \ag_{\tx{i}}, \ag_{\tx{j}}) \\
 & \text{ended\ without\ agreement.} \\
\end{cases}
\]
for $l\in \{1,2\}$, where $\util_1$, $\util_2$, $\rv_1$, and $\rv_2$ are the utility functions and reservation values of domain $\dom_d$.

Pay attention to the fact that the numbers $\util_{1}^{d,\tx{i},\tx{j},r}$, $\util_{2}^{d,\tx{i},\tx{j},r}$, $\util_{1}^{d,\tx{j},\tx{i},r}$, and $\util_{2}^{d,\tx{j},\tx{i},r}$ are, in general, all different numbers. For example, suppose we have agent $\ag_{\tx{3}}$ and $\ag_{\tx{4}}$ negotiating over domain $\dom_7$ and that in this domain the two agents are referred to as the `buyer' and the `seller' respectively. Then we can distinguish between two different scenarios:
\begin{itemize}
\item The scenario $(\prot, \dom_7, \ag_{\tx{3}}, \ag_{\tx{4}})$ in which agent $\ag_{\tx{3}}$ is the buyer and $\ag_{\tx{4}}$ is the seller.
\item The scenario $(\prot, \dom_7, \ag_{\tx{4}}, \ag_{\tx{3}})$ in which agent $\ag_{\tx{4}}$ is the buyer and $\ag_{\tx{3}}$ is the seller.
\end{itemize}
If we assume that each of these scenarios is repeated only once, then we obtain four numbers: the two utility values obtained by the two respective agents from the negotiation in the first scenario ($\util_{1}^{7,\tx{3},\tx{4},1}$ and $\util_{2}^{7,\tx{3},\tx{4},1}$), and the two utility values obtained by the two agents from the negotiation in the second scenario  ($\util_{1}^{7,\tx{4},\tx{3},1}$ and $\util_{2}^{7,\tx{4},\tx{3},1}$). That is:
\begin{itemize}
\item $\util_{1}^{7,\tx{3},\tx{4},1}$ is the utility obtained by agent $\boldsymbol{\ag_{\tx{3}}}$ when it acted as the \textbf{buyer} in domain $\dom_7$ against agent $\ag_{\tx{4}}$. 
\item $\util_{2}^{7,\tx{3},\tx{4},1}$ is the utility obtained by agent $\boldsymbol{\ag_{\tx{4}}}$ when it acted as the \textbf{seller} in domain $\dom_7$ against agent $\ag_{\tx{3}}$. 
\item $\util_{1}^{7,\tx{4},\tx{3},1}$ is the utility obtained by agent $\boldsymbol{\ag_{\tx{4}}}$ when it acted as the \textbf{buyer} in domain $\dom_7$ against agent $\ag_{\tx{3}}$. 
\item $\util_{2}^{7,\tx{4},\tx{3},1}$ is the utility obtained by agent $\boldsymbol{\ag_{\tx{3}}}$ when it acted as the \textbf{seller} in domain $\dom_7$ against agent $\ag_{\tx{4}}$. 
\end{itemize}

For any pair of agents $\ag_{\tx{i}}$, $\ag_{\tx{j}}$ and any domain $\dom_d$ we can now calculate the average utility $\ts_{\tx{i}}^{d,\tx{j}}$ obtained by agent $\ag_{\tx{i}}$ against opponent $\ag_{\tx{j}}$ in domain $\dom_d$ (i.e. averaged over all repetitions of the scenarios $(\prot, \dom_d, \ag_{\tx{i}}, \ag_{\tx{j}})$ and $(\prot, \dom_d, \ag_{\tx{j}}, \ag_{\tx{i}})$):
\begin{equation}\label{eq:average_score}
\ts_{\tx{i}}^{d,\tx{j}} \ \ := \ \ \frac{1}{2 \numReps^{d,\tx{i},\tx{j}}} \sum_{r=1}^{\numReps^{d,\tx{i},\tx{j}}} \util_{1}^{d,\tx{i},\tx{j},r} + \frac{1}{2 \numReps^{d,\tx{j},\tx{i}}} \sum_{r=1}^{\numReps^{d,\tx{j},\tx{i}}} \util_{2}^{d,\tx{j},\tx{i},r}
\end{equation}
Note that this formula can also be applied to the case that $i=j$.

Then, for each agent $\ag_{\tx{i}}$, we can calculate its \textbf{tournament score} $\ts_{\tx{i}}$ by averaging over all domains and all opponents (including itself):
\begin{equation}\label{eq:tournament_score}
\ts_{\tx{i}} \ \ := \ \ \frac{1}{|\Dom| \cdot |\Ag|} \sum_{d=1}^{|\Dom|} \sum_{\tx{j}=1}^{|\Ag|} \ts_{\tx{i}}^{d,\tx{j}}
\end{equation}
We can now rank all agents in the tournament based on their tournament scores. The agent with the highest tournament score can be considered the best.

For example, suppose that we have implemented an agent called \textit{IlPadrino} and we have tested it against 3 benchmark agents, called \textit{CrazyAgent}, \textit{RandomAgent}, and \textit{MegaBarter3000}. Then we can display the results of a tournament between these agents in a table such as Table~\ref{tab:tournament_1}.

\begin{table}
\begin{center}
\begin{tabular}{|l|c|}
\hline
\textbf{Agent} & \textbf{Tournament Score} ($\ts_{\tx{i}}$) \\
\hline
MegaBarter3000 &  0.729 \\
\hline
IlPadrino & 0.665 \\
\hline
CrazyAgent & 0.604 \\
\hline
RandomAgent & 0.488 \\
\hline
\end{tabular}
\caption{Results of a fictional tournament between four fictional agents.}\label{tab:tournament_1}
\end{center}
\end{table}

\begin{exercise}\label{ex:run_tournament}
\textbf{Run a Tournament.} Implement code that runs a tournament among the agents that you implemented in the previous exercises (if you didn't do those exercises you can use the agents from the folder `Solutions to Exercises'). Also, make sure the tournament involves both example domains that are given with the framework, as well as the negotiation domain that you created yourself in Exercise~\ref{ex:nego_domain} (if you did that exercise). \newline

Hint: the NegoSim framework already contains a function to run a single negotiation. So, all you have to do is create a loop that iterates over all possible negotiation scenarios, and that calls this function in each iteration. \newline

Make sure the results of all the individual negotiations in this tournament are stored in a text file (or a .csv file) on your hard disk. 
Each line should correspond to the result of one negotiation, and should contain the following information: 
\begin{itemize}
\item The names of the two agents in that negotiation.
\item The name of the domain used in that negotiation.
\item Whether or not the agents came to an agreement.
\item The utility values obtained by the two respective agents (their reservation values in case there was no agreement).
\end{itemize}
For example, this text file could look as follows: \\
\ \\
\begin{tabular}{llllll}
Agent 1; & Agent2; & Domain; & Agree?; & util1; & util2; \\
IlPadrino; & IlPadrino; & CarSale; & YES; & 0.8; & 0.54; \\
IlPadrino; & CrazyAgent; & CarSale; & NO; & 0.0; & 0.0; \\
IlPadrino; & RandomAgent; & CarSale; & NO; & 0.0; & 0.0; \\
IlPadrino; & MegaBarter3000; & CarSale; & YES; & 0.74; & 0.67; \\
\dots
\end{tabular}
\end{exercise}

%
%Agent 1; & Agent2; & domain; & repetition; & agreement?; & util1; & util2 \\
%IlPadrino; IlPadrino; CarSaleDomain; 1; YES; 0.8; 0.54; \\
%IlPadrino; CrazyAgent; CarSaleDomain; 1; NO; 0.0; 0.54; \\
%IlPadrino; RandomAgent; CarSaleDomain; 1; NO; 0.80; 0.54; \\
%IlPadrino; MegaBarter; CarSaleDomain; 1; YES; 0.80; 0.54; \\

\begin{exercise}\label{ex:calculate_tournament_scores}
\textbf{Calculate Tournament Scores.}
Implement a program that does the following:
\begin{enumerate}
\item Read the text file from Exercise~\ref{ex:run_tournament}.
\item Based on the contents of that file, calculate the tournament score of each agent.
\item Display a table such as Table~\ref{tab:tournament_1}, showing the names and scores of the agents, sorted in order of decreasing tournament score. 
\end{enumerate}
\end{exercise}

\subsection{Storing your Data}

Note that in Exercises \ref{ex:run_tournament} and \ref{ex:calculate_tournament_scores} we asked you to implement two \textit{separate} programs. One for running the tournament and one for calculating the tournament scores. There are very good reasons for this.

It is a common mistake among beginning researchers to just write one single program that runs the tournament and then immediately calculates the tournament scores, without storing the results of the individual negotiations. This is a big mistake, because after running your first tournament it often turns out that you need to increase the number of repetitions, or that you need to add more agents or domains to your experiment, or that you may want to analyze one particular aspect of the data that you did not think of before. If you didn't store the individual results, it means that you would then have to run the entire tournament all over again, which may cost a lot of time. On the other hand, if you did store the results, you can simply run a number of extra negotiations and add the results of those negotiations to the data that you already had.

\subsection{Agreement Rate and Utility-Under-Agreement}
While the tournament score $\ts_{\tx{i}}$ can be used to determine \textit{which} agents performed well and which performed poorly, we typically also want to know \textit{why} some agent did or did not perform well. Recall from Chapter \ref{sec:negotiation_strategies} that negotiating well comes down to striking the right balance between being \hard{} and being \soft{}. An agent that is \textit{too} \hard{} will not make many agreements, while an agent that is \textit{too} \soft{} will only make agreements with low utility. To measure whether our agent is too \hard{} or too \soft{}, we can use the following two quantities:

\begin{itemize}
\item \textbf{Agreement Rate}: the percentage of all negotiations that our agent was involved in that ended in agreement.
\item \textbf{Utility-under-Agreement}: the average utility obtained by our agent among only those negotiations that ended in agreement.
\end{itemize}
A \hard{} agent would typically score a low agreement rate, but a high utility-under-agreement. Reversely, a \soft{} agent would typically score a high agreement rate, but a low utility-under-agreement.

Let us make this precise. Let $\ind_{agr(d,\tx{i},\tx{j},r)}$ be the `indicator function' that has value 1 if the $r$-th repetition of scenario $(\prot, \dom_d, \ag_{\tx{i}}, \ag_{\tx{j}})$ ended with agreement, and 0 otherwise.
\[
\ind_{agr(d,\tx{i},\tx{j},r)} := 
\begin{cases}
1 & \text{If\ the\ $r$-th\ repetition\ of\ scenario\ } (\prot, \dom_d, \ag_{\tx{i}}, \ag_j) \\
 & \text{ended\ with\ agreement}. \\
0 & \text{If\ the\ $r$-th\ repetition\ of\ scenario\ } (\prot, \dom_d, \ag_{\tx{i}}, \ag_{\tx{j}}) \\
 & \text{ended\ without\ agreement.} \\
\end{cases}
\]
Furthermore, let $\numSucc^{d,\tx{i},\tx{j}}$ represent the number of times a negotiation in the scenario $(\prot, \dom_d, \ag_{\tx{i}}, \ag_{\tx{j}})$ ended with agreement:
\[\numSucc^{d,\tx{i},\tx{j}} \ \ := \ \ \sum_{r=1}^{\numReps^{d,\tx{i},\tx{j}}} \ind_{agr(d,\tx{i},\tx{j},r)} \]

Then we can calculate the agreement rate of agent $\ag_{\tx{i}}$, for a given opponent and a given domain as:
\[\ar_{\tx{i}}^{d,\tx{j}} \ \ := \ \ \frac{1}{2}\cdot \frac{\numSucc^{d,\tx{i},\tx{j}}}{\numReps^{d,\tx{i},\tx{j}}} + \frac{1}{2}\cdot \frac{\numSucc^{d,\tx{j},\tx{i}}}{\numReps^{d,\tx{j},\tx{i}}} \]
and the agreement rate of agent $\ag_{\tx{i}}$ for the entire tournament is obtained by averaging this quantity over all domains and opponents:
\[\ar_{\tx{i}} \ \ := \ \ \frac{1}{|\Dom| \cdot |\Ag|} \sum_{d=1}^{|\Dom|} \sum_{\tx{j}=1}^{|\Ag|} \ar_{\tx{i}}^{d,\tx{j}}\]
This quantity will of course always be a number between 0 and 1, but it is custom to present it as a percentage, so 0.65 would be presented as 65\%.

Similarly, we can define the average \textit{utility-under-agreement} $\uua_{\tx{i}}$ using the following two equations:
\[\uua_{\tx{i}}^{d,\tx{j}} \ \ := \ \ \frac{1}{2\cdot \numSucc^{d,\tx{i},\tx{j}}} \sum_{r=1}^{\numReps^{d,\tx{i},\tx{j}}} \util_{1}^{d,\tx{i},\tx{j},r}\ind_{agr(d,\tx{i},\tx{j},r)} + \frac{1}{2\cdot \numSucc^{d,\tx{j},\tx{i}}} \sum_{r=1}^{\numReps^{d,\tx{j},\tx{i}}} \util_{1}^{d,\tx{j},\tx{i},r}\ind_{agr(d,\tx{i},\tx{j},r)}\]

\[\uua_{\tx{i}} \ \ := \ \ \frac{1}{|\Dom| \cdot |\Ag|} \sum_{d=1}^{|\Dom|} \sum_{\tx{j}=1}^{|\Ag|} \uua_{\tx{i}}^{d,\tx{j}}\]

Now, we can present the results of our experiment including these values, such as in Table~\ref{tab:tournament_2}. Note that this table is much more informative than Table~\ref{tab:tournament_1}. While the first table only showed us that our agent `IlPadrino' was second-best, after MegaBarter3000, we can now also see \textit{why} it performed worse than MegaBarter3000. To see this, let us analyze each agent one by one.

%That is, we see that our agent scores the lowest utility-under-agreement of all agents, but at the same time it scores the highest agreement rate. This clearly indicates that our agent is too \soft{}. On the other hand, CrazyAgent suffers from the opposite problem. It has a very low agreement rate, but a very high utility-under-agreement, indicating that it is a very \hard{} agent. Furthermore, we see that MegaBarter3000 scores neither the highest nor the lowest utility-under-agreement, and the same holds for its agreement rate. However, it does score the highest overall tournament score. This indicates it strikes the best balance between being \hard{} and being \soft{} among the four agents. Finally, we note that RandomAgent scores low on both quantities. This means that it suffers from more fundamental problems rather than just being too \hard{} or too \soft{}. One possible explanation is that it mainly proposes offers that are really bad for both himself and his opponent at the same time.

\begin{itemize}
\item \textbf{IlPadrino}: We see that IlPadrino scores the lowest utility-under-agreement of all agents, but at the same time it scores the highest agreement rate. This clearly indicates that it is too \soft{}.
\item \textbf{CrazyAgent}: On the other hand, CrazyAgent suffers from the opposite problem. It has a very low agreement rate, but a very high utility-under-agreement, indicating that it is a very \hard{} agent.
\item \textbf{MegaBarter3000}: We see that MegaBarter3000 scores neither the highest nor the lowest utility-under-agreement, and the same holds for its agreement rate. However, it does score the highest overall tournament score, which is the score that actually matters. Therefore, we can say that it is the best of the four agents, because it strikes the best balance between being \hard{} and being \soft{}.
\item \textbf{RandomAgent}: Finally, we note that RandomAgent scores low on all quantities. This means that it suffers from more fundamental problems rather than just being too \hard{} or too \soft{}. One possible explanation is that it mainly proposes offers that are really bad for both himself and his opponent at the same time.
\end{itemize}

\begin{table}
\begin{center}
\begin{tabular}{|l|c|c|c|}
\hline
 & \textbf{Tournament} & \textbf{Utility-under-} & \textbf{Agreement} \\
 \textbf{Agent}			   & \textbf{Score} ($\ts_{\tx{i}}$) & \textbf{Agreement} ($\uua_{\tx{i}}$) & \textbf{Rate} ($\ar_{\tx{i}}$) \\
\hline
MegaBarter3000 & 0.729 & 0.81 & 90\% \\
\hline
IlPadrino &  0.665 & 0.70 & 95\% \\
\hline
CrazyAgent & 0.604 & 0.93 & 65\% \\
\hline
RandomAgent & 0.488 & 0.75 & 65\% \\
\hline
\end{tabular}
\caption{Results of a fictional tournament between four fictional agents. This time presented together with the Utility-under-Agreement and the Agreement Rate of each agent.}\label{tab:tournament_2}
\end{center}
\end{table}

\begin{exercise}\label{ex:calculate_other_scores}
\textbf{Calculate Agreement Rates and Utility-under-Agreement.}
Modify your code of Exercise \ref{ex:calculate_tournament_scores} so that it also calculates the $\uua_{\tx{i}}$ and $\ar_{\tx{i}}$ of every agent, and so that it displays a table such as  Table~\ref{tab:tournament_2}.
\end{exercise}

\subsection{Abuse of $\uua_{\tx{i}}$ and $\ar_{\tx{i}}$}\label{sec:abuse}
We should stress the fact that the utility-under-agreement $\uua_{\tx{i}}$ and agreement rate $\ar_{\tx{i}}$ should only be used for \textit{diagnostic} purposes. That is, they should only be used to answer the question \textit{why} a particular agent did or did not perform well. These quantities, however, should \textit{not} be used to determine \textit{whether} or not it performed well. Instead, the tournament score is the only of the three measures that should be used for that.
 
In fact, it is a common mistake among beginning researchers to argue that they implemented a strong agent based only on the observation that it scored a high utility-under-agreement in their experiments. It is, however, easy to see that a high value of $\uua_{\tx{i}}$ by itself does not say anything about the quality of the agent, because implementing an agent that scores high $\uua_{\tx{i}}$ is entirely trivial. For example, we can simply implement an agent that only proposes or accepts any offers with utility value $\util_1(\off) \geq 0.99$ (assuming the maximum utility is $1.0$) and that rejects any other offers. Clearly, when such an agent comes to an agreement, it will always be one with a very high utility for himself, and this agent would be guaranteed to achieve a $\uua_{\tx{i}}$ of at least $0.99$. But that is completely useless because this agent would probably almost never make any agreements at all, and therefore would likely end up with a very low average utility overall.

For the same reason, we cannot use the agreement rate as a performance measure for our agent. After all, it is also trivial to implement an agent that simply always accepts every proposal, and which is therefore guaranteed to achieve an agreement rate of 100\%.

\begin{observation}
The agreement rate and the utility-under-agreement should never be used as measures of performance of an agent. They should only be used as an aid to understand \underline{why} the agent performed well or not.
\end{observation}

\subsection{The Importance of Self-Play}\label{sec:self_play}
One aspect that is often overlooked in experiments, is the importance of agents negotiating not only against other agents, but also against \textit{themselves}. Many experiments described in the literature do not involve such self-play (that is, they set $\numReps^{d,\tx{i},\tx{j}} = 0$ whenever $\tx{i}=\tx{j}$). However, we argue that it is in fact extremely important that any new negotiation strategy is not only tested against a number of benchmark agents, but also against itself.

The reasoning behind this is as follows. Imagine a world in which many people use negotiation algorithms to do their negotiations for them. Obviously, those people would want to use the best such algorithms available. Now, suppose that we have invented a new algorithm that is very strong against most other existing algorithms, but that performs poorly when pitted against itself. Then initially, when we just start using our new algorithm, it may obtain very good results. However, sooner or later other people will  also discover our algorithm and will start using it as well. But then it also becomes more and more likely that our agent will encounter opponents that use exactly the same algorithm, or a very similar one. And since our algorithm performs poorly against such opponents, eventually, its results will deteriorate.

\begin{observation}
The stronger a negotiation algorithm performs, the more likely it is that, in a real-world setting, it would encounter itself as its opponent. Therefore, any `strong' algorithm must necessarily also be strong against itself.
\end{observation}

A good example of a strategy that may perform well against other agents, but not against itself, is a very \hard{} time-based agent.

Another important reason for self-play, is that it is a requirement for a so-called \textit{empirical game-theoretical analysis}, which we will discuss later on.

\subsection{The Importance of Benchmark vs. Benchmark Negotiations}

Another mistake that beginning researchers sometimes tend to make, is that, when testing a new agent, they only run negotiations between their new agent and some existing benchmark agents, without running any negotiations between the benchmark agents themselves.
We will here argue that it is extremely important that you also include the results of mutual negotiations among the benchmark agents in your data.

The reason for this is the same as the reason why we argued against the use of the utility-under-agreement as a performance measure, in Section \ref{sec:abuse}.
To see this, suppose again that we implement a very \hard{} agent that only proposes or accepts any offers with utility value $\util_1(\off) \geq 0.90$ (assuming the maximum utility is $1.0$) and that rejects any other offers.  Clearly, whenever this agent makes an agreement, it will be one with very high utility for itself, and therefore most likely with very low utility for its opponent. However, it is then also very likely that the negotiations will fail very often, so the tournament score of our agent would actually be very low. 

For example, say that when our agent $\ag_{\tx{1}}$ negotiates against $\ag_{\tx{2}}$, then on average, if there is an agreement, $\ag_{\tx{1}}$ scores 0.92 and $\ag_{\tx{2}}$ scores $0.2$. However, they only come to an agreement in 10\% of all negotiations, so their overall average utilities would be 0.092 and 0.02 (assuming they both have a reservation value of 0). Similarly,  when our agent $\ag_{\tx{1}}$ negotiates against $\ag_3$, then on average, if there is an agreement, $\ag_{\tx{1}}$ scores 0.94 and $\ag_{\tx{2}}$ scores $0.22$. However, they only come to an agreement in 10\% of all negotiations, so their overall average utilities would be 0.094 and 0.022.

Now, if we stop here, we might falsely conclude that $\ag_{\tx{1}}$ is the best among the three agents, since it clearly scored higher utilities than the other two. However, if we also let the other two agents negotiate against each other, they might reach an agreement in 100\% of the negotiations, and they might on average achieve utility values of 0.8 and 0.78 respectively. If we now calculate the tournament scores of each agent, we get the results as displayed in Table~\ref{tab:bench_vs_bench}. We see that our agent scores, by far, the lowest utility. Now, if you look at this table, ask yourself: which agent would you choose to do your negotiations?

\begin{table}
\begin{center}
\begin{tabular}{|l|c|c|c|}
\hline
 & $\ts_{\tx{i}}$ & $\uua_{\tx{i}}$ & $\ar_{\tx{i}}$ \\
 \hline
$\ag_{\tx{2}}$ & 0.410 & 0.75 & 55\% \\
\hline
$\ag_{\tx{3}}$ & 0.401 & 0.73 & 55\% \\
\hline
$\ag_{\tx{1}}$ & 0.093 & 0.93 & 10\% \\
\hline
\end{tabular}
\caption{Results of a fictional tournament in which $\ag_{\tx{1}}$ is a very \hard{} agent.}\label{tab:bench_vs_bench}
\end{center}
\end{table}

%Note that the mistakes we discussed previously in Sections \ref{sec:abuse} and \ref{sec:self_play}, as well as the one we discussed here, in the end all come down to the same important observation: you can easily implement an agent that strongly exploits its opponents, but that does not make it a ``strong'' agent in general. 

%This is for the simple reason that if you want to compare the average utility of your agent with the average utility obtained by some other agent, then these two agents must have negotiated against the same opponents.

%\subsection{Choosing the Benchmark Agents}
%\later{Write some paragraphs about how to select benchmark agents and domains.}

\section{Empirical Game-Theoretical Analysis}\label{sec:egta}
In the previous section we discussed how we can compare agents experimentally, through the use of a `tournament evaluation'. While this is by far the most commonly used method, it suffers from one important problem. Namely, that it is very sensitive to the presence of weak agents. We therefore here present an alternative evaluation method that avoids this problem.

To explain this, imagine a tournament between an extremely \hard{} agent, a very \soft{} agent, and an intermediate agent. In this tournament the \hard{} agent might obtain a very high score because it is able to fully exploit the \soft{} agent, while the \soft{} agent will achieve a very low score.  This may make it \textit{seem} that the \hard{} agent is very strong. However, one could argue that this outcome is unrealistic, because in a real-world situation it would unlikely to encounter a very poorly performing \soft{} agent, and without the presence of the \soft{} agent the \hard{} agent would achieve a lot less utility.

This problem can be partially mitigated by organizing the tournament over multiple rounds. That is, after we have obtained the results, we remove the worst performing agents, and run another tournament, but this time with only the top performing agents from the first round.

However, there exists a more principled approach to this, known as `\textit{empirical game-theoretical analysis}' (EGTA) \cite{Wellman2025EGTAsurvey}. In the following subsections we will first explain this through an example, then  we will define it formally, and finally we will compare the advantages and disadvantages of EGTA with those of tournament evaluation.

\subsection{Example}

Suppose we have again the same four agents as before: \textit{MegaBarter3000}, \textit{IlPadrino}, \textit{CrazyAgent} and \textit{RandomAgent}. Furthermore, assume that we have run a tournament in which all agents negotiate against each other (including against themselves), as in the previous section, and that we have stored the results of all the negotiations in this tournament in a database.

Now, imagine that there are two people, called Alice and Bob, that have to negotiate about something, but they will not do these negotiations themselves, but instead they each choose one of those four agents to do the negotiations on their behalf. Note that it is perfectly possible that they each choose the same agent, so in that case there would be two copies of the same agent negotiating against each other. 

Now, the question is, which agents should Alice and Bob choose, assuming they have access to the full database with all the results of the tournament?

Let's assume that, initially, Alice is a bit naive and simply chooses the agent that achieved the highest tournament score in our tournament. Let's say that that is \textit{MegaBarter3000}. Bob, however, is not so naive. He also sees that \textit{MegaBarter3000} scored the highest tournament score, but when he looks a bit closer to the data, he realizes that \textit{MegaBarter3000} mainly achieved high utility against  \textit{CrazyAgent} and \textit{RandomAgent}, which each achieved a very low tournament score. On the other hand, in the negotiations between  \textit{MegaBarter3000} and \textit{IlPadrino}, it was actually \textit{IlPadrino} that performed much better. Therefore, assuming that Alice chooses \textit{MegaBarter3000}, Bob is smart and chooses \textit{IlPadrino}. If they now let these two agents negotiate each other, then they will likely achieve a deal that is much better for Bob than for Alice.

Luckily for Alice, however, she also realizes this, just in time before they start their negotiations. So, knowing that Bob will choose \textit{IlPadrino}, she now carefully examines the data to determine which of the four agents performs best against \textit{IlPadrino}. She learns from this that the agent that scores the highest utility against \textit{IlPadrino}, is \textit{IlPadrino} itself. Therefore, she also chooses \textit{IlPadrino}. So, in the end, Alice and Bob both end up choosing \textit{IlPadrino}.

We learn from this example that the agent that scores the highest average utility in the tournament is not necessarily the best choice.

You may have noticed that Alice and Bob were following the same pattern of reasoning as what we discussed in Section \ref{sec:pure_nash_equilibria}. That is, Alice and Bob are essentially playing a normal-form game, in which their `actions' are the 4 respective agents, and their utility functions are given by the scores that the agents obtained in the tournament. In the example, \textit{IlPadrino} was a best response against \textit{MegaBarter3000}, and \textit{IlPadrino} was also a best response against itself. Therefore, the action profile (\textit{IlPadrino}, \textit{IlPadrino}) was a pure Nash equilibrium of the game. So, the essence of an empirical game-theoretical evaluation, is to model the choice of agents as a normal-form game and then determine its Nash equilibria.

\subsection{Formal Definition}
We will now formally define the procedure described in the example above.

Suppose we have a finite set of agents $\Ag = \{\ag_{\tx{1}}, \ag_{\tx{2}}, \dots \ag_{\tx{\numAgents }}\}$, and some finite set of domains $\Dom$, so we have $|\Dom|\times|\Ag|\times|\Ag|$ possible scenarios. Furthermore, suppose that each scenario $(\prot, \dom_d, \ag_{\tx{i}}, \ag_{\tx{j}})$ is repeated a certain number of times, denoted $\numReps^{d,\tx{i},\tx{j}}$.

Then, after running the tournament, we calculate for each pair of agents $(\ag_{\tx{i}}, \ag_{\tx{j}})$ the average utility $\ts_{\tx{i}}^{\tx{j}}$ achieved by agent $\ag_{\tx{i}}$ against agent $\ag_{\tx{j}}$, and the average utility $\ts_{\tx{j}}^{\tx{i}}$ achieved by agent $\ag_{\tx{j}}$ against agent $\ag_{\tx{i}}$.

\begin{eqnarray}
\ts_{\tx{i}}^{\tx{j}}  & := & \frac{1}{|\Dom|} \sum_{d=1}^{|\Dom|} \ts_{\tx{i}}^{d,\tx{j}}\\
 & = & \frac{1}{|\Dom|}\sum_{d=1}^{|\Dom|} \Big( \frac{1}{2\numReps^{d,\tx{i},\tx{j}}} \sum_{r=1}^{\numReps^{d,\tx{i},\tx{j}}} \util_{1}^{d,\tx{i},\tx{j},r} + \frac{1}{2\numReps^{d,\tx{j},\tx{i}}} \sum_{r=1}^{\numReps^{d,\tx{j},\tx{i}}} \util_{2}^{d,\tx{j},\tx{i},r} \Big)\label{eq:_agent_agent_score}
\end{eqnarray}

This yields $|\Ag|\times|\Ag|$ numbers which can be organized in a square matrix, where the cell in row $i$ and column $j$ contains the number $\ts_{\tx{i}}^{\tx{j}}$. This can be seen as the payoff matrix of a symmetric normal-form game (recall from Section \ref{sec:symmetric_equilibria} that for symmetric games the payoff matrix only needs to contain one number in each cell, which is the utility of the row-player).

An example of such a payoff matrix is presented in Table \ref{tab:egta}. Each number represents the average utility obtained by the agent in the row-header, averaged over all scenarios in which it negotiated against the agent indicated in the column-header. For example, we see that \textit{IlPadrino} obtained an average utility of 0.91 over all the sessions in which it negotiated against \textit{MegaBarter300}. On the other hand, the average utility that \textit{MegaBarter300} obtained in those same negotiations (against \textit{IlPadrino}), was only 0.23. 

We then need to determine the pure symmetric Nash equilibria of this game (recall that in Section \ref{sec:symmetric_equilibria} we argued that for symmetric games we are only interested in \textit{symmetric} equilibria), and among those equilibria we then determine which one achieves the highest utility (since we are talking about \textit{symmetric} equilibria, both players achieve the same utility). Note that the fact that this equilibrium is pure, means that in this equilibrium each player selects exactly one agent. Furthermore, the fact that it is symmetric, means that both players choose the same agent. Therefore, a pure symmetric Nash equilibrium consists of exactly one agent. So, according to the empirical game-theoretical analysis, the best agent is the one that forms the pure symmetric Nash equilibrium with highest utility.

In order to find, for any given agent, the best response against that agent, we need to look at the column corresponding to that agent and then find the highest utility in that column. For example, in Table \ref{tab:egta}, we see that in the column representing \textit{MegaBarter3000}, the highest utility is 0.91, which is achieved by \textit{IlPadrino}. Therefore, \textit{IlPadrino} is the best response against \textit{MegaBarter3000}. Similarly, to find the best response against \textit{CrazyAgent}, we see that the highest utility in that column is 0.80, which is obtained by \textit{MegaBarter3000}, so therefore we conclude that \textit{MegaBarter3000} is the best response against \textit{CrazyAgent}. To make this easier to see the highest value in each column is highlighted in bold.

To find the pure symmetric Nash equilibria, we just have to see for which agents the best response falls on the diagonal of the matrix. In Table \ref{tab:egta} we see that this is the case for IlPadrino: the highest value in that column is 0.82, which is achieved by IlPadrino itself. Therefore, the profile (\textit{IlPadrino}, \textit{IlPadrino}) forms a pure symmetric Nash equilibrium. 

%Unfortunately, however, it may also happen that there is no pure symmetric Nash equilibrium at all (which means there must be a \textit{mixed} symmetric Nash equilibrium), or there may be multiple pure symmetric equilibria that each obtain exactly the same utility. In those cases,  empirical game theoretical evaluation does not return a unique optimal agent.

\begin{table}
\small
\begin{tabular}{l|c|c|c|c}
\ & MegaBarter3000 & IlPadrino & CrazyAgent & RandomAgent\\
\hline
MegaBarter3000 & 0.53 & 0.23 & \textbf{0.80} & \textbf{0.90} \\
\hline
IlPadrino & \textbf{0.91} & \textbf{0.82} & 0.20 & 0.10 \\
\hline
CrazyAgent & 0.40 & 0.23 & 0.30 & 0.35 \\
\hline
RandomAgent & 0.30 & 0.25 & 0.27 & 0.28 \\
\end{tabular}
\normalsize
\caption{Each cell contains the average score obtained by the agent in the row-header, when negotiating against the agent in the column-header. For example, IlPadrino scores an average utility of 0.91 when negotiating against MegaBarter3000. In each column, the highest value is highlighted in bold, indicating the best response against the agent in the column header.}\label{tab:egta}
\end{table}

Unfortunately, however, empirical game-theoretical evaluation does have a number of disadvantages. Namely:
\begin{enumerate}
\item It allows us only to find the `best' agent, but it does not allow us to determine a further ranking among the agents. That is, it doesn't allow us to tell which agent is the second best or the third best.
\item The optimal symmetric Nash equilibrium may not be unique.
\item The optimal pure symmetric Nash equilibrium may actually be dominated by \textit{mixed} symmetric Nash equilibrium.
\item EGTA requires a lot more data, compared to tournament evaluation, to get statistically significant results.
\item It is based on the assumption that the players are fully rational and have full knowledge of the payoff matrix which, one can argue, is not entirely realistic.
\end{enumerate}

The first of these points is not a big issue if we are only interested in proving that our agent is the best. However, if, for example, we want to run a competition that awards a prize for the second and third-best agents, then EGTA cannot help us with that. Also, if our agent is not the best, then EGTA cannot really tell us how great the difference between our agent and the best agent is.

The second issue happens when we have two agents that each score exactly the same utility when negotiating against themselves. This is not a very big problem, because the chance of that happening is not very big, and moreover, the same problem can occur in a tournament evaluation just as well. In fact, this can happen in literally any evaluation method, because it is always possible that two agents are simply equally strong.

The third and fourth issues are more problematic ones. 

Suppose that (\textit{IlPadrino}, \textit{IlPadrino}) is the optimal \textit{pure} symmetric Nash equilibrium, but that there also exists a \textit{mixed} symmetric Nash equilibrium, for which the two players achieve even higher utilities. For example, a strategy profile in which both players select \textit{MegaBarter3000} with 60\% probability and \textit{CrazyAgent} with 40\% probability. It is then questionable to say that \textit{IlPadrino} is the game-theoretically best agent. After all, each player would be better off selecting the mixed strategy. But then there is no longer one unique agent that can be considered the best. 

Ideally, we should therefore also check if there are any such \textit{mixed} symmetric Nash equilibria and, if yes, check that they do not dominate the optimal pure symmetric equilibrium. This can be done using software libraries such as Gambit \cite{gambit},  but unfortunately this gets computationally very costly if there are many agents.

If it happens that the game does have one or more \textit{mixed} symmetric Nash equilibria that dominate the pure ones, then we cannot really conclude anything from the evaluation. In this case, the EGTA does not consider any single agent to be the best, and instead the optimal strategy for Alice and Bob would be to flip a coin to choose between the agents that make up the optimal mixed symmetric Nash equilibrium.

The fourth issue is caused by the fact that for a tournament evaluation we only need to determine $|\Ag|$ different numbers (one tournament score $\ts_{\tx{i}}$ for each agent), while an EGTA requires $|\Ag| \times |\Ag|$ different numbers (one $\ts_{\tx{i}}^{\tx{j}}$ for every pair of agents $(\ag_{\tx{i}}, \ag_{\tx{j}})$) and for each of these numbers the standard error needs to be sufficiently low. This means that we need to run a lot more negotiations for EGTA than for tournament evaluation, which can be very time-consuming.

%We need to accurately determine $|\Ag| \times |\Ag|$ different numbers, while a tournament evaluation only requires $|\Ag|$ different numbers. 

Finally, the fifth issue is more of theoretical problem rather than a practical one. The problem here, is that if humans are not fully rational, or they do not have full knowledge of the payoff matrix, then they may not make optimal choices, and it may therefore be that the agent that forms a symmetric Nash equilibrium is actually a sub-optimal choice if our opponent is irrational. For example, if Bob is irrational and chooses \textit{CrazyAgent}, then Alice would actually be better of with her original choice \textit{MegaBarter3000} than with \textit{IlPadrino}. 

\begin{exercise}\label{ex:perform_EGTA}
\textbf{Perform EGTA.}
Implement a program that, similarly to Exercise \ref{ex:calculate_tournament_scores}, does the following:
\begin{enumerate}
\item Read the text file from Exercise~\ref{ex:run_tournament}.
\item Based on the contents of that file, calculate the values $\ts_{\tx{i}}^{\tx{j}}$ for every pair of agents $(\ag_{\tx{i}}, \ag_{\tx{j}})$.
\item Display them in a table such as Table~\ref{tab:tournament_4}.
\item Determines and outputs the pure symmetric Nash equilibrium with highest utility values.
\end{enumerate}
\end{exercise}

\subsection{EGTA vs. Tournament Evaluation}
In a certain sense, game-theoretical evaluation and  tournament evaluation can be seen as two extreme ends of a spectrum. While game-theoretical evaluation is based on the assumption that all people are fully rational and have full knowledge of the experimental results, tournament evaluation is based on the opposite assumption that other people are completely irrational, or have absolutely no knowledge at all about the payoff matrix. 

To see this, assume that Alice indeed does not apply any form of rationality. That is, she picks her agent completely at random. That means that for each of the four agents there is a 25\% chance that Alice will pick that agent. Therefore, the optimal choice for Bob is to pick the agent that maximizes his expected utility against Alice's `uniform' mixed strategy. For example, if Bob picks an agent $\ag_{\tx{i}}$, then his expected utility would be:
\[E(\util_{\tx{i}}) \quad = \quad \frac{1}{4}\ts_{\tx{i}}^{\tx{1}} + \frac{1}{4}\ts_{\tx{i}}^{\tx{2}}   + \frac{1}{4}\ts_{\tx{i}}^{\tx{3}}  + \frac{1}{4}\ts_{\tx{i}}^{\tx{4}}  \]
but it is easy to see that this is exactly the tournament score $\ts_{\tx{i}}$ of agent $\ag_{\tx{i}}$. In other words, Bob should pick the agent with the highest tournament score. So, we see that tournament evaluation indeed selects the agent that is optimal under the assumption that our opponent is not rational at all, or has no knowledge whatsoever about the payoff matrix. 

\begin{opinion}
In my opinion, neither of the two methods is based on entirely realistic assumptions. On the one hand, people are often not perfectly rational and in a real-world situation they would probably not have perfect knowledge about how well every possible agent performs against every possible other agent.

On the other hand, however, I also don't think it's realistic to assume people would have absolutely no knowledge at all about how well each agent performs. Typically, if one agent clearly under-performs compared to other agents, then sooner or later people would stop using that agent. This is demonstrated by the fact that for many real-world applications their popularity follows a `\textit{power law distribution}', which means that just a few apps are significantly more popular than others. For example, if you search the Internet for data about the most popular messaging apps, dating apps, web browsers, or any other type of application, then in almost all cases you will find a histogram that follows such a pattern. So, people clearly do not select their applications purely at random.

In other words, neither tournament evaluation, nor empirical game-theoretical evaluation gives us the definitive answer to the question which agent is the `best'.  I therefore recommend to always perform \textit{both} evaluations. In the ideal case both methods yield the same result. If they give different results, however, then the answer to the question which agent is the best just depends on which set of assumptions you would consider more realistic in a specific domain of application.
\end{opinion}

\section{Sequential Elimination Ranking}
Previously, we have discussed the two most common methods to evaluate negotiation algorithms. We have argued that neither of the two are clearly the best, and that they are respectively based on two strictly opposing sets of assumptions, neither of which is entirely realistic.

Therefore, we would here like to propose a third, alternative method that lies somewhere in between tournament evaluation and EGTA. We will call it `Sequential Elimination Ranking'.  It should be noted that, to the best of our knowledge, this method has never actually been applied in the literature on automated negotiation. However, similar methods have been studied and applied extensively in other research topics, such as \textit{computational social choice}.
\later{Cite some work from social choice theory}

This method is based on the idea is that in a real-world situation people would be less likely to choose weaker agents. So, the final score of an agent should be weighted more heavily by the utility it obtained against the stronger opponents than by the utility it obtained against weaker opponents.

\later{Change the notation? Instead of $\ts^n$ use something else.}

As usual, we assume we have some set of negotiation domains $\Dom$, and some set of agents $\Ag$ of size $\numAgents$ (i.e. $\numAgents := |\Ag|$). However, for reasons that will become clearly imminently, we will here denote the set of agents as $\Ag_\numAgents$ instead of $\Ag$. Furthermore, for any $i \in \{1, 2, \dots, \numAgents\}$ we will use the notation $\ag_{[i]}$ to denote the agent that ranks in $i$-th place according to the Sequential Elimination Ranking.

Sequential Elimination Ranking works as follows.
\begin{enumerate}
\item First, we run a tournament, just like for tournament evaluation, in which all agents $\ag \in \Ag_n$ negotiate against each other (including against themselves), over all domains $\dom \in \Dom$.
\item We then calculate, for each pair of agents $(\ag_{\tx{i}}, \ag_{\tx{j}})$ the average utility obtained by agent $\ag_{\tx{i}}$ against agent $\ag_{\tx{j}}$, which we denote as $\ts(\ag_{\tx{i}}, \ag_{\tx{j}})$. That is:
\[\ts(\ag_{\tx{i}}, \ag_{\tx{j}})  =  \frac{1}{|\Dom|}\sum_{d=1}^{|\Dom|} \Big( \frac{1}{2\numReps^{d,\tx{i},\tx{j}}} \sum_{r=1}^{\numReps^{d,\tx{i},\tx{j}}} \util_{1}^{d,\tx{i},\tx{j},r} + \frac{1}{2\numReps^{d,\tx{j},\tx{i}}} \sum_{r=1}^{\numReps^{d,\tx{j},\tx{i}}} \util_{2}^{d,\tx{j},\tx{i},r} \Big)\]
Note that this is the same as Equation~(\ref{eq:_agent_agent_score}), except that we here use the notation $\ts(\ag_{\tx{i}}, \ag_{\tx{j}})$ instead of $\ts_{\tx{i}}^{\tx{j}}$.
\item Next, we can use these values to calculate for each agent $\ag \in \Ag_{\numAgents}$ its tournament score, which we here denote as $\ts^\numAgents(\ag)$:
\[\ts^\numAgents(\ag)\ \ := \ \ \frac{1}{\numAgents} \sum_{\ag' \in \Ag_\numAgents}\ts(\ag, \ag')\]
\item We then determine the agent with the lowest tournament score, and we set its \textbf{rank} equal to $\numAgents$. In other words, this agent ends the tournament in last place. We denote this agent by $\ag_{[\numAgents]}$.
\begin{eqnarray*}
\ag_{[\numAgents]} & := & \argmin \{\ts^\numAgents(\ag)\mid \ag \in \Ag_\numAgents \}\\
\end{eqnarray*}
\item We now regard $\ag_{[\numAgents]}$ as `eliminated' from the tournament, and we define the set $Ag_{\numAgents-1}$ as the set of all remaining agents:
\[\Ag_{\numAgents-1}  \ \ :=  \ \ \Ag_\numAgents \setminus \{\ag_{[\numAgents]}\}\]
\item Now, we recalculate the scores of all the remaining agents $\ag \in \Ag_{\numAgents-1}$, by calculating the average of all the utilities they obtained against the other remaining agents (i.e. we are no longer counting the utilities they obtained against the eliminated agent). We will denote this new score as  $\ts^{\numAgents-1}(\ag)$:
\[\ts^{\numAgents-1}(\ag) \ \ := \ \ \frac{1}{\numAgents-1} \sum_{\ag' \in \Ag_{\numAgents-1}}\ts(\ag, \ag')\]
\item We now again determine the agent with the lowest score (among the remaining agents), which we will denote as $\ag_{[\numAgents-1]}$. This agent ends the tournament in second-last place.
\begin{eqnarray*}
\ag_{[\numAgents-1]} & := & \argmin \{\ts^{\numAgents-1}(\ag)\mid \ag \in \Ag_{\numAgents-1} \}\\
\end{eqnarray*}
\item This agent can now also be considered eliminated and we remove it from the set of remaining agents.
\[\Ag_{\numAgents-2} \ \ := \ \ \Ag_\numAgents \setminus \{\ag_{[\numAgents]}, \ag_{[\numAgents-1]}\}\]
\item We keep doing this over and over until only one agent is left. That is, in each iteration we calculate the scores of all remaining agents, by averaging only over the utilities obtained against those same remaining agents themselves and then determine the agent with the lowest score, which will be next to be eliminated.
\item Finally, the winner of the tournament is $\ag_{[1]}$, the second-best agent is $\ag_{[2]}$, etcetera.
\end{enumerate}

Formally, for any $i \in \{1,2, \dots, \numAgents\}$
the agent $\ag_{[i]}$ is defined by the following recursive set of equations:
\begin{eqnarray*}
\ag_{[i]} & := & \argmin \{\ts^i(\ag)\mid \ag \in \Ag_{i} \}\\
\ts^{i}(\ag) & := & \frac{1}{i}\sum_{\ag' \in \Ag_{i}}\ts(\ag, \ag') \\
\Ag_{i} & := & \Ag \setminus \{\ag_{[i+1]}, \ag_{[i+2]}, \dots, \ag_{[\numAgents]}\}\\
\end{eqnarray*}

The motivation for this method is as follows. Imagine a society in which many people use negotiation algorithms to do their negotiations for them. Initially there are, say, 10 different algorithms available, and initially each of them is equally popular. However, as time passes and people start using them more and more often, it quickly becomes obvious that some of these algorithms are stronger than others. Therefore, people will quickly abandon the weakest ones. As those weaker algorithms start disappearing, this will influence the results obtained by the other algorithms. That is, some algorithms that are very good at exploiting the weaker ones and that initially obtained good results, may now slowly start performing worse, because of the increasing lack of weaker opponents to exploit. So, again, as time passes, the weaker algorithms among those that are left will now also start declining in popularity. This process will continue until slowly but surely only one algorithm is left at the end, which can therefore be considered the `best'. 

Note that, on the one hand, this system does not assume that people simply choose their agents randomly, but on the other hand it also does not assume that all people have perfect knowledge of the full payoff matrix. Instead, regarding knowledge, it merely assumes that people know, at any moment, which agent is at that moment the weakest one available. Furthermore, regarding to rationality, it only assumes that nobody wants to use the worst agent available. So, indeed, we see that the assumptions underlying this method lie somewhere in between the assumptions underlying tournament evaluation and the assumptions underlying EGTA.

Moreover, this system is very robust against the presence of unrealistically weak agents. For example, if we add one extremely weak agent to the set of agents, then that agent will probably be eliminated in the first iteration and after that will no longer have any influence on the other agents' scores. Thus, the rankings of the other agents remain unaffected.

%Or alternatively, the popularity of the algorithms stabilizes into a mixture that will likely take the shape of a `power-law distribution', in which some algorithms are much more popular than others

\begin{exercise}\label{ex:sequential_elimination}
\textbf{Sequential Elimination Ranking.}
Implement a program that does the following:
\begin{enumerate}
\item Read the text file from Exercise~\ref{ex:run_tournament}.
\item Based on the contents of that file, calculate the value of $\ts(\ag_{\tx{i}}, \ag_{\tx{j}})$ for every pair of agents $(\ag_{\tx{i}}, \ag_{\tx{j}})$.
\item Calculate for each agent its rank according to the sequential elimination ranking, as explained in this section.
\item Display the results in a table.
\end{enumerate}
\end{exercise}

\section{Statistics}\label{sec:statistics}
In the previous sections of this chapter we have discussed various methods to compare agents. Each of these methods involved calculating the average utility of each agent over a some set of negotiation scenarios.

However, just like in any other experimental science, we have to keep in mind that such averages may be heavily affected by statistical noise, especially if the agents are, to some extent, non-deterministic. This means that if we repeat the entire tournament a second time, then the agents may obtain different  scores than in the first tournament,  which might  lead to to  different conclusions. Obviously, this is highly undesirable in a scientific study.

Luckily, we can avoid this problem, by making sure that the agents' utilities are averaged over a very large number of negotiations. The more negotiations we perform, the less likely it will be that a repetition of the experiment will lead to different conclusions. In this section we will therefore discuss how to ensure that we have run enough negotiations to be confident that our conclusions are reliable.

Before we continue, we should first note that the utility values achieved by the agents are actually influenced by \textit{two} different types of statistical noise.
\begin{enumerate}
\item \textbf{Scenario Noise:} Noise due to the variation of opponents and domains. That is, our agent may perform very well against one specific opponent, or on one specific domain, but may perform poorly against another opponent or on another domain.
\item \textbf{Randomization Noise:} Purely random noise due to the non-determinism of one or both of the agents. That is, even if we repeat exactly the same scenario, our agent may still achieve different utility values in each negotiation if some of its actions, or some of its opponent's actions, are randomized.
\end{enumerate}

In the rest of this section we will present several mathematical equations. However, we will not go into a   full derivation of those equations, because that would require a much more in-depth study of the topic of statistics, which is a vast field in itself. For more details we therefore refer to the many text books that have been written on this topic, such as \cite{diez2012openintro} and \cite{freedman2007statistics}.  Furthermore, we will restrict our discussion purely to the statistical analysis of \textit{tournament evaluation} and ignore the two other evaluation methods we discussed.

\subsection{Random Variables}
In statistics, any process that returns random outcomes is called a `\textit{random variable}'. A typical example of a random variable is a standard six-sided die, which can return any integer between 1 and 6, each with an equal probability of $\frac{1}{6}$. Every time this process returns a new outcome (e.g. every time we throw a die), we say we \textbf{draw an observation} from that variable. 

Formally, we can define a \textbf{finite real-valued random variable} $\mathcal{X}$ as a  pair $(\varDomain_{\rvX}, P_{\rvX})$ where $\varDomain_{\rvX}$ is some finite set of real numbers $\varDomain_{\rvX} = \{x_1, x_2, \dots, x_K\} \subset \mathbb{R}$, and $P_{\rvX}$ is a probability distribution over this set. That is, $P_{\rvX} : \varDomain_{\rvX} \rightarrow \mathbb{R}$ such that:
\[\forall x\in \varDomain_{\rvX} : \ \ P_{\rvX}(x) \geq 0 \quad \text{and} \quad \sum_{x\in \varDomain_{\rvX}} P_{\rvX}(x) = 1\]
Then, the \textbf{mean} of ${\rvX}$, denoted $\mu_{\rvX}$, is  defined as:
\begin{equation}\label{eq:mean}
\mu_{\rvX} \ \ := \ \ \sum_{x\in \varDomain_{\rvX}} P_{\rvX}(x) \cdot x
\end{equation}
the \textbf{variance} of ${\rvX}$, denoted $\mathit{Var}_{\rvX}$, is defined as:
\begin{equation}\label{eq:variance}
\mathit{Var}_{\rvX}  \ \ := \ \  \sum_{x\in \varDomain_{\rvX}} P_{\rvX}(x) \cdot (\mu_{\rvX} - x)^2
\end{equation}
and the \textbf{standard deviation} of ${\rvX}$, denoted $\sigma_{\rvX}$, is defined as:
\begin{equation}\label{eq:std_err}
\sigma_{\rvX}  \ \ := \ \ \sqrt{\mathit{Var}_{\rvX}}
\end{equation}

In the context of a negotiation tournament, for any negotiation over some negotiation scenario $(\prot, \dom_d, \ag_{\tx{i}}, \ag_{\tx{j}})$ we can see the utility  $\util_1^{d,\tx{i},\tx{j},r}$ obtained by agent $\ag_{\tx{i}}$, as an observation from a random variable. Similarly, the utility $\util_2^{d,\tx{i},\tx{j},r}$ obtained by the other agent $\ag_{\tx{j}}$ can also be seen as an observation drawn from another random variable. Therefore, any negotiation scenario is associated with two random variables, which we will denote as $\rvX_1^{d,\tx{i},\tx{j}}$ and $\rvX_2^{d,\tx{i},\tx{j}}$.

In general, whenever we have two real-valued random variables, $\rvX_1$ and $\rvX_2$, then we can define a new random variable, denoted as $\rvX_1 + \rvX_2$, which corresponds to the process that we draw one observation from $\rvX_1$ and one observation from $\rvX_2$, and then return the sum of the two observations. 
For example, if $\rvX_1$ and $\rvX_2$ each represent an ordinary 6-sided die, then an observation of the variable $\rvX_1 + \rvX_2$ represents the process of throwing two dice at the same time, and recording the sum of the two numbers shown by the two dice. This means that an observation of the variable $\rvX_1 + \rvX_2$ can be any integer between 2 and 12, so we have $S_{\rvX_{1} + \rvX_2} =  \{2,3,4, \dots, 11, 12\}$

Similarly, if $\rvX$ is a real-valued random variable and $c$ is any arbitrary real number, then we can define a new random variable $c \rvX$. An observation from $c \rvX$ is defined as an observation from $\rvX$, multiplied by $c$. For example, if $c=2$ and $\rvX$ represents a 6-sided die, then $c \rvX$ can be any \textit{even} number between 2 and 12, so we have $S_{2\rvX} =  \{2,4,6,8,10,12\}$

Note, therefore, that $\rvX + \rvX$ is \textit{not} the same as  $2 \rvX$, because in the first case we are drawing \textit{two} observations from $\rvX$, which may be different from each other, while in the second case we just draw one observation from $\rvX$ which is then multiplied by 2. For example, if $\rvX$ is a 6-sided die, then in the first case we could throw a 3 and then a 6, so our observation would be the number $3+6=9$, while in the second case we can only observe even numbers.

Of course, these operations can also be combined to create `linear combinations' of random variables, such as $a\rvX_1 + b\rvX_2$ or $\frac{1}{R}\sum_{r=1}^R{\rvX_r}$. It can be shown that, for any two \textit{independent} variables $\rvX_1$ and $\rvX_2$ the variance of the linear combination $a \rvX_1 + b \rvX_2$ satisfies:
\begin{equation}\label{eq:variance_of_sum}
\mathit{Var}_{a \rvX_1 + b \rvX_2} \ \ = \ \ a^2 \mathit{Var}_{\rvX_1} + b^2 \mathit{Var}_{\rvX_2}
\end{equation}
%However, if they are not independent, we have to use the more general equation:
%\begin{equation}
%\mathit{Var}_{a \rvX_1 + b \rvX_2} \ \ = \ \ a^2 \mathit{Var}_{\rvX_1} + b^2 \mathit{Var}_{\rvX_2} + 2ab \cdot Cov(\rvX_1, \rvX_2)
%\end{equation}
%where $Cov(\rvX_1, \rvX_2)$ is the covariance between $\rvX_1$ and $\rvX_2$:
%\begin{equation}
%Cov(\rvX_1, \rvX_2) \ \ = \ \ \sum_{x\in S_{\rvX_1}}\sum_{y\in S_{\rvX_2}}P(x,y) \cdot (\mu_1 - x)\cdot (\mu_2 - y)
%\end{equation}

\subsection{Standard Error}\label{sec:standard_error}

Whenever we estimate the mean of a random variable, we typically also want to know \textit{how accurately} we have estimated this mean. A commonly used quantity to measure the accuracy of an estimated mean, is the \textit{standard error}, which we will discuss in this section.

In the following subsections we will derive the equation to calculate the standard error on the tournament score of a negotiating agent. To explain it properly, we will have to do this step by step. First, in Section \ref{sec:std_err_single_scen} we will discuss how we can calculate the standard error over just one single scenario. Next, in Section \ref{sec:std_err_multiple_scen} we will show how to generalize this to a \textit{fixed} set of multiple scenarios. This generalization only takes into account randomization noise, but not scenario noise. Finally, in Section \ref{sec:std_err_multiple_variable_scen}, we will generalize this further to include possibility that we may use a different set of negotiation scenarios every time we repeat the experiment. This means that our expression for the standard error will also include scenario noise. Our final expression for the standard error will be given by the square root of Equation (\ref{eq:standard_error_final}).

%\later{
%We will first discuss randomization noise. That is, in Section \ref{sec:std_err_single_scen} we will assume we are measuring the performance of our agent only on one specific scenario. Next, in Section \ref{sec:std_err_multiple_scen} we will show how to generalize this to a \textit{fixed} set of multiple scenarios (meaning we are still only including randomization noise). And then in Section \ref{sec:std_err_multiple_variable_scen} we will generalize this further to include possibility that we may use a different set of scenarios every time we repeat the experiment, hence including scenario noise.
%}

\subsubsection{Standard Error for a Single Scenario}\label{sec:std_err_single_scen}

Suppose that have a random variable $\rvX$, but we do not know its probability distribution $P_{\rvX}$ and we also don't know its mean, variance or standard deviation. In that case, we can \textit{estimate} these quantities, by drawing a sequence of observations $(\obs_1, \obs_2, \dots, \obs_\numReps)$ from $\rvX$, where each $o_i$ is an element of $\varDomain_{\rvX}$ that was randomly selected (with replacement) with probability $P_{\rvX}(\obs_i)$. Such a sequence of observations is called a \textbf{sample} and we will denote its size by $\numReps$. We can then calculate an \textit{estimated} mean $\est{\mu}_{\rvX}$, \textit{estimated} variance $\est{\mathit{Var}}_{\rvX}$ and \textit{estimated} standard deviation $\est{\sigma}_{\rvX}$ of $\rvX$ as follows:
\begin{eqnarray}
\est{\mu}_{\rvX} &=& \frac{1}{\numReps} \sum_{i=1}^\numReps o_i \label{eq:estimate_mu}\\
\est{\mathit{Var}}_{\rvX} &=& \frac{1}{\numReps-1} \sum_{i=1}^\numReps (\est{\mu}_{\rvX} - o_i )^2\label{eq:estimate_var}\\
\est{\sigma}_{\rvX} &=& \sqrt{\est{\mathit{Var}}_{\rvX}} \ \ = \ \ \frac{\sqrt{\sum_{i=1}^\numReps (\est{\mu}_{\rvX} - o_i )^2}}{\sqrt{\numReps-1}}\label{eq:estimate_sigma}
\end{eqnarray}
For a linear combination of random variables, the equation for its estimated variance becomes:
\begin{equation}\label{eq:variance_of_sum_estimated}
\est{\mathit{Var}}_{a \rvX_1 + b \rvX_2} \ \ = \ \ a^2 \est{\mathit{Var}}_{\rvX_1} + b^2 \est{\mathit{Var}}_{\rvX_2}
\end{equation}

The estimated variance $\est{\mathit{Var}}_{\rvX}$ and standard deviation $\est{\sigma}_{\rvX}$ are also known as the \textbf{sample variance} and \textbf{sample standard deviation}. On the other hand, the \textit{actual} variance $\mathit{Var}_{\rvX}$ and standard deviation $\sigma_{\rvX}$ (given by Equations (\ref{eq:variance}) and (\ref{eq:std_err})) are known as the \textbf{population variance} and \textbf{population standard deviation}. 

Note that the expression for the sample variance has the number $\numReps - 1$ in the denominator instead of $\numReps$. This is a correction for the fact that the expression does not use the true mean $\mu_{\rvX}$, but the estimated mean $\est{\mu}_{\rvX}$. For a mathematical derivation of this fact we refer to any standard text book on statistics.

Of course, the estimated mean $\est{\mu}_{\rvX}$ is only an approximation of the real value of $\mu_{\rvX}$ and will typically not be exactly equal to $\mu_{\rvX}$. In fact, every time we draw a new sample from $\rvX$ and re-calculate $\est{\mu}_{\rvX}$ we will likely obtain a different value.
Now, a key insight, is that the value $\est{\mu}_{\rvX}$ that we obtain from these calculations  can itself  also be seen as an observation from a random variable. Let us denote this random variable by $\rvY$. That is:
\[\rvY \ \ := \ \ \frac{1}{\numReps}\sum_{r=1}^\numReps {\rvX}\]
So, a \textit{single} observation from $\rvY$ is, by definition, obtained from drawing $\numReps$ observations from $\rvX$ and then calculating their average.

The point of this, is that one can mathematically prove that the mean of $\rvY$ equals the mean of $\rvX$ (i.e. $\mu_{\rvX} = \mu_{\rvY}$), but the standard deviation of $\rvY$ is smaller than that of $\rvX$. In particular: $\sigma_{\rvY} = \frac{\sigma_{\rvX}}{\sqrt{\numReps}}$. The random variable $\rvY$ is therefore said to be an \textit{estimator} for $\mu_{\rvX}$. 

Note that the larger we choose the sample size $\numReps$, the smaller $\sigma_{\rvY}$, and therefore the more likely it is that $\est{\mu}_{\rvX}$ will be close to $\mu_{\rvX}$. The standard deviation of $\rvY$ can therefore be seen as a measure of accuracy of our estimation of $\mu_{\rvX}$ and is known as the \textbf{standard error} $\se_{\mu_{\rvX}}$ on our estimation of $\mu_{\rvX}$:
\[\se_{\mu_{\rvX}} \quad := \quad \sigma_{\rvY} \quad = \quad \frac{\sigma_{\rvX}}{\sqrt{\numReps}}\]

%In general, if we draw $\numReps$ observations from a random variable $\rvX$, which we'll denote as $(x_1, x_2, \dots, x_\numReps)$, then we can calculate its estimated mean $\est{\mu}_{\rvX}$ as:
%\begin{equation}\label{eq:estimate_mu}
%\est{\mu}_{\rvX} = \frac{1}{\numReps}\sum_{r=1}^{\numReps} x_i
%\end{equation}
%and its estimated standard deviation $\est{\sigma}_X$ as:
%\begin{equation}\label{eq:estimate_sigma}
%\est{\sigma}_X = \frac{\sqrt{(\mu_{\rvX} - x_1)^2 + (\mu_{\rvX} - x_2)^2 + \dots + (\mu_{\rvX} - x_\numReps)^2}}{\sqrt{\numReps-1}}
%\end{equation}
%
% Note: this is the *sample* standard deviation as opposed to the 
% *population* standard deviation.
%

Roughly speaking, the interpretation of the standard error is that  there is a 95\% probability that the true mean $\mu_{\rvX}$ lies somewhere between ${\est{\mu}_{\rvX} - 2\cdot \se_{\mu_\rvX}}$ and $\est{\mu}_{\rvX} + 2\cdot \se_{\mu_{\rvX}}$. For example, if $\est{\mu}_{\rvX} = 10$ and $\se_{\mu_{\rvX}} = 1$  then there is a 95\% probability that $\mu_{\rvX}$ lies in the interval $[8\ , 12]$.

In practice, however, we often can't directly calculate $\se_{\mu_{\rvX}}$ because we do not know the exact value of $\sigma_\rvX$. Instead, we can only calculate the \textit{estimated} standard error for $\mu_{\rvX}$, denoted $\est{\se}_{\mu_{\rvX}}$, by using the estimated standard deviation $\est{\sigma}_{\rvX}$ as defined by Equation~(\ref{eq:estimate_sigma}):
\[\est{\se}_{\mu_{\rvX}} \ \ := \ \ \frac{\est{\sigma}_{\rvX}}{\sqrt{\numReps}}\]

It is important to understand that, for any given random variable $\rvX$, its standard deviation $\sigma_{\rvX}$ is fixed. It is a property of that random variable, so we can \textit{estimate} it, but it is not something that we can \textit{change}. On the other hand, the standard error on $\mu_{\rvX}$ is not a property of $\rvX$, but rather, it is a property of our experiment. It is a measure of how accurately we have estimated the mean of $\rvX$, which depends on how many observations we have drawn from $\rvX$. The larger the number of observations $\numReps$, the smaller the standard error. This means we can make the standard error as small as we like, as long as we draw enough observations from $\rvX$.

Another very important thing to remember is not to confuse the standard \textit{error} on $\mu_{\rvX}$  with the standard \textit{deviation} on $\rvX$. \textit{\underline{This is a very common mistake}}. Many beginning researchers erroneously report their estimated mean $\est{\mu}_{\rvX}$ of $\rvX$ together with the standard deviation $\sigma_{\rvX}$ as a measure of accuracy. However, this is incorrect because, as mentioned above, the standard deviation is a fixed property of $\rvX$ and therefore does not say anything about the accuracy of the experiment. Instead, we should report the standard \textit{error}, which is the standard deviation $\sigma_{\rvY}$ of $\rvY$, and which decreases with the number of observations $\numReps$.

\subsubsection{Standard Error for Multiple Scenarios}\label{sec:std_err_multiple_scen}
As mentioned above, every negotiation scenario $(\prot, \dom_d, \ag_{\tx{i}}, \ag_{\tx{j}})$ is associated with two random variables $\rvX_1^{d,\tx{i},\tx{j}}$ and $\rvX_2^{d,\tx{i},\tx{j}}$. However, whenever we run a tournament we are not just interested in one single negotiation scenario, but rather we want to calculate the average utility of our agent over many different scenarios. So, if our agent is $\ag_{\tx{1}}$, then we are interested in all random variables of the form $\rvX_1^{d,\tx{1},\tx{j}}$ or $\rvX_2^{d,\tx{j},\tx{1}}$ and, instead of estimating the mean of one such variable, we aim to estimate the \textit{average of the means} of \textit{all} those variables.

Specifically, if there are $\numDomains$ domains and $\numAgents$ agents (including our own agent) then the calculation of our agent's tournament score would  involve a total $2 \numDomains \numAgents$ of these variables. To see this, note that there are $\numAgents-1$ opponents, and for each domain and each opponent there are 2 possible scenarios (one in which our agent has utility function $\util_1$ and one in which it has utility function $\util_2$), yielding $2\numDomains (\numAgents-1)$ possible scenarios involving our agent. Furthermore, on top of that there are also $\numDomains$ scenarios (one for each domain) in which our agent negotiates against itself. In such scenarios there are two outcomes for our agent (one for the copy of our agent that had utility function $\util_1$ and one for the copy that had utility function $\util_2$). Yielding a total of $2 \numDomains \cdot (\numAgents-1) + 2\numDomains = 2\numDomains\numAgents $ random variables for the entire tournament.

For example, if there are 2 domains and 3 agents, and we want to calculate the tournament score of agent $\ag_{\tx{1}}$ then this involves the following 12 random variables:
\[\rvX_{1}^{1,\tx{1},\tx{1}}, \rvX_{2}^{1,\tx{1},\tx{1}}, \rvX_{1}^{1,\tx{1},\tx{2}}, \rvX_{1}^{1,\tx{1},\tx{3}}, \rvX_{2}^{1,\tx{2},\tx{1}}, \rvX_{2}^{1,\tx{3},\tx{1}}\]
%%%%%%%%%%%%%%%%%%%%%%%%%
\[\rvX_{1}^{2,\tx{1},\tx{1}}, \rvX_{2}^{2,\tx{1},\tx{1}}, \rvX_{1}^{2,\tx{1},\tx{2}}, \rvX_{1}^{2,\tx{1},\tx{3}}, \rvX_{2}^{2,\tx{2},\tx{1}}, \rvX_{2}^{2,\tx{3},\tx{1}}\]
Note that these are indeed all the variables of the form $\rvX_1^{d,\tx{1},\tx{j}}$ or $\rvX_2^{d,\tx{j},\tx{1}}$, for any domain $\dom_d$ and opponent $\ag_{\tx{j}}
$, and that they all return utility values for agent $\ag_{\tx{1}}$.

In the rest of this section we will assume we are calculating the tournament score of some arbitrary agent $\ag_{\tx{i}}$, and to simplify the notation 
we will just denote its corresponding variables as:
\[\rvX_{\tx{i},1}, \quad \rvX_{\tx{i},2},  \quad \dots, \quad \rvX_{\tx{i},\numVars}\]
where $\numVars = 2 \numDomains\numAgents$. Each random variable will, in general, have a different mean and a different standard deviation, so we have $\numVars$ different unknown means $\mu_{\tx{i},1}, \mu_{\tx{i},2}, \dots, \mu_{\tx{i},\numVars}$ and $\numVars$ different unknown standard deviations $\sigma_{\tx{i},1}, \sigma_{\tx{i},2}, \dots, \sigma_{\tx{i},\numVars}$. 

Now, when we run an experiment and we calculate the tournament score of agent $\ag_{\tx{i}}$, we first calculate for each scenario the average utility the agent achieved on that scenario, and then we calculate the average over all scenarios. This can be modeled by means of the following two random variables:
\begin{equation}\label{eq:Y_i_s}
\rvY_{\tx{i},s} \ \ := \ \ \frac{1}{\numReps_s}\sum_{r=1}^{\numReps_s} \rvX_{\tx{i},s}
\end{equation}
\begin{equation}
\rvZ_{\tx{i}} \ \ := \ \ \frac{1}{\numVars} \sum_{s=1}^{\numVars} \rvY_{\tx{i},s}
\end{equation}
Here, an observation of $\rvY_{\tx{i},s}$ represents the average utility obtained by agent $\ag_{\tx{i}}$ over all repetitions in the scenario corresponding to variable $\rvX_{\tx{i},s}$, and a single observation from $\rvZ_{\tx{i}}$ represents the average utility obtained by $\ag_{\tx{i}}$ in the entire tournament. In other words, the tournament score $\ts_{\tx{i}}$ of agent $\ag_{\tx{i}}$ is, by definition, a single observation from the variable $\rvZ_{\tx{i}}$. 

Note that the means of $\rvX_{\tx{i},s}$ and $\rvY_{\tx{i},s}$ are equal:
\[\mu_{\rvX_{\tx{i},s}} = \mu_{\rvY_{\tx{i},s}}\]
we will therefore use the notation $\mu_{\tx{i},s}$ as a shorthand for $\mu_{\rvX_{\tx{i},s}}$ and $\mu_{\rvY_{\tx{i},s}}$. Furthermore, the mean $\mu_{\rvZ_{\tx{i}}}$ of $\rvZ_{\tx{i}}$ is given by:
\begin{equation}\label{eq:mu_Z_i}
\mu_{\rvZ_{\tx{i}}} = \frac{1}{\numVars}\sum_{s=1}^\numVars \mu_{\tx{i},s}
\end{equation}
Of course, every time we repeat the experiment, we may obtain a  different tournament score for $\ag_{\tx{i}}$, but the higher the number of repetitions $\numReps_s$ for each scenario, the closer the observed tournament scores will be to the mean of $\rvZ_{\tx{i}}$, which can therefore be seen as an measure of the true strength of the agent.

\begin{observation} The mean of $\rvZ_{\tx{i}}$ represents the \underline{true} strength of agent $\ag_{\tx{i}}$, while the tournament score $\ts_{\tx{i}}$ is only a noisy approximation of the agent's strength.
\end{observation} 

This means that the (estimated) standard deviation of $\rvZ_{\tx{i}}$ can be used as a measure of how accurately we have estimated the true strength of $\ag_{\tx{i}}$. This is, therefore, the (estimated) standard error of our experiment.
\[\est{\se}_{\mu_{\rvZ_{\tx{i}}}} \quad := \quad \est{\sigma}_{\rvZ_{\tx{i}}} \quad = \quad \sqrt{\est{\mi{Var}}_{\rvZ_{\tx{i}}}}\]

So, to calculate the standard error we need to estimate $\mi{Var_{\rvZ_{\tx{i}}}}$. To do this, we first estimate the variance of each individual variable $\rvX_{\tx{i},s}$, using Eq.~(\ref{eq:estimate_var}). Then, using Eq.~(\ref{eq:variance_of_sum_estimated}) and Eq.~(\ref{eq:Y_i_s}) we find that the estimated variance of $\rvY_{\tx{i},s}$ is given by:
\begin{eqnarray*}
\est{\mi{Var}}_{\rvY_{\tx{i},s}} & = & \frac{1}{\numReps_s} \cdot \est{\mi{Var}}_{\rvX_{\tx{i},s}}\\
& = & \frac{1}{\numReps_s \cdot (\numReps_s-1)}\sum_{r=1}^{\numReps_s} (\est{\mu}_{\tx{i},s} - \util_{\tx{i},s,r} )^2
\end{eqnarray*}

% $\sigma_{\rvX_{\tx{i},s}}$, as in Eq.~(\ref{eq:estimate_sigma}) and then for each $Y_{\tx{i},s}$ it standard deviation is given by $ \sigma_{Y_{\tx{i},s}} = \frac{\sigma_{\rvX_{\tx{i},s}}}{\sqrt{\numReps_s}}$
%
%
%for each $\rvX_s$, the estimated standard error $\est{\se}_s$ on the mean $\mu_s$ as $\est{\se}_s = \frac{\est{\sigma}_s}{\sqrt{\numReps_s}}$, 
%where $\numReps_s$ is the number of observations from $\rvX_s$. However, in practice it is often easier to first calculate the square of the estimated standard error, as $\est{\se}_s^2 = \frac{1}{\numReps_s}\est{\mi{Var}}_s$. Thus, we have:
%\[\est{\se}_s^2 \quad = \quad \frac{1}{\numReps_s \cdot (\numReps_s-1)}\sum_{r=1}^{\numReps_s} (\est{\mu}_s - \obs_{s,r} )^2\]
where each $\util_{\tx{i},s,r}$ is the $r$-th observation from variable $\rvX_{\tx{i},s}$, and $\est{\mu}_{\tx{i},s}$ is calculated as:
\begin{equation}\label{eq:mu_s}
\est{\mu}_{\tx{i},s} = \frac{1}{\numReps_s}\sum_{r=1}^{\numReps_s}\util_{\tx{i},s,r}
\end{equation}
Note that we are now using the symbol $\util$ for observations from $\rvX_{\tx{i},s}$, to stress the fact that they are indeed utility values obtained by our agent in the respective negotiations it is participating in.

%\[\est{\se}_s = \frac{1}{\sqrt{\numReps^s}} \cdot \frac{1}{\sqrt{\numReps^s-1}}\sqrt{\sum_{i=1}^\numReps (\est{\mu}_{\rvX} - o_i )^2}\]

Similarly, we see that, if all random variables are mutually independent, the estimated variance of $\rvZ_{\tx{i}}$ can then be calculated as:
\begin{eqnarray}
\label{eq:x}
\est{\mi{Var}}_{\rvZ_{\tx{i}}}  &=& \frac{1}{\numVars^2}\sum_{s=1}^\numVars \est{\mi{Var}}_{\rvY_{\tx{i},s}}\\
\label{eq:partial_standard_error}
&=& \frac{1}{\numVars^2}\sum_{s=1}^\numVars  \frac{1}{\numReps_s \cdot (\numReps_s-1)}\sum_{r=1}^{\numReps_s} (\est{\mu}_{\tx{i},s} - \util_{\tx{i},s,r} )^2
\end{eqnarray}
Unfortunately, however, the assumption that all variables $\rvX_k^{d,\tx{i}, \tx{j}}$ are mutually independent does not hold in our context. Actually, it does hold for most variables, but not for those that correspond to the self-play scenarios. Specifically, for any domain $\dom_d$ and any agent $\ag_{\tx{i}}$, the variables $\rvX_1^{d,\tx{i}, \tx{i}}$ and $\rvX_2^{d,\tx{i}, \tx{i}}$ are not independent from each other, because their values are drawn from the same negotiation scenario $(\prot, \dom_d, \ag_{\tx{i}}, \ag_{\tx{i}})$. To correct for this, we need to add, for every domain $\dom_d$, a term involving the \textit{covariance} between these two variables. This yields the following expression:
\begin{multline}\label{eq:std_err_with_dependence}
\est{\mi{Var}}_{\rvZ_{\tx{i}}} = \frac{1}{\numVars^2}\Big(\sum_{s=1}^{\numVars}  \frac{1}{\numReps_{s} \cdot (\numReps_{s}-1)}\sum_{r=1}^{\numReps_{s}} (\est{\mu}_{\tx{i},s} - \util_{\tx{i},s,r} )^2 \quad +\\ 
\frac{2}{\numReps^{d,\tx{i},\tx{i}} \cdot (\numReps^{d,\tx{i},\tx{i}}-1)}\sum_{d=1}^{|\Dom|} \sum_{r=1}^{R^{d,\tx{i},\tx{i}}}(\est{\mu}_{1}^{d,\tx{i},\tx{i}} - \util_1^{d,\tx{i},\tx{i},r})(\est{\mu}_{2}^{d,\tx{i},\tx{i}} - \util_2^{d,\tx{i},\tx{i},r})\Big)
\end{multline}
where $\numReps^{d,\tx{i},\tx{i}}$ denotes the number of repetitions of the self-play scenario $(\prot, \dom_d, \ag_{\tx{i}},\ag_{\tx{i}})$.

If we take the square root of this expression, then we obtain the estimated standard error $\est{\sigma}_{\rvZ_{\tx{i}}}$, which is a measure of how close the agent's tournament score $\ts_{\tx{i}}$ is to the true mean $\mu_{\rvZ_{\tx{i}}}$. If $\est{\sigma}_{\rvZ_{\tx{i}}}$ is very low, it means that if we repeat the entire tournament again, \textit{on the same set of scenarios}, then we will probably obtain a tournament score that is very close to the one we obtained the first time. 

However, there is still a problem, namely that our calculations above only take into account randomization noise, but not scenario noise. This means that even if $\est{\sigma}_{\rvZ_{\tx{i}}}$ is very low, our agent might still get an entirely different tournament score if we repeat the tournament \textit{on a different set of scenarios}. 

This is fine if we were don't care about other scenarios, and only care about exactly those scenarios that we used for our tournament. In reality, however, we are usually interested in the performance of our agent, \textit{in general}. Therefore, the set of domains that we have used for our experiments should be seen as just a randomly selected subset from the set of \textit{all} possible negotiation domains that we are interested in, and similarly, the set of opponents used in our experiments should only be considered as a randomly selected subset of all the possible opponents our agent might encounter in reality. This means we still need to adapt our calculation of the standard error to also include scenario noise.

\subsubsection{Standard Error with Scenario Noise}\label{sec:std_err_multiple_variable_scen}
In order to include scenario noise, we can model our calculation of the tournament score $\ts_{\tx{i}}$ as a two-step process. That is, we can imagine that there exists a very large set of possible scenarios, $\Scen$. This set may contain billions of possible scenarios, or may even be infinite,  so we cannot test our agent on all of them. Therefore, the first step of our experiment consists of randomly choosing some limited subset $\{\scen_1, \scen_2, \scen_3, \dots \}$ of scenarios from $\Scen$. 
Then, in the second step, we run a tournament over only that selected subset.

Now, recall that when we calculate the standard error, we aim to answer the question how likely it would be that we find the same tournament score for our agent if we repeated the \textit{entire} experiment. In this case, `repeating the experiment' means that we also repeat the first step in which we randomly selected the negotiation scenarios. So, the second repetition of our experiment would involve a different set of scenarios than the first repetition.

As before, the utility values obtained by $\ag_{\tx{i}}$ in these negotiations can be seen as observations drawn from corresponding random variables, and we will again denote these variables as $\rvX_{\tx{i},1}, \rvX_{\tx{i},2}, \dots, \rvX_{\tx{i},\numVars}$ with $\numVars= 2 \numDomains \numAgents$.

%If each scenario is repeated multiple times, this means we are drawing several observations from each of these variables. After the tournament has finished we can calculate our agent's tournament score by first calculating the estimated mean $\est{\mu}_s$ of each of these variables, and then calculating the average of all these estimated means.

Now, to calculate the standard error, we should not only take into account the variance of each individual variable $\rvX_{\tx{i},s}$, but also the variance \textit{among} the means of the various random variables $\rvX_{\tx{i},s}$. After all, if we were to repeat this entire experiment, we would get different outcomes, not only because for each scenario the negotiations on that scenario might yield different outcomes, but also because we are now using entirely different scenarios.

For now, let us ignore again the fact that some variables may be mutually dependent. It can then be shown mathematically that the correct formula for the variance of $\rvZ_{\tx{i}}$ is given by:
\begin{equation}\label{eq:ideal_standard_error}
\mi{Var}_{\rvZ_{\tx{i}}} \quad = \quad \frac{1}{\numVars^2}\sum_{s=1}^\numVars \mi{Var}_{\rvY_{\tx{i}}} + \frac{1}{\numVars^2} \sum_{s=1}^\numVars (\mu_{\rvZ_{\tx{i}}} - \mu_{\tx{i},s})^2
\end{equation}
%where $\mu_s$ is the mean of random variable $\rvX_{\tx{i},s}$, and $\mu_{Z_{\tx{i}}}$ is the mean of $Z_{\tx{i}}$
The first sum in this expression represents the randomization noise, while the second sum represents the scenario noise.

The problem with this equation, however, is that it requires knowledge of the \textit{exact} means $\mu_{\tx{i},s}$ and $\mu_{\rvZ_{\tx{i}}}$. Of course, in reality we don't know these numbers, because if we did know them, then we wouldn't need to perform our experiment in the first place. 

A naive idea to solve this, would be to simply replace the values of $\mu_{\tx{i},s}$ and $\mu_{\rvZ_{\tx{i}}}$ by their estimations $\est{\mu}_{\tx{i},s}$ and $\est{\mu}_{\rvZ_{\tx{i}}}$. While this idea is in principle correct, it turns out that if we do that, then we have to remove the first sum $\frac{1}{\numVars^2}\sum_{s=1}^\numVars \mi{Var}_{\rvY_{\tx{i}}}$ from the equation. As usual it would go too far to give a mathematical proof of this claim here, but the basic idea is that the agents' randomization noise is already represented in the fact that we are using \textit{estimations} $\est{\mu}_{\tx{i},s}$, rather than the true means $\mu_{\tx{i},s}$ themselves. To see this, note that if all agents were purely deterministic, then every $\est{\mu}_{\tx{i},s}$ would be exactly equal to $\mu_{\tx{i},s}$. Therefore, the difference between each $\est{\mu}_{\tx{i},s}$ and each $\mu_{\tx{i},s}$ indeed represents the non-determinism of the agents.

So, the correct mathematical equation for the \textit{estimated} variance of $\rvZ_{\tx{i}}$ (ignoring possible dependency between variables) is:
\begin{equation}\label{eq:full_standard_error}
\est{\mi{Var}}_{\rvZ_{\tx{i}}} \quad = \quad \frac{1}{\numVars \cdot (\numVars-1)}\cdot \sum_{s=1}^\numVars (\est{\mu}_{Z_{\tx{i}}} - \est{\mu}_{\tx{i},s})^2
\end{equation}
where, as before, each $\est{\mu}_{\tx{i},s}$ is given by Eq.~(\ref{eq:mu_s}), and $\est{\mu}_{\rvZ_{\tx{i}}}$ by:
\[\est{\mu}_{\rvZ_{\tx{i}}} = \frac{1}{\numVars}\sum_{s=1}^\numVars \est{\mu}_{\tx{i},s}\]

Finally, we still need to take into account that for each domain $\dom_d$, the means $\est{\mu}_1^{d,\tx{i},\tx{i}}$ and $\est{\mu}_2^{d,\tx{i},\tx{i}}$ are not independent. The final correct formula to calculate the standard error is therefore given by the square root of the following expression:
%\begin{eqnarray*}
%\est{se}_{tot}^2 & = & \frac{1}{\numVars \cdot (\numVars-1)}\cdot \sum_{s=1}^\numVars (\est{\mu}_{tot} - \est{\mu}_s)^2 \quad +\\
%&  &  \quad 2 \cdot \sum_{d=1}^{|\Dom|}(\est{\mu}_{tot} - \est{\mu}_1^{d,\tx{i},\tx{i}})(\est{\mu}_{tot} - \est{\mu}_2^{d,\tx{i},\tx{i}})
%\end{eqnarray*}
%
\begin{equation}\label{eq:standard_error_final}
\est{\mi{Var}}_{\rvZ_{\tx{i}}}  \ \ = \ \ \frac{1}{\numVars \cdot (\numVars-1)}\cdot \Big( \sum_{s=1}^\numVars (\est{\mu}_{\rvZ_{\tx{i}}} - \est{\mu}_{\tx{i},s})^2  +  2 \cdot \sum_{d=1}^{|\Dom|}(\est{\mu}_{\rvZ_{\tx{i}}} - \est{\mu}_1^{d,\tx{i},\tx{i}})(\est{\mu}_{\rvZ_{\tx{i}}} - \est{\mu}_2^{d,\tx{i},\tx{i}})\Big)
\end{equation}
\later{I have verified this equation experimentally, but I'm not sure why it's correct theoretically.}

As a final remark, we should mention that the above calculations were all based on the assumption that in the first step of the experiment we picked the scenarios \textit{randomly}. Of course, in reality  when we run an experiment, we often do not \textit{really} pick the scenarios randomly. Instead, we often just use some given set of agents and domains that happen to be available to us, or we \textit{manually} pick a subset of those agents and domains. Nevertheless, this does not change the fact that we should model this selection of scenarios \textit{as if} it was done randomly, because only in that way we can properly take into account that the results can be different if the agent encounters different domains or opponents. The fact that our choice is not truly random doesn't matter, as long as we make sure that the domains and agents we pick are \textit{representative} for the set of all agents and domains that we are interested in.

\subsubsection{Calculating the Standard Error -- an Example}

Now let us look at an example. Suppose we have a very small tournament with only $\numDomains = 2$ domains and $\numAgents = 3$ agents, so we have $\numDomains \times \numAgents^2 = 2\times 3\times 3 = 18$ scenarios, and suppose that each scenario is repeated three times, so there are $3 \times 18 = 54$ negotiations. The three agents are called \textit{IlPadrino}, \textit{MegaBarter3000}, and \textit{CrazyAgent}, which we may alternatively refer to as $\ag_{\tx{1}}$, $\ag_{\tx{2}}$, and $\ag_{\tx{3}}$, respectively. Our goal is to calculate the tournament score and  standard error of $\ag_{\tx{1}}$, a.k.a. \textit{IlPadrino}.

While there are 18 scenarios in this tournament, we are for now only interested in the tournament score of  \textit{IlPadrino}, so we are only interested in the scenarios involving that agent. There are 10 such scenarios and, as explained above, these correspond to $2\numDomains \numAgents = 2\times 2\times 3 = 12$ different `variables' that involve \textit{IlPadrino} (there are 2 scenarios in which \textit{both} roles are assumed by \textit{IlPadrino}, and 8 scenarios in which \textit{IlPadrino} negotiates against a different agent, so that makes $2\times 2 + 8\times 1 = 12$ random variables).

Now suppose the outcomes of our tournament are as given in Table~\ref{tab:tournament_3}. Since we are for now only interested in the 10 scenarios that involve \textit{IlPadrino} we have omitted the other scenarios from this table. Furthermore, since we only need the utilities obtained by \textit{IlPadrino}, we have highlighted those in boldface.
\begin{table}
\begin{center}
\begin{small}
\begin{tabular}{lllcc}
\hline
\textbf{Agent 1} & \textbf{Agent 2} & \textbf{Domain} & $\mathbf{\util_1}$ & $\mathbf{\util_2}$\\
\hline
IlPadrino & IlPadrino & CarSale & \textbf{0.724} & \textbf{0.600}\\
IlPadrino & IlPadrino & CarSale & \textbf{0.714} & \textbf{0.487}\\
IlPadrino & IlPadrino & CarSale & \textbf{0.791} & \textbf{0.541}\\
\hline
IlPadrino & MegaBarter3000 & CarSale & \textbf{0.539} & 0.485\\
IlPadrino & MegaBarter3000 & CarSale & \textbf{0.520} & 0.468\\
IlPadrino & MegaBarter3000 & CarSale & \textbf{0.521} & 0.498\\
\hline
IlPadrino & CrazyAgent & CarSale & \textbf{0.751} & 0.682\\
IlPadrino & CrazyAgent & CarSale & \textbf{0.787} & 0.697\\
IlPadrino & CrazyAgent & CarSale & \textbf{0.762} & 0.703\\
\hline
MegaBarter3000 & IlPadrino & CarSale & 0.586 & \textbf{0.746} \\
MegaBarter3000 & IlPadrino & CarSale & 0.671 & \textbf{0.693} \\
MegaBarter3000 & IlPadrino & CarSale & 0.646 & \textbf{0.631} \\
\hline
CrazyAgent & IlPadrino & CarSale & 0.770 & \textbf{0.422} \\
CrazyAgent & IlPadrino & CarSale & 0.776 & \textbf{0.426} \\
CrazyAgent & IlPadrino & CarSale & 0.704 & \textbf{0.407} \\
\hline
IlPadrino & IlPadrino & Cinema & \textbf{0.366} & \textbf{0.445}\\
IlPadrino & IlPadrino & Cinema & \textbf{0.342} & \textbf{0.432}\\
IlPadrino & IlPadrino & Cinema & \textbf{0.320} & \textbf{0.513}\\
\hline
IlPadrino & MegaBarter3000 & Cinema & \textbf{0.604} & 0.754\\
IlPadrino & MegaBarter3000 & Cinema & \textbf{0.608} & 0.733\\
IlPadrino & MegaBarter3000 & Cinema & \textbf{0.517} & 0.642\\
\hline
IlPadrino & CrazyAgent & Cinema & \textbf{0.532} & 0.444\\
IlPadrino & CrazyAgent & Cinema & \textbf{0.532} & 0.444\\
IlPadrino & CrazyAgent & Cinema & \textbf{0.532} & 0.444\\
\hline
MegaBarter3000 & IlPadrino & Cinema & 0.461 & \textbf{0.730} \\
MegaBarter3000 & IlPadrino & Cinema & 0.357 & \textbf{0.758} \\
MegaBarter3000 & IlPadrino & Cinema & 0.534 & \textbf{0.701} \\
\hline
CrazyAgent & IlPadrino & Cinema & 0.643 & \textbf{0.602} \\
CrazyAgent & IlPadrino & Cinema & 0.630 & \textbf{0.602} \\
CrazyAgent & IlPadrino & Cinema & 0.535 & \textbf{0.537} \\
\hline
\end{tabular}
\caption{Outcomes of all negotiations involving IlPadrino from a fictional tournament. The results are grouped by scenario, with 3 repetitions for each scenario. The utility values obtained by IlPadrino are indicated in bold.}\label{tab:tournament_3}
\end{small}
\end{center}
\end{table}

To calculate the tournament score and standard error of \textit{IlPadrino}, we first start by calculating the average utility of each instance of \textit{IlPadrino} in each scenario (i.e. the estimated mean of each random variable), according to:
\begin{equation}\label{eq:mu_scenario}
\est{\mu}_l^{d,\tx{i},\tx{j}}  \quad := \quad \frac{1}{\numReps^{d,\tx{i},\tx{j}}}\sum_{r=1}^{\numReps^{d,\tx{i},\tx{j}}} \util_l^{d,\tx{i},\tx{j},r}
\end{equation}
Using the numbers from Table \ref{tab:tournament_3} we get:
\begin{eqnarray*}
\est{\mu}_1^{1,\tx{1},\tx{1}} &=& \frac{1}{3}(0.724 + 0.714 + 0.791) = 0.743 \\
\est{\mu}_2^{1,\tx{1},\tx{1}} &=& \frac{1}{3}(0.600 + 0.487 + 0.541) = 0.543 \\
\est{\mu}_1^{1,\tx{1},\tx{2}} &=& \frac{1}{3}(0.539 + 0.520 + 0.521) = 0.527 \\
\est{\mu}_1^{1,\tx{1},\tx{3}} &=& \frac{1}{3}(0.751 + 0.787 + 0.762) = 0.767 \\
\est{\mu}_2^{1,\tx{2},\tx{1}} &=& \frac{1}{3}(0.746 + 0.693 + 0.631) = 0.690 \\
\est{\mu}_2^{1,\tx{3},\tx{1}} &=& \frac{1}{3}(0.422 + 0.426 + 0.407) = 0.418 \\
\est{\mu}_1^{2,\tx{1},\tx{1}} &=& \frac{1}{3}(0.366 + 0.342 + 0.320) = 0.343 \\
\est{\mu}_2^{2,\tx{1},\tx{1}} &=& \frac{1}{3}(0.445 + 0.432 + 0.513) = 0.463 \\
\est{\mu}_1^{2,\tx{1},\tx{2}} &=& \frac{1}{3}(0.604 + 0.608 + 0.517) = 0.576 \\
\est{\mu}_1^{2,\tx{1},\tx{3}} &=& \frac{1}{3}(0.532 + 0.532 + 0.532) = 0.532 \\
\est{\mu}_2^{2,\tx{2},\tx{1}} &=& \frac{1}{3}(0.730 + 0.758 + 0.701) = 0.730 \\
\est{\mu}_2^{2,\tx{3},\tx{1}} &=& \frac{1}{3}(0.602 + 0.602 + 0.537) = 0.580 \\
\end{eqnarray*}

Next, we calculate the tournament score of $\ag_{\tx{1}}$ as the total estimated mean:
\begin{eqnarray*}
\ts_{\tx{1}} = \est{\mu}_{\rvZ_{\tx{1}}} &=& \frac{1}{2 \numDomains  \numAgents}\sum_{d=1}^\numDomains \sum_{\tx{j}=1}^\numAgents \est{\mu}_1^{d,\tx{1},\tx{j}} + \est{\mu}_2^{d,\tx{j},\tx{1}}\\
%%%%%%%%%%%%%%%%%%%%%%%%
 &=& \frac{1}{2\cdot 2 \cdot 3}\sum_{d=1}^2 \sum_{\tx{j}=1}^3 \est{\mu}_1^{d,\tx{1},\tx{j}} + \est{\mu}_2^{d,\tx{j},\tx{1}}\\
 %%%%%%%%%%%%%%%%%%%%%%%%
&=& \frac{1}{12} \cdot  \Big(\est{\mu}_1^{1,\tx{1},\tx{1}} + \est{\mu}_2^{1,\tx{1},\tx{1}} + \est{\mu}_1^{1,\tx{1},\tx{2}} + \est{\mu}_1^{1,\tx{1},\tx{3}} + \est{\mu}_2^{1,\tx{2},\tx{1}} + \est{\mu}_2^{1,\tx{3},\tx{1}} \\
%%%%%%%%%%%%%%%%%%%%%%%%
& \ & \quad \quad \quad  + \est{\mu}_1^{2,\tx{1},\tx{1}} + \est{\mu}_2^{2,\tx{1},\tx{1}} + \est{\mu}_1^{2,\tx{1},\tx{2}} + \est{\mu}_1^{2,\tx{1},\tx{3}} + \est{\mu}_2^{2,\tx{2},\tx{1}} + \est{\mu}_2^{2,\tx{3},\tx{1}}\Big)\\
%%%%%%%%%%%%%%%%%%%%%%%%
 &=& \frac{1}{12} \cdot  \Big(0.743 + 0.543 + 0.527 + 0.767 + 0.690 + 0.418 \\
 %%%%%%%%%%%%%%%%%%%%%%%%
& \ & \quad \quad \quad  + 0.343 + 0.463 + 0.576 + 0.532 + 0.730 + 0.580\Big)\\
%%%%%%%%%%%%%%%%%%%%%%%%
&=& 0.576
\end{eqnarray*}

\normalsize

%Our goal is to calculate the standard error using Eq.~(\ref{eq:standard_error_final}). To do this, we first need to calculate the covariances of the self-play variables:
%\begin{eqnarray*}
%
%\end{eqnarray*}

Then, using Eq.~(\ref{eq:standard_error_final}) we get:
%\[\se_{tot}^2 = frac{1}{12^2}\cdot \Big()\]

\begin{eqnarray*}
\est{\mi{Var}}_{\rvZ_{\tx{1}}} &=& \frac{1}{12 \cdot (12-1)}\cdot \Big((0.576 - 0.743)^2 + (0.576 - 0.543)^2 + (0.576 -  0.527)^2 + \\
  & &  \quad \quad (0.576 - 0.767)^2 + (0.576 - 0.690)^2 + (0.576 - 0.418)^2 + \\
    & &  \quad \quad (0.576 - 0.343)^2 + (0.576 - 0.463)^2 + (0.576 -0.576 )^2 + \\
    & &  \quad \quad  (0.576 - 0.532)^2 + (0.576 - 0.730)^2 + (0.576 - 0.580 )^2 + \\
   & &  \quad \quad 2\cdot (0.576-0.743)\cdot (0.576-0.543) + \\ 
   & & \quad \quad 2\cdot (0.576-0.343)\cdot (0.576-0.463)
    \Big)\\
    &=& 1.817 \times 10^{-3}
\end{eqnarray*}
And from this it finally follows that the total estimated standard error of our agent is:
\[\est{\sigma}_{\rvZ_{\tx{1}}} \quad = \quad \sqrt{\est{\mi{Var}}_{\rvZ_{\tx{1}}}}\quad = \quad \sqrt{1.817 \times 10^{-3}}  \quad = \quad  0.0426\]

Now, we still need to repeat these calculations for every other agent as well, and then finally we can present the results of the entire tournament as in Table \ref{tab:tournament_4}. Note that we have multiplied the utility values and the standard errors by a factor of 1,000. This is purely for the purpose of readability, because 576 $\pm$ 42.6 is a bit easier to read than 0.576 $\pm$ 0.0426
\begin{table}
\begin{center}
\begin{tabular}{|l|c|c|c|}
\hline
 & \textbf{Tournament} & \textbf{Utility-under-} & \textbf{Agreement} \\
 \textbf{Agent}			   & \textbf{Score} ($\ts_{\tx{i}}$) & \textbf{Agreement} ($\uua_{\tx{i}}$) & \textbf{Rate} ($\ar_{\tx{i}}$) \\
\hline
MegaBarter3000 & 717 $\pm$ 30.0 & 815 & 88\% \\
\hline
IlPadrino &  576 $\pm$ 42.6 & 619 & 93\% \\
\hline
CrazyAgent & 511 $\pm$ 23.7 & 824 & 62\% \\
\hline
\end{tabular}
\caption{Results of a fictional tournament between three fictional agents. This time presented together with the standard errors. We have multiplied the utility values and standard errors by a factor of $1,000$ for the purpose of readability.}\label{tab:tournament_4}
\end{center}
\end{table}

\begin{exercise}\label{ex:standard_error}
\textbf{Standard Error} Modify your code of Exercise \ref{ex:calculate_tournament_scores} so that it also calculates, for each agent, the standard error on its tournament score, according to Equation~(\ref{eq:full_standard_error}) and then displays a table such Table \ref{tab:tournament_4}.
\end{exercise}

\subsubsection{Decreasing the Standard Error}
A common situation that we may encounter when doing experiments, is that although our agent's tournament score is higher than that of other agents, the difference is not large enough compared to the standard error to conclude that our agent is indeed truly better.

A naive solution would be to simply increase the number of repetitions, in order to achieve more accurate results. However, this can only decrease the standard error to a limited extent.  To see this, note that in  Equation~(\ref{eq:full_standard_error}) or Equation~(\ref{eq:standard_error_final}) this would allow the estimated means $\est{\mu}_{\tx{i},s}$ and $\est{\mu}_{\rvZ_{\tx{i}}}$ to get very close to the true means $\mu_{\tx{i},s}$ and $\mu_{\rvZ_{\tx{i}}}$, 
but this would still not reduce the total standard error to zero, because the $\est{\mu}_{\tx{i},s}$ would still be different from $\est{\mu}_{\rvZ_{\tx{i}}}$. Therefore, a much more effective way to reduce statistical noise, is to increase the number of negotiation scenarios, instead (note that this increases the value of $\numVars$ and therefore decreases the standard error).

%Even if it allows us to completely remove the noise from the agents' non-deterministic behavior (reflected by the sum $\sum_{s=1}^\numVars \se_s^2$), it cannot remove the noise due to the agents' different behavior on different domains (reflected by the sum $\sum_{s=1}^\numVars (\mu_{tot} - \mu_s)^2$). Similarly, 

%we can see from Equation~(\ref{eq:full_standard_error}) or Equation~(\ref{eq:standard_error_final}) that it would allow the estimated means $\est{\mu}_s$ to get very close to the true means $\mu_s$, but this would still not reduce the total standard error to zero. 

Of course, in some cases it may happen that we simply can't increase the number of scenarios, because we only have a limited number of domains and agents  available to us. In that case, increasing the number of repetitions is the only solution. If this does not decrease the standard errors enough to get statistically significant results, then, as an alternative, we can still use Equation~(\ref{eq:std_err_with_dependence}) instead. Using this equation is not \textit{wrong} per se, but rather, it changes the  \textit{interpretation} of our results.

Specifically, if the standard errors of the agents are very small according to Equation~(\ref{eq:std_err_with_dependence}), it means that if we repeat the experiment on \textit{exactly the same set of scenarios}, then the agents' tournament scores for the second repetition will likely not be very different from their tournament scores in the first repetition. On the other hand, if the standard errors are very small according to Equation~(\ref{eq:standard_error_final}), it means that the agents' tournament scores will likely not change much  \textit{even if} we repeat the experiment \textit{on an entirely different set of scenarios} (as long as in both experiments we selected a set of scenarios that is \textit{representative} for the hypothetical set of all possible scenarios $\Scen$ we might be interested in). This is of course a much stronger statement, and therefore it makes scientifically more sense to use Equation~(\ref{eq:standard_error_final}).

\subsection{Statistical Significance}\label{sec:statistical_significance}
Calculating the standard error on the tournament score is a useful way to get an idea of how accurately we have determined an agent's strength. However, this is not sufficient to determine whether or not any difference between two agents is statistically significant. In this section we will therefore explain how to perform a proper statistical test. 

We should note again, however, that this is a vast topic, so we can't go into full detail here. For a full understanding of statistical tests we therefore refer the reader to a more specialized textbook on statistics, such as \cite{diez2012openintro} or \cite{freedman2007statistics}.

In the rest of this section we will assume that we have calculated the tournament scores of agents $\ag_{\tx{i}}$ and $\ag_{\tx{j}}$ and that this was higher for $\ag_{\tx{i}}$ than for $\ag_{\tx{j}}$, so $\ts_{\tx{i}} > \ts_{\tx{j}}$. We now want to determine whether that difference is statistically significant. In order to do this we need to perform a so-called \textit{paired t-test}.

\subsubsection{Formulating the Problem}
We have previously discussed that $\ts_{\tx{i}}$ can be seen as an observation from a random variable $\rvZ_{\tx{i}}$ and moreover that the tournament score can also be seen as an approximation $\est{\mu}_{\rvZ_{\tx{i}}}$ of the mean  $\mu_{\rvZ_{\tx{i}}}$ of $\rvZ_{\tx{i}}$.

The idea in this section, is that we will now regard the \textit{difference} $\ts_{\tx{i}} - \ts_{\tx{j}}$ as an observation from a random variable that we will denote $\rvZ_{\tx{i},\tx{j}}$. This observation can be seen as an estimation $\est{\mu}_{\rvZ_{\tx{i},\tx{j}}}$ of the true mean $\mu_{\rvZ_{\tx{i},\tx{j}}}$ of this random variable, which represents the `true' difference in strength between the two agents. So, we have $\est{\mu}_{\rvZ_{\tx{i},\tx{j}}} = \ts_{\tx{i}} - \ts_{\tx{j}}$, and since we have $\ts_{\tx{i}} > \ts_{\tx{j}}$ it follows that $\est{\mu}_{\rvZ_{\tx{i},\tx{j}}} > 0$.

Our \textbf{observation} that $\est{\mu}_{\rvZ_{\tx{i},\tx{j}}}$ is greater than 0 \textit{suggests} that $\ag_{\tx{i}}$ is stronger than $\ag_{\tx{j}}$. However, we can only say that this is really true if in fact we have $\mu_{\rvZ_{\tx{i},\tx{j}}} > 0$. This is therefore our \textbf{hypothesis}.

%The idea is that we view the quantity $\ts_{\tx{i}} - \ts_{\tx{j}}$ once again as an observation from a random variable $Y_{tot}^\Delta$, and this observation can be seen as an estimation $\est{\mu}_{tot}^\Delta$ of the true mean $\mu_{tot}^\Delta$ of that random variable. 

%In other words, given our \textbf{observation}  $\est{\mu}_{\rvZ_{\tx{i},\tx{j}}}$, we would like to prove the hypothesis that 
%
%
%Since $\ts_{\tx{i}} > \ts_{\tx{j}}$ we have that $\est{\mu}_{\rvZ_{\tx{i},\tx{j}}} = \ts_{\tx{i}} - \ts_{\tx{j}} > 0$. This is our \textbf{observation}. We now want to know whether it also holds that $\mu_{\rvZ_{\tx{i},\tx{j}}} > 0$. This is our \textbf{hypothesis}.

Now, ideally, \textit{we would like to prove that, given our observation, we can conclude that our hypothesis is true. }

%In general, in statistics, the hypothesis that we aim to prove is (for some strange reason) known as \textit{the alternative hypothesis}.

Unfortunately, however, in the world of statistics we can never really prove anything with 100\% certainty, so instead we could reformulate our goal as follows: 

\textit{``We would like to prove that, given our observation, there is a very high probability that our hypothesis is true."}

Once again, however, it turns out that, in general, this question is impossible to answer. To understand why, it may be helpful to look at the following two problems:
\begin{enumerate}
\item If we flip a coin 10 times, and we know that the coin is  \textit{fair} (i.e. for each flip the probability of getting `heads' is exactly 50\%), then what is the probability that we will observe `heads' 7 times and `tails' 3 times? 
\item If we flip a coin 10 times and we observe `heads' 7 times  and `tails' 3 times,  then what is the probability that the coin is fair? 
\end{enumerate}
The first of these two problems is easy to solve. It can be calculated in a straightforward manner using only basic probability theory. The second problem, however, is impossible to answer without any further information. In principle it can be solved with Bayes' rule (Section \ref{sec:bayesian_in_general}), but that would require  knowing a prior probability for the `fairness' of the coin. That is, we would first have to know what the probability would be that the coin is fair if we don't have any observations. Without such extra information, however, it is not even a mathematically well-defined problem.

The difference between these two problems is that in the first case we know the parameters of the probability distribution, and we use them to calculate the probability of getting a certain observation, while in the second case we are given the observation, and with that we are hoping to figure out the probability that the parameters have a certain value.

Our goal that we described above was also formulated as a problem of the second type. Luckily, however, we can reformulate it again, so that it becomes a problem of the first type:

\textit{``We would like to prove that, if our hypothesis is \underline{not} true, then the probability of getting our observation is very \underline{small}."}

This probability is called the $\boldsymbol{p}$-\textbf{value}. The lower this value, the more confident we can be that our hypothesis is true. In order to determine if it is low \textit{enough}, it is common in the scientific literature to apply a threshold of 0.05. So, if the $p$-value is below 0.05, then we can say that our conclusion is statistically significant. This threshold is also known as the $\boldsymbol{\alpha}$-\textbf{value}.

The \textit{negation} of our hypothesis is usually called the \textbf{null hypothesis}, and our hypothesis itself (which we are trying to prove) is also known as the \textbf{alternative hypothesis}. So, in our example, the null hypothesis can be formulated as ``\textit{agent $\ag_{\tx{i}}$ is \underline{not} better than agent $\ag_{\tx{j}}$}" or, mathematically, as $\mu_{\rvZ_{\tx{i},\tx{j}}}\leq 0$.

If the $p$-value is below the $\alpha$-value, we say the \textbf{null hypothesis is rejected}. This statement may at first sound a bit confusing, because it's a kind of double negation. It means the \textit{null hypothesis} can be considered \textit{false}, which in turn means the \textit{alternative hypothesis} (which we wanted to prove) can be considered to be \textit{true}.

In order to calculate the $p$-value, we need to have some probability distribution for our random variable $\rvZ_{\tx{i},\tx{j}}$. The $p$-value is then defined as the probability that we draw an observation from $\rvZ_{\tx{i},\tx{j}}$ that is equal to, or higher than, the actual observation $\est{\mu}_{\rvZ_{\tx{i},\tx{j}}}$ that we made. 

We will assume that $\rvZ_{\tx{i},\tx{j}}$ has a Gaussian probability distribution. This assumption is safe, since any observation from $\rvZ_{\tx{i},\tx{j}}$ is obtained by calculating and subtracting averages over many (mostly independent) observations from random variables $\rvX_l^{d,\tx{i},\tx{j}}$. It is well-known that, no matter what kind of probability distribution those variables have, as long as there are enough of them, the distribution of $\rvZ_{\tx{i},\tx{j}}$ will indeed approximate the Gaussian distribution.

Next, we need to know the mean $\mu_{\rvZ_{\tx{i},\tx{j}}}$ of that Gaussian distribution. Recall that we are trying to prove the hypothesis that $\mu_{\rvZ_{\tx{i},\tx{j}}} > 0$, and as explained above, to do this we have to calculate the $p$-value under the assumption that this hypothesis is \textit{not} true. So, we have to assume that $\mu_{\rvZ_{\tx{i},\tx{j}}} \leq 0$. Now, in order to be absolutely sure that our conclusions are valid, we have to make the worst-case assumption, which means we have to assume $\mu_{\rvZ_{\tx{i},\tx{j}}} = 0$. Indeed, note that the higher the mean $\mu_{\rvZ_{\tx{i},\tx{j}}}$ the more likely it is that we make the an observation $\est{\mu}_{\rvZ_{\tx{i},\tx{j}}} > 0$, and thus the higher the $p$-value. So, if our $p$-value stays below the $\alpha$-value \textit{even if} we chose the highest value of $\mu_{\rvZ_{\tx{i},\tx{j}}}$, then we are sure that it is also below the $\alpha$-value for any other value of $\mu_{\rvZ_{\tx{i},\tx{j}}}$.

The final ingredient we need, is the standard deviation $\sigma_{\rvZ_{\tx{i},\tx{j}}}$ of $\rvZ_{\tx{i},\tx{j}}$. We will discuss how to estimate $\sigma_{\rvZ_{\tx{i},\tx{j}}}$ next, but for now, let us assume that we already know its exact value. In that case, it follows directly that we can calculate the $p$-value as follows.
\begin{equation}\label{eq:p_value_1}
p \quad = \quad \int_{\est{\mu}}^\infty \mc{N}(x | \mu, \sigma) dx \quad = \quad \frac{1}{\sqrt{2\pi}\sigma}\int_{\est{\mu}}^\infty e^{-\frac{(x-\mu)^2}{2\sigma^2}} dx
\end{equation}
where $\mc{N}$ denotes the Gaussian distribution, and we used $\mu$, $\est{\mu}$ and $\sigma$ as shorthands for $\mu_{\rvZ_{\tx{i}, \tx{j}}}$, $\est{\mu}_{\rvZ_{\tx{i}, \tx{j}}}$ and $\sigma_{\rvZ_{\tx{i}, \tx{j}}}$

Then, with our assumption that $\mu=0$, and with a change of variables $x' := \frac{x}{\sigma}$ and defining the $\boldsymbol{z}$-\textbf{statistic}: $z := \frac{\est{\mu}}{\sigma}$ this can be rewritten as:
\begin{equation}\label{eq:p_value_2}
p \quad = \quad \frac{1}{\sqrt{2\pi}}\int_{z}^\infty e^{-\frac{x'^2}{2}} dx'
\end{equation}

The advantage of this expression is that it contains only one parameter: the $z$-statistic. So, the $p$-value only depends on the value of $z$.

\subsubsection{Estimating the Standard Deviation}
In reality, however, we typically wouldn't know the standard deviation of $\rvZ_{\tx{i},\tx{j}}$, so instead we have to estimate it, which we will explain here.

In Section \ref{sec:standard_error} we have seen how the standard deviation of $\rvZ_{\tx{i}}$ could be calculated by modeling $\rvZ_{\tx{i}}$ as a sum over variables $\rvY_{\tx{i},s}$ and then using the observations of those variables to calculate $\est{\sigma}_{\rvZ_{\tx{i}}}$. So, here we will do something similar. That is, we define:
\[\rvY_{\tx{i}, \tx{j},s} := \rvY_{\tx{i},s} - \rvY_{\tx{j},s}\]
\[\rvZ_{\tx{i}, \tx{j}} := \frac{1}{\numVars} \sum_{s=1}^{\numVars} \rvY_{\tx{i},\tx{j},s}\]
with $\rvY_{\tx{i},s}$ and $\rvY_{\tx{j},s}$ defined as before (See Eq.~(\ref{eq:Y_i_s})). That is, an observation of $\rvY_{\tx{i},s}$ is defined as the average utility obtained by agent $\ag_{\tx{i}}$ over some number of repetitions $R_s$ of some scenario $sc_s$.

However, we have to be careful here, because the two agents $\ag_{\tx{i}}$ and $\ag_{\tx{j}}$ do not participate in exactly the same scenarios. While we mentioned that each random variable of agent $\ag_{\tx{i}}$ can be denoted as $\rvX_{\tx{i},s}$ for some integer $s$, we never specified how to decide which random variable gets assigned which integer $s$. So, if we don't do this carefully, the two variables $\rvX_{\tx{i},s}$ and $\rvX_{\tx{j},s}$ may correspond to two entirely different and unrelated scenarios. For example, they could be representing negotiations on two different negotiation domains. But it doesn't make any sense to compare the utility of $\ag_{\tx{i}}$ on one domain with the utility of $\ag_{\tx{j}}$ on a totally different domain. We therefore have to make sure that for each integer $s$, the variables $\rvX_{\tx{i},s}$ and $\rvX_{\tx{j},s}$ can be compared to each other. 

To do this, we first have to make a careful distinction between three different types of scenario:
\begin{enumerate}
\item Scenarios in which agent $\ag_{\tx{i}}$ or agent $\ag_{\tx{j}}$ was negotiating against a \textit{third} agent $\ag_{\tx{l}}$ (with $\ag_{\tx{i}} \neq \ag_{\tx{l}}$  and $\ag_{\tx{j}} \neq \ag_{\tx{l}}$). \\
That is, scenarios of the form $(\prot, \dom, \ag_{\tx{i}}, \ag_{\tx{l}})$, $(\prot, \dom, \ag_{\tx{j}}, \ag_{\tx{l}})$, $(\prot, \dom, \ag_{\tx{l}}, \ag_{\tx{i}})$, or $(\prot, \dom, \ag_{\tx{l}}, \ag_{\tx{j}})$.
\item Scenarios in which the two agents $\ag_{\tx{i}}$ and $\ag_{\tx{j}}$ were negotiating directly against \textit{each other}. \\
That is, scenarios of the form $(\prot, \dom, \ag_{\tx{i}}, \ag_{\tx{j}})$  or  $(\prot, \dom, \ag_{\tx{j}}, \ag_{\tx{i}})$.
\item Scenarios in which either agent $\ag_{\tx{i}}$ or $\ag_{\tx{j}}$ was negotiating against \textit{itself}. \\
That is, scenarios of the form $(\prot, \dom, \ag_{\tx{i}}, \ag_{\tx{i}})$  or  $(\prot, \dom, \ag_{\tx{j}}, \ag_{\tx{j}})$.
\end{enumerate}

For the first type of scenario we calculate, for each domain $\dom_d$ and each opponent $\ag_{\tx{l}}$, the average utility obtained by $\ag_{\tx{i}}$ against that opponent and the average utility obtained by $\ag_{\tx{j}}$ against that same opponent, and we then calculate the difference. In fact, we do that twice, once for those negotiations in which $\ag_{\tx{l}}$ had utility $\util_1$ and once for those negotiations in which $\ag_{\tx{l}}$ had utility $\util_2$. In other words, we calculate the following two quantities:
\[\diff_{\tx{i},\tx{j}}^{d,\boldsymbol{\cdot}, \tx{l}}\ \ := \ \ \est{\mu}_1^{d,\tx{i}, \tx{l}} - \est{\mu}_1^{d,\tx{j}, \tx{l}}\] 

%\[\diff_{\tx{i},\tx{j}}^{d,\tx{l}, \boldsymbol{\cdot}}\ \  := \ \  \frac{1}{\numReps^{d,\tx{l},\tx{i}}}\sum_{r=1}^{\numReps^{d,\tx{l},\tx{i}}}\util_2^{d,\tx{l},\tx{i},r} - \frac{1}{\numReps^{d,\tx{l},\tx{j}}} \sum_{r=1}^{\numReps^{d,\tx{l},\tx{j}}}\util_2^{d,\tx{l},\tx{j},r}\]
\[\diff_{\tx{i},\tx{j}}^{d,\tx{l}, \boldsymbol{\cdot}}\ \ := \ \ \est{\mu}_2^{d,\tx{l}, \tx{i}} - \est{\mu}_2^{d,\tx{l}, \tx{j}}\] 
where each $\est{\mu}$ on the right-hand side is calculated with Equation (\ref{eq:mu_scenario}).

Next, for the second type of scenario, we calculate, for each domain $\dom_d$, the difference between the average utility obtained by $\ag_{\tx{i}}$ and the average utility obtained by $\ag_{\tx{j}}$. This is also done twice: once for the case that $\ag_{\tx{i}}$ had utility $\util_1$ and one for the case that $\ag_{\tx{i}}$ had utility $\util_2$:

\[\diff_{\tx{i},\tx{j}}^{d,\tx{i},\tx{j}} \ \ := \ \  \est{\mu}_1^{d,\tx{i},\tx{j}} - \est{\mu}_2^{d,\tx{i},\tx{j}}\] 

%
%\[\diff_{\tx{i},\tx{j}}^{d,\tx{j},\tx{i}} \ \ := \ \  \frac{1}{\numReps^{d,\tx{j},\tx{i}}}\sum_{r=1}^{\numReps^{d,\tx{j},\tx{i}}}(\util_2^{d,\tx{j},\tx{i},r}-\util_1^{d,\tx{j},\tx{i},r})\] 

\[\diff_{\tx{i},\tx{j}}^{d,\tx{j},\tx{i}} \ \ := \ \  \est{\mu}_2^{d,\tx{j},\tx{i}} - \est{\mu}_1^{d,\tx{j},\tx{i}}\] 
Finally, we calculate, for each domain $\dom_d$ the difference between the agents' average utilities in the  `self-play' scenarios:
%\[\diff_{\tx{i},\tx{j}}^{d,sp_1} \ \ := \ \  \frac{1}{\numReps^{d,\tx{i},\tx{i}}}\sum_{r=1}^{\numReps^{d,\tx{i},\tx{i}}}\util_1^{d,\tx{i},\tx{i},r} - \frac{1}{\numReps^{d,\tx{j},\tx{j}}}\sum_{r=1}^{\numReps^{d,\tx{j},\tx{j}}}\util_1^{d,\tx{j},\tx{j},r}\] 

\[\diff_{\tx{i},\tx{j}}^{d,sp_1} \ \ := \ \  \est{\mu}_{1}^{d,\tx{i},\tx{i}} - \est{\mu}_{1}^{d,\tx{j},\tx{j}}\] 

%\[\diff_{\tx{i},\tx{j}}^{d,sp_2} \ \ := \ \  \frac{1}{\numReps^{d,\tx{i},\tx{i}}}\sum_{r=1}^{\numReps^{d,\tx{i},\tx{i}}}\util_2^{d,\tx{i},\tx{i},r} - \frac{1}{\numReps^{d,\tx{j},\tx{j}}}\sum_{r=1}^{\numReps^{d,\tx{j},\tx{j}}}\util_2^{d,\tx{j},\tx{j},r}\]

\[\diff_{\tx{i},\tx{j}}^{d,sp_2} \ \ := \ \  \est{\mu}_{2}^{d,\tx{i},\tx{i}} - \est{\mu}_{2}^{d,\tx{j},\tx{j}}\] 
To better understand these expressions, note that each of them consists of an average utility of $\ag_{\tx{i}}$ minus an average utility of $\ag_{\tx{j}}$. The expressions just differ in the question for which scenarios those averages are calculated. Furthermore, it may help to note that for each of these quantities the subscript indices $\tx{i}$ and $\tx{j}$ always refer to the two agents $\ag_{\tx{i}}$ and $\ag_{\tx{j}}$ that we are comparing, while the superscript indices refer to the scenarios for which each quantity is calculated.

The idea is that, for each of these equations, the two means on the right-hand side can be seen as observations from two variables $\rvY_{\tx{i},s}$ and  $\rvY_{\tx{j},s}$ respectively, and therefore the number on the left-hand side can be seen as an observation from a variable $\rvY_{\tx{i},\tx{j},s}$.

We now have a set of in total $2 \numDomains \numAgents$ different numbers, with $\numDomains = |\Dom|$ and $\numAgents = |\Ag|$. For example, if we have 2 domains and 4 agents then we have to calculate are 16 different numbers. Specifically, if we are comparing the scores of agents $\ag_{\tx{1}}$ and $\ag_{\tx{2}}$, then we have $\tx{i}=1$, $\tx{j}=2$. Furthermore, since there are two other agents, $\ag_{\tx{3}}$ and $\ag_{\tx{4}}$, we then have $\tx{l} \in \{3,4\}$ and since there are two domains we have $d \in \{1,2\}$. Therefore, we have to calculate the following 16 numbers:
\renewcommand{\arraystretch}{2}
\begin{center}
\begin{tabular}{cccc}
$\diff_{\tx{1},\tx{2}}^{1,\boldsymbol{\cdot},\tx{3}}$ \ \ & \ \ $\diff_{\tx{1},\tx{2}}^{1,\boldsymbol{\cdot},\tx{4}}$ & $\diff_{\tx{1},\tx{2}}^{2,\boldsymbol{\cdot},\tx{3}}$ \ \  & \ \ $\diff_{\tx{1},\tx{2}}^{2,\boldsymbol{\cdot},\tx{4}}$    \\
%%%%%%%%%%%%%%
$\diff_{\tx{1},\tx{2}}^{1,\tx{3}, \boldsymbol{\cdot}}$\ \  & \ \ $\diff_{\tx{1},\tx{2}}^{1,\tx{4},\boldsymbol{\cdot}}$ \ \ & \ \ $\diff_{\tx{1},\tx{2}}^{2,\tx{3},\boldsymbol{\cdot}}$ \ \ & \ \ $\diff_{\tx{1},\tx{2}}^{2,\tx{4},\boldsymbol{\cdot}}$\\
%%%%%%%%%%%%%%
$\diff_{\tx{1},\tx{2}}^{1,\tx{1},\tx{2}}$ \ \ & \ \ $\diff_{\tx{1},\tx{2}}^{1,\tx{2},\tx{1}}$ \ \ & \ \ $\diff_{\tx{1},\tx{2}}^{2,\tx{1},\tx{2}}$ \ \ & \ \ $\diff_{\tx{1},\tx{2}}^{2,\tx{2},\tx{1}}$\\
%%%%%%%%%%%%%%
$\diff_{\tx{1},\tx{2}}^{1,sp_1}$ \ \  & \ \ $\diff_{\tx{1},\tx{2}}^{1,sp_2}$ \ \ & \ \ $\diff_{\tx{1},\tx{2}}^{2,sp_1}$ \ \ & \ \  $\diff_{\tx{1},\tx{2}}^{2,sp_2}$ 
\end{tabular}
\end{center}
\renewcommand{\arraystretch}{1}
To simplify notation we will now instead denote these numbers as:
\[\diff_{\tx{i},\tx{j},1}, \quad \diff_{\tx{i},\tx{j},2}, \quad \dots, \quad \diff_{\tx{i},\tx{j},16}\]
Each of these numbers $\diff_{\tx{i},\tx{j},s}$ can  be seen as an observation from a different random variable $\rvY_{\tx{i},\tx{j},s}$. 

It is easy to check that if we calculate the estimated mean of $\rvZ_{\tx{i},\tx{j}}$, i.e. the average of these 16 observations, we just get the difference between the two agents' tournament scores:
\[\est{\mu}_{\rvZ_{\tx{i},\tx{j}}} \quad = \quad \frac{1}{2 \numDomains \numAgents}\sum_{s=1}^{2 \numDomains \numAgents} \diff_{\tx{i},\tx{j},s} \quad = \quad \ts_{\tx{i}} - \ts_{\tx{j}}\]

We are now ready to calculate the estimated standard deviation of $\rvZ_{\tx{i},\tx{j}}$. As before, however, we have to take into account that not all of these observations are independent. Specifically,  for each domain $D_d$ the values of $\diff_{\tx{i},\tx{j}}^{d,sp_1}$ and $\diff_{\tx{i},\tx{j}}^{d,sp_2}$ are mutually dependent, because they are both calculated from the same two scenarios $(\prot, D_d, \ag_{\tx{i}}, \ag_{\tx{i}})$ and $(\prot, D_d, \ag_{\tx{j}}, \ag_{\tx{j}})$. We therefore use an expression similar to Eq.~(\ref{eq:standard_error_final}):

\begin{equation}\label{eq:standard_error_diff}
\est{\mi{Var}}_{\rvZ_{\tx{i},\tx{j}}} \ \ = \ \ \frac{1}{\numVars \cdot (\numVars-1)}\cdot \Big( \sum_{s=1}^\numVars (\est{\mu}_{\rvZ{\tx{i},\tx{j}}} - \diff_{\tx{i},\tx{j},s})^2 \ + \  2 \cdot \sum_{d=1}^{|\Dom|}(\est{\mu}_{\rvZ{\tx{i},\tx{j}}} - \diff_{\tx{i},\tx{j}}^{d,sp_1})\cdot (\est{\mu}_{\rvZ{\tx{i},\tx{j}}} - \diff_{\tx{i},\tx{j}}^{d,sp_2})\Big)
\end{equation}
with $\numVars = 2\numDomains\numAgents = 2 \cdot  |\Dom|\cdot |\Ag|$

%Next, our goal is to calculate the sample standard deviation and standard error. \essential{However, there is one complication that we have to take into account, which is that our observations $\diff_1, \quad \diff_2, \quad \dots, \quad \diff_{16}$ are not all independent. Specifically, for each domain $D_d$ the values of $\diff_{\tx{i},\tx{j}}^{d,sp_1}$ and $\diff_{\tx{i},\tx{j}}^{d,sp_2}$ are mutually dependent, because they are each calculated from the same two scenarios $(\prot, D_d, \ag_{\tx{i}}, \ag_{\tx{i}})$ and $(\prot, D_d, \ag_{\tx{j}}, \ag_{\tx{j}})$. This means we cannot use the usual formula for the sample standard deviation (Eq.~(\ref{eq:estimate_sigma})).
%%\[\sigma \quad = \quad\sqrt{\frac{\sum_{k=1}^{2 \numDomains \numAgents} (\mu - \diff_k)^2}{2 \numDomains \numAgents-1}}\]
%Instead, we have to add a correction term for every domain, involving the \textit{covariance} of the two dependent variables.} That is:
%\[\est{\sigma} \quad = \quad\sqrt{\frac{1}{2 \numDomains \numAgents-1}\sum_{k=1}^{2 \numDomains \numAgents} (\est{\mu} - \diff_k)^2 + \frac{1}{\numDomains \numAgents}\sum_{d=1}^\numDomains (\est{\mu} - \diff_{\tx{i},\tx{j}}^{d,sp_1}) \cdot (\est{\mu} - \diff_{\tx{i},\tx{j}}^{d,sp_2})}\]
%Once again, we will not go into the details of how this formula is derived.

Finally, the estimated standard deviation of $\rvZ_{\tx{i},\tx{j}}$ can be calculated by taking the square root:
\[\est{\sigma}_{\rvZ_{\tx{i},\tx{j}}} \quad = \quad \sqrt{\est{\mi{Var}}_{\rvZ_{\tx{i},\tx{j}}} }\]

\subsubsection{Performing the Test}

We are now finally ready to calculate the $p$-value of our hypothesis. Essentially, we are going to do the same as in Equations (\ref{eq:p_value_1}) and (\ref{eq:p_value_2}), except that this time we will be using the \textit{estimated} standard deviation $\est{\sigma}_{\rvZ_{\tx{i},\tx{j}}}$ instead of the true standard deviation $\sigma_{\rvZ_{\tx{i},\tx{j}}}$. This means that instead of the $z$-statistic we are now calculating what is called the `\textbf{t-statistic}':
\[\tstat \quad = \quad \frac{\est{\mu}_{\rvZ_{\tx{i},\tx{j}}}}{\est{\sigma}_{\rvZ_{\tx{i},\tx{j}}}}\]
A major consequence of this difference, is that we can now no longer assume that this statistic is drawn from a Gaussian distribution. Instead, the $t$-statistic is drawn from what is known as the `\textbf{t-distribution with} $2\numDomains \numAgents-1$ \textbf{degrees of freedom}' (denoted $\tdist_{2\numDomains \numAgents-1}$). This is because $\est{\sigma}_{\rvZ_{\tx{i},\tx{j}}}$ is not a constant, but rather a quantity that was calculated from observations of a number of random variables. The $t$-distribution is very similar to the Gaussian distribution but has `heavier' tails. As usual, we refer to more specialized text books on statistics, such as \cite{diez2012openintro} and \cite{freedman2007statistics}, for more details on this.

So, we now need to calculate the $p$-value as follows: 
\[p \quad = \quad  \int_{\tstat}^\infty \tdist_{2\numDomains \numAgents-1}(x) dx\]
Now, unfortunately, the $t$-distribution has a very complicated expression, so we can't expect to calculate this integral analytically. Luckily, however, we don't have to care about this, because most programming languages and statistics programs already have this function built-in. For example, if $\tstat=3.1$ and $ 2\numDomains \numAgents -1 = 15$, then, in Excel, we can calculate the $p$-value with the following formula:
\[\text{=T.DIST.RT(3.1,\ 15)}\]
and in Python we can calculate it as follows (after installing the scipy package):
\begin{lstlisting}[language=Python]
from scipy import stats
p_value = 1 - stats.t.cdf(3.1, 15)
\end{lstlisting}

Now, if our $p$-value is smaller than or equal to the $\alpha$-value (which is typically set to 0.05), then we say that the null hypothesis is rejected, which is another way of saying that our conclusion that agent $\ag_{\tx{i}}$ is better than agent $\ag_{\tx{j}}$, is statistically significant.

On the other hand, if $p > \alpha$, this does \textit{not} mean that we can draw the opposite conclusion that $\ag_{\tx{j}}$ is better than $\ag_{\tx{i}}$. It just means that we do not have sufficient evidence to say confidently that $\ag_{\tx{i}}$ is better than $\ag_{\tx{j}}$.

\begin{exercise}\textbf{Paired t-test.}
Modify your code of the exercises \ref{ex:calculate_tournament_scores} and \ref{ex:standard_error} so that it also calculates, for each pair of agents, the $p$-value for the hypothesis that the agent with the higher tournament score is indeed stronger than the other.
\end{exercise}

\subsubsection{Combining Multiple Tests}

In the previous sections we have compared the scores of two agents, and determined whether the difference was statistically significant. Of course, `statistically significant' is still no guarantee that $\ag_{\tx{i}}$ is truly better than $\ag_{\tx{j}}$. After all, as we mentioned above, in statistics we can never be 100\% sure of anything. The best we can do is to say that it is \textit{unlikely} that $\ag_{\tx{i}}$ is \textit{not} better than $\ag_{\tx{j}}$, because if that were the case, then the probability of obtaining the data we observed would have been smaller than 5\%.

This is fine if we are just comparing two agents, but this becomes problematic when we are comparing our agent against multiple benchmark agents (which we would normally do). The problem is that if we compare our agent $\ag_{\tx{1}}$ against multiple other agents, then even though for each \textit{individual} opponent there is only a small chance that we draw a false conclusion, these probabilities add up, yielding a relatively large possibility that \textit{at least one} of our conclusions is false.

For example, suppose that we test our agent against four different opponents, so for each of these four opponents we test the null hypothesis that that opponent is better than or equal to our agent $\ag_{\tx{1}}$, and suppose that for each of these hypotheses we apply an $\alpha$-value of 0.05. Furthermore assume the worst case-scenario that all null hypotheses are in reality true (so in reality our agent does not outperform any of the benchmark agents). Then the probability that we will falsely reject at least one null hypothesis would be:
\[P(\text{at\ least\ one\ of\ the\ null\ hypotheses\ is\ rejected}) = 1-(1-\alpha)^4 \approx 4 \cdot \alpha\]
In other words, the fact that we are testing \textit{multiple} hypotheses increases the chance that we will draw a false conclusion from 5\% to 20\%, which is too high. In order to compensate for this, we therefore need to apply a correction to our acceptance threshold. To this end, we define the \textbf{global} $\alpha$-value, which we'll denote as $\al_G$, to be the maximum probability of error that we tolerate for the \textit{entire} experiment. That is, the maximum probability that we draw at least one false conclusion.  This value would typically be set to 0.05. Furthermore, we define for every single null hypothesis $\hypo_i$ a \textbf{local} $\alpha$-value, which we'll denote as $\al_i$, which represents the maximum $p$-value for which we reject that individual null hypothesis. 

A commonly used and simple method to determine the local $\alpha$-values is the \textbf{Bonferroni correction}. 
\begin{definition}
Suppose we are testing $n$ hypotheses and that we are given a global $\alpha$-value $\alpha_G$. Then, we say that we have applied a \textbf{Bonferroni correction} if the local $\alpha$-values $\alpha_i$ are set as:
\[\alpha_i = \frac{\alpha_G}{n}\]
\end{definition}

%This means that we divide the desired alpha-level by the number of hypotheses that we are testing, and each individual null hypothesis is only rejected if its $p$-level is below that corrected value.

For example, suppose if we are testing 4 different null hypotheses $\hypo_1$, $\hypo_2$, $\hypo_3$, and $\hypo_4$, and for each hypothesis $\hypo_i$ we have calculated a corresponding $p$-value $p_i$. Furthermore, suppose we have a global $\alpha$-level of 0.05. Then, applying a Bonferroni correction means that any individual null hypothesis $\hypo_i$ is rejected if and only if $p_i \leq \frac{0.05}{4} = 0.0125$.

While this method is simple, it turns out that it is actually overly strict, in the sense that it reduces the individual $\alpha$-levels too much, which means that we might throw away some results that are actually statistically significant. A better, but somewhat more complicated, solution is the so-called \textbf{Holm-Bonferroni procedure}. 

\begin{definition}
Suppose we have $n$ null hypotheses, denoted $\hypo_1, \hypo_2, \dots \hypo_n$, and suppose that for each null hypothesis $\hypo_i$ we have calculated a corresponding $p$-value $p_i$. Furthermore, assume that these hypotheses are sorted in order of increasing $p$-value, so we have: $p_1 \leq p_2 \leq \dots \leq p_{n-1} \leq p_n$. Finally, assume we are given some global $\alpha$-value $\alpha_G$. Then, according to the \textbf{Holm-Bonferroni procedure}:
\begin{itemize}
\item Hypothesis $\hypo_1$ is rejected iff $p_1 \leq \frac{\alpha_G}{n}$.
\item Hypothesis $\hypo_2$ is rejected iff $\hypo_1$ is rejected and $p_2 \leq \frac{\alpha_G}{n-1}$.
\item Hypothesis $\hypo_3$ is rejected iff $\hypo_1$ and $\hypo_2$ are both rejected and $p_3 \leq \frac{\alpha_G}{n-2}$.
\item etcetera...
\end{itemize}
So, in general, the Holm-Bonferroni procedure rejects a null hypothesis $\hypo_i$ iff all null hypotheses $\hypo_j$ with $j<i$ have been rejected and $p_i \leq \frac{\alpha_G}{n+1-i}$.
\end{definition}

\section{Summary of Chapter}

\begin{itemize}
\item In order to compare the strengths of various negotiating agents, we should run a tournament in which all these agents negotiate against each other (as well as against themselves), over a large set of different negotiation domains.
\item We have discussed three different ways to analyze the results of such a tournament:
\begin{enumerate}
\item Tournament Evaluation
\item Empirical Game-Theoretical Analysis (EGTA)
\item Sequential Elimination Ranking
\end{enumerate}
\item Tournament Evaluation makes most sense if we can assume that the people who are going to use the agents have no knowledge whatsoever about the strength of each agent, or completely ignore such knowledge.
\item EGTA makes most sense if we assume that other people are perfectly rational and have detailed knowledge of how well each agent performs against every other agent.
\item Sequential Elimination Ranking forms a middle ground between the other two methods.
\item Tournament evaluation ranks each agent based on their tournament score, which is the average utility of each agent, over the entire tournament. 
\item In addition to the tournament score, we may also calculate the utility-on-agreement and the agreement rate. These measures may help to determine \textit{why} each agent performs well or poorly, but should not be used as performance measures themselves.
\item To assess how accurately we have calculated an agent's tournament score, we can calculate its standard error, by taking the square root of Equation (\ref{eq:standard_error_final}).
\item If the standard error is too high, then:
\begin{enumerate}
\item We should increase the number of negotiation domains.
\item If, for whatever reason, we can't increase the number of negotiation domains, then instead we can increase the number of repetitions of each scenario.
\item If that still does not decrease the standard error enough, then we can take the square root of Eq.~(\ref{eq:std_err_with_dependence}) instead of  Eq.~(\ref{eq:standard_error_final}). However, this changes the interpretation of the results. If the standard error is low according to Eq.~(\ref{eq:std_err_with_dependence}) we can only conclude the tournament score remains stable if we repeat the tournament on exactly the same set of scenarios, but this does not say anything about what happens if we repeat the tournament on a different set of scenarios.
\end{enumerate} 
\item Whenever we conclude that one agent is stronger than another agent, we should test if that conclusion is statistically significant, by performing a paired t-test with Holm-Bonferroni correction.
\end{itemize}

\chapter{Multilateral Negotiations}\label{sec:multilateral}
In this section we will generalize our discussion of automated negotiation to scenarios with more than two agents. That is, we will discuss so-called \textit{multilateral} negotiations.

In the case of bilateral negotiations, the question how negotiations work is fairly straightforward: once an agent accepts a proposal by the other, it becomes a binding agreement. Of course, there exist other protocols for bilateral negotiations, besides the AOP, in which agreements may be defined in a slightly different way, but the basic idea remains the same.  

However, when we generalize this to multiple agents, there are many possible ways in which the negotiations could be structured. In particular, we mention the following types of multilateral negotiations:
\begin{enumerate}
\item \textbf{All-Together negotiations}: Negotiations in which all agents together take part in one single negotiation, and for which at the end either all agents come to an agreement together, or the negotiations fail completely and there is no agreement at all.
\item \textbf{One-to-Many negotiations}:
One agent takes part in multiple bilateral negotiations at the same time, each with a different opponent.
\item \textbf{Many-to-Many negotiations}:
The set of agents can be divided in two disjoint subsets, often referred to as the `buyers' and the `sellers'. Each buyer takes part in multiple bilateral negotiations at the same time, each with a different seller, and each seller also takes part in multiple bilateral negotiations at the same time, each with a different buyer.
\item \textbf{All-Subsets negotiations}: For any subset of the agents (with size greater than 1) there can be a separate negotiation taking place at the same time. So, there can be up to $2^\numAgents - (\numAgents+1)$ separate negotiations taking place in parallel.
\end{enumerate}
We will discuss these types in more detail in the following sections. We should stress, however, that these are by no means the \textit{only} possible types of multilateral negotiation. 

%In all cases, we will define a multilateral negotiation scenario (with $\numAgents$ agents) as follows.
%\begin{definition}\label{def:scenario_multi}
%A multilateral negotiation \textbf{scenario} $\scen$ as a tuple $\scen = (\prot, \dom, \ag_1, \ag_2, \dots \ag_n)$, where $\prot$ is a multilateral negotiation protocol, $\dom$ is a negotiation domain for $\numAgents$ agents, and each $\ag_i$ is a negotiating agent.
%\end{definition}
%\essential{This is not really correct because in one-to-many agents the buyer agent is denoted $\ag_0$}

%Note that, except from all-together negotiations, all these types of negotiations in fact involve

Whenever there are multiple separate negotiations that take place in parallel, we will refer to those parallel negotiations as \textbf{negotiation threads}.

\section{All-Together Negotiations}
All-Together negotiations form the most straightforward generalization from bilateral to multilateral negotiations. The idea is that all agents take part in one single negotiation which may either end with a single agreement for all agents, or with no agreement at all.

An example could be a negotiation between multiple founders of a new start-up company about how they will divide the shares of that company among themselves. Each founder brings his own essential expertise into the business, so they can't start the company without all of them on board.

In the literature, the most commonly used negotiation protocol for this type of negotiation is the Stacked Alternating Offers Protocol (SAOP)~\cite{Aydogan2017SAOP}, which is a straightforward generalization from the ordinary Alternating Offers Protocol as defined in Section~\ref{sec:aop}. To explain it, imagine that all agents are sitting around a round table. Let's assume, without loss of generality, that to the left of $\ag_1$ sits agent $\ag_2$, that to the left of $\ag_2$ sits $\ag_3$, etcetera, and to the left of $\ag_\numAgents$ sits $\ag_1$.

\begin{enumerate}
\item One agent, say $\ag_1$, starts by proposing any offer $\off$ from the offer space. We then say that that offer $\off$ has become the \textit{standing} offer.
\item  Next, agent $\ag_2$ can either \textit{accept} the standing offer $\off$, or \textit{propose} another offer $\off'$. If she accepts $\off$, then that offer remains the standing offer. If she proposes another offer $\off'$, then that offer becomes the new standing offer. 
\item Then, agent $\ag_3$ has the same choice: either accept the standing offer, or propose a new offer (which then becomes the new standing offer).
\item etcetera...
\item After the last agent's turn, it is again $\ag_1$'s turn, which may accept the standing offer or again propose a new offer. 
\item Next, it is again $\ag_2$'s turn, etcetera...
\end{enumerate}
This continues until one of the following  happens:
\begin{enumerate}
\item In some turn an agent proposes some offer $\off_k$, and then in the next $\numAgents-1$ turns all other agents accept that offer one by one. 
\item A given deadline $\dead$ has passed.
\item A maximum number of turns $\maxRounds$ have passed.
\end{enumerate}
In the first case, the negotiations end with the offer $\off_k$ as the agreement, and every agent $\ag_i$ receives its corresponding utility value $\util_i(\off_k)$. In the other two cases the negotiations end without agreement, and every  agent $\ag_i$ receives its respective reservation value $\rv_i$. Note that the ordinary AOP is just a special case of the SAOP, with $\numAgents=2$.

Now, it is important to understand that in the SAOP, whenever some agent $\ag_j$ makes a proposal, this should be seen as a proposal to \textit{all} other agents, and not just as a proposal only to the next agent $\ag_{j+1}$. It just happens that $\ag_{j+1}$ will be the first agent to be able to react to this proposal. For this reason, in the SAOP, all actions are public. This means that when $\ag_j$ proposes or accepts an offer, this action is observed by \textit{all} other agents. This allows all agents to keep track of what is going on, and to build a separate opponent model of every other agent. 

For negotiations under the SAOP we define a negotiation action in the same way as for the AOP (Def.~\ref{def:action}), except that the agent index $i$ can now be any value between 1 and $\numAgents$.
\begin{definition}\label{def:multilateral_action}
Let $\numAgents$ be the number of agents in the negotiation. Then we define a \textbf{negotiation action} to be a tuple 
\[(i, \actype, \off, t) \ \ \in  \ \  \{1,2, \dots, \numAgents\}\times \{\prop, \acc\} \times \Off \times \mathbb{R}^+\]
where all symbols have the same meaning as in Definition~\ref{def:action}.
\end{definition}

Similarly, a negotiation history $\hist$ for the SAOP is defined in exactly the same way as for the AOP (Def.~\ref{def:history}), except that the actions $\ac_j$ in the history are now defined as in Def.~\ref{def:multilateral_action}. However, we should specify that the delay $\delay_j$ is to be interpreted as the time it takes for the message to arrive at the \textit{next} agent $ag_{i_{j+1}}$, even though the message will also be sent to all other agents. For example, suppose $\ac_{10} = (2,\prop, \off, t)$. That is, the 10th action in the history consists of agent $\ag_2$ proposing offer $\off$ at time $t$. Then $\delay_{10}$ represents the time it takes for that message to arrive at agent $\ag_3$. All other agents may receive the message at a slightly earlier or slightly later time. 

We can now formally define the SAOP by means of the following two definitions.
\begin{definition}\label{def:saop}
We say a negotiation history $\hist$ satisfies the SAOP (with deadline $\dead$ and maximum number of rounds $\maxRounds$) if and only if all of the following conditions hold:
\begin{enumerate}
%\item For any two consecutive negotiation actions $\ac_j = (i_j, \actype_j, \off_j, t_j)$ and $\ac_{j+1} = (i_{j+1}, \actype_{j+1}, \off_{j+1}, t_{j+1})$ in $h$, we have:
%\[i_{j+1} = 
%\begin{cases}
%i_j + 1 & \text{if\ } i_j < \numAgents\\
%1  & \text{if\ } i_j = \numAgents
%\end{cases}
%\]
\item For any two consecutive negotiation actions $\ac_j = (i, \actype, \off, t)$ and $\ac_{j+1} = (i', \actype', \off', t')$ in $h$, we have:
\[i' = 
\begin{cases}
i + 1 & \text{if\ } i < \numAgents\\
1  & \text{if\ } i = \numAgents
\end{cases}
\]
%\item For any two consecutive negotiation actions $\ac_j = (i_j, \actype_j, \off_j, t_j)$ and $\ac_{j+1} = (i_{j+1}, \actype_{j+1}, \off_{j+1}, t_{j+1})$ in $h$, we have:
%\[t_j + \delay_j \ \ < \ \ t_{j+1}\]
\item For any two consecutive negotiation actions $\ac_j = (i, \actype, \off, t)$ and $\ac_{j+1} = (i', \actype', \off', t')$ in $h$, we have:
\[t + \delay_j \ \ < \ \ t'\]
where $\delay_j$ is the delay between $\ac_j$ and $\ac_{j+1}$ in $\hist$.
\item The first negotiation action must be a proposal (i.e. it cannot be an acceptance): $\actype_1 = \prop$.
\item If $(i, \actype, \off, t)$ and $(i', \actype', \off', t')$ are two consecutive actions of the negotiation history, and $\actype' = \acc$, then we must have $\off = \off'$.
\item If the history contains $\numAgents -1$ \underline{consecutive} acceptances, then they must be exactly the $\numAgents-1$ last actions of the history.
\item For all negotiation actions $(i, \actype, \off, t)$ in $\hist$ we have $t\leq \dead$.
\item The history $\hist$ can contain at most $\maxRounds$ negotiation actions.
\end{enumerate}
\end{definition}
The first condition states that the agents take turns in a round-the-table manner. The second condition states that an agent can only propose or accept an offer after it has received the action from the previous agent. The fourth condition states
that one can only accept the offer that was proposed last, and not any offer that was proposed earlier. The fifth condition states that the negotiations are over as soon as a proposal has been accepted by all other agents consecutively. The sixth condition states that the negotiations are over when the deadline $\dead$ has passed, and the last condition states that the negotiations are over as soon as $\maxRounds$ turns have passed.

\begin{definition}\label{def:agreement}
Let $\hist$ be a negotiation history that satisfies the SAOP and let $\ac_k$ be the last negotiation action in this history. Furthermore, let  $\ac_j$ be the last \underline{proposal} in this history (so all following actions are acceptances). Then, the SAOP defines that the negotiation has ended in \textbf{agreement} if $j + \numAgents - 1 = k$ and $t_k + \epsilon_k < \dead$ (where $\numAgents$ is the number of agents). In that case we say that $\off_k$ is the \textbf{accepted offer}. Otherwise, we say the negotiations have \textbf{failed}.
\end{definition}

\begin{exercise}\label{ex:saop}
\textbf{Implement the SAOP protocol.} Make a copy of the file \textbf{nego\_simulator.py} and adapt it to execute the SAOP protocol, for any arbitrary number of agents $\numAgents$, instead of just the bilateral alternating offers protocol. \newline

Note that there are two different ways you can do this: 
\begin{enumerate}
\item Whenever an agent $\ag_i$ returns an action, immediately pass that action to every other agent $\ag_j$, so that they can all update their opponent models of $\ag_i$ before the next agent's turn.
\item Whenever an agent $\ag_i$  returns an action, only pass that action to the next agent $\ag_{i+1}$, together with all $\numAgents-2$ other actions that were chosen in the previous turns, so that $\ag_{i+1}$ can update its opponent models of all other agents, before choosing its next action.
\end{enumerate}
Also note that you will need to create a new base class for the agents in this protocol, because the Agent class provided with the framework is only designed for bilateral negotiations.
\end{exercise}

Most negotiation strategies that we discussed in Chapter~\ref{sec:negotiation_strategies} can be adapted for multilateral negotiations under the SOAP in a relatively simple way.

For example, a time-based strategy that doesn't require opponent modeling (e.g. based on Equation~(\ref{eq:time_based_min})) does not really need any adaptation at all. On the other hand, if we have a time-based agent based on Equation~(\ref{eq:time_based_max}) then, to adapt it for multilateral negotiations, we have to replace the maximization over $\est{\util}_2$ by something that takes into account the estimated utility of multiple opponents. For example, if our agent is $\ag_1$, then, for any offer $\off \in \Off$, we could define $\est{\util}_{opp}(\off)$ as the \textit{minimum} estimated utility of $\off$ among all our opponents:
\begin{equation}\label{eq:util_opp}
\est{\util}_{opp}(\off) \ \ := \ \ \min\ \{\est{\util}_i(\off) \mid i\in \{2, 3, \dots, \numAgents\} \}
\end{equation}
where $\est{\util}_i$ denotes the estimation that our agent $\ag_1$ makes about the utility function $\util_i$ of opponent $\ag_i$.

We can then generalize Eq.~(\ref{eq:time_based_max}) to:
\begin{equation}\label{eq:time_based_multilateral}
\offNext \quad = \quad \argmax_{\off\in \Off }\ \{\ \est{\util}_{opp}(\off) \mid  \util_1(\off) \geq \asp(t) \ \land \ \off \not\in \Pro_{t}\}
\end{equation}
The reason that we chose the \textit{minimum} operator in Eq.~(\ref{eq:util_opp}), as opposed to, for example, the \textit{maximum} or the \textit{average}, is that a deal can only be made if \textit{all}  agents agree with it. Therefore, to maximize the probability that the proposal becomes an agreement, we have to make sure that it is acceptable \textit{even} to the agent that benefits the least from it.

Recall that for adaptive strategies, we do not only need an opponent \textit{utility} modeling algorithm, but we also need an opponent \textit{strategy} modeling algorithm, in order to determine our target value. For this, we can follow a similar approach. That is, we keep a separate opponent model for each opponent, and for each of them we try to estimate the offer $\off_i^*$ with highest utility for us that agent $\ag_i$ would still be willing to propose or accept at the end of the negotiations. Then, we set our target value $\targ$ equal to:
\[\targ \ \ = \ \ \min \{\util_1(\off_i^*) \mid i\in \{2, 3, \dots, \numAgents\} \}\]
Again, we need to take the \textit{minimum} here, because we aim to concede far enough so that \textit{every} opponent would be willing to accept our final proposal.

To adapt MiCRO for the SAOP protocol, we need to make two adaptations~\cite{Aguilera2025MulitlateralMiCRO}. Recall that the bilateral version of MiCRO will propose a new offer if and only if it has so far not proposed more unique offers than its opponent. The first adaptation we have to make, is that we should no longer just count the number of \textit{proposed} offers by each agent, but also the number of \textit{accepted} offers by each agent. So, let $m_i$ denote the number of unique offers that were either proposed \textit{or accepted} by agent $\ag_i$.

The second adaptation we have to make is that, just as for  time-based- and adaptive- strategies, we have to aggregate these numbers $m_i$ over all opponents. As with these other strategies, we apply the \textit{minimum} operator for this. So, a multilateral MiCRO agent $\ag_1$ would propose a new offer only if:
\[m_1 \leq \min \{m_2, m_3, \dots, m_{\numAgents}\}\]

In the case of bilateral negotiations, it was not necessary for MiCRO to count acceptances, because in such scenarios the negotiations are over as soon as any of the agents makes an acceptance. However, in the case of the SAOP, the negotiations continue. Furthermore, recall from our discussion in Section \ref{sec:proposals_and_acceptances} that  `accepting' an offer is essentially the same as `proposing' an offer. Both actions are just a way for an agent to signal that he agrees with that offer. The only difference is that a proposal is the first in a sequence of `endorsements' of the same offer, while all following endorsements are called `acceptances'. So, the distinction between proposals and acceptances is mainly just a naming convention, rather than a real semantic difference.

%If MiCRO doesn't count acceptances, then it might appear to MiCRO that some opponent is unwilling to make any concessions, because he is never making any new proposals, while in reality that opponent is conceding by accepting the offers he receives.

\begin{exercise}\label{ex:saop_agents}
\textbf{Implement SAOP Agents}. Implement new variants of the agents you have implemented before, but now for multilateral negotiations under the SAOP protocol.
\end{exercise}

\later{Discuss classic tit-for-tat strategies?}

\later{Add algorithm?}

\section{One-to-Many Negotiations}
A second model of multilateral negotiation that has been studied extensively in the literature, is that of a one-to-many negotiation. In this type of negotiation, one agent is conducting multiple \textit{bilateral} negotiations at the same time, each with a different opponent. As mentioned above, we will refer to each of these bilateral negotiations as a \textit{negotiation thread}.

For example, we can imagine a customer that aims to buy a car, and is conducting negotiations over the Internet with multiple potential sellers at the same time.

In this model the agent that is conducting  multiple negotiations is often referred to as the `buyer' while its various opponents are often referred to as the `sellers'. We will here follow this same convention. However, It should be understood that this is really just a naming convention, and that in reality this same model can just as well be used to model a single seller negotiating with multiple buyers.

When studying this model, authors usually focus on the implementation of the buyer agent. This is because each individual seller is only involved in a single bilateral negotiation, so it is the implementation of the buyer that makes this specific model interesting.

In this section, we will use the convention that the buyer agent is denoted as $\ag_0$, and the sellers are denoted as $\ag_1,  \ag_2, \dots, \ag_\numAgents$. 
\begin{definition}
Let $\numAgents$ be a positive integer. Then, a \textbf{one-to-many negotiation domain} with $\numAgents$ sellers, consists of:
\begin{itemize}
\item $\numAgents$ offer spaces: $\Off_1, \Off_2, \dots, \Off_\numAgents$, one for each seller $\ag_i$.
\item $\numAgents$ `seller' utility functions: $\util_i \ : \ \Off_i \rightarrow \mathbb{R}$
\item A utility function $\util_0$ for the buyer.
\begin{equation}\label{eq:util_buyer_multiple_sellers}
\util_0 \ : \ \prod_{i=1}^\numAgents \big( \Off_i \cup \{\dis\} \big) \rightarrow \mathbb{R}
\end{equation}
where $\dis$ represents `disagreement' (i.e. a negotiation ending without agreement).
\item $\numAgents + 1$ reservation values $\rv_i \in \mathbb{R}$, one for each agent, where the reservation value for the buyer is defined as $\rv_0 := \util_0(\dis, \dis, \dots, \dis)$.
\end{itemize}
\end{definition}
For example, suppose the buyer is negotiating with 3 sellers, and that he comes to an agreement $\off \in \Off_1$ with seller $\ag_1$, and to another agreement  $\off' \in \Off_3$ with seller $\ag_3$. Furthermore, assume that his negotiation with seller $\ag_2$ ends without any agreement. Then, the buyer's final utility is given by $\util_0(\off, \dis, \off')$.

Note that the various offer spaces $\Off_i$ do not necessarily have to be all different. It is perfectly possible that the buyer is negotiating over exactly the same offer space with several different sellers, or over partially overlapping offer spaces.

Within the one-to-many negotiation model we can further distinguish between three more specific models (but once again these are not necessarily the \textit{only} possibilities):
\begin{enumerate}
\item \textbf{Single Agreement:} The buyer only aims to come to \textit{one} agreement in total. 
\item \textbf{Many Agreements:} The buyer aims to come to a separate agreement in each  negotiation thread. 
\item \textbf{One Agreement Per Subset:} This model is a hybrid of the previous two, in which the set of sellers is partitioned into several disjoint subsets, and the buyer only aims to come to one agreement for every such subset. %\later{So, when the buyer comes to an agreement with some seller $\ag_i$, he ends all negotiations with sellers $\ag_j$ that are in the same subset as $\ag_i$, but continues negotiating with all other sellers.}
\end{enumerate}

An example of the single-agreement model could be the scenario we mentioned above, where a customer is negotiating with multiple car sellers at the same time, to buy a car. Of course, the buyer only needs to buy one car, so as soon as he makes a deal with one of the sellers, he has no reason to continue negotiating with the other sellers.

An example of the many-agreements model could be the scenario in which a customer is buying different types of furniture for a flat. He negotiates with one seller to buy a carpet, while he negotiates with another seller to buy a closet, and he negotiates with yet another seller to buy a couch. The buyer's goal is to buy one of each product, so once the buyer has made an agreement in one negotiation thread, he still needs to continue negotiating with the other sellers to buy the other products.

The one-agreement-per-subset model can also be exemplified with the furniture-buying scenario, but this time for each product the buyer may be in a negotiation with several sellers at the same time. For example, he might be negotiating the purchase of a carpet with three different carpet sellers, and at the same time be negotiating the purchase of a closet with several other sellers, and be negotiating the purchase of a couch with yet another set of sellers. Once he buys a carpet he immediately breaks off the negotiations with all other carpet sellers, but continues negotiating with the closet sellers and the couch sellers.

\subsection{Dependencies in One-to-Many Negotiations}

The many-agreements model is most interesting if the utility obtained by the buyer depends on the \textit{combination} of agreements made in the various different negotiation threads. For example, the buyer may want his carpet and his couch to have matching colors. Therefore, once he buys a couch, this will affect his preferences over the set of possible carpets. Without such interdependencies, the buyer would literally just be conducting multiple bilateral negotiations that are completely unrelated to one another, and thus such a model would not add anything interesting to what we already discussed in the previous chapters. For this reason we defined the buyer's utility function in Eq.~(\ref{eq:util_buyer_multiple_sellers}) as a function over the set of all possible combinations of negotiation outcomes.

%\later{The agent's reservation value is then defined as the utility it receives in case it does not make any agreements at all, in any negotiation:
%\[\rv_0 := \util_0(\dis, \dis, \dots, \dis)\]}

In the case of a single-agreement model, we still use the same expression for the buyer's utility function, because even though he aims to achieve only one agreement, in many such models studied in the literature it is still \textit{possible} for the buyer to make multiple agreements, and therefore his utility function must be defined over all such possible outcomes. The fact that the agent \textit{aims} to make just one agreement, just means that, in general, the buyer receives \textit{less} utility if he makes more than one agreement, than if he makes just one agreement.

%\essential{This text below needs to be adapted, because it may be the case that it is *possible* for the agent to make multiple agreements, but he is only *interested* in making at most one agreement.}
%In the single-agreement model, on the other hand, it does not make sense to say that the utility of an agreement in one negotiation depends on agreements in the other negotiations, since there can be only one agreement. So, the buyer's utility is defined over the \textit{union} of all offer spaces.
%\[\util_0 \ : \ \bigcup_{i=1}^\numAgents \Off_i  \rightarrow \mathbb{R}\]
%The buyer's reservation value $\rv_0$ is, in this case, given separately, just as in the case of bilateral negotiations.

%While there cannot be any dependency between agreements in the single-agreement case,

Also note that even in the single-agreement model the various concurrent bilateral negotiation threads still depend on each other, but in a different way. Namely, in the sense that any proposal the buyer receives in one negotiation may affect the buyer's strategy in other negotiations. For example, if one car seller offers a price of \$3000, then the buyer will no longer have any reason to propose or accept any price higher than \$3000 to any of the other car sellers (assuming they all sell comparable cars).

%if the buyer is not trying to buy `matching products',

\subsection{One-to-Many Protocols}\label{sec:one-to-many_protocols}
Let us now discuss the various negotiation protocols that have been proposed in the literature for one-to-many negotiations.

In the many-agreements model, each separate bilateral negotiation thread can simply follow the alternating offers protocol. However, in the single-agreement model this is not ideal. The reason for this is the following. If the buyer makes several different proposals at the same time, to different sellers, then it may happen that more than one seller replies with an acceptance. This means the buyer could end up with multiple agreements, and this would be very hard to prevent. 

Therefore, when studying one-to-many single-agreement negotiations (or one-agreement-per-subset negotiations), authors typically make use of some alternative protocol. However, there  does not seem to be any single protocol that is generally accepted in the literature as the default protocol for this type of negotiations. Therefore, we will just discuss several different protocols that have been proposed by various authors.

Nguyen and Jennings~\cite{Nguyen2004MultipleConcurrentNegotiations} studied a model in which each negotiation thread between the buyer $\ag_0$ and a seller $\ag_i$ follows a small variation of the alternating offers protocol. The difference between their protocol and the standard AOP, is that in their protocol, whenever an offer is accepted, this is only considered a \textit{binding} agreement for the seller, but not for the buyer. This means that the buyer can make a separate agreement in each parallel negotiation thread, and then in the end, when all negotiations are finished, select one \textit{final} agreement among the accepted offers, and cancel all the other agreements he made in the other negotiation threads.

Later, however, the same authors argued that this model is unfair for the sellers~\cite{Nguyen2005ManagingCommitments}. So, they adapted their protocol by including `\textit{decommitment fees}'. Such a decommitment fee is a price the buyer must pay for every agreement he makes, but not selects as the final agreement. This  fee increases over time, so the later the buyer decides to cancel the agreement, the higher the fee.

An, Gatti and Lesser~\cite{An2009ExtendingAOP} proposed a protocol that, besides the standard  `propose' and `accept' actions, also includes an `exit' action and a `confirm' action. As in the ordinary AOP, after an agent $\ag_i$ (which can be either the buyer or a seller) proposes an offer, his opponent $\ag_j$ can reply to this with either a counter proposal or an `accept'. However, if she replies with an `accept', then this is not yet a binding agreement. Instead, the agent $\ag_i$ (who \textit{proposed} the offer) should now reply to this acceptance, but can only choose between `confirm' or `exit'. If he chooses `confirm', then the agreement becomes binding, but if he chooses `exit' then the negotiations between those two agents stop immediately, without agreement (but the other bilateral negotiation threads still continue). It is not hard to see that, as long as the buyer is careful enough not to send out more than one accept or confirm message at the same time, he can  guarantee himself to never reach more than one agreement.

%Note that this protocol still allows the buyer to make multiple agreements. However, the idea is that the buyer is only \textit{interested} in making one agreement, which is reflected in his utility function (making multiple agreements yields \textit{less} utility than making a single agreement).

Williams et al.~\cite{Williams2012NegotiatingConcurrently} proposed a combination of the previous two protocols. That is, just like An, Gatti and Lesser~\cite{An2009ExtendingAOP} they include `confirm' and `exit' actions in their protocol, which can be used in the same way as before, and with the same consequences, except that even after sending or receiving a `confirm' message, an agent may still cancel the agreement by sending a `decommit' message, which, however, forces that agent to pay a decommitment fee. 

Alrayes, Kafali, and Stathis \cite{Alrayes2018concurrent} proposed an even more elaborate protocol, in which an offer can either be definitively accepted (meaning one cannot decommit from it) or, instead, the buyer and seller may agree to `reserve' an offer, which means that either party can still decommit from it, but at the cost of a decommitment fee.

\section{Many-to-Many Negotiations}
We will now briefly describe Many-to-Many negotiations. We will keep this discussion short, because not much changes with respect to One-to-Many negotiations.

\begin{definition}
Let $b$ and $s$ be two positive integers. Then, a \textbf{many-to-many negotiation domain} with $b$ buyers and $s$ sellers, consists of:
\begin{itemize}
\item $b \times s$ offer spaces: $\Off_{1,1}, \Off_{1,2}, \dots, \Off_{b,s}$.
\item $b$ `buyer' utility functions $\util_i$:
\begin{equation}
\util_i \ : \ \prod_{j=1}^s \big( \Off_{i,j} \cup \{\dis\} \big) \rightarrow \mathbb{R}
\end{equation}
\item $s$ `seller' utility functions $\util_j$:
\begin{equation}
\util_j \ : \ \prod_{i=1}^b \big( \Off_{i,j} \cup \{\dis\} \big) \rightarrow \mathbb{R}
\end{equation}
where $\dis$ represents `disagreement' (i.e. a negotiation ending without agreement).
\end{itemize}
\end{definition}
As before, the various offer spaces $\Off_{i,j}$ do not necessarily have to be all different. The buyers and sellers also have reservation values, but they are defined by their utility functions:
\[\rv_i \ \ := \ \ \util_i(\dis, \dis, \dots, \dis)\]
\[\rv_j \ \ := \ \ \util_j(\dis, \dis, \dots, \dis)\]

Most of the protocols described in Section \ref{sec:one-to-many_protocols} can also be applied to many-to-many negotiations, without much modification.

A more general variant of this model, which has been used in various editions of the Supply Chain Management League (SCML) of the ANAC competition, divides the agents into more than two `layers'~\cite{Mohammad2019SCML}. That is, the agents are partitioned into disjoint subsets $\Ag  = \Ag_1 \cup \Ag_2 \cup \dots \cup \Ag_l$. Each agent in the \textit{first} layer $\Ag_1$ negotiates in a separate bilateral negotiation with every agent in the second layer $\Ag_2$. Similarly, each agent in the \textit{last} layer $\Ag_l$ negotiates with each agent in the second-last layer $\Ag_{l-1}$. Finally, for each \textit{intermediate} layer $\Ag_j$ (i.e. with $j \neq 1$ and $j\neq l$) each agent in that layer is involved in a separate bilateral negotiation with every agent from the previous layer $\Ag_{j-1}$ as well as with every agent from the next layer $\Ag_{j+1}$.

This model represents a supply chain, in which the agents in $\Ag_1$ are the suppliers of some source material which they sell to the agents in $\Ag_2$. The agents in $\Ag_2$ then process this material to produce some intermediate product which they sell to the agents in $Ag_3$. Those agents, in turn, process that product further to create yet another product which they sell to the agents in $\Ag_4$, etcetera. The agents in the last layer $Ag_l$, finally, are the consumers of the end product.

\section{All-Subsets Negotiations}

In all-subsets negotiations, all restrictions on who negotiates with whom are removed. Any agent can be negotiating with any other agent, in groups of arbitrary size.

A good example of this type of negotiation is a group of children that are exchanging collectible football cards. Any child may approach any other child to propose an exchange of cards. Moreover, it is even possible that multiple children agree to exchange cards in a circular fashion (e.g. Alice gives a card to Bob, Bob gives a card to Charles, and Charles gives a card to Alice). Another example could be the negotiations among political parties to form a coalition. Any party may propose to form a coalition with any subset of other political parties. Also, the negotiations that take place among players of the board game \textit{Diplomacy} are of this type.

For this type of negotiations, de Jonge and Sierra~\cite{deJonge2015nb3} proposed the \textit{unstructured negotiation protocol} (UNP), which is sometimes also called the unstructured \textit{communication} protocol. The idea behind the UNP is to impose the fewest constraints possible on the negotiators. That is:
\begin{itemize}
\item Any subset of 2 or more agents may be engaged in a separate parallel negotiation thread.
\item Within each such negotiation thread, the agents may make more than one agreement. So, after making an agreement, the negotiators may continue negotiating in order to come to a better deal.
\item Agents do not have to wait for their turn.  That is, after making a proposal, the agent is allowed to make another proposal immediately afterwards, without having to wait for any of the other agents to reply to the first proposal.
\item Agents are allowed to accept \textit{any} previous proposals, instead of only the \textit{last} one.
\end{itemize}

In a nutshell: at any time any agent may propose or accept any offer. In principle, the offer becomes a binding agreement whenever all agents that are involved in it have accepted it. However, there are two exceptions. Firstly, after proposing or accepting an offer, the same agent may withdraw his support for that offer. If he does so before all other agents have accepted it, then it can no longer become a binding agreement. However, if he does so \textit{after} the offer became a binding agreement, then he is too late, and the agreement remains  binding anyway. Secondly, if a certain offer $\off$ is incompatible with other offers that have already become binding agreements before, then $\off$ cannot become a binding agreement, even if all agents accept it.

In the UNP, once an offer becomes a binding agreement, this can never be reverted, even if the agents all agree that they want to cancel the agreement. At first sight, this may seem limiting and at odds with the philosophy of the UNP to impose as few restrictions as possible. However, the idea is that in reality this is not really a restriction, because instead of canceling the agreement, the agents can simply come to a \textit{new} agreement that undoes the \textit{effects} of the original agreement. For example, suppose Alice and Bob agree to exchange two football cards. That is, Alice gives a card $x$ to Bob, and in return Bob gives a card $y$ to Alice. Now, if later on they both regret this decision, then they can simply make a new agreement in which Alice gives card $y$ back to Bob, and Bob gives card $x$ back to Alice. 

We will discuss the UNP in more detail in the following sections. Note that we here define the UNP in a slightly different way from its original definition in \cite{deJonge2015nb3}.

\subsection{Negotiation Domains in the UNP}
As mentioned above, in the UNP each agent may negotiate with any non-empty subset of the other agents. This means that for each offer $\off$ in the offer space, we need to be able to tell which agents are participating in that specific offer. For example, if $\off$ represents an exchange of football cards in which agent $\ag_1$ gives a card to $\ag_3$, and $\ag_3$ gives a card to $\ag_7$, and $\ag_7$ gives a card to $\ag_1$, then we have $\pa(\off) = \{1, 3, 7\}$.

In general, whenever we talk about the `participating agents' of an offer, we mean the set of agents that need to accept it in order for it to become an agreement. Therefore, to formally define a negotiation domain for the UNP, we need to include a function $\pa$ in the definition, called the `participating agents function', which assigns to each offer $\off\in \Off$, the set of agents that are participating in it.

Furthermore, we mentioned that the UNP allows agents to make \textit{multiple} binding agreements. However, in some negotiation domains, it may happen that certain offers are incompatible with other offers. 

For example, suppose agent $\ag_i$ has a budget of \$10,000. First he agrees to buy car for \$6,000 (represented by offer $\off$). Next, he agrees to buy another car for \$3,000 (represented by offer $\off'$). After this, he may want to buy another car for \$3,000 (represented by offer $\off''$), but that will not be possible because he no longer has enough money left for that.  So, we say that $\off'$ is compatible with $\off$, but $\off''$ is not compatible with $\off$ and $\off'$.

In order to formally define a negotiation domain, we therefore have to be able to specify which offers are compatible with which other offers. For this we need to define a function $\feas: 2^\Off \rightarrow 2^\Off$ that, given any subset of offers $S \subseteq \Off$, maps it to the set of all offers that are compatible with $S$. In our example, the fact that $\off'$ was compatible with $\off$, would be denoted as $\off' \in \feas(\{\off\})$ and the fact that $\off''$ was not compatible with $\off$ and $\off'$, would be denoted as $\off'' \not \in \feas(\{\off, \off'\})$.

Together, this leads to the following definition.
\begin{definition}
Let $[\numAgents]$ denote the set of all integers from 1 to $\numAgents$. A \textbf{negotiation domain for the UNP} (with $\numAgents$ agents) consists of:
\begin{itemize}
\item An offer space $\Off$.
\item a \textbf{participating agents function}: $\pa : \Off \rightarrow 2^{[\numAgents]}$
\item A \textbf{feasibility function}: $\feas :  2^\Off \rightarrow 2^\Off$.
\item $\numAgents$ utility functions:
$\util_i : 2^\Off \rightarrow \mathbb{R}$
\end{itemize}
where $\pa$ must satisfy $\pa(\off) \geq 2$ for every $\off \in \Off$. 
\end{definition}

The agents also have reservation values, but they are defined implicitly as $\rv_i := \util_i(\emptyset)$.

\subsection{Negotiation Actions in the UNP}
%Although one could still say that the negotiations are divided into separate threads, one for each subset of agents, the UNP does not explicitly model the negotiations in this way. Instead, it assumes that each negotiation message includes the set of agents to which it is directed. We therefore define a negotiation action for the UNP as follows.

In the UNP, a negotiation action is defined as follows.
\begin{definition}\label{def:action_unp}
A \textbf{negotiation action for the UNP} is defined as a tuple:
\[(i, \actype, \off, t) \ \ \in  \ \  \{1,2\} \times\{\acc, \rej\} \times \Off \times \mathbb{R}^+\]
where $i$ represents the index of the agent performing the action,  $\actype$ represents the \textbf{type} of the action, which can be either the symbol $\acc$ (`accept'), or the symbol $\rej$ (`reject'), $\off$ is the offer that is being accepted or rejected, and $t$ is the time at which the agent accepts or rejects the offer.
\end{definition}

Whenever an agent accepts or rejects an offer $\off$, this message will be sent to all agents participating in the offer $\off$ (i.e. all agents $\ag_i$ with $i \in \pa(\off)$).

Note that the UNP does not define a ‘propose’ action. As we already discussed in Section~\ref{sec:proposals_and_acceptances}, this does not matter because there is no strict need to distinguish between `propose' or `accept' actions. So, while the UNP \textit{formally} does not include any `propose' actions,  \textit{informally}, we can still say that an agent is \textit{proposing} an offer $\off$, if he is the first agent to send and `accept' message regarding to that offer. Alternatively, we may in that case just say that that agent is \textit{accepting} the offer, even though he is the first to suggest offer $\off$. The main advantage of this is that we don't need to define any rules about when an agent is allowed to use the `propose' action and when it is allowed to use the `accept' action.

On the other hand, the UNP does define a `reject' action which is not present in the AOP or SAOP. The reject action has two purposes. Firstly, it allows the \textit{receiver} of a proposal to explicitly reject it. This may help the other agents to understand the desires of the rejecting agent, but this is purely optional. If an agent is not interested in a proposal, she may just as well completely ignore it and never reply to it. So, there is no formal difference between \textit{rejecting} a proposal and simply \textit{ignoring} it. The second, and more important, use of the `reject' action, however, is for an agent to withdraw his \textit{own} earlier proposal. That is, an agent $\ag_i$ may first propose some offer $\off$, but later change his mind and then reject this same offer himself, in order to avoid it from becoming a binding agreement. Note that in the AOP or SAOP such a withdrawal of a proposal is usually not necessary, because in those protocols one can only accept the \textit{last} offer anyway, so the moment that another agent replies to the proposal of $\off$ with counter proposal $\off'$, the first proposal $\off$ can no longer be accepted, so it is already withdrawn automatically. In the UNP, on the other hand, agents are allowed to accept any earlier proposals, so if the proposer changes his mind, he needs to be able to withdraw his earlier proposal explicitly.

%Normally, when some agent $\ag_i$ makes a proposal $(i, \recip, \acc, \off, t)$, the set $\recip$ should at least contain the set of participating agents $\pa(\off)$. After all, if $\recip$ is a strict subset of $\pa(\off)$ it means that not all agents participating in the offer receive the message, which means they cannot reply to it, which in turn means it can never become a  agreement. Such a proposal would therefore be rather pointless. Nevertheless, the UNP does allow agents to send proposals with $\recip \subsetneq \pa(\off)$, because the authors wanted to impose a minimal set of restrictions on the agents. They argued that it is simply the agents' own responsibility to select the correct set of recipients.
%
%Furthermore, the UNP also allows $\recip$ to be a strict \textit{superset} of $\pa(\off)$. In that case, it means that some agents will receive this message even though they are not participating in the offer. While this means they could reply to the proposal, such a reply from a non-participating agent would  not have any effect on the question whether it will become a binding agreement or not. Again, however, the UNP allows this, to impose a minimum set of restrictions. In addition, one could argue that it could be useful to include non-participating agents in a proposal, as a means to inform them of the fact that some other agents are discussing the offer $\off$, so as to put them under pressure to propose a better alternative.

\subsection{The Notary Agent}
The fact that the UNP allows agents to withdraw their own proposals, plus the fact that there are many different agents that can accept or reject any offers in any random order, plus the fact that different agents may receive the same messages at different times, makes that the UNP can easily get very messy. Specifically, each agent could end up with a different view of the order in which the various actions were taken, which means they could disagree about which offers have become binding agreements. 

To deal with this problem, the authors introduced the concept of a `notary' agent to keep track of the agreements. As explained in Section \ref{sec:notary_agents}, such notary agents are implicitly also present in most implementations of the alternating offers protocol. It is just that the increased complexity of the negotiations under the UNP now makes the necessity of a notary agent much more obvious.

So, for the implementation of the UNP, we assume that there is a special agent called the `notary' which does not participate in the negotiations, but that does receive every message sent by every negotiating agent. The main task of the notary is to establish a unique chronological order between the messages. That is, while different agents may have each received the various negotiation messages in a different order, the question whether or not an offer has become a binding agreement is determined by the order in which the \textit{notary} has received those messages. 

Furthermore, the notary may also be used to verify that any accepted agreement is consistent with earlier agreements, to inform the agents that an agreement has been made (whenever this happens), or to ensure that each message is sent to the correct recipients.

%Now, the question whether any offer is considered a binding agreement, is determined by the order in which the notary receives the agents' `accept' and `reject' messages. If, for any offer $\off$, all participating agents have accepted it, the notary may check if this offer is consistent with all binding agreements that have so far been made. If yes, then $\off$ also becomes a binding agreement, and then the notary will send a message to each agent participating in it, to notify them of this fact. If not, then offer $\off$ does not become a binding agreement.

\later{We might explain here that it can become a binding agreement later, after some other offers become binding agreements, but that requires at least one more agent to accept. This might be slightly different from the protocol that was originally proposed.}

\subsection{Formal Definition of the UNP}
As we have seen before, to define a negotiation protocol, we need to define two things: the set of histories that obey the protocol and, for any such history, which are the binding agreements.

A negotiation history $\hist$ for the UNP is defined in exactly the same way as for the AOP (Def.~\ref{def:history}), except that the actions $\ac_j$ in the history are now defined as in Def.~\ref{def:action_unp}. Furthermore, we should remark that any delay $\delay_j$ is to be interpreted as the time it takes for the message $\ac_j$ to arrive at the \textit{notary} agent (which may differ from the time it takes to arrive at any of the other agents).

Now, defining which such histories obey the UNP is simple: \textit{any} such history obeys the UNP, except that we require that an agent cannot send more than one message at the same time  and that, of course, the history has to take place within a given deadline.
\begin{definition}
A history $\hist$ satisfies the UNP (with deadline $T$) iff the following constraint holds:
\begin{itemize}
\item If $(i, \actype, \off, t) \in \hist$, and $(i', \actype', \off', t') \in \hist$, then we must have either $i\neq i'$ or $t\neq t'$.
\item For all negotiation actions $(i, \actype, \off, t) \in \hist$ we have $t\leq \dead$.
\end{itemize}
\end{definition}
Of course, in practice you can't really send two messages at exactly the same time anyway, so we impose this first constraint purely for mathematical reasons, to ensure the formalization is consistent with reality. It doesn't really impose any restriction on the agents themselves.

Let $\hist$ be some negotiation history that satisfies the UNP. Then we use the notation $\obsHist{not}$ to denote the sequence of actions as observed by the notary. For example, if 
\[\hist = \Big( (i_1, \actype_1, \off_1, t_1), \delay_1, (i_2, \actype_2, \off_2, t_2), \delay_2,  \dots, (i_k, \actype_k, \off_k, t_k), \delay_k \Big)\]
Then $\obsHist{not}$ would be:
\[\obsHist{not} = \Big((i_1, \actype_1, \off_1, t_1 + \delay_1),  (i_2, \actype_2, \off_2, t_2 + \delay_2),  \dots, (i_k, \actype_k, \off_k, t_k + \delay_k)\Big)\]
except, however, that the actions in $\obsHist{not}$ may appear in a different order than in $\hist$, because the delays may cause one message to arrive later than another message, even though it was sent earlier. For example, if $t_1 < t_2$, but $t_2 + \delay_2 < t_1 + \delay_1$, then we would have:
\[\obsHist{not} = \Big((i_2, \actype_2, \off_2, t_2 + \delay_2), (i_1, \actype_1, \off_1, t_1 + \delay_1),    \dots, (i_k, \actype_k, \off_k, t_k + \delay_k)\Big)\]

Now, in order to define how offers become binding agreements in the UNP, we first have to define the following concepts.
\begin{definition}
Let $\hist$ be some negotiation history (for the UNP) and $\obsHist{not}$ the corresponding history as observed by the notary. We use the notation $\mi{rej}_h(i, \off, t)$ to denote that agent $\ag_i$ has rejected the offer $\off$ at some time  \underline{after} the given time $t$ (from the point of view of the notary):
\[\mi{rej}_h(i, \off, t) \quad \leftrightarrow \quad \exists  t': \ \ (i,\rej, \off, t') \in \obsHist{not} \ \ \land \ \ t < t'\]
\end{definition}

\begin{definition}
Let $\hist$ be some negotiation history (for the UNP) and $\obsHist{not}$ the corresponding history as observed by the notary. Given such a history, we say that an offer $\off$ is \textbf{accepted by agent} $\ag_i$ iff there is an action $(i, \acc, \off, t) \in \hist$ and it was not rejected by that same agent afterwards. We will denote this as $acc_\hist(i, \off)$.
\[acc_\hist(i, \off) \quad \leftrightarrow \quad \exists t: (i,\acc,\off,t)\in \obsHist{not} \ \ \text{and} \ \  \lnot \mi{rej}_h(i,\off, t)\]
\end{definition}

\begin{definition}
Let $\hist$ be some negotiation history (for the UNP). Given such a history, we say that an offer $\off$ has been \textbf{fully accepted} iff it is accepted by all agents participating in $\off$. We denote this as $acc_\hist(\off)$. So:

\[acc_\hist(\off) \quad \leftrightarrow \quad \forall i\in \pa(\off): \ \ acc_\hist(i,\off)\]

\end{definition}

An offer $\off$ becomes a binding agreement if it is fully accepted and it is compatible with all previous binding agreements.

\begin{definition}
Let $\hist$ be some negotiation history (for the UNP) and $\hslash$ the corresponding history as observed by the notary (so we here use the notation $\hslash$ as an alternative to $\obsHist{not}$). Then, we define $Agr_{\hslash}$ to denote the set of offers that are considered officially binding agreements after that history. Of course, at the start of the negotiations no agreement has been made, so we define: 
\[
Agr_\emptyTuple := \emptyset
\]
where $\emptyTuple$ denotes the empty history. Now, let $\hslash$ be non-empty, let $\ac_k$ denote the last action of $\hslash$, and let $\hslash'$ denote the prefix of $\hslash$ before $\ac_k$ (so we have $\hslash = \hslash' \circ \ac_k$). Furthermore, let $\off$ be the offer accepted or rejected in $\ac_k$. Then we define:
\[
Agr_{\hslash} :=
\begin{cases}
Agr_{\hslash'} \cup \{\off\} & \text{if\ } \ acc_\hist(\off)\ \  \text{and} \ \ \off \in \feas(Agr_{\hslash'})\\
Agr_{\hslash'} & \text{otherwise}
\end{cases}
\]
\end{definition}

\subsection{Some Final Remarks about the UNP}
The UNP as defined here is slightly different from how it was defined in the original paper~\cite{deJonge2015nb3}. Here, we impose the constraint that when an agent accepts or rejects an offer $\off$, this message must be sent to all participating agents $\pa(\off)$ of that offer. In~\cite{deJonge2015nb3}, on the other hand, the authors allowed the sender to choose any set of recipients. The idea behind this, was that in practice it is hard to enforce the rules of a protocol onto the negotiating agents, so they wanted to impose as few restrictions on them as possible.

However, this makes the formalization somewhat more complicated, and moreover, it may lead to problems because it means an agent could withdraw his own earlier proposal without informing all the other participating agents. This, in turn could at some point cause the other agents to falsely believe the offer has become an agreement.

To avoid these problems, we therefore chose here to assume each message is always sent to all participating agents. We don't think that this is really much of a restriction, because we could for example implement the notary such that whenever it receives a message, it automatically forwards it to all participating agents. In this way we are guaranteed that our assumption holds, without imposing any constraint on the negotiating agents.

\chapter{Large and Complex Negotiation Domains}

Until now, we have always assumed that the negotiation domains were small and simple enough that we did not have to care about the computational complexity of negotiation. Most algorithms discussed so far, for example, were based on the assumption that we could simply iterate over all offers in the negotiation domain, and that the utility value of any offer could be calculated easily. In this chapter, however, we will look at what happens when we cannot make such assumptions anymore. 

A real-world negotiation domain can easily get extremely large. This is because the size of a multi-issue domain increases exponentially with the number of issues. For example, a negotiation domain with only 20 issues and with 10 options for each issue, already contains a total of $10^{20}$ possible offers, which is an astronomically large number (a \textit{hundred billion billion}). In such a large domain, even a simple time-based negotiation algorithm would fail if it uses a naive brute-force search to implement Equation~(\ref{eq:time_based_max}) or~(\ref{eq:time_based_min}), because that would be extremely time-consuming. Similarly, the MiCRO strategy would fail on such a domain, because it requires the agent to create a sorted list of all possible offers, and a list of $10^{20}$ offers would obviously never fit in our computer's memory.

A second reason why a real-world negotiation domain could be too complex for the algorithms we have discussed so far in this book, is that the agents' utility functions may be computationally complex. That is, for any given offer $\off$ it could be very time consuming to calculate its utility value $\util_1(\off)$. This is not the case for the linear utility functions we have mostly discussed so far, but if we allow the agents to have \textit{non-linear} utility functions, then this may certainly happen. We will give some examples of such negotiation domains later on in this chapter.

So, there are two main reasons why negotiations may be computationally complex: 
\begin{enumerate}
\item The offer space is very large, so one cannot apply brute-force search to find the next offer to propose.
\item For any single offer $\off$, the calculation of its utility value $\util_1(\off)$ is a computationally complex problem (e.g. NP-hard).
\end{enumerate}
It was therefore proposed that the BOA model should be extended to a BOASE model, where the letter $S$ stands for `search', and the letter $E$ stands for `evaluation'~\cite{deJonge2022multiObjectiveVRP}. In other words, when implementing a negotiating agent for highly complex negotiation domains, we do not only need to choose a bidding strategy, an opponent modeling algorithm and an acceptance strategy, but we also need to choose a search algorithm to \textit{find} the optimal offers to propose (according to the bidding strategy), as well as an evaluation algorithm to calculate the utility values of the offers.

So, in this chapter we will discuss what happens if:
\begin{itemize}
\item The size of the offer space becomes very large.
\item The utility functions become non-linear.
\item Calculating the utility value for any given deal is a computationally complex problem.
\end{itemize}

\section{Large Linear Domains}\label{sec:large_linear_domains}
We will start by only increasing the size of the negotiation domains. That is, we will here use the same type of linear negotiation domains that we introduced in Section \ref{sec:basic_negotiations}, except that we now assume there are many issues (say, 20 or more), yielding extremely large offer spaces. As explained above, such large negotiation domains require us to implement a specific \textit{search} algorithm. We will discuss two such search techniques, both introduced by Koça et al. \cite{koça2024}. Namely, the BIDS algorithm and the IPS algorithm.

Since the utility functions are still assumed linear, we don't have to care about \textit{evaluation} here. The utility values can still be calculated in an entirely straightforward way.

\subsection{The BIDS Algorithm}\label{sec:bids}
Suppose we have a very large negotiation domain with, for example, 20 issues and 10 options per issue, and we want to implement a time-based negotiation algorithm, based on an equation similar to Eq.~(\ref{eq:time_based_min}). To be precise, we will here assume our agent aims to propose an offer $\offNext$ for which the utility value $\util_1(\offNext)$ is as close as possible to the aspiration level $\asp(t)$:
\begin{equation}\label{eq:time_based_closest}
\offNext \quad = \quad \argmin_{\off \in \Off} \ \{ \ |\util_1(\off) - \asp(t)| \ \}
\end{equation}
In order to find this offer, we can use the BIDS algorithm (\textit{Bidding under Diversified Search}).

We assume our agent $\ag_1$ has a linear utility function $\util_1$. In order to simplify the notation we will here express it in the form of Eq.~(\ref{eq:linear_util_func_no_weights}), without weights. That is:
\[u_1(\off) = \sum_{j=1}^m \eval_1^j(\offComp{j})\]
where $\numIssues$ denotes the number of issues in the domain, and the offer is given as $\off = (\offComp{1}, \offComp{2}, \dots, \offComp{\numIssues})$. As explained in Section \ref{sec:linear_util_functions} the assumption that the utility function does not have weights is not a restriction, because any linear utility function with weights (of the form of Eq.~(\ref{eq:linear_util_func})) can be re-written into a form without weights, by `absorbing' the weights into the evaluation functions.

Now, let $k$ be an integer such that $0 < k \leq \numIssues$. Then, for any offer $\off$, we define $\partoff{k}$ to be the tuple consisting of only the first $k$ options of $\off$. We'll call this a \textbf{partial offer}. For example, if $\off = (\text{The Godfather}, \text{Rialto}, \text{Fri}, \text{20:00})$, then $\partoff{2} =  (\text{The Godfather}, \text{Rialto})$. We will use $\Partoff{k}$ to denote the set of all partial offers of size $k$:
\[\Partoff{k} \ \ := \ \ \issue{1} \times \issue{2} \times \dots \times \issue{k}\]
Of course, for $k = \numIssues$ we have $\Partoff{\numIssues} = \Off$, and for any offer $\off$ we have $\partoff{\numIssues} = \off$. 

For any partial offer $\partoff{k}$ we define its partial utility value $\util_1(\partoff{k})$ to be the sum of its first $k$ evaluations:
\begin{equation}\label{eq:util_partial_offer}
\util_1(\partoff{k}) \ \ := \ \ \sum_{j=1}^k  \eval_1^j(\offComp{j})
\end{equation}
For any partial offer $\partoff{k-1}$ of size $k-1$ and any option $x$ of issue $\issue{k}$ we use the notation $\partoff{k-1} \circ x$ to denote the partial offer of size $k$ resulting from concatenating the option $x$ to $\partoff{k-1}$. For example, if  we have $\partoff{2} =  (\text{The Godfather}, \text{Rialto})$ and $x = \text{Fri}$, then $\partoff{2} \circ x = (\text{The Godfather}, \text{Rialto}, \text{Fri})$

Let $\bids_k$ denote the function that, given any aspiration value $\asp \in \mathbb{R}$ returns the partial offer $\partoff{k}$ for which the utility value $\util_1(\partoff{k})$ is closest to $\asp$. That is:
\[\bids_k(\asp)  \ \ := \ \  \argmin_{\partoff{k} \in \Partoff{k}} \ \{\ |\util_1(\partoff{k}) - \asp| \ \}\] 
So, to implement our agent according to Eq.~(\ref{eq:time_based_closest}), we need to be able to find, for any given aspiration value $\asp$ the offer $\bids_\numIssues(\asp)$.

%Note that for any $k$, we can calculate $\bids_k(\asp)$ in a recurrent manner. That is, for any option $\offComp{k} \in \issue{k}$ we can calculate $\bids{k-1}(\asp - \eval_1^{k}(\offComp{k})$

%Let $\bidsUtil_k(\asp)$ denote the utility of the partial offer that comes closest to $\asp$. That is:
%\begin{eqnarray*}
%\bidsUtil_k(\asp) &:=& \util_1(\bids_k(\asp))\\
%&=&  \min \{|\util_1(\partoff{k}) - \asp| \mid \partoff{k} \in \Partoff{k}\}\\
%&\approx& \asp
%\end{eqnarray*}

%Note that $\bids_k$ satisfies the following recurrence relation:
%
%\begin{eqnarray*}
%\bids_k(\asp) \ \ = \ \ \arg \min_{\offComp{k} \in \issue{k}} \{\bidsUtil_{k-1}(\lambda - \eval_1^{k}(\offComp{k})) + \eval_1^{k}(\offComp{k})\}\\
%\bids_k(\asp) \ \ = \ \ 
%\end{eqnarray*}

%That is, to calculate $\bids_k(\asp)$, we can first take an arbitrary option $\offComp{k}$ from issue $\issue{k}$, calculate its evaluation $\eval_1^k(\offComp{k})$, which we subtract from $\asp$, and then find the partial offer of size $k-1$ for which the utility comes closest to $\asp - \eval_1^k(\offComp{k})$ Then, if we extend $\bids_{k-1}$ with $\offComp{k}$ we know that its utility will come close to $\asp$. Then, if we repeat this for all options $\offComp{k} \in \issue{k}$ and find the one that comes closest to $\asp$. 

To explain how the BIDS algorithm works, let us first suppose that we already have an algorithm that gives us, for any arbitrary value of $\asp$ the partial offer $\bids_{\numIssues-1}(\asp)$. Then, we can calculate $\bids_{\numIssues}(\asp)$ as follows:
\begin{enumerate}
\item Pick an option $x$ from issue $\issue{\numIssues}$.
\item Define $\asp_x := \asp - \eval_1^{\numIssues}(x)$
\item Obtain $\partoff{\numIssues-1}_x := \bids_{\numIssues-1}(\asp_x)$.
\item Define $\partoff{\numIssues}_x$ by concatenating $\partoff{\numIssues-1}_x$ with $x$. That is:
\[\partoff{\numIssues}_x \quad := \quad  \partoff{\numIssues-1}_x \circ x\]
Note that, by Eq.~(\ref{eq:util_partial_offer}), this implies $\util_1(\partoff{\numIssues}_x) = \util_1(\partoff{\numIssues-1}) + \eval_1^\numIssues(x)$ and since $\util_1(\partoff{\numIssues-1}_x) \approx \asp_x$ it follows that $\util_1(\partoff{\numIssues}_x) \approx \asp$.
\item Repeat steps 1--4 for all options $x \in \issue{\numIssues}$.
\item We now have a number of offers $\partoff{\numIssues}_x$, one for each option $x$ in $\issue{\numIssues}$. Pick the one for which the utility is closest to $\asp$, and return it. 
\end{enumerate}
So, we see that given an algorithm for $\bids_{\numIssues-1}$, we can calculate $\bids_\numIssues(\asp)$ by making $|\issue{\numIssues}|$ calls to that algorithm. 

Of course, this only works if we already have an algorithm for $\bids_{\numIssues-1}$, so we we are now left with the question how to implement that algorithm. Now, to answer that question, we simply assume that we have an algorithm  for $\bids_{\numIssues-2}$ and apply exactly the same approach as above, which allows us to calculate $\bids_{\numIssues-1}(\asp)$ by making $|\issue{\numIssues-1}|$ calls to $\bids_{\numIssues-2}$, which, in turn, implies that we can calculate $\bids_{\numIssues}(\asp)$ by making $|\issue{\numIssues-1}|\times |\issue{\numIssues}|$ calls to $\bids_{\numIssues-2}$. 

Now, to implement $\bids_{\numIssues-2}$, we can repeat the same argument, etcetera, etcetera, and in the end we conclude that we only need an algorithm for $\bids_1$ and that we can calculate $\bids_{\numIssues}(\asp)$ by making $|\issue{2}| \times |\issue{3}| \times \dots \times |\issue{\numIssues}|$ calls to $\bids_1$. Now, the good news is that $B_1$ is very easy to implement, as a brute-force search over $\issue{1}$. The bad news, however, is that the number of calls we have to make to $\bids_1$ increases exponentially with the number of issues, which means that in this way we have not gained anything with respect to a direct brute-force search over the entire offer space $\Off$,

%Similarly, if we have an algorithm for $\bids_{\numIssues-2}$, then $\bids_{\numIssues-1}(\asp)$ can be calculated by making $|\issue{\numIssues-1}|$ calls to the algorithm for $\bids_{\numIssues-2}$, in exactly the same way as explained above. Combining this, we observe that $\bids_\numIssues(\asp)$ can be calculated by making $|\issue{\numIssues-1}| \times |\issue{\numIssues}|$ calls to an algorithm for $\bids_{\numIssues-2}$, etcetera. In the end, this means that in theory we only need an algorithm for $\bids_1$ that needs to be called a total of $|\issue{1}| \times |\issue{2}| \times \dots \times |\issue{\numIssues}|$ times. 

However, there is a simple trick that allows us to strongly reduce the number of computations. That is, instead of calculating the exact values $\bids_k(\asp)$, we only calculate \textit{approximate} values $\bids_k(\tilde{\asp})$ where $\tilde{\asp}$ is, for example, $\asp$ rounded off to 2 decimals.

To see why this makes a big difference, suppose that there are two options $x, y \in \issue{\numIssues-1}$ and two options  $x', y' \in \issue{\numIssues-2}$ such that:
\[\eval_1^{\numIssues-1}(x) = 0.020 \quad \eval_1^{\numIssues-2}(x') = 0.030\]
\[\eval_1^{\numIssues-1}(y) = 0.024 \quad \eval_1^{\numIssues-2}(y') = 0.027\]
This means that during the calculation of $\bids_\numIssues(\asp)$, at some moment, we need to calculate:
\[\bids_{\numIssues-2}\big(\asp - \eval_1^{\numIssues-1}(x) - \eval_1^{\numIssues-2}(x')\big) \ \ = \ \  \bids_{\numIssues-2}(\asp - 0.050)\]
and at some later moment we need to calculate: 
\[\bids_{\numIssues-2}\big(\asp - \eval_1^{\numIssues-1}(y) - \eval_1^{\numIssues-2}(y')\big) \ \ = \ \  \bids_{\numIssues-2}(\asp - 0.051)\]
Now, if we round these numbers off, then, instead of calculating the partial offer $\bids_{\numIssues-2}(\asp - 0.051)$ the algorithm can simply return the partial offer ${\bids_{\numIssues-2}(\asp - 0.050)}$ that it already calculated before. This saves a lot of computations. While this means that the algorithm will no longer be 100\% accurate, it makes a \textit{huge} difference in computation time.

%The BIDS algorithm therefore assumes, for each value of $k \in \{0,1,\dots \numIssues\}$, a finite set of values for the utility $\util_1(\partoff{k})$ of any partial offers.

To make this precise, let $\util_1^{min,k}$ and $\util_1^{max,k}$ denote the minimum and maximum possible utility values for any partial offer of size $k$:
\[\util_1^{min,k} \ \ := \ \ \min \ \{ \util_1(\partoff{k}) \mid \partoff{k} \in \Partoff{k} \} \]
\[\util_1^{max,k} \ \ := \ \ \max \ \{ \util_1(\partoff{k}) \mid \partoff{k} \in \Partoff{k} \} \]
Furthermore, let $\precision$ be some positive integer called the \textit{precision} parameter. Then, for each value of $k$ we can define the \textit{range} $\range{\precision}{k}$ as a discretization of the interval $[\util_1^{min,k} , \util_1^{max,k}]$ into $\precision$ equally spaced points:
\[\range{\precision}{k} \ \ := \ \ \{ \ (\util_1^{max,k}-\util_1^{min,k})\cdot \frac{j}{\precision-1} +\util_1^{min,k} \ \mid \ j= 0, 1, 2, \dots, \precision-1 \ \}\]
For example, if $\util_1^{min,k} = 20$, $\util_2^{max,k} = 80$, and $q=7$, then we have:
\[\range{\precision}{k} = \{20, 30, 40, 50, 60, 70, 80\}\]
The idea behind the BIDS algorithm is that it creates a table of size $\numIssues \times \precision$ that stores, for each $k \in \{1, 2, \dots, \numIssues\}$ and each $\tilde{\asp} \in \range{\precision}{k}$ the (partial) offer $\bids_{k}(\tilde{\asp})$. The algorithm starts by filling the first row of the table with all such partial offers for $k=1$. Then, it uses those partial offers to calculate the partial offers for $k=2$ as explained above, which will fill the second row, etcetera, until the entire table is filled. This is displayed in Algorithm~\ref{alg:bids}.  

The key point, is that once it has filled row $k-1$ of the table, it can determine any (partial) offer $\bids_{k}(\asp)$ in $O(|\issue{k}|)$ time, because instead of \textit{calculating} the partial offer  $\bids_{k-1}(\asp_x)$ for the \textit{exact} value $\asp_x$ it can just look up the partial offer $\bids_{k-1}(\tilde{\asp}_x)$ from the table for the \textit{approximate} value $\tilde{\asp}_x$. So, if each issue has $d$ options, then the full table can be calculated in $O(\numIssues \times \precision \times d)$ time, which is a lot faster than the $O(d^\numIssues )$ time a brute-force calculation would require.

Of course, as always, there is no free lunch. As mentioned above, the algorithm is not perfectly accurate, but at least we can control its accuracy. The higher the value of $\precision$, the more accurate the algorithm will be, but also the slower it will run.

So, at the start of the negotiations, our agent will first run the BIDS algorithm to fill the table. Then, each time it is our agent's turn to propose a new offer, it first calculates its aspiration level $\asp(t)$ for that time $t$. Next, it determines the value $\tilde{\asp}$ in $R_{\precision,\numIssues}$ that is closest to $\asp(t)$, and finally looks up the offer $\bids_{\numIssues}(\tilde{\asp})$ from the table and proposes it (unless, of course, the agent's acceptance strategy decides that it's better to accept the previously received offer).

Unfortunately, one disadvantage of BIDS, is that it doesn't allow us to take into account the constraint that the agent should only consider offers it hasn't proposed yet ($\off \not \in \Pro_{t}$), which we see in Equation (\ref{eq:time_based_min}). This is not a very big problem, because the alternating offers protocol does allow the agent to simply repeat the same offer, but it is sub-optimal, because it means the agent misses the opportunity to propose some other offer that the opponent might be more willing to accept.

\later{Can we adapt it to follow exactly Eq.~(\ref{eq:time_based_min})?}

\begin{algorithm}
\caption{The BIDS algorithm}\label{alg:bids}
\begin{algorithmic}[1]
	\Statex \textbf{Input:} 
	\Statex $\Off$ \Comment{The offer space (consisting of $\numIssues$ issues $\issue{k}$).}
	\Statex $\util$ \Comment{The agent's own utility function (consisting of one }
	\Statex	\Commentt{evaluation function $\eval^k$ for every issue $\issue{k}$)}
	\Statex $q$ \Comment{The precision parameter (a positive integer).}
    \Statex
    \State $\bids \leftarrow$ Initialize a table  of size $\numIssues \times \precision$.
    \State $\util^{max}[0] \leftarrow 0$ 
    \State $\util^{min}[0] \leftarrow 0$ 
    \For{$k \in \{1, 2, \dots, \numIssues\}$}
    		\comment{Determine max. and min. util. that a partial offer of size $k$ may have.}
    		\State $\util^{max}[k] \leftarrow \util^{max}[k-1] + \max\ \{\eval^k(x) \mid x \in \issue{k} \}$
    		\State $\util^{min}[k] \leftarrow \util^{min}[k-1] + \min\ \{\eval^k(x) \mid x \in \issue{k} \}$
    		\Statex
    		\For{$l \in \{0,1, \dots, \precision-1\}$}
    			\comment{Each index $l$ corresponds to a value of $\lambda$ between $\util^{min}[k]$ and $\util^{max}[k]$}
    			\State $\lambda \leftarrow (\util^{max}[k] - \util^{min}[k])\frac{l}{\precision-1} + \util^{min}[k]$ 
    				\Statex
    				\For{$x \in \issue{k}$}
    					\If{$k = 1$}
    						\State $\partoff{k} \leftarrow x$
    					\Else
    					\comment{Subtract the evaluation of option $x$ from $\asp$}
    					\State $\asp' \leftarrow \asp -  \eval^k(x)$
\comment{Convert the result back to an index (i.e. re-scale it to a number between 0 and $\precision-1$ and then round it off to an integer).}    					
    					\State $l' \leftarrow round((\precision -1) \cdot \frac{\asp' - \util^{min}[k-1]}{\util^{max}[k-1] - \util^{min}[k-1]})$
    					\Statex
    					\comment{Get the partial offer of size $k-1$ for which the utility is closest to $\asp'$:}
    					\State $\partoff{k-1} \leftarrow 	\bids[k-1,l']$
			\Statex    			
    			\comment{Extend it with option $x$:}
    					\State $\partoff{k} \leftarrow \partoff{k-1} \circ x$
    					\EndIf
    			\Statex  
    			\comment{If the utility of the offer $\partoff{k}$ is closer to $\asp$ than the offer currently stored in the table (or if this table entry is still empty), then update the table with this new offer:}
    			\If{$\bids[k,l] = null$ or $|\util_1(\partoff{k} ) - \asp| < |\util_1(\bids[k,l]) - \asp|$}
    				\State $\bids[k,l] \leftarrow \partoff{k}$ 
    				\EndIf
    			\EndFor
    		\EndFor
    \EndFor
\end{algorithmic}
\end{algorithm}

\begin{exercise}\textbf{BIDS.} Implement the BIDS algorithm, as displayed in Algorithm~\ref{alg:bids}. Furthermore, implement a time-based or adaptive agent that uses the BIDS algorithm to find its offers to propose and test it on a very large domain with, say, 20 issues and 10 options per issue.
\end{exercise}

\subsection{Iterated Pareto Search}

Next, let us suppose again we have a large negotiation domain with, for example, 20 issues and 10 options per issue, but this time we want to implement a time-based negotiation algorithm based on an equation similar to Eq.~(\ref{eq:time_based_max}). Specifically, we assume it uses the following equation to determine its next offer to propose $\offNext$.
\begin{equation}\label{eq:time_based_ips}
\offNext \quad = \quad \argmax_{\off\in \Off }\ \{\ \est{\util}_2(\off) \mid  \util_1(\off) \geq \asp(t) \ \}
\end{equation}
This means we now have to take \textit{two} utility functions into consideration: our own agent's utility function $\util_1$ and the estimation of our opponent's utility function $\est{\util}_2$, which may be obtained from some opponent modeling algorithm as discussed in Section~\ref{sec:opp_util_modeling} (in case we actually have the opponent's \textit{true} utility function $\util_2$, everything discussed here stays the same, except that we can replace $\est{\util}_2$ with $\util_2$). 

For this, our agent can use an algorithm called \textit{Iterated Pareto Search} (IPS). The idea behind this approach is based on two important observations. The first observation, is that Eq.~(\ref{eq:time_based_ips}) is equivalent to the following equation:
\begin{equation}\label{eq:time_based_ips_rewritten}
\offNext \quad = \quad \argmin_{ \off\in \pareto{\Off} }\ \{\ \util_1(\off) \ \mid \ \util_1(\off) \geq \asp(t) \ \}
\end{equation}
where $\pareto{\Off}$ denotes the Pareto-set (i.e. the set of all Pareto-optimal offers). So, instead of searching for the offer with maximum opponent utility, we can search for the offer with minimum utility for ourselves, among those offers that are Pareto-optimal.

The second observation, is that the Pareto-set is typically \textit{much} smaller than the full offer space. For example, it is perfectly possible that a randomly generated offer space with $10^{20}$ offers, only has a few hundred Pareto-optimal offers. So, once we have obtained the Pareto-set, even a very naive and straightforward implementation of  Eq.~(\ref{eq:time_based_ips_rewritten}) would take very little time.

So, the idea is that at any time $t$, our agent will first use the IPS algorithm to determine the Pareto-set (based on its estimation $\est{\util}_2$ of the opponent's utility function at that time), and then pick the offer $\off$ from this set for which his own utility $\util_1(\off)$ is closest to (but still greater than or equal to) its current aspiration value $\asp(t)$.

We will discuss two versions of IPS: `\textit{exact IPS}' and `\textit{discretized IPS}'.

\subsubsection{Pareto-optimality}
Before we can explain the IPS algorithm, let us first extend our previous definition of Pareto-optimality for \textit{offers} (Def.~\ref{def:pareto_optimal}), to \textit{partial} offers and \textit{options}.  

Just as in Section \ref{sec:bids} we will assume the utility functions of both agents are given in the form of Eq.~(\ref{eq:linear_util_func_no_weights}), without weights. To simplify notation we will use the symbols $\util_2$ and $\eval_2^j$ even though in practice the implementation would typically use the \textit{estimated} utility function $\est{\util}_2$ and evaluation functions $\est{\eval}_2^j$.

\begin{definition}
For any issue $\issue{j}$, and any pair of options $x, x' \in \issue{j}$ of that issue, we say that $x$ is \textbf{dominated} by $x'$, denoted $\domin(x,x')$, iff the following both hold:
\[\forall i\in \{1,2\}: \quad \eval_i^{j}(x) \leq \eval_i^{j}(x') \]
\[\exists i\in \{1,2\}: \quad \eval_i^{j}(x) < \eval_i^{j}(x') \]
and we say an option $x \in \issue{j}$ is \textbf{Pareto-optimal} iff it is not dominated by any other option of the same issue:
\[\not \exists \ x' \in \issue{j}: \quad \domin(x,x')\]
We will use $\pareto{\issue{j}}$ to denote the set of Pareto-optimal options of $\issue{j}$.

Similarly, let $\partoff{k}$ and ${\partoff{k}}'$ denote two partial offers of size $k$. Then we say that $\partoff{k}$ is \textbf{dominated} by ${\partoff{k}}'$, denoted $\domin(\partoff{k},{\partoff{k}}')$ iff the following both hold:
\[\forall i\in \{1,2\}: \quad \util_i(\partoff{k}) \leq \util_i({\partoff{k}}') \]
\[\exists i\in \{1,2\}: \quad \util_i(\partoff{k}) < \util_i({\partoff{k}}') \]
and we say a partial offer $\partoff{k}$ is \textbf{Pareto-optimal} iff it is not dominated by any other partial offer of the same size:
\[\not \exists \ {\partoff{k}}' \in \Partoff{k}: \quad \domin({\partoff{k}}, {\partoff{k}}')\]
We will use $\pareto{\Partoff{k}}$ to denote the set of Pareto-optimal partial offers of size $k$.

More generally, if $S \subseteq \Partoff{k}$ is any subset of the set of partial offers of size $k$, then we say a partial offer $\partoff{k} \in S$ is \textbf{Pareto-optimal in} $\mathbf{S}$ iff it is not dominated by any other partial offer in $S$:
\[\not \exists \ {\partoff{k}}' \in S: \quad \domin({\partoff{k}}, {\partoff{k}}')\]
We will use $\pareto{S}$ to denote the set of  partial offers that are Pareto-optimal in $S$. 
\end{definition}

\subsubsection{Exact IPS}\label{sec:ips}
The exact IPS algorithm allows us to find the Pareto-set $\pareto{\Off}$ efficiently, as follows:
\begin{enumerate}
\item Determine the Pareto-set $\pareto{\issue{1}}$ of issue $\issue{1}$.
\item Determine the Pareto-set $\pareto{\issue{2}}$ of issue $\issue{2}$.
\item Create the Cartesian product $\pareto{\issue{1}} \times \pareto{\issue{2}}$. 
\item Determine the set of Pareto-optimal partial offers within that Cartesian product. That is, determine the set $\pareto{ \ \pareto{\issue{1}} \times \pareto{\issue{2}}\ }$.
Thanks to Theorem \ref{thm:pareto} (see below) and the observation that ${\Partoff{1} = \issue{1}}$ we know that this is exactly equal to the set $\pareto{\Partoff{2}}$.
\item Determine the Pareto-set $\pareto{\issue{3}}$ of issue $\issue{3}$.
\item Create the Cartesian product  $\pareto{\Partoff{2}} \times \pareto{\issue{3}}$.
\item Determine the set of Pareto-optimal partial offers within this set. That is, determine $\pareto{ \ \pareto{\Partoff{2}} \times \pareto{\issue{3}}\ }$. Again, thanks to Theorem~\ref{thm:pareto}, we know that this is exactly $\pareto{\Partoff{3}}$.
\item Keep repeating this until we have the set $\pareto{\Partoff{\numIssues}}$, which is the set we are looking for.
\end{enumerate}

\subsubsection{Example}

Let us look at an example with 3 issues and 4 options per issue, as displayed in Table~\ref{tab_ips_1}. Each column corresponds to an issue $\issue{j}$ and a cell in row $l$ and column $j$ corresponds to option $l$ of issue $j$, which we here denote as $x_{j,l}$. The numbers in that cell represent the evaluations $\eval_1^{j}(x_{j,l})$ and $\eval_2^j(x_{j,l})$ of that option for agents $\ag_1$ and $\ag_2$ respectively.

\begin{table}
\begin{center}
\begin{tabular}{l|c|c|c|c|}
 & $\issue{1}$ & $\issue{2}$ & $\issue{3}$ \\
 \hline
 $x_{j,1}$ & $(0.48\ ,\ 0.50)$ & $(0.28\ ,\ 0.35)$ & $(0.60,0.40)$  \\
 $x_{j,2}$ & $(0.30\ ,\ 0.20)$ & $(0.20\ ,\ 0.25)$ & $(0.40,0.50)$  \\
 $x_{j,3}$ & $(0.60\ ,\ 0.30)$ & $(0.24\ ,\ 0.50)$ & $(0.30,0.30)$  \\
 $x_{j,4}$ & $(0.54\ ,\ 0.25)$ & $(0.40\ ,\ 0.35)$ & $(0.20,0.20)$\\
\end{tabular}
\caption{Example domain with 3 issues and 4 options per issue.}\label{tab_ips_1}
\end{center}
\end{table}

The algorithm now proceeds as follows:
\begin{enumerate}
\item For issue $\issue{1}$, note that option $x_{1,2}$ has utility vector $(0.30\ ,\ 0.20)$, which is dominated by both $x_{1,1}$ (with utility vector $(0.48\ ,\ 0.50)$) and $x_{1,3}$ (with utility vector $(0.60\ ,\ 0.30)$), and that option $x_{1,4}$ (with utility vector $(0.54\ ,\ 0.25)$) is also dominated by $x_{1,3}$. However, $x_{1,1}$ does not dominate $x_{1,3}$ and $x_{1,3}$ does not dominate $x_{1,1}$. Therefore, we have $\pareto{\issue{1}} = \{x_{1,1} , x_{1,3}\}$
\item For issue $\issue{2}$, note that option $x_{2,1}$ is dominated by $x_{2,3}$ and that option $x_{2,2}$ is dominated by every other option. So, we have $\pareto{\issue{2}} = \{x_{2,3} , x_{2,4}\}$.
\item The Cartesian product $\pareto{\issue{1}} \times \pareto{\issue{2}}$ is given by:
\[\{\ (x_{1,1},x_{2,3})\ ,\ (x_{1,1} , x_{2,4})\ ,\ (x_{1,3}, x_{2,3})\ ,\ (x_{1,3} , x_{2,4}) \  \}\]
\item The utility vectors of these partial offers are:
\[\vec{\util}(x_{1,1}, x_{2,3}) \ \ = \ \ (0.48\ ,\ 0.50) + (0.24\ ,\ 0.50) \ \ = \ \  (0.72\ ,\ 1.00)\]
\[\vec{\util}(x_{1,1} , x_{2,4})\ \ = \ \  (0.48\ ,\ 0.50) + (0.40\ ,\ 0.35)\ \ = \ \  (0.88\ ,\ 0.85)\]
\[\vec{\util}(x_{1,3}, x_{2,3})\ \ = \ \  (0.60\ ,\ 0.30) + (0.24\ ,\ 0.50) \ \ = \ \  (0.84\ ,\ 0.80)\]
\[\vec{\util}(x_{1,3} , x_{2,4}) \ \ = \ \  (0.60\ ,\ 0.30) + (0.40\ ,\ 0.35)\ \ = \ \ (1.00\ ,\ 0.65)\]
We see that here the only partial offer that is dominated, is $(x_{1,3},x_{2,3})$ which is dominated by $(x_{1,1},x_{2,4})$. So, the Pareto-set consists of the three other partial offers:
\begin{eqnarray*}
\pareto{\Partoff{2}} &=& \pareto{\ \pareto{\issue{1}} \times \pareto{\issue{2}}\ }\\
&=&  \{\ (x_{1,1} , x_{2,3})\ ,\ (x_{1,1} , x_{2,4})\ ,\ (x_{1,3} , x_{2,4}) \  \}\\
\end{eqnarray*}
\item Next, we note that $\pareto{\issue{3}}$ equals $\{x_{3,1} , x_{3,2}\}$
\item and therefore that
\begin{eqnarray*}
\pareto{\Partoff{2}} \times \pareto{\issue{3}}  &=& \{\ (x_{1,1} , x_{2,3} ,  x_{3,1})\ ,\ (x_{1,1} , x_{2,4} , x_{3,1})\ ,\ (x_{1,3} , x_{2,4} , x_{3,1}) \ ,\\
& & \quad (x_{1,1} , x_{2,3} , x_{3,2})\ ,\ (x_{1,1} , x_{2,4} ,  x_{3,2})\ ,\ (x_{1,3} , x_{2,4}, x_{3,2}) \ \}
\end{eqnarray*}
\item The utility vectors of these partial offers are:
\[\vec{\util}(x_{1,1}, x_{2,3}, x_{3,1}) \ \ = \ \ (0.72\ ,\ 1.00) + (0.60\ ,\ 0.40) \ \ = \ \  (1.32\ ,\ 1.40) \]
\[\vec{\util}(x_{1,1} , x_{2,4}, x_{3,1}) \ \ = \ \  (0.88\ ,\ 0.85) + (0.60\ ,\ 0.40)\ \ = \ \ (1.48\ ,\ 1.25 ) \]
\[\vec{\util}(x_{1,3} , x_{2,4}, x_{3,1}) \ \ = \ \  (1.00\ ,\ 0.65) + (0.60\ ,\ 0.40)\ \ = \ \ (1.60\ ,\ 1.05 )\]
\[\vec{\util}(x_{1,1}, x_{2,3}, x_{3,2}) \ \ = \ \ (0.72\ ,\ 1.00) + (0.40\ ,\ 0.50) \ \ = \ \ (1.12\ ,\ 1.50)  \]
\[\vec{\util}(x_{1,1} , x_{2,4}, x_{3,2}) \ \ = \ \  (0.88\ ,\ 0.85) + (0.40\ ,\ 0.50)\ \ = \ \ (1.28\ ,\ 1.35) \]
\[\vec{\util}(x_{1,3} , x_{2,4}, x_{3,2}) \ \ = \ \  (1.00\ ,\ 0.65) + (0.40\ ,\ 0.50)\ \ = \ \ (1.40\ ,\ 1.15) \]
\item Finally, removing the dominated elements from this set, we find that the Pareto-set $\pareto{\Off}$ is given by:
\begin{eqnarray*}
\pareto{\Off} &=& \{\ (x_{1,1}, x_{2,3}, x_{3,1}) \ , \ (x_{1,1} , x_{2,4}, x_{3,1}) \\
& & \quad (x_{1,3} , x_{2,4}, x_{3,1})\ , \ (x_{1,1}, x_{2,3}, x_{3,2})\ \}\\
\end{eqnarray*}
\end{enumerate}

\subsubsection{Correctness of the Algorithm}
The correctness of the IPS algorithm is based on Theorem~\ref{thm:pareto} below. In order to prove this theorem, we first need the following three lemmas.

%\begin{lemma}\label{lem:pareto}
%Let $\dom$ be some bilateral negotiation domain with $\numIssues$ issues and linear utility functions.
%Then, for any integer $k$ with $1\leq k \leq \numIssues$ we have:
%\[\pareto{\Partoff{k}} \ \ \subseteq \ \  \pareto{\issue{1}} \times  \pareto{\issue{2}} \times \dots \times  \pareto{\issue{k}} \]
%\end{lemma}
%\begin{proof}
%Suppose the opposite: that there exists a Pareto-optimal element $\partoff{k} = (x_1, x_2, \dots, x_k)$ in $\Partoff{k}$ that is not contained in $\pareto{\issue{1}} \times  \pareto{\issue{2}} \times \dots \times \pareto{\issue{k}}$. This means there is some integer $j$ with $1\leq j \leq k$ such that the option $x_j$ of $\partoff{k}$ is not contained in $\pareto{\issue{j}}$. But that means there exists an option $x_j' \in \issue{j}$ that dominates $x_j$. 
%
%Now, let us define ${\partoff{k}}'$ to be equal to $\partoff{k}$, except that $x_j$ has been replaced by $x_j'$. Then, by the assumption that the utility functions are linear, it follows directly that $\partoff{k}$ is dominated by ${\partoff{k}}'$, which contradicts the assumption that $\partoff{k}$ was Pareto-optimal.
%\end{proof}

\begin{lemma}\label{lem:pareto1}
Let $\dom$ be some bilateral negotiation domain with $\numIssues$ issues of finite size and linear utility functions.
Then, for any integer $k$ with $1 < k \leq \numIssues$ we have:
\[\pareto{\Partoff{k}} \ \ \subseteq \ \  \pareto{\Partoff{k-1}}   \times \pareto{\issue{k}} \]
\end{lemma}
\begin{proof}
Suppose the contrary: that there exists a Pareto-optimal element $\partoff{k} = (x_1, x_2, \dots, x_k) \in \pareto{\Partoff{k}}$ that is not contained in $\pareto{\Partoff{k-1}} \times \pareto{\issue{k}}$. This means that either $(x_1, x_2, \dots, x_{k-1}) \not \in \pareto{\Partoff{k-1}}$, or that $x_k \not \in \pareto{\issue{k}}$.

In the first case, this means that there is some partial offer ${\partoff{k-1}}' \in \pareto{\Partoff{k-1}}$ that dominates $(x_1, x_2, \dots, x_{k-1})$. However, since the utility functions are linear, that would mean that ${\partoff{k-1}}' \circ x_k$ dominates $\partoff{k}$, which is in contradiction to our assumption that $\partoff{k}$ was Pareto-optimal. 

In the second case, this means there exists an option $x_k' \in \issue{j}$ that dominates $x_k$. But in that case $(x_1, x_2, \dots, x_{k-1}, x_k')$ would dominate $\partoff{k}$, which again contradicts the assumption that $\partoff{k}$ was Pareto-optimal.
\end{proof}

\begin{lemma}\label{lem:pareto2}
Let $\dom$ be some bilateral negotiation domain with $\numIssues$ issues of finite size and linear utility functions.
Then, for any integer $k$ with $1 < k \leq \numIssues$ we have:
\[\pareto{\Partoff{k}} \ \ \subseteq \ \  \pareto{\  \pareto{\Partoff{k-1}}   \times \pareto{\issue{k}}\ }\]
\end{lemma}
\begin{proof}
To simplify the notation, define:
\[A \ := \ \Partoff{k} \quad \quad B \ := \ \pareto{\Partoff{k-1}}   \times \pareto{\issue{k}}\]
Note that we have $B \subseteq A$ and, by Lemma \ref{lem:pareto1}, we have $\pareto{A} \subseteq B$. Now, our goal is to prove: 
\[\pareto{A} \ \ \subseteq \ \ \pareto{B}\]
To prove this, let us assume the contrary: that there exists a partial offer $\partoff{k} \in \pareto{A}$ that is not contained in $\pareto{B}$. Now, since $\pareto{A} \subseteq B$, we have that $\partoff{k} \in B$, so the fact that $\partoff{k} \not \in \pareto{B}$ means there exists at least one other partial offer ${\partoff{k}}' \in B$ that dominates $\partoff{k}$. But since $B \subseteq A$, we have that ${\partoff{k}}' \in A$. So, $\partoff{k}$ and ${\partoff{k}}'$ are both contained in $A$. But then the fact that  $\partoff{k}$ is dominated by ${\partoff{k}}'$ contradicts the assumption that $\partoff{k} \in \pareto{A}$.
%
%Note that $\pareto{B}$ is obtained from $B$ by removing its dominated elements. However, since all elements of $B$ are also elements of $A$, any such elements that are removed, are elements of $A$ that are dominated in $A$, and thus they are not elements of $\pareto{A}$. 
%
%In other words, the elements that are removed from $B$ to obtain $\pareto{B}$ are not contained in $\pareto{A}$ and therefore $\pareto{A} \subseteq \pareto{B}$.
%
%More formally:
%
%Let $b \in B \setminus \pareto{B}$. This means it is an element of $B$ that is dominated by some other element in $B$. But since $B \subseteq A$, we have that $b$ is also an element of $A$ that is dominated by some other element in $A$ and thus, $b \in A \setminus \pareto{A}$.
%
%\essential{finish this}
\end{proof}

%\begin{lemma}\label{lem:pareto2}
%If a partial offer is Pareto-optimal in $\pareto{\issue{1}} \times  \pareto{\issue{2}} \times \dots \times  \pareto{\issue{k}}$, then it is also Pareto-optimal in $\Partoff{k}$.
%\end{lemma}
%\begin{proof}
%Note that we can reformulate the lemma as: ``If some partial offer $\off$ is dominated by some other offer $\off'$ in $\Partoff{k}$ then it is also dominated by some offer in  $\pareto{\issue{1}} \times  \pareto{\issue{2}} \times \dots \times  \pareto{\issue{k}}$."
%
%To prove this statement, let us first assume that $\off'$ is Pareto-optimal. In that case we know by Lemma \ref{lem:pareto}
%that $\off' \in \pareto{\issue{1}} \times  \pareto{\issue{2}} \times \dots \times  \pareto{\issue{k}}$, an therefore the statement obviously holds.
%
%If $\off'$ is not Pareto-optimal, then we can simply choose some other partial offer $\off''$ that dominates $\off'$ and therefore also dominates $\off$. If $\off''$ is Pareto-optimal, then the lemma is again proven with the same argument as before. If not, then we can repeat our procedure and find another offer $\off'''$ that dominates $\off''$ etcetera, until we find one that is Pareto-optimal.
%\end{proof}

\begin{lemma}\label{lem:pareto3}
Let $\dom$ be some bilateral negotiation domain with $\numIssues$ issues of finite size.
Then, for any integer $k$ with $1 < k \leq \numIssues$ we have: 
\[\pareto{\ \pareto{\Partoff{k-1}} \times  \pareto{\issue{k}}\ } \ \ \subseteq \ \ \pareto{\Partoff{k}}\]
\end{lemma}
\begin{proof}
To prove this lemma, first notice that it is logically equivalent to: \\
\ \\
``If a partial offer is \textit{not} Pareto-optimal in $\Partoff{k}$, then it is also not Pareto-optimal in $\pareto{\Partoff{k}} \times  \pareto{\issue{k}}$." \\
\ \\
This, in turn, can be reformulated as: \\
\ \\
``If some partial offer ${\partoff{k}}$ in $\Partoff{k}$ is dominated by some other offer ${\partoff{k}}'$ in $\Partoff{k}$ then ${\partoff{k}}$ is also dominated by some offer in  $\pareto{\Partoff{k-1}} \times  \pareto{\issue{k}}$." \\
\ \\
To prove this statement, let us first assume that ${\partoff{k}}'$ is Pareto-optimal in $\Partoff{k}$. In that case we know by Lemma \ref{lem:pareto1}
that ${\partoff{k}}' \in \pareto{\Partoff{k-1}} \times \pareto{\issue{k}}$, and therefore the statement obviously holds.

If ${\partoff{k}}'$ is not Pareto-optimal in $\Partoff{k}$, then we can simply choose some other partial offer ${\partoff{k}}''$ in $\Partoff{k}$ that dominates ${\partoff{k}}'$ and therefore also dominates ${\partoff{k}}$. If ${\partoff{k}}''$ is Pareto-optimal, then the lemma is again proven with the same argument as before. If not, then we can repeat our procedure and find another partial offer ${\partoff{k}}'''$ in $\Partoff{k}$ that dominates ${\partoff{k}}''$ etcetera. Since $\Off$ is finite we must at some point encounter a partial offer that is not dominated, and thus Pareto-optimal.
\end{proof}

\begin{theorem}\label{thm:pareto}
Let $\dom$ be some bilateral negotiation domain with $\numIssues$ issues of finite size and linear utility functions.
Then, for any integer $k$ with $1 < k \leq \numIssues$ we have
\[\pareto{\Partoff{k}}\ \ = \ \  \pareto{\ \pareto{\Partoff{k-1}} \times  \pareto{\issue{k}}\ } \]
\end{theorem}
\begin{proof}
This is just a combination of Lemmas \ref{lem:pareto2} and \ref{lem:pareto3}. 
\end{proof}

\subsubsection{Discretized IPS}\label{sec:ips_discretized}
The exact IPS algorithm is by itself already a lot faster than a brute-force search. However, if we `discretize' the set of possible utility values, similar to what we did for the BIDS algorithm, then we can make it even faster, albeit at the expense of reduced accuracy.

One way to do this is as follows. For every integer $k$ with $1 \leq k \leq \numIssues$, we define a set of numbers $\range{\precision}{k}$ of size $\precision$, in the same way as for the BIDS algorithm. Next, we define a function $\close:\Partoff{k} \rightarrow \range{\precision}{k}$ as follows:\footnote{We are somewhat sloppy with our notation here, because technically speaking, the $\argmin$ operator returns a \textit{set} of elements rather than a single element. However, here, in most cases that set will contain only one element, so there is no confusion. In some cases it may return a set of two elements, if it happens that $\util_1(\partoff{k})$ lies exactly in the middle of two points in $\range{\precision}{k}$. In that case we can just arbitrarily choose any of these two elements.}
\[\close(\partoff{k}) \ \ := \ \ \argmin_{\tilde{\util} \in \range{\precision}{k}} \ \{\ |\util_1(\partoff{k}) - \tilde{\util}| \ \}\]
So, $\close(\partoff{k})$ is the element of $\range{\precision}{k}$ that is closest to $\util_1(\partoff{k})$. We then use this to define, for each value $\tilde{\util} \in \range{\precision}{k}$, a `cluster' of partial offers $\cluster{\tilde{\util},k}$, consisting of those partial offers for which $\tilde{\util}$ is the closest element of $\range{\precision}{k}$:
\[\cluster{\tilde{\util},k} \ \ := \ \ \{\partoff{k}\in \Partoff{k} \mid \close(\partoff{k}) = \tilde{\util} \}\]
This is also illustrated in Figure~\ref{fig:ips}.

Then, the discretized version of IPS (for some given precision parameter~$\precision$), works as follows: 
\begin{enumerate}
\item Determine the Pareto-set $\pareto{\issue{1}}$ of issue $\issue{1}$.
\item Determine the Pareto-set $\pareto{\issue{2}}$ of issue $\issue{2}$.
\item Create the Cartesian product $\pareto{\issue{1}} \times \pareto{\issue{2}}$. 
\item Determine the set $\pareto{ \ \pareto{\issue{1}} \times \pareto{\issue{2}}\ }$, which equals $\pareto{\Partoff{2}}$
\item For each value $\tilde{\util}\in \range{\precision}{2}$ determine the cluster $\cluster{\tilde{\util},2}$.
\item For each such cluster, if it is non-empty, choose exactly one partial offer $\partoff{2}$ from that cluster. We then define the approximate Pareto-set $\approxPareto{\Partoff{2}}$ to be the collection of those chosen partial offers (see Figure~\ref{fig:ips2}). Note that its size will be at most $\precision$.
\item Determine the Pareto-set $\pareto{\issue{3}}$ of issue $\issue{3}$.
\item Create the Cartesian product  $\approxPareto{\Partoff{2}} \times \pareto{\issue{3}}$.
\item Determine the set of Pareto-optimal partial offers within this set. That is, $\pareto{ \ \approxPareto{\Partoff{2}} \times \pareto{\issue{3}}\ }$. 
\item For each $\tilde{\util}\in \range{\precision}{3}$ determine the cluster $\cluster{\tilde{\util},3}$, and for each such cluster (if  non-empty), pick one element, to form the approximate Pareto set $\approxPareto{\Partoff{3}}$.
\item Keep repeating this until we have the set $\approxPareto{\Partoff{\numIssues}}$.
\end{enumerate}

%for every integer $k$ with $1\leq k \leq \numIssues$, after calculating $\pareto{\Partoff{k}}$, 
%
%Then, for every such $k$, after calculating  $\pareto{\Partoff{k}}$, we calculate for each partial offer  $\partoff{k} \in \Partoff{k}$, the value $\tilde{\util} \in \range{\precision}{k}$ that is closest to its true utility value $\util_1(\partoff{k})$. Note that this can be modeled as  a function $\close:\Partoff{k} \rightarrow \range{\precision}{k}$:
%\[\close(\partoff{k}) \ \ := \ \ \argmin_{\tilde{\util} \in \range{\precision}{k}} \ \{\ |\util_1(\partoff{k}) - \tilde{\util}| \ \}\]
%So, $\close(\partoff{k})$ is the element of $\range{\precision}{k}$ that is closest to $\util_1(\partoff{k})$. This, in turn, defines for each value $\tilde{\util} \in \range{\precision}{k}$ a `cluster' of partial offers $\cluster{\tilde{\util}}$, consisting of those partial offers for which $\tilde{\util}$ is the closest element of $\range{\precision}{k}$:
%\[\cluster{\tilde{\util}} := \{\partoff{k}\in \Partoff{k} \mid \close(\partoff{k}) = \tilde{\util} \}\]
%This is also illustrated in Figure~\ref{fig:ips}.

%\[\cluster{\tilde{\util}} := \{\ \partoff{k}\in \Partoff{k} \mid \forall \util' \in \range{\precision}{k}: \ |\util_1(\partoff{k}) - \tilde{\util}| \leq |\util_1(\partoff{k}) - \util'| \ \}\]

%Finally, we can create the approximate Pareto-set $\approxPareto{\Partoff{k}}$ by choosing, for each $\tilde{\util} \in \range{\precision}{k}$ exactly one partial offer from the cluster $\cluster{\tilde{\util}}$ (if it is non-empty).

\begin{figure}
\tikzset{every picture/.style={line width=1pt}} %set default line width to 1pt        
\pgfplotsset{width=8cm}
\begin{center}
\begin{tikzpicture}[scale=0.9]
\begin{axis}[xmin=0, xmax=1,ymin=0,ymax=1,xlabel={Utility of Agent 1}, ylabel={Utility of Agent 2}, 
		clip=false, %ensures that you can draw outside the boundaries.
		xtick pos=left,
		ytick pos=left		
		]
	%%OFFERS:
    \filldraw [black] (0.0,1.0) circle [radius=2pt]
    					  (0.05,0.98) circle [radius=2pt]
    					  (0.12,0.97) circle [radius=2pt]
    					  (0.18,0.965) circle [radius=2pt]
                      (0.26,0.95) circle [radius=2pt]   
                      (0.31,0.94) circle [radius=2pt] 
                      (0.342,0.926) circle [radius=2pt]   
                      (0.422,0.906) circle [radius=2pt]  
                      (0.49,0.866) circle [radius=2pt]
                     (0.573,0.82) circle [radius=2pt]	 
                     (0.64,0.766) circle [radius=2pt] 
                     (0.71,0.71) circle [radius=2pt]
                     (0.766,0.64) circle [radius=2pt]   
                     (0.82,0.573) circle [radius=2pt]  
                      (0.866,0.5) circle [radius=2pt] 
                      (0.899,0.422) circle [radius=2pt] 
                      (0.94,0.342) circle [radius=2pt]   
                      (0.966,0.26) circle [radius=2pt] 
                       (0.98,0.15) circle [radius=2pt]  
                     (1.0,0.0) circle [radius=2pt]
                     ;
%%%%%%%%%%%
% ELLIPSES:
	\draw[draw=red,rotate around={-20:(0.0,1.0)}] (0.0,1.0) ellipse (0.10 and 0.025);
	\draw[draw=red,rotate around={-10:(0.19,0.96)}] (0.19,0.96) ellipse (0.09 and 0.030);
	\draw[draw=red,rotate around={-17:(0.4,0.9)}] (0.4,0.9) ellipse (0.12 and 0.030);
	\draw[draw=red,rotate around={-30:(0.6,0.8)}] (0.6,0.8) ellipse (0.08 and 0.030);
	\draw[draw=red,rotate around={-50:(0.81,0.57)}] (0.81,0.57) ellipse (0.2 and 0.045);
	\draw[draw=red,rotate around={-78:(0.97,0.16)}] (0.97,0.16) ellipse (0.18 and 0.045);
%%%%%%%%%%%
\node[red] at (0.03,1.06) {$\cluster{0.0,k}$};
\node[red] at (0.23,1.05) {$\cluster{0.2,k}$};
\node[red] at (0.48,0.95) {$\cluster{0.4,k}$};
\node[red] at (0.68,0.85) {$\cluster{0.6,k}$};
\node[red] at (0.89,0.65) {$\cluster{0.8,k}$};
\node[red] at (1.09,0.20) {$\cluster{1.0,k}$};
%%%%%%%%%%%%%%%%%%%%%%%
% DASHED VERTICAL LINES
 	\draw[black, dashed] (0.2,0) -- (0.2,1);
 	\draw[black, dashed] (0.4,0) -- (0.4,1);
 	\draw[black, dashed] (0.6,0) -- (0.6,1);
   \draw[black, dashed] (0.8,0) -- (0.8,1); %node[anchor=south,color=black] {$\util_1(\off) = 0.5$}; 
\end{axis}
\end{tikzpicture}
\caption{Pareto-frontier of some set of partial offers $\Partoff{k}$. The vertical dashed lines correspond to the elements of the range ${\range{q}{k} = \{0 , 0.2 , 0.4 , 0.6, 0.8, 1.0\} }$. The red ovals represent the clusters $\cluster{\tilde{\util},k}$ corresponding to each of the elements of $\range{q}{k}$.}\label{fig:ips}
\end{center}
\end{figure}

\begin{figure}
\tikzset{every picture/.style={line width=1pt}} %set default line width to 1pt        
\pgfplotsset{width=8cm}
\begin{center}
\begin{tikzpicture}[scale=0.9]
\begin{axis}[xmin=0, xmax=1,ymin=0,ymax=1,xlabel={Utility of Agent 1}, ylabel={Utility of Agent 2}, 
		clip=false, %ensures that you can draw outside the boundaries.
		xtick pos=left,
		ytick pos=left		
		]
	%%OFFERS:
    \filldraw [black] (0.0,1.0) circle [radius=2pt]
    					  %(0.05,0.98) circle [radius=2pt]
    					  %
    					  %(0.12,0.97) circle [radius=2pt]
    					  (0.18,0.965) circle [radius=2pt]
                      %(0.26,0.95) circle [radius=2pt]   
                      %
                      %(0.31,0.94) circle [radius=2pt] 
                      %(0.342,0.926) circle [radius=2pt]   
                      (0.422,0.906) circle [radius=2pt]  
                      %(0.49,0.866) circle [radius=2pt]
                      %  
                     (0.573,0.82) circle [radius=2pt]	 
                     %(0.64,0.766) circle [radius=2pt] 
                     % 
                     %(0.71,0.71) circle [radius=2pt]
                     %(0.766,0.64) circle [radius=2pt]   
                     (0.82,0.573) circle [radius=2pt]  
                      %(0.866,0.5) circle [radius=2pt] 
                      %(0.899,0.422) circle [radius=2pt] 
                      %
                      %(0.94,0.342) circle [radius=2pt]   
                      %(0.966,0.26) circle [radius=2pt] 
                       %(0.98,0.15) circle [radius=2pt]  
                     (1.0,0.0) circle [radius=2pt]
                     ;

%%%%%%%%%%%%%%%%%%%%%%%
% DASHED VERTICAL LINES
 	\draw[black, dashed] (0.2,0) -- (0.2,1);
 	\draw[black, dashed] (0.4,0) -- (0.4,1);
 	\draw[black, dashed] (0.6,0) -- (0.6,1);
   \draw[black, dashed] (0.8,0) -- (0.8,1); %node[anchor=south,color=black] {$\util_1(\off) = 0.5$}; 
\end{axis}
\end{tikzpicture}
\caption{The approximate Pareto-frontier, after choosing one partial offer for each cluster in Figure \ref{fig:ips}.}\label{fig:ips2}
\end{center}
\end{figure}

\subsubsection{Disadvantages of IPS}
One disadvantage of IPS, is that, just as for BIDS, it does not allow us to take into account the constraint that the agent should only consider offers it hasn't proposed yet ($\off \not \in \Pro_{t}$).

Another disadvantage, is that it needs a very accurate opponent modeling algorithm to estimate the opponent's utility function $\util_2$. Note that for a brute-force implementation of Eq.~(\ref{eq:time_based_ips}) it doesn't actually matter if $\est{\util}_2$ is a very accurate estimation of $\util_2$, as long as it correctly estimates the \textit{ordering} of the offers. That is, as long as $\est{\util}_2$ satisfies: 
\[\forall \off, \off' \in \Off: \quad \est{\util}_2(\off) \leq \est{\util}_2(\off') \ \  \leftrightarrow \ \ \util_2(\off) \leq \util_2(\off')\]
we will get the correct offer $\off_{next}$, even if the actual values of $\est{\util}_2(\off)$ are entirely wrong. However, this no longer holds with IPS, because it needs the correct values of $\util_2(\off)$ to find the correct Pareto-sets at each step of the algorithm.

\begin{exercise}\textbf{IPS.} Implement the exact IPS algorithm as explained in Section \ref{sec:ips}, implement the discretized version of the IPS algorithm as explained in Section~\ref{sec:ips_discretized} and also implement an algorithm that finds the Pareto-frontier in a brute-force manner. 

Pick a value of $q$ (e.g. $q=100$), and compare the speed of the three algorithms, by letting each of them run on a domain of 5 issues and 5 options per issue. Next, increase the number of issues of the domain and repeat the experiment. Keep doing this and determine, for each of the three algorithms, how large the number of issues can get before the algorithm takes more than a minute to run. 

Finally, implement a time-based or adaptive agent that uses the IPS algorithm to find its offers to propose.
\end{exercise}

\section{Non-Linear Utility Functions}

So far, we have almost always assumed the utility functions were linear, as defined by Eq.~(\ref{eq:linear_util_func_no_weights}) or (\ref{eq:linear_util_func}). It is now time to discuss alternative negotiation domains in which this no longer holds.

While linear negotiation domains have been studied much more extensively, one could argue that such domains are actually rather unrealistic. After all, they are based on the assumption that there is no dependency between the various issues, while it seems reasonable to say that in many real-world scenarios people's preferences over one issue do depend on the options chosen for one or more other issues.

For example, in our cinema example, suppose that cinema \textit{Paradiso} is very far from Alice's home. This means that if she agrees with Bob to see a movie there, then she would prefer to see the movie at an earlier time, because it would take her  a lot of time to get home after the movie. On the other hand, if they agree to see the movie at cinema \textit{Rialto}, which is closer to Alice's home, but farther away from her work, then she might prefer to go to the cinema at a later time, because she would need more time to travel to the cinema after work. So, indeed, we see that her most preferred option for one issue (the time) depends on the option they choose for another issue (the cinema). We say there is an \textit{interdependency} between the issues. This cannot be modeled with a linear utility function over those issues.

Another common reason why linear utility functions may fail to describe real-world situations, is because there may exist \textit{hard constraints} among the issues. That is, certain combinations of options may simply be impossible. For example, suppose the negotiators are able to choose between two different cinemas: \textit{Rialto} and \textit{Paradiso}, and three different movies: \textit{The Godfather}, \textit{Casablanca}, and \textit{The Big Lebowski}. However, in general this does not mean that they can freely choose any arbitrary combination of those options, because, for example, \textit{The Godfather} might only screen in cinema \textit{Paradiso}, while \textit{The Big Lebowski} might only screen in cinema \textit{Rialto}. 

Note that a hard constraint can be seen as an extreme case of an interdependency.

In the following we will discuss several ways to model non-linear utility functions.

\subsection{Utility Functions with Hard Constraints}
Our first type of non-linear utility function, consists of a linear utility function with hard constraints imposed on top of it. That is, we first define, for each agent $\ag_i$ a \textit{linear} utility function $\util_i'$ in the standard way, according to Eq.~(\ref{eq:linear_util_func}), and then we define some subset $\infeas_i \subseteq \Off$ as the set of \textbf{infeasible offers} for that agent. We can then combine these, to define a \textit{non}-linear utility function $\util_i$ as:
\begin{equation}\label{eq:util_with_hard_constraints}
\util_i(\off) \ \ := \ \  
\begin{cases}
\rv_i & \text{if\ } \off \in \infeas_i\\
\util_i'(\off) & \text{otherwise}
\end{cases}
\end{equation}
where $\rv_i$ is the agent's reservation value.

\later{change the notation for $\infeas$?}

There can be several reasons why an offer $\off$ might be infeasible for an agent $\ag_i$. For example:
\begin{itemize}
\item The offer contains a combination of options that is impossible for that particular agent (e.g. on weekdays Alice cannot arrive at \textit{Rialto} before 18:00, because it is too far from her work).
\item The offer contains a combination of options that is simply impossible in general (e.g. \textit{The Godfather} is not being screened in \textit{Rialto}).
\item The offer contains a combination of options that is disliked so much by Alice, that she prefers not to make any agreement at all.
\end{itemize}
In principle, in either of these cases it should be sufficient to define the utility $\util_i(\off)$ of an infeasible offer $\off \in \infeas_i$ as being equal to the agent's reservation value $\rv_i$, as we have done in Eq.~(\ref{eq:util_with_hard_constraints}). After all, a rational agent would never agree with an offer that only yields the reservation value, so this should be enough to prevent the agents from making such agreements. However, in case we want to explicitly model the fact that the offer $\off$ is even \textit{worse} than disagreement, then we can just as well set $\util_i(\off)$ equal to any arbitrary value lower than $\rv_i$.

\subsection{Soft Interdependencies}
Our second type of non-linear utility function, involves what we call \textit{soft} interdependencies. We say there exists a soft interdependency whenever some combination of options does influence an agent's utility, but not by that much that it makes the offer infeasible for that agent.

One way to include soft interdependencies into a utility function, is to introduce \textit{second order} evaluation functions. By this, we mean an evaluation function $\eval_i^{j,k}: \issue{j} \times \issue{k} \rightarrow \mathbb{R}$ that assigns evaluations to \textit{pairs} of options. Then, the utility function can be defined as:
%\[\util_i(\off) \ \ = \ \ \sum_{j=1}^\numIssues \eval_i^j(\off) + \sum_{j=1}^\numIssues \sum_{k=j+1}^\numIssues \eval_i^{j,k}(\off) \]
\begin{equation}\label{eq:util_first_and_second_order}
\util_i(\offComp{1}, \offComp{2}, \dots, \offComp{\numIssues}) \ \ = \ \ \sum_{j=1}^\numIssues \eval_i^j(\offComp{j}) + \sum_{j=1}^\numIssues \sum_{k=j+1}^\numIssues \eval_i^{j,k}(\offComp{j}, \offComp{k})
\end{equation}
for any offer $\off = (\offComp{1}, \offComp{2}, \dots, \offComp{\numIssues}) \in \Off$.
So, the second-order evaluation function $\eval_i^{j,k}$ represents the interdependency between issues $\issue{j}$ and $\issue{k}$. 

For example, if agent $\ag_i$ likes some option $\offComp{j} \in \issue{j}$ and also likes some option $\offComp{k} \in \issue{k}$, but he does not like the \textit{combination} of those two options, then $\eval_i^j(\offComp{j})$ and $\eval_i^k(\offComp{k})$ would each have a high value, but the term $\eval_i^{j,k}(\offComp{j}, \offComp{k})$ would be negative. On the other hand, if the agent  likes these two options \textit{only} when they come together, then $\eval_i^j(\offComp{j})$ and $\eval_i^k(\offComp{k})$ would each have a low value, but the term $\eval_i^{j,k}(\offComp{j}, \offComp{k})$ would be high.

Note that any utility function of the form of Eq.~(\ref{eq:util_first_and_second_order}), which includes both first-order and second-order evaluation functions, can be written equivalently as a function with only second-order evaluation functions. After all, if we define:
\[{\eval_i^{j,k}}'(\offComp{j}, \offComp{k}) \quad := \quad  {\eval_i^{j,k}}(\offComp{j}, \offComp{k}) + \frac{1}{m-j}v_i^{j}(\offComp{j}) \]
Then Eq.~(\ref{eq:util_first_and_second_order}) can be rewritten as:
\[\util_i(\offComp{1}, \offComp{2}, \dots, \offComp{\numIssues}) \ \ = \ \ \sum_{j=1}^\numIssues \sum_{k=j+1}^\numIssues {\eval_i^{j,k}}'(\offComp{j}, \offComp{k})\]

Of course, second-order evaluation functions can only capture the interdependencies between \textit{pairs} of two issues, but not interdependencies between three or more issues.
However, we can easily generalize this approach to also allow for \textit{third}-order evaluation functions ${\eval_i^{j,k,l}: \issue{j} \times \issue{k} \times \issue{l} \rightarrow \mathbb{R}}$, or evaluation functions of any arbitrary order.

The higher the order of evaluation functions we include, the more general the utility functions we can describe. However, including higher-order evaluation functions also means that the description of such utility functions becomes larger, because we need to specify more parameters. In the extreme case that we even allow $\numIssues^{th}$-order evaluation functions (where $\numIssues$ is the total number of issues) then we could literary describe any arbitrary utility function (as long as each issue is finite), but it would essentially just be one giant table that lists every single offer with its utility value. So, in practice this would only work for small offer spaces.

An alternative can be to only specify first-order evaluation functions, but allow products between them. For example:
\begin{equation}\label{eq:util_first_with_products}
\util_i(\offComp{1}, \offComp{2}, \dots, \offComp{\numIssues}) \ \ = \ \ \sum_{j=1}^\numIssues \eval_i^j(\offComp{j}) + \sum_{j=1}^\numIssues \sum_{k=j+1}^\numIssues \eval_i^{j}(\offComp{j}) \cdot \eval_i^{k}(\offComp{k}) 
\end{equation}
Note that this is essentially a special case of Eq.~(\ref{eq:util_first_and_second_order}), because we can rewrite Eq.~(\ref{eq:util_first_with_products}) in the form of Eq.~(\ref{eq:util_first_and_second_order}) by defining
\[\eval_i^{j,k}(\offComp{j}, \offComp{k}) \ \ := \ \ \eval_i^{j}(\offComp{j}) \cdot \eval_i^{k}(\offComp{k})\]
The advantage of Eq.~(\ref{eq:util_first_with_products}) is that we need fewer parameters, because we only need to define the first-order evaluation functions, but the disadvantage is that it is less general than Eq.~(\ref{eq:util_first_and_second_order}).

%there is no benefit at all anymore, because we would literally need to specify an evaluation function for every single offer $\off \in \Off$, which means we have just created a giant table that lists the utility value of each offer.

%Note that, in principle, Eq.~\ref{X} can be rewritten in terms of only second-order evaluation functions, because we can define:
%\[{\eval_i^{j,k}}'(\offComp{j},\offComp{k}) \ \ := \ \ \eval_i^{j,k}(\offComp{j},\offComp{k}) + \eval_i^j(\offComp{j}) + \eval_i^k(\offComp{k})\]

\subsection{The Hypercube Model}
A third way to model non-linear utility functions, which has actually been studied more often in the literature, is through what we will here call the \textit{hypercube model} \cite{ito2008nonlinear}. To introduce this model we first need the following definition.
\begin{definition}
Let $\Off = \issue{1} \times \issue{2} \times \dots \times \issue{\numIssues}$ be an offer space composed of $\numIssues$ issues. Then a \textbf{hypercube} $\hypCube \subseteq \Off$ is a Cartesian product ${\hypCube = \subIssue{1} \times \subIssue{2} \times \dots \times \subIssue{\numIssues}}$ where each $\subIssue{i}$ is a non-empty subset of $\issue{i}$ (possibly equal to $\issue{i}$). Its \textbf{volume} $\vol(\hypCube)$ is defined as:
\[\vol(\hypCube) \ \ := \ \ |\subIssue{1}| \times |\subIssue{2}| \times \dots \times |\subIssue{\numIssues}| \]
and its associated `membership function' $\delta_{\hypCube} : \Off \rightarrow \{0,1\}$ is defined as:
\[
\delta_{\hypCube}(\off) = 
\begin{cases}
1 & \text{ if } \off \in \hypCube \\
0 & \text{ if } \off \not \in \hypCube
\end{cases}
\]
\end{definition}

\begin{definition}
A negotiation domain $\dom$ with offer space $\Off$ is called a \textbf{hypercube domain} if, for each agent $\ag_i$ there is a set of hypercubes $\HypCube_i$, such that its utility function $\util_i$ is given by:
\[\util_i(\off) = \sum_{\hypCube \in \HypCube_i} w_{\hypCube} \cdot \delta_{\hypCube}(\off)\]
with $w_{\hypCube} \in \mathbb{R}$ for each $\hypCube \in \HypCube_i$. 
%A pair $(\hypCube, w_{\hypCube})$ is called a \textbf{constraint}.
\end{definition}
The hypercube model has been used, for example, in the ANAC  competition of 2014~\cite{anac2014}.

\later{include an image of a hypercube utility function for a 2-dimensional offer space.}

One advantage of the hypercube model, is that it allows agents to propose many offers at the same time. That is, Ito et al.~\cite{ito2008nonlinear} defined a \textbf{bid} as a set of hypercubes such that their intersection is non-empty. They then allowed the agents to propose bids, instead of single offers,  meaning that the agent would essentially propose all offers that lie in the intersection of the hypercubes of a bid, at once.
\begin{definition}\label{def:bid}
In the context of the hypercube model, a \textbf{bid} $\bid$  for agent $\ag_i$ is defined as a subset $\bid \subseteq \HypCube_i$ of that agent's set of hypercubes $\HypCube_i$, such that the intersection of the hypercubes in $\bid$ is non-empty:
\[\bid \subseteq \HypCube_i  \quad \text{such that} \quad \bigcap_{\hypCube \in \bid} \hypCube \neq \emptyset\]
\end{definition}
Note that each hypercube $\hypCube \in \HypCube_i$ is a subset of $\Off$, so a bid $\bid$ is a set of subsets of $\Off$. However, a bid also corresponds directly to a subset of $\Off$, by considering the intersection of all hypercubes in $\bid$. We will denote this intersection by $\overline{\bid}$:
\[\overline{\bid} \ \ := \ \ \bigcap_{\hypCube \in \bid} \hypCube\]
So, we have $\overline{\bid} \subseteq \Off$, and whenever an agent proposes a bid $\bid$, this essentially means that he is proposing all offers $\off \in \overline{\bid}$ at once.

Furthermore, for any offer $\off$ and agent $\ag_i$ we can define the corresponding bid $\bid_{\off}^i$ as the set of all hypercubes in $\HypCube_i$, that contain $\off$:
\[\bid_{\off}^i \ \ := \ \ \{\hypCube \in \HypCube_i \mid \off \in \hypCube\}\]
Note that all offers contained in the intersection of these hypercubes have the same utility as $\off$:
\[\forall \off' \in \overline{\bid}_{\off}^i : \quad \util_i(\off') = \util_i(\off)\]
We may therefore write $\util_i(\bid_{\off}^i)$ instead of $\util_i(\off)$. 

\subsection{Search Algorithms for Non-linear Utility Functions}
\later{Rename this to `Search Algorithms for Hypercube domains'?}

In Section \ref{sec:large_linear_domains} we discussed search algorithms that allows us to find the next offer to propose, from an offer space that may be astronomically large. However, the two algorithms we discussed there (BIDS and IPS) both depended on the assumption that the utility functions were linear. So, for negotiation domains that are both large and non-linear, we need alternative search algorithms.

The topic of `search and optimization' is a vast research field within the field of artificial intelligence, so there are thousands of algorithms that one can choose from. Common examples are Branch \& Bound, Genetic Algorithms, Simulated Annealing, and Ant Colony Optimization. Each of these algorithms aims to find a solution $x$ that maximizes some objective $f(x)$ over a large set of candidate solutions $X$.

We should keep in mind, however, that in the field of automated negotiation,  we typically do not aim to find the offer with highest utility, but rather, we aim to find the offer that strikes a certain balance between our own utility and our opponent's utility. In the literature, several approaches have been suggested to deal with this, such as:
\begin{enumerate}
\item Combine the agents' utility functions $\util_1$ and $\util_2$ into a single objective, such as the Nash product $f := (\util_1-\rv_1)\cdot (\util_2 - \rv_2)$ or the utility sum $f := \util_1 + \util_2$.
\item Optimize only $\util_1$, but restrict the search to 
a set of offers that are in the neighborhood of some offer that has been proposed by the opponent.
\item Use a specialized algorithm for \textit{multi-objective optimization problems} (MOOP), that aims to find the set of all Pareto-optimal offers, rather than a single optimal offer.
\end{enumerate}

We will now discuss several specific search approaches that have been proposed in the literature. For now, we will only focus on algorithms designed for hypercube domains.

\subsubsection{Simulated Annealing}
Ito et al.~\cite{ito2008nonlinear} described an approach in which each agent $\ag_i$ used simulated annealing~\cite{kirkpatrick1983SA} to find offers that maximize (approximately) that agent's own utility $\util_i(\off)$. So, this would return a set of offers $\off_{i,1}, \off_{i,2}, \off_{i,3} \dots$ corresponding to the local maxima found by the simulated annealing. Then, each agent would propose the corresponding bids $B_{\off_{i,1}}^i, B^i_{\off_{i,2}}, B^i_{\off_{i,3}}, \dots$. Next, a neutral \textit{mediator} agent would use a branch \& bound approach to select for each negotiating agent $\ag_i$ one of its bids $\bid_i' \in \{B^i_{\off_{i,1}}, B^i_{\off_{i,2}}, B^i_{\off_{i,3}}, \dots\}$, such that their intersection is non-empty:
\[\bigcap_{i=1}^\numAgents \overline{\bid}_i' \ \ = \ \ \bigcap_{i=1}^\numAgents \bigcap_{\hypCube \in \bid_i'} \hypCube \ \ \neq \ \ \emptyset \]
and such that their utility-sum $\sum_{i=1}^\numAgents \util_i(\bid_i')$ is maximized. Finally, any arbitrary offer $\off$ in this intersection can be selected as the final agreement. It doesn't matter which offer, because they all have the same utility vector anyway.

\subsubsection{Combinatorial Optimization}
Marsa-Maestre et al.~\cite{marsaMaestre2009nonlinear} experimented with various combinatorial optimization techniques designed to find, among any set of hypercubes $\HypCube$, the non-empty subset that maximizes the agent' utility $\util_1$. However, to ensure a proper trade-off between $\util_1$ and $\util_2$, they assigned a quality measure $Q(\hypCube)$ to each hypercube $\hypCube \in \HypCube_1$, based on the weight $w_{\hypCube}$ as well as the volume of the hypercube:
\[Q(\hypCube) \ \ := \ \ w_{\hypCube}^\alpha \cdot \vol(\hypCube)^\beta\]
where $\alpha \in \mathbb{R}$ and $\beta \in \mathbb{R}$ are two tunable parameters. The idea being that the greater the volume, the higher the probability that it will contain offers that are acceptable to the other agent(s), so the higher the quality of the hypercube. Then, they randomly selected a subset $\HypCube_1' \subseteq \HypCube_1$ from the agent's set of hypercubes $\HypCube_1$, but such that hypercubes with higher quality value $Q(\hypCube)$ had a bigger chance of getting selected. Finally they would run their optimization algorithms only on that set of hypercubes $\HypCube_1'$, and then their agent would propose the bid returned by those algorithms.

\subsubsection{Genetic Algorithms}
The GANGSTER algorithm (Genetic Algorithm NeGotiator Subject To alternating offERs)~\cite{deJonge2016gangster} was developed for the ANAC 2014 competition, and used genetic algorithms to search for good offers to propose, from a large bilateral hypercube domain. For more details about genetic algorithms in general, we refer to~\cite{Holland1992GA}.

In order to determine the best offer to propose, GANGSTER considered a trade-off between three quantities:
\begin{enumerate}
\item \textbf{Utility} $\util_1(\off)$: the agent's own utility value for the offer $\off$.
\item \textbf{Distance} $dist_t(\off)$: the distance between offer $\off$ and any offer $\off'$ already proposed by the opponent:
\[dist_t(\off) \ \ := \ \  \min \{\dist(\off, \off') \mid \off'\in \Rec_t\}\]
\item \textbf{Diversity} $div_t(\off)$: the distance between offer $\off$ and any offer $\off'$ already proposed by our own agent $\ag_1$ itself:
\[div_t(\off) \ \ := \ \  \min \{\dist(\off, \off') \mid \off'\in \Pro_t\}\]
\end{enumerate}
where $\dist(\off, \off')$ is the Manhattan distance between offers $\off = (\offComp{1}, \offComp{2}, \dots, \offComp{\numIssues})$ and $\off' = (y_1, y_2, \dots, y_\numIssues)$:
\[\dist(\off, \off') \ \ := \ \ \sum_{j=1}^\numIssues |x_j - y_j|\]
which is defined because the ANAC 2014 competition only involved domains in which all options of all issues were integers between 0 and 9.

The GANGSTER algorithm used these quantities to choose its next offer to propose $\off_{next}$ according to:
\[\off_{next} \ \  = \ \ \argmax_{\off \in \offSet} \{div_t(\off) \mid \util_1(\off) \geq \asp_1(t) \ \land \ dist_t(\off) \leq \asp_1'(t) \}\]
where $\offSet \subseteq \Off$ is the set of offers returned by the agent's genetic algorithms, and $\asp_1$ and $\asp_1'$ are two different aspiration functions used by agent $\ag_1$.

The idea is that a low value of $dist_t(\off)$ can be seen as a proxy for a high opponent utility value $\util_2(\off)$. After all, it is likely that the offers $\off'$ so far proposed by the opponent yield high utility to that opponent, and therefore any offer $\off$ that is very similar to $\off'$, would likely also yield high utility to that opponent. 

Furthermore, proposing offers with high `diversity' ensures that our agent proposes offers that lie far away from each other, thereby increasing the probability that some of them lie in regions of high opponent utility.

GANGSTER used two separate genetic algorithms: a \textit{global} genetic algorithm, that used $\util_1$ as a fitness function to search for good offers among the entire offer space, and a \textit{local} genetic algorithm, that also used $\util_1$  as its fitness function, but that sampled its initial population exclusively from the neighborhood around the last received proposal. The point of this local algorithm, was to find offers that satisfy $dist_t(\off) \leq \asp_1'(t)$. The agent would run both of these genetic algorithms at the start of every new turn, before proposing a new offer.

%  there are several solution for this:
% 1) use multi-objective optimization techinques to find the pareto-set.
% 2) somehow combine the utility functions into a single objective (e.g. Nash product or utility-sum).
	% -- Ito, Klein, Hattori
% 3) used a local search in the neighborhood of the opponent's proposals.
	% -- GANGSTER

%\subsubsection{Genetic Algorithms}
%A genetic algorithm is a standard technique to find some optimal solution from a large space of solutions with non-linear utility functions.
%
%In general, it works as follows: start with a random `population' of offers. For example, 20 randomly selected offers. Then, a new population is generated using two operators: cross-over and mutation. Next, for each offer in this new population, we calculate its utility value. Then, we select a number of offers from the original and new population together, based on their utility values. That is, the higher the utility value, the higher the chance of `survival'. This is repeated several times, either until we no longer see improvement, or until some deadline passes.

\section{Computationally Complex Utility Functions}

%\later{It is often claimed that any arbitrary utility function can be modeled in the hypercube model. \textit{In theory} this claim is indeed correct, at least, if the offer space has finite size. Nevertheless, one can argue that this claim is an exaggeration, because in practice it may be entirely unrealistic to specify the utility functions in this model.}
%
%The reason for this, is that when the utility functions are given in some other form, transforming them into the hypercube model, may be a computationally complex problem (e.g. NP-hard). 
%
%In this section we will give several examples of negotiation domains with such utility functions.
%
%For example, in \cite{deJonge2015nb3} the `negotiating salesmen problem' was defined. This is a variant of the well-known traveling salesman problem, involving negotiation.
%
%
%The traveling salesman problem is one of the most famous problems for which it is believed that it cannot be solved in an  \textit{efficient} manner. The traveling salesman problem i

% Diplomacy
% vehicle routing
% Colored Trails
% General Game Playing
% MCNS

COMING SOON!

%\section{Negotiation and Logic}

%\nocite{*}								%display all references from the bib file, even if they haven't been cited in the text.
\bibliographystyle{plain}		%default bibliography style
\bibliography{intro_to_nego}			%get references from mybib.bib

\end{document}